\documentclass[aps,twocolumn,nofootinbib,preprintnumbers,superscriptaddress]{revtex4-1}
\usepackage{psfrag,graphicx}

\usepackage{mathrsfs}

\usepackage{amsfonts}
\usepackage{amsmath}
\usepackage{empheq}		
\usepackage{cancel}     
\usepackage[toc,page]{appendix}

\usepackage{amssymb}
\usepackage{graphicx}
\usepackage{epsfig}
\usepackage{float}
\usepackage{color}
\usepackage{epstopdf}
\usepackage{esint}

\usepackage{graphicx}
\RequirePackage{CJKutf8}

\usepackage[colorlinks=true,  
            citecolor=red,
            urlcolor= blue,                     %
            filecolor=black,
            linktocpage=true,  
            linkcolor=blue,
             ]{hyperref}

\newcommand{\beq}{\begin{equation}}
\newcommand{\eeq}{\end{equation}}
\newcommand{\bq}{\begin{equation}}
\newcommand{\eq}{\end{equation}}
\newcommand{\ba}{\begin{array}}
\newcommand{\ea}{\end{array}}
\newcommand{\beqa}{\begin{eqnarray}}
\newcommand{\eeqa}{\end{eqnarray}}
\newcommand{\bsel}{{\begin{subequations}\begin{empheq}}}
\newcommand{\bse}{{\begin{subequations}\begin{empheq}[left={\ii}\empheqlbrace]{align}}}
\newcommand{\ese}{{\end{empheq}\end{subequations}}}
\def\bc{\begin{center}}
\def\ec{\end{center}}
\def\bnum{\begin{enumerate} }
\def\enum{\end{enumerate}}
\def\nn{\nonumber}
\def\ii{\!\!\!\!\!\!}  
\def\3i{\!\!\!}
\def\2i{\!\!}

\def\ea{{e_a}}

\def\ec{{e_c}}

\def\log{\ln}

\def\nn{\nonumber}

\def\[{\left[}
\def\]{\right]}
\def\({\left(}
\def\){\right)}

\def\>{\rightarrow}

\def\Diracslash#1{\not{\hbox{\kern-4pt $#1$}}}

\def\Dslash{\not{\hbox{\kern-4pt $D$}}}
\def\pslash{\not{\hbox{\kern-4pt $p$}}}
\def\qslash{\not{\hbox{\kern-4pt $q$}}}

\def\lv{\not{\hbox{\kern-4pt $L$}}}
\def\lsim{\mathrel{\raise.3ex\hbox{$<$\kern-.75em\lower1ex\hbox{$\sim$}}}}
\def\gsim{\mathrel{\raise.3ex\hbox{$>$\kern-.75em\lower1ex\hbox{$\sim$}}}}
\def\ifmath#1{\relax\ifmmode #1\else $#1$\fi}

\evensidemargin=\oddsidemargin

\def\dr{\mathrm{d}}

\numberwithin{equation}{section}

\allowdisplaybreaks
\begin{document}
\begin{CJK}{UTF8}{gbsn}

\begin{titlepage}
\begin{flushright}
\end{flushright}

\begin{center}
 \vspace*{10mm}

{\LARGE\bf
Topological Non-Fermi Liquid
}\\
\medskip
\bigskip\vspace{0.6cm}
{\large {Rong-Gen Cai$^{\star}$,} \, {Yong-Hui Qi$^{\dag}$,} \, {Yue-Liang Wu$^{\star,\ddag}$,}\, {Yun-Long Zhang$^{\star}$} }
\\[7mm]
{\it
$^{\dag}$Center for High Energy Physics, Peking University, Beijing, 100871,\\
$^{\star}$State Key Laboratory of Theoretical Physics, Kavli Institute for Theoretical Physics China, \\
Institute of Theoretical Physics, Chinese Academy of Sciences, Beijing, 100190, \\
$^{\ddag }$University of Chinese Academy of Sciences, Beijing, 100049, P.R.China\\ 
}
\vspace*{0.3cm}
 {\tt cairg@itp.ac.cn,~yhqi@pku.edu.cn,~ylwu@itp.ac.cn,~zhangyl@itp.ac.cn}
\bigskip\bigskip\bigskip

{
\centerline{\large\bf Abstract}
\begin{quote}
In this paper we investigate the $(2+1)$-dimensional topological non-Fermi liquid in strongly correlated electron system, which has a holographic dual description by Einstein gravity in $(3+1)$-dimensional anti-de Sitter (AdS) space-time. In a dyonic Reissner-Nordstrom black hole background, we consider a Dirac fermion coupled to the background $U(1)$ gauge theory and an intrinsic chiral gauge field $b_M$ induced by chiral anomaly. UV retarded Green's function of the charged fermion in the UV boundary from AdS$_4$ gravity is calculated, by imposing in-falling wave condition at the horizon. We also obtain IR correlation function of the charged fermion at the IR boundary arising from the near horizon geometry of the topological black hole with index $k=0,\pm 1$. By using the UV retarded Green's function and IR correlation function, we analyze the low frequency behavior of the topological non-Fermi liquid at zero and finite temperatures, especially the relevant non-Fermi liquid behavior near the quantum critical region. In addition, we find that when $k=\pm 1$, the effective mass of bulk Dirac fermion at IR fixed point is topologically quantized, consequently the Fermi momentum presents a discrete sparse distributed pattern, in contrary to the Ricci flat case with $k=0$. As demonstrated examples, we calculate the spectral functions of topological non-Fermi liquid due to anomalous Hall effect (AHE) ($|\vec{b}|\ne 0, b_0=0$) and chiral magnetic effect (CME) ($|\vec{b}| = 0, b_0\ne 0$), which can be tested in topological materials such as topological insulator and topological metal with strong couplings.
\end{quote}}
\end{center}
\end{titlepage}

\tableofcontents

\section{Introduction}
\label{sec:Intro}

One of the cornerstones of condensed matter physics is Landau's theory of condensed state~\cite{Landau}, which underlies our understanding on almost all normal state metal, conductor, semiinsulator (or semimetal), as well as low temperature superconductor and superfluid, etc. For example, the electronic state of  metal is well described\footnote{At the phenomenological level (with pheno input), the theory can be used to predict essentially all the low energy behavior of the system, especially to successfully explain almost all metallic states in nature.} by Fermi liquid theory~\cite{Anderson}. The metallic states have a free stable renormalization group (RG) fixed point~\cite{Wilson:1971bg,Polyakov}, they are manifested by Landau quasi-particles as poles in a retarded Green's function of the fermion operator near the Fermi surface $\lambda\sim \lambda_F$ in the complex frequency plane. The retarded Green's function describers the causal response of the system to an adding fermion and has the form:
\beqa
G^R(\omega,\lambda) = \frac{{\mathcal Z}}{\omega - v_F(\lambda-\lambda_F) +  \Sigma(\omega,\lambda)} + \ldots  \label{Eq:GR_FL}
\eeqa
where the dots represent incoherent contributions. ${\mathcal Z}$ is the residue of the pole in momentum space,
it can be interpreted as the overlap between the quasi-particle state and the external electron state generated by acting the fermion operator on the vacuum. $v_F=\lambda_F/m_\star$ is the Fermi velocity and $m_\star$ is the effective mass of quasi-particle or hole,
accounting to the interactions between many bodies, which are not necessarily weak\footnote{The theory does not require that the interactions between fundamental constitute particles to be weak, e.g., for heavy fermion compounds in the $3$d transition element series and the $4$f rare earth series in the periodic table, the effective mass of the quasi-particle can be as large as $m_\star\sim (10^2-10^3)\, m_e$, which indicates that the interactions between the atoms are not weak.}.
The momentum locating at the surface of the sphere $\lambda_F\equiv \lambda$ is the Fermi surface\footnote{By definition, a Fermi surface $\lambda=\lambda_F$ can be defined as the surface in momentum space where there are gapless excitations.}, which characterizes the ground state of an interacting fermionic system. The lower energy excitation near the Fermi surface behaves as weakly interacting particle or ``anti-particle", namely quasi-particle or hole, which satisfies the Fermi statistics, and follows the dispersion relations $\epsilon(\lambda)= v_F(\lambda-\lambda_F)+\ldots$ near Fermi surface. They can be distinguished by the sign of $\lambda-\lambda_F$, indicating that it is sightly outside or inside the Fermi surface.
$\Sigma(\omega, \lambda)$ is the self energy, whose imaginary part corresponds to the decay width\footnote{After Fourier transformation, the retarded Green's function in momentum space can be transformed into that in coordinate space, i.e., $G^R(\omega, \lambda) \to G^R(t, x) \sim e^{-i\epsilon(\lambda)t - \frac{\Gamma}{2}t}$, where $\Gamma$ represents the decay width of the quasi-particle.} of the quasi-particle $\Gamma \propto \text{Im}\Sigma \sim \epsilon(\lambda)^2 \sim \omega^2 \overset{\omega\to 0}{\ll} \omega \sim \epsilon(\lambda) $. This implies that at low energy, the quasi-particle near the Fermi surface has a long lifetime and thus approximates a single particle picture. To be brief, the weakly interacting quasi-particles picture can be described as the low energy collective excitation of a strongly interacting many body system.

The retarded Green's function can be measured by the Angular Resolved Photon Emission Spectra (ARPES) experiment, where the incident photons knock out electrons from the sample and the {intensity of the electron beam} $I(\omega,\lambda)$ is proportional to $A(\omega,\lambda)f(\omega,\lambda)$, where $f(\omega,\lambda)$ is the Fermi-Dirac distribution and $A(\omega,\lambda)$ is the {electron spectral function} defined by\footnote{The spectral function of an operator is a measure of the density of states, which couples to the operator.}
\beqa
A(\omega,\lambda) = \frac{1}{\pi} \text{Im} G^R(\omega,\lambda). \nn
\eeqa
Near the Fermi surface ($\lambda \approx \lambda_F$), the surface of the normal Fermi liquid is approached when $\lambda \to \lambda_F$, $\omega\to 0$, then $A(\omega,\lambda) \xrightarrow[]{}  {\mathcal Z} \delta(\omega-v_F(\lambda-\lambda_F))$, where the residue ${\mathcal Z}>0$ is finite on Fermi surface. Intuitively, the spectral density becomes arbitrarily sharp, which means that the existence of the Fermi surface in momentum space where the gapless excitation is located at, results in a {non-analytic behavior of electron spectral function} obtained from APRES.

While it is phenomenologically found that some system's low energy properties are significantly different from those predicted by the Fermi liquid theory, which implies that the Landau's theory breaks down for these systems. For example, the strange metal phases of high critical temperature $T_c$ cuprates superconductors and heavy fermion material close to a quantum critical point\cite{Hertz:1976zz,Si:2011hh}, which are all universally named non-Fermi liquids~\cite{Varma:1989zz,Hewson,Stewart:2001zz}. In the mysterious Non-Fermi liquid electronic system, data from ARPES experiments indicate that there are gapless excitations, namely, a Fermi surface still exists, but the quasi-particle picture breaks down. The electron spectral function $A(\omega,\lambda)$ is not prior as expected, it exhibits non-analyticity at $\omega\to 0, \lambda\to \lambda_F$, i.e., $A(\omega,\lambda)\to 0$ around Fermi surface with decay width of the excitation, $\Gamma \propto \text{Im}\Sigma \sim \omega$ instead of $\omega^2$ for a normal Fermi liquid. As a result, such an excitation can no longer be considered as a quasi-particle. Meanwhile when the Fermi surface is approached $\lambda\to \lambda_F$, the residue of the low energy excitation or pole in the complex momentum plane, scales like ${\mathcal Z}\sim [\log(\lambda-\lambda_F)]^{-1}\xrightarrow[]{} 0$, thus vanishes logarithmically. Therefore, the singularity of electron spectral function $A(\omega,\lambda)$ near the Fermi surface is much soft than that ${\mathcal Z} \sim \delta(\lambda-\lambda_F)$ for a normal Fermi liquid. All of the evidence show that it is a {non-Fermi liquid with a Fermi surface without long lived quasi-particles}, the excitation exhibits a much broader peak than that of a normal Fermi liquid, where the self-energy is given by $\Sigma(\omega,\lambda)=c \, \omega \log\omega + z\, \omega$, where $c$ is a real constant and $z$ is a complex constant.

In particular, when one is approaching the quantum critical point (QCP)~\cite{Senthil:2008}, the non-Fermi liquid system is becoming strongly coupled, and the physics around the region is characterized effectively by scaling law of critical parameters, the dynamics of which can be described by a conformal field theory.

The non-Fermi liquid behavior has successfully been described through the gauge-gravity duality~\cite{Polchinski:1992ed,Lee:2008xf,Liu:2009dm,Cubrovic:2009ye,Basu:2009qz,Faulkner:2010tq,Faulkner:2010zz,Faulkner:2011tm,Sachdev:2011ze}, which is a breakthrough of the {AdS/CFT} correspondence~\cite{Maldacena:1997re,Gubser:1998bc,Witten:1998qj}, where a higher dimensional gravity theory can be used as a tool to explore physical observables such as mass spectrum, correlation length and critical exponents of a field theory lived on its boundary. In small curvature and low energy limit, all known gravity theories reduce to the universal classical Einstein gravity. Through studying the weakly interacting gravity, one can extract universal properties of a large class of strongly coupled quantum field theories lives on the boundary~\cite{Henningson:1998cd,Henneaux:1998ch}. It not only reveals the correspondence relation between strongly coupled conformal field theory and weakly interacting classical gravity, but also provides a strongly powerful tool in dealing with strongly coupled quantum many body physics~\cite{Son:2002sd,Iqbal:2009fd,Hartnoll:2009sz,McGreevy:2009xe}. The near horizon region behavior of the classical Einstein gravity with well understanding behavior at long distance, has provide a phenomenological understanding on the IR CFT close to the QCP for non-Fermi liquid~\cite{Faulkner:2009wj,Hartnoll:2009qx,Iqbal:2011ae}

The non-Fermi liquid phases are classified by scaling dimension of spinor operator of CFT at IR fixed point: (1)When the operator is {relevant} in the IR CFT, the quasi-particle is {unstable}, thus there is no quasi-particle description. As the physical consequence, its width is linearly proportional to its energy and the quasi-particle residue vanishes when approaching the Fermi surface. (2)When the operator is {irrelevant}, the quasi-particle becomes stable, scaling towards the Fermi surface with a non-vanishing quasi-particle residue. In some special cases, the ordinary Landau's Fermi liquid deduced too. (3)When the operator is {marginal}, the spectrum function has the form of a ``marginal non-Fermi liquid". The most intriguing aspect of the non-Fermi liquid is that its low energy behavior is controlled by the IR fixed point, which exhibits non-analytic behavior only in the time direction. In particular, single particle spectral function and charge transport can be characterized by the scaling dimension of the fermionic operator at IR fixed point. Its dynamics is described by IR conformal quantum mechanics, i.e., CFT$_1$, dual to the near horizon behavior of a black hole. It has been shown numerically in literatures that the Dirac fermions in RN AdS system has non-Fermi liquid behavior~\cite{Liu:2009dm}. Physical observables such as Fermi momentum, decay width and correlation functions are theoretically predictable and phenomenologically detectable.

Recently, some novel strongly correlated electron systems, namely, topological metals~\cite{Wan:2011,Haldane:2014}, have been suggested in $5$d transition metal oxides, topological semi-metallic phases, i.e., topological Dirac and Weyl semimetals arises from the interplay of strong correlation of electrons and strong spin-orbit interactions~\cite{Moon:2012rx}, so that novel quantum effects such as anomalous Hall effect (AHE)~\cite{Haldane:1988,Nagaosa:2003,Nagaosa:2010ahe,Chang:2013} and a magnetic field induced charge density wave are present~\cite{Yang:2011}. Three dimensional Weyl semimetal phase have been realized by utilizing multilayer structure~\cite{Burkov:2011} with Topological Insulators (TIs)~\cite{Hasan:2010xy,Qi:2008ew} and the metallic phase is unusual, which is characterized by a finite anomalous Hall conductivity and topologically protected edge states. Weyl semimetal~\cite{Zyuzin:2012vn,Grushin:2012mt} is considered as a platform toward interacting topological states of matter. The topological structure leads to anomalous transport phenomena such as chiral magnetic effect (CME)~\cite{Fukushima:2008xe,Kharzeev:2012dc,Landsteiner:2013sja}, negative magneto-resistivity~\cite{Son:2012bg} and anomalous Hall effect~\cite{Chen:2013rza,Jho:2013jd}. It seems that chiral anomaly is responsible for anomalous transport phenomena in Weyl metral~\cite{Liu:2012hk,Zyuzin:2012tv}, and can be described by an axion field theory~\cite{Goswami:2012db}. Naively speaking, a Weyl metallic state is expected to appear by applying magnetic field externally or intrinsically into the Dirac metal~\cite{Gorbar:2013qsa}. Recently, the topological metal has been observed in experiments~\cite{Haldane:2004zz,Xu:2015,Xu:2015cga,Lu:2015zre,Lv:2015pya}.

All these new experiments reveals a fact that chiral anomaly~\cite{Adler:1969gk,Bell:1969ts} plays a universal and significant physical role in understanding the IR physics of strongly correlated electron systems with novel topological phases in numerous solid state crystals~\cite{Nagaosa:1999uc}. For example, Halperin-Lee-Read composite fermion charge liquid state of a half-filled Landau level, can be viewed as a topological non-Fermi liquid~\cite{Barkeshli:2012hs,Metlitski:2014zsa}. Although a plenty of works on non-Fermi liquid theory through gauge-gravity duality have been established, there are few to seriously consider the non-Fermi liquid with topological phase, e.g., AHE, CME and topological magnetoelectric effect. This motivates us to study these physical effects originated from chiral anomaly in the non-Fermi liquid at or near QCP, which can be described by a CFT dual to Reissner-Nordstr\"om (RN) AdS black holes from classical Einstein gravity. In this paper, we aim at exploring the universal non-Fermi liquid behavior in topological phases due to chiral anomaly.

This paper is organized as follows.

In section~\ref{sec:Intro}, we firstly introduce the basic experimental observations on non-Fermi liquid behavior in heavy fermion system near quantum critical point. Secondly we introduce the non-Fermi liquid theory, our motivation and promising results. As supplementary material, in appendix~\ref{app:ChiralAnomaly_ChiralGaugeField}, we briefly discuss the chiral anomaly and chiral gauge field in both $(3+1)$-dimensional Euclidean and Minkowski space-time, as well as universal topological effects in topological metals such as CME and AHE due to $\theta$ vacuum term.

In section~\ref{sec:ChargedSpinor-CurvedSpaceTime}, we establish the general Lagrangian of charged fermions in curved space-time, with chiral gauge field introduced, which couples with the chirality operators in the bulk. The bulk Dirac equation with chiral gauge field in arbitrary dimensional space-time can be simplified by choosing proper representations for bulk Dirac fermions, with the assumption of spatial homogeneousness of the space-time. For more detail, refer to appendix~\ref{app:DiracRep_RotationalInvariance}.

In section~\ref{sec:BulkDiracEOM-SpatialRotaionalInvariance}, we study the bulk Dirac equation in high dimension with spatial rotational invariance. The generic Dirac equation in the transverse (spatial and time) sector and radial sector is separated. The generic solution to the radial sector of the Dirac equation is summarized in appendix~\ref{app:radial_Dirac_EOMs}.

In section~\ref{sec:UVGR-AdS(d+1)-SpatialRotationalInvariance}, we investigate the UV retarded Green's function from CFT dual to high dimensional bulk gravity and the bulk fermion wave functions with the assumption of rotational invariance. The UV retarded Green's functions can be obtained directly by solving flow equations, from which, the properties of the retarded Green's function are deduced in Sec.~\ref{sec:GR-prop}. Fermi surface of non-Fermi liquid from AdS$_4$ gravity, including both zero frequency and finite frequency cases, can be obtained from the UV retarded Green's function.

In section~\ref{sec:IRgR-AdS(d+1)}, we mainly study the retarded Green's function of Dirac fermion at IR fixed point on the near horizon boundary. The IR correlation functions for charged fermion at both zero and finite temperatures, in the bulk gravity with Ricci flat hypersurface are studied in Sec.~\ref{sec:flat-k=0}.  As a demo, we consider the non-Fermi liquid of AHE in retarded Green's function at the low frequency. The parameters to characterize the NFL behavior, i.e., residue, $Z_k$, Fermi velocity $v_F$ and self energy, $\Sigma(\omega,\lambda)$ can be obtained by fitting the exact retarded Green's function obtained through solving flow equation.

The basic results on conformal invariance of the near horizon boundary in topological charged black holes with Ricci flat or Ricci topological hypersurface in AdS$_4$ are summarized in appendix~\ref{app:TCBH_AdS}.

Further discussions on topological hypersurface with $k \ne 0$ case and its relevant physical implications on topological non-Fermi liquid are given in Sec.~\ref{sec:concl-dis}.

In this paper, we use the Greek symbols $\mu,\nu,...$ to denote indices of the boundary space-time coordinates $x^{\mu}\sim (t,x^{i})$,
with $i=1,2,...,d-1$ denoting the indices of pure spatial coordinates $x^{i}$.
The capital  letters  $M,N,...$ denote the indices of bulk space-time coordinates $x^{M}\sim (r,x^{\mu})$,
and small letters $a,b,...$ denote tangent space indices.

\section{Charged Dirac Fermion in Curved Space-Time}
\label{sec:ChargedSpinor-CurvedSpaceTime}

Consider a boundary theory with spinor operator ${\mathcal O}(t,\vec{x})$ dual to a bulk Dirac field $\psi(u,t,\vec{x})$ ($u\sim 1/r$) in curved space-time background $g_{MN}$. The action of this bulk fermion field coupled with a massless $U(1)$ gauge field $A_M$ is
\beqa
\ii\ii
S=-\int d^{d+1}x\sqrt{-g}i(\bar\psi \Gamma^M D_M \psi - m_D \bar\psi \psi) + S_{\text{bd}}, \label{Eq:S_fermion}
\eeqa
where the boundary term $S_{\text{bd}}$ is included to ensure a well defined variational principle for the bulk fermion~\cite{Henneaux:1998ch}, and $D_M$ is the covariant derivative in bulk space-time with minimal couplings to both of the background gauge field and the spin connection of the charged background geometry
\footnote{$b_\mu$ can be viewed as the other independent gauge field, which distinguishes itself from gauge field $A_M$ minimally coupled to the fermion.
In condensed matter physics, $b_\mu$ can be realized through imposing external source, or is an induced effective gauge field in the system, e.g., the chemical potential in CME as the time component of $b_\mu$. In some literature~\cite{Nagaosa:1999uc}, one may use the conventional notation $a_\mu$ instead of $b_\mu$.},
\beqa
 D_M &=& \partial_M - \frac{i}{4}{\omega^{\,ab}}_{M}\sigma_{ab} - i q A_M - i b_M \Gamma^{2p+1}, \nn\\
 \sigma_{ab} & \equiv & i \Gamma_{ab} = \frac{i}{2}[\Gamma_a, \Gamma_b], \label{Eq:DM-Lorentz-rep}
\eeqa
where ${\omega^{\,ab}}_{M}$ is a representation independent field known as spin connection~\cite{Brill:1957fx}, and $\sigma_{ab}$ is the matrix representing the generator of the $(d+1)$-dimensional homogeneous Lorentz group in the spinor representation of $SO(d,1)$. The $q$ is the charge of spinor operator ${\mathcal O}$ under the $U(1)$ current $J^\mu$. By setting the charge $q$ as a unit, one can study the physics of the probe changes as one varies the gauge coupling $g$, which is equivalent to rescaling of the gauge field. $p\equiv [{D}/{2}]$ and $\Gamma^{2p+1}$ is the Chirality operator defined in $2p$-even dimensions, e.g., if $D$ is even, $\Gamma^{2p+1}=\Gamma^{D+1}$ is the Chirality operator in $D$-even dimensions; while if $D$ is odd, $\Gamma^{2p+1}=\Gamma^{D}$ is the chirality operator defined in $(D-1)$-even dimensions. In the large $N$ limit, the 't Hooft coupling $g_S\equiv g^2N$ is fixed, thus the increasing of the gauge couplings is equivalent to the decreasing of $N^2$ strongly coupled degrees of freedom of the boundary field theory~\cite{McGreevy:2009xe}.

It is straightforward to read off Lagrangian density from the action given in Eq.(\ref{Eq:S_fermion}),
\beqa
{\mathcal L}_f = i\sqrt{-g}(\bar\psi \Gamma^M D_M \psi - m_D \bar\psi \psi)
\eeqa
from which one obtains the conjugate momentum of the bulk Dirac fermion
\beqa
\pi_f = \frac{\partial {\mathcal L}_f}{ \partial(\partial_t{\psi}) } = i \sqrt{-g} \bar\psi \Gamma^{t}.
\eeqa
Thus the Hamiltonian density is
\beqa
{\mathcal H}_f =\sqrt{-g} i \bar\psi \Gamma^{t} \dot{\psi} - {\mathcal L}_f.
\eeqa

It is worthy to notice that the bulk spinor $\psi$ has $(d+1)$-components for $d$-odd and $d$-components for $d$-even, while the spinor operator in the boundary ${\mathcal O}$ has only half components. Here
\beqa
 \bar\psi &=& \psi^\dag \Gamma^{\underline{t}}, \quad \Gamma^M D_M = \frac{1}{2}(\overrightarrow{\cancel{D}}-\overleftarrow{\cancel{D}}), \label{Eq:psibar-Dslash}\\
 \overrightarrow{\cancel{D}} & = & e_c^{\,\, M}\Gamma^c \bigg(\partial_M + \frac{1}{4}\omega_{ab\,M} \Gamma^{ab}-iqA_M - ib_M \Gamma^{2p+1} \bigg), \nn
\eeqa
where $M$ and $a,b$ are used to denote the indexes of bulk space-time and tangent space (local Lorentz space-time or local laboratory coordinates), respectively.
For tangent space, we use a underline to label the corresponding coordinates, i.e., $a=(\underline{r}, \underline{t},\underline{i})$.
The Greece latters $\mu$, $\nu$ etc., are used to denote indexes along the hypersurface or cutoff brane, i.e., $x^M\sim(r,x^\mu)$, $x^\mu\sim(t,x^i)$.
The Gamma matrices in bulk space-time are linear combination of the Gamma matrix in the tangent space time, $\Gamma^{M}= e_a^{\,\,M}\Gamma^{a}$, where $\Gamma^a$ are Gamma matrices satisfy Grassman algebra $\{ \Gamma^a, \Gamma^b  \}=2\eta^{ab}$. To be concrete,
\beqa
(\Gamma^{\underline{t}})^2 = -1, \quad (\Gamma^{\underline{r}})^2 = 1, \quad (\Gamma^{\underline{x}})^2 = 1. \label{Eq:Grassmanian-Algebra}
\eeqa
We have introduced vierbein $e_a^{\,M}$, $a$ here is used to denote tangent space index as the $a$ above,
which are a set of $(d+1)$-dimensional orthogonal normal vector bases and the label $a$ telling the direction of the vector.
The spinor has a Lorentz transformation rule locally.

The spin connection $\omega_{ab M}$ is a $1$-form defined as
\beqa
&& \omega_{ab\,M} = g_{NL} e_{a}^{\,L} \nabla_M e_b^{\,N} = e_{aN} (\partial_M e_{b}^{\ N} + \Gamma_{M L}^{N}e_b^{ \ L }),\nn\\
&& {\omega^{ab}}_{M} = \eta^{ac}\eta^{bd} \omega_{cd\,M}.   \label{Eq:omega-spin-connection}
\eeqa
It is worth emphasizing that the spin connection is antisymmetric in $a$ and $b$, since $\eta^{ab}=g^{MN}e^a_{\,M}e^b_{\,N}$ is a quantity with vanishing covariante derivative ($\nabla_M \eta^{ab}= {\omega^{ab}}_M+ {\omega^{ba}}_M=0$).

\section{Dirac Fermions in Bulk with Spatial Rotational Invariance}
\label{sec:BulkDiracEOM-SpatialRotaionalInvariance}

\subsection{Dirac equation in $(3+1)$-dimensions}

Consider a generic diagonal background metric with rotationally invariance along $x^i$ directions,
\beqa
\ii \dr s^2  =  - g_{tt}(r)\dr t^2 + g_{rr}(r)\dr r^2 + g_{xx}(r)\dr x^i \dr x_i, \label{Eq:ds2-generic-(3+1)D}
\eeqa
where $i = 1, 2$. According to  Eq.(\ref{Eq:Dirac-Psi-k}), the most general bulk Dirac equation in $(3+1)$-dimensions can also be re-expressed as
\beqa
\ii
&& \ii \bigg(\frac{\sqrt{g_{xx}}}{\sqrt{g_{tt}}} \Gamma^{\underline{t}} ( \partial_t \!-\! i q A_t \!+\! i b_0 \Gamma^{\underline{5}} )
\!+\!  \frac{\sqrt{g_{xx}}}{\sqrt{g_{rr}}} \Gamma^{\underline{r}} (\partial_r - i b_r \Gamma^{\underline{5}}) \nn\\
&& \!-\! m_D  \sqrt{g_{xx}}  \!+\!   \Gamma^{\underline{j}} (\partial_j \!-\! i q A_j \!-\! i b_j \Gamma^{\underline{5}} )  \bigg)\Psi \!=\! 0,\qquad
\label{Eq:Dirac-Psi-k-1}
\eeqa
where $j=x,y$ and $ \Psi \equiv (-gg^{rr})^{\frac{1}{4}}\psi$.
The EOMs can be simplified by absorbing $b_r$ into the radial sector of the wave function, which gives
\beqa
\ii
&& \bigg[ (\partial_r - \sqrt{g_{rr}} m_D \Gamma^{\underline{r}})  + i \frac{\sqrt{g_{rr}}}{\sqrt{g_{xx}}}\Gamma^{\underline{r}}\Gamma^{\underline{\mu}}(K_\mu - B_\mu \Gamma^{\underline{5}}) \bigg] \Psi = 0, \ii \nn \\
&& K_\mu  = (-K_0, K_j) \equiv \bigg(-\frac{\sqrt{g_{xx}}}{\sqrt{g_{tt}}}( \omega +  q A_t ), ( k_j  - q A_j)  \bigg),  \nn \\
&& B_\mu  = (-B_0, B_j) \equiv \bigg(-\frac{\sqrt{g_{xx}}}{\sqrt{g_{tt}}} b_0 , b_j \bigg), \label{Eq:Dirac-Psi-k-2}
\eeqa
where $b_r$ has been absorbed into the phase redefinition of the Dirac field through
\beqa
\Psi \equiv (-gg^{rr})^{\frac{1}{4}}e^{-i b(r) \Gamma^{\underline{2p+1}} } \psi,
\eeqa
where the phase factor $b(r)$ is a function defined in Eq.(\ref{Eq:b(r)}). In the chosen representation in Eq.(\ref{Eq:Gamma-rep-1}) with chirality operator given in Eq.(\ref{Eq:Gamma-rep-1_Chirality}), the second phase factor becomes
\beqa
&& \exp{(-ib(r)\Gamma^{\underline{5}})} = \exp{\bigg( -b(r)  \left(
                                                                                                       \begin{array}{cc}
                                                                                                         0 & -\sigma^2 \\
                                                                                                         \sigma^2 & 0 \\
                                                                                                       \end{array}
                                                                                                     \right)
   \bigg)} \nn\\
&=& \left(
                                                                      \begin{array}{cccc}
                                                                        \cosh{b(r)} & 0 & 0 & i \sinh{b(r)} \\
                                                                        0 & \cosh{b(r)} & -i\sinh{b(r)} & 0 \\
                                                                        0 & i\sinh{b(r)} & \cosh{b(r)} & 0 \\
                                                                        -i\sinh{b(r)} & 0 & 0 & \cosh{b(r)} \\
                                                                      \end{array}
                                                                    \right). \nn
\eeqa

By assuming the rotational invariance in the $2$-dimensional spatial sector, i.e., $x$-$y$ plane, perpendicular to the $r$ direction, the bulk Dirac equation in $(3+1)$-dimensions can be simplified in analogy to those in $(2+1)$-dimensions. Due to the rotational invariance in the $2$-dimensional spatial directions, one does not lose generality by setting
\beqa
k_1 = k_x, ~ b_1 = b_x ; \quad k_2=k_y, ~ b_2=b_y. \label{Assump:rotational-invariance}
\eeqa
While in our case, since we are considering the chiral gauge field in the bulk, we would like to consider both the electric and magnetic field at the same time. Assume that the static magnetic field pass through the $x$-$y$ plane, i.e., $h_\mu=(h_r,h_t,\vec{h})=(h,0,h_x,h_y,\textbf{0}_{d-3})$,
\beqa
\vec{A} =  \frac{1}{2} (\vec{h} \times \vec{x}),
\eeqa
namely,
\beqa
&& A_x(y) = \frac{1}{2}(h_y r - h_r y ) = - \frac{1}{2} h y, \nn\\
&& A_y(x) =  \frac{1}{2}( h_r x - h_x r) = \frac{1}{2} h x.
\eeqa
For simplicity and without loss of generality, one can choose the nonvanishing elements of the gauge field as
\beqa
A_x(y) = - h y,  \quad A_y(x) = 0.
\eeqa
While equivalently, one can also take
\beqa
A_x(y) = 0, \quad A_y(x) = h x.
\eeqa
Thus, in $(3+1)$-dimensional space-time case, we choose the static gauge field as
\beqa
A_M = (A_r, A_\mu) = \bigg(0, A_t(r), -hy, 0 \bigg), \label{Eq:A_M}
\eeqa
which gives
\beqa
E_M = (E_r, E_\mu) = (\partial_r A_t(r),0,0,0), \quad h_M = (h,0,0,0). \nn
\eeqa

\subsection{Dirac equations in bulk with spatial rotational invariance}

In the assumption that the rotational invariance in the $2$-dimensional space as shown in Eq.(\ref{Assump:rotational-invariance}), e.g., except $r$ direction, the bulk Dirac equations in $(3+1)$-dimensions can be simplified. Under the representation in Eq.(\ref{Eq:Gamma-rep-1}) and chirality operator in Eq.(\ref{Eq:Gamma-rep-1_Chirality}), the Dirac equation in Eq.(\ref{Eq:Dirac-Psi-k-1}) becomes
\beqa
( D_{L}(r) + D_{T}(r)  )\Psi \!=\! 0, \quad \Psi \equiv (-gg^{rr})^{\frac{1}{4}}\psi.
\eeqa
where
\beqa
D_{L}(r) & \equiv & \frac{\sqrt{g_{xx}}}{\sqrt{g_{tt}}} \Gamma^{\underline{t}} ( \partial_t \!-\! i q A_t \!+\! i b_0 \Gamma^{\underline{5}} ) \nn\\
&& +  \frac{\sqrt{g_{xx}}}{\sqrt{g_{rr}}} \Gamma^{\underline{r}} (\partial_r - i b_r \Gamma^{\underline{5}}) \!-\! m_D  \sqrt{g_{xx}} , \nn\\
D_{T}(r) & \equiv & \Gamma^{\underline{1}} (\partial_x \!-\! i b_x \Gamma^{\underline{5}} ) + \Gamma^{\underline{2}} (\partial_y \!-\! i b_y \Gamma^{\underline{5}} ). \qquad
\eeqa

Although the matrices $D_1$ and $D_2$ do not commute, it is possible to construct an anti-unitary matrix $U$, i.e., $U^2=-1$, $U^{-1}=U^\dag$ and $U^\dag = - U$ which in the representation in Eq.(\ref{Eq:Gamma-rep-1}), turns out to be
\beqa
U = \left(
     \begin{array}{cc}
       -i\sigma^2 & 0 \\
       0 & -i\sigma^2 \\
     \end{array}
   \right), \label{Eq:transU}
\eeqa
thus
\beqa
\ii && \ii U \Gamma^{\underline{t}} = \left(
     \begin{array}{cc}
     - i \sigma^3 & 0 \\
     0 & - i \sigma^3  \\
     \end{array}
   \right), ~ U \Gamma^{\underline{r}} = \left(
     \begin{array}{cc}
     - \sigma^1 & 0 \\
     0 & - \sigma^1  \\
     \end{array}
   \right),  \nn\\
\ii && \ii U \Gamma^{\underline{1}} = \left(
     \begin{array}{cc}
     i \textbf{1}_2  & 0 \\
     0 & - i \textbf{1}_2  \\
     \end{array}
   \right) , ~ U \Gamma^{\underline{2}} = \left(
     \begin{array}{cc}
     0 & - i\textbf{1}_2 \\
     - i\textbf{1}_2 & 0  \\
     \end{array}
   \right) ,  \nn\\
\ii && \ii \Gamma^{\underline{t}}\Gamma^{\underline{5}} = \left(
     \begin{array}{cc}
       0 & -i\sigma^3 \\
     i\sigma^3 & 0 \\
     \end{array}
   \right), ~ \Gamma^{\underline{r}}\Gamma^{\underline{5}} = \left(
     \begin{array}{cc}
       0 & -\sigma^1 \\
     \sigma^1 & 0 \\
     \end{array}
   \right), \nn\\
\ii && \ii \Gamma^{\underline{1}}\Gamma^{\underline{5}} = \left(
     \begin{array}{cc}
       0 & -i\textbf{1}_2 \\
     -i\textbf{1}_2  & 0 \\
     \end{array}
   \right) , ~ \Gamma^{\underline{2}}\Gamma^{\underline{5}} = \left(
     \begin{array}{cc}
     -i \textbf{1}_2  & 0 \\
     0 & i \textbf{1}_2  \\
     \end{array}
   \right), \qquad \label{Eq:UGamma_trx}
\eeqa
and one can check that the matrix has the following properties
\beqa
 [U, \Gamma^{\underline{5}}] =0,
\eeqa
and
\beqa
&& \{ U, \Gamma^{\underline{t}} \} =0, \quad  \{ U, \Gamma^{\underline{r}} \}  = 0, \nn\\
&& \{ U, \Gamma^{\underline{t}}\Gamma^{\underline{2p+1}}  \} =0, \quad \{ U, \Gamma^{\underline{r}}\Gamma^{\underline{2p+1}} \} =0,   \nn\\
&& [U, \Gamma^{\underline{1}}] =0, \quad  [U, \Gamma^{\underline{2}}] = 0,\nn\\
&& [ U, \Gamma^{\underline{1}}\Gamma^{\underline{2p+1}}  ] =0, \quad [ U, \Gamma^{\underline{2}}\Gamma^{\underline{2p+1}}  ] =0.
\eeqa
where $[,]$ and $\{,\}$ means commutator and anti-commutator, respectively.
Therefore, one obtains
\beqa
\{ U, D_{L}(r) \} =0, \quad  [U, D_2 (r)] =0,
\eeqa
which implies that
\beqa
\ii\ii\ii U^{-1}D_{L}(r)U = - D_{L}(r) \quad U^{-1} D_{T}(r) U = D_{T}(r).
\eeqa
thus $UD_{L}(r)$,$UD_{T}(r)$ are commuting Hermitian operators,
\beqa
[UD_{L}(r), UD_{T}(r)]=0,
\eeqa
which means that they have the same eigenvectors, i.e., the Dirac equation becomes
\beqa
UD_{L}(r)\Psi = - UD_{T}(r)\Psi = L\Psi,   \label{Eq:UD1-UD2-L}
\eeqa
where $L$ is a real eigenvalue.

\subsection{Spatial coordinate dependence}
We will try to solve the spatial coordinate dependence of the equation firstly, namely, the second sector of Eq.(\ref{Eq:UD1-UD2-L}).

\subsubsection{The case without external magnetic field}

\begin{widetext}
\beqa
U D_{T}(r)\Psi = [U\Gamma^{\underline{1}} (\partial_x \!-\! i b_x \Gamma^{\underline{5}}  ) + U\Gamma^{\underline{2}} (\partial_y \!-\! i b_y \Gamma^{\underline{5}} )] \Psi =  - L\Psi. \label{Eq:DiracEOMs_D2Psi}
\eeqa
By using Eq.(\ref{Eq:UGamma_trx}), we obtain
\beqa
U D_{T}(r) =  \left(
     \begin{array}{cc}
     i \textbf{1}_2 \partial_x + i b_y \sigma^2 & i b_x \sigma^2 - i\textbf{1}_2 \partial_y \\
     i b_x \sigma^2 - i\textbf{1}_2 \partial_y & - i \textbf{1}_2 \partial_x - i b_y \sigma^2   \\
     \end{array}
   \right). \label{Eq:UD2}
\eeqa
By doing Fourier transformation and considering the assumption in Eq.(\ref{Assump:rotational-invariance}), one obtains
\beqa
\ii
\psi(r,x^\mu) = (-gg^{rr})^{-\frac{1}{4}}  \int d \omega \int dk_i e^{-i\omega t + i k_i x^i }\Psi(r,k^\mu), \label{Eq:psi-Psi-Fourier}
\eeqa
where $x^\mu=(t,x^i)=(t,x,y)$ and $k^\mu = (\omega, k^i) = (\omega,k_x,k_y)$.

Thus in the momentum space, by making the replacement $\partial_x \to i k_x$, and by using the notation for bulk Dirac spinor $\Psi$, in terms of two Weyl fermions, as shown in Eq.(\ref{Eq:Psi}),
\beqa
\Psi\equiv \left(
             \begin{array}{c}
               \Psi_+ \\
               \Psi_- \\
             \end{array}
           \right),
\eeqa
the Dirac equation in Eq.(\ref{Eq:DiracEOMs_D2Psi}) becomes
\beqa
 \left(
     \begin{array}{cc}
     - k_x \textbf{1}_2 + i b_y \sigma^2 & i b_x \sigma^2 - i\textbf{1}_2 \partial_y \\
     i b_x \sigma^2 - i\textbf{1}_2 \partial_y &  k_x \textbf{1}_2 - i b_y \sigma^2   \\
     \end{array}
   \right) \left(
             \begin{array}{c}
               \Psi_+ \\
               \Psi_- \\
             \end{array}
           \right)   = - \lambda \textbf{1}_4  \left(
             \begin{array}{c}
               \Psi_+ \\
               \Psi_- \\
             \end{array}
           \right).
\eeqa
One can make a unitary transformation $U_1$ according to the convention used above and it acts on all coordinate sector, namely, $t,r,x,y$,
\beqa
 \left(      \begin{array}{c}
               \tilde\Psi_+ \\
               \tilde\Psi_- \\
             \end{array}
           \right) = U_1 \left(
             \begin{array}{c}
               \Psi_+ \\
               \Psi_- \\
             \end{array}
           \right), \quad U_1  \equiv \frac{1}{\sqrt{2}}(1- i \sigma^1) = \frac{1}{\sqrt{2}}\left(
                                                                                           \begin{array}{cc}
                                                                                             1 & -i \\
                                                                                             -i & 1 \\
                                                                                           \end{array}
                                                                                         \right),  \label{Eq:transU1_tPsi-Psi} \label{Eq:MAMinv}
\eeqa
which gives
\beqa
U_1 \left(
     \begin{array}{cc}
     (\lambda - k_x) \textbf{1}_2 + i b_y \sigma^2 & i b_x \sigma^2 - i\textbf{1}_2 \partial_y \\
     i b_x \sigma^2 - i\textbf{1}_2 \partial_y &  (\lambda + k_x) \textbf{1}_2 - i b_y \sigma^2   \\
     \end{array}
   \right) U_1^{-1}  \left(      \begin{array}{c}
               \tilde\Psi_+ \\
               \tilde\Psi_- \\
             \end{array}
           \right)  = 0. \nn
\eeqa
By substituting transformation matrix $U_1$ into the field equations above, one obtains
\beqa
\left(
     \begin{array}{cc}
     \lambda \textbf{1}_2  & i (b_x+i b_y) \sigma^2 - i \textbf{1}_2 (\partial_y + k_x ) \\
     i (b_x-i b_y) \sigma^2 - i \textbf{1}_2 (\partial_y - k_x ) &  \lambda  \textbf{1}_2    \\
     \end{array}
   \right)  \left(      \begin{array}{c}
               \tilde\Psi_+ \\
               \tilde\Psi_- \\
             \end{array}
           \right)  = 0,  \label{Eq:M(UD2-L)Minv_Psi12}
\eeqa
from which, we obtain two coupled 1st order ODEs,
\beqa
\lambda \textbf{1}_2 \tilde\Psi_+ + [i (b_x+i b_y) \sigma^2 - i \textbf{1}_2 (\partial_y + k_x ) ]\tilde\Psi_- = 0, \quad \lambda \textbf{1}_2 \tilde\Psi_- + [i (b_x-i b_y) \sigma^2 - i \textbf{1}_2 (\partial_y - k_x ) ]\tilde\Psi_+ = 0. \nn
\eeqa
Thus one can obtain two decoupled 2nd order ODEs,
\beqa
&&   [ -\lambda^2 \textbf{1}_2 -(b_x^2+b_y^2)\textbf{1}_2+ 2\sigma^2 (b_x \partial_y - i b_y k_x) + \partial_y(b_x - i b_y)\sigma^2  -(\partial_y^2 - k_x^2)\textbf{1}_2  ] \tilde\Psi_+ = 0, \nn\\
&&   [ -\lambda^2 \textbf{1}_2 -(b_x^2+b_y^2)\textbf{1}_2+ 2\sigma^2 (b_x \partial_y - i b_y k_x) + \partial_y(b_x + i b_y)\sigma^2  -(\partial_y^2 - k_x^2)\textbf{1}_2  ] \tilde\Psi_- = 0.
\eeqa
We assume that $b_x, b_y$ are two constants, namely $\partial_y b_i =0$ with $i=x,y$. Thus, one has
\beqa
   [ \lambda^2 \textbf{1}_2 + (b_x^2+b_y^2)\textbf{1}_2 - 2\sigma^2 (b_x \partial_y - i b_y k_x) + (\partial_y^2 - k_x^2)\textbf{1}_2  ] \tilde\Psi_\beta = 0,
\eeqa
with $\beta=\pm$, and
\end{widetext}
\beqa
\tilde\Psi_{\beta} \equiv \tilde{R}_\beta(r) \otimes \left(\begin{array}{c}
                                                                                             \chi_1(y) \\
                                                                                             \chi_2(y)  \\
                                                                                           \end{array}
                                                                                         \right), \label{Eq:tPsi-chi_12-y}
\eeqa
where $\tilde{R}_\beta=(\tilde{F}_\beta,\tilde{G}_\beta)$ defined in Eq.(\ref{Eq:Rt_alpha}), are the radial sector of the wave function, we will solve it in the following sections.

One can introduce other unitary transformation $U_2$, which only acts on the wave functions in the spatial directions, namely, the transverse sector $x^i=(x,y)$, which are transverse to the time and radial directions $t$ and $r$,
\beqa
\tilde\Psi_\beta = U_2 \tilde{\tilde\Psi}_\beta,
\eeqa
where $\beta=\pm$ and
\beqa
\ii\ii\ii U_2 = \frac{1}{\sqrt{2}}\left(
                                                                                           \begin{array}{cc}
                                                                                             -i & i \\
                                                                                             1 & 1 \\
                                                                                           \end{array}
                                                                                         \right), ~ U_2^{-1}\sigma^2 U_2 = \left(
                                                                                           \begin{array}{cc}
                                                                                             1 & 0 \\
                                                                                             0 & -1 \\
                                                                                           \end{array}
                                                                                         \right) = \sigma^3. ~ \label{Eq:transU2_tPsi_chi}
\eeqa
Thus
\beqa
\tilde{\tilde\Psi}_\beta = U_2^{-1}\tilde\Psi_\beta.
\eeqa
By using the properties of the transformation matrix $U_2$,
\beqa
U_2^{-1}\sigma^2 U_2 = \sigma^3, \quad U_2^{-1}\sigma^1 U_2 = -\sigma^2, \quad U_2^{-1}\sigma^3 U_2 = -\sigma^1, \nn
\eeqa
one obtains
\beqa
&& [ \lambda^2 \textbf{1}_2 + (b_x^2+b_y^2)\textbf{1}_2 - 2\sigma^3 (b_x \partial_y - i b_y k_x) \nn\\
&& + (\partial_y^2 - k_x^2)\textbf{1}_2  ] \tilde{\tilde\Psi}_\beta = 0, \nn
\eeqa
where $\beta = \pm $.
By assuming that
\beqa
\tilde{\tilde\Psi}_{\beta}
= \tilde{R}_\beta(r)\otimes \left(
                    \begin{array}{c}
                      \chi_+(y) \\
                      \chi_-(y) \\
                    \end{array}
                  \right),   \label{Eq:ttPsi-chi_pm-y}
\eeqa
one obtains
\beqa
 && [ \lambda^2  + (b_x^2+b_y^2) - 2(b_x \partial_y - i b_y k_x) + (\partial_y^2 - k_x^2)  ] \chi_+ = 0, \nn\\
 && [ \lambda^2  + (b_x^2+b_y^2) + 2(b_x \partial_y - i b_y k_x) + (\partial_y^2 - k_x^2)  ] \chi_- = 0, \nn
\eeqa
which gives
\beqa
\chi_+ &=& e^{b_x y} [C_{1} e^{ - y\sqrt{(k_x - ib_y)^2-\lambda^2}} + C_{2} e^{ y\sqrt{(k_x -i b_y)^2 -\lambda^2} }], \nn\\
\chi_- &=& e^{-b_xy} [D_{1} e^{ -y \sqrt{(k_x + ib_y)^2-\lambda^2}} + D_{2} e^{y\sqrt{(k_x +i b_y)^2 -\lambda^2} }], \nn
\eeqa
where $C_1$, $C_2$, $D_1$ and $D_2$ are all constants.

By imposing the infinite boundary conditions as
\beqa
\chi_\pm \sim e^{\pm ik_y y}, \quad k_y = -i b_x \pm \sqrt{\lambda^2 - (k_x-ib_y)^2}, \nn
\eeqa
we obtain pseudo eigenvalues (since they are not real) as
\beqa
\lambda_{\pm}^2 = (k_x \mp i b_y)^2 + (k_y \pm i b_x)^2,
\eeqa
which are real only in the case when $k_y=0, b_y=0$ (or $k_x=0,b_x=0$), then the eigenvalue becomes degenerate as
\beqa
\ii\lambda_+ = \lambda_- \equiv \lambda = \pm \sqrt{k_x^2 - b_x^2} \equiv \pm \sqrt{k^2 - b^2}, \label{Eq:lambda-k-b}
\eeqa
since $x$ and $y$ directions have no anistropy in the absence of external field, e.g., by making the replacement,
\beqa
k_x \to -i \partial_x, \quad \partial_y \to  i k_y,
\eeqa
the equations of motion above become
\beqa
 && [ \lambda^2  + (b_x^2+b_y^2) - 2(i b_x k_y -  b_y \partial_x) + (-k_y^2 + \partial_x^2)  ] \chi_+ = 0, \nn\\
 && [ \lambda^2  + (b_x^2+b_y^2) + 2(i b_x k_y -  b_y \partial_x) + (-k_y^2 + \partial_x^2)  ] \chi_- = 0. \nn
\eeqa
Therefore, in the $k_y=b_y=0$ case, the equations of motion become
 \beqa
( \lambda^2  + b_x^2  +  \partial_x^2  ) \chi_\pm = 0,
\eeqa
which gives
\beqa
\chi_\pm &=& C_{1,2}e^{i\sqrt{\lambda^2+b_x^2}x}+D_{1,2}e^{-i\sqrt{\lambda^2+b_x^2}x} \nn\\
&\equiv & C_{1,2}e^{ik_x x}+D_{1,2}e^{-ik_x x},
\eeqa
thus the eigenvalues are
\beqa
\ii\ii\ii \lambda = \pm \sqrt{k_x^2 - b_x^2} \equiv \pm \sqrt{k^2 - b^2}, \quad k\equiv k_x, \, b\equiv b_x.
\eeqa
Otherwise, one can make a replacement for the chiral gauge field from the beginning, which is allowed in Euclidean space for the chiral gauge field,
\beqa
b_x \to - i b_x, \quad b_y \to - i b_y,
\eeqa
so that the wavefunction is completely in the oscillator region,
\beqa
\chi_+ &\sim & 
 e^{-ib_x y}  e^{ - y\sqrt{(k_x - b_y)^2-\lambda^2}} = e^{i k_y y}, \nn\\
\chi_- &\sim & 
 e^{ib_xy} e^{ -y \sqrt{(k_x + b_y)^2-\lambda^2}} = e^{-i k_y y},
\eeqa
where we have just kept the physical wavefunctions by imposing $C_2=0$ and $D_2=0$, which are exponentially divergent in the infinite boundary, and the eigenvalues are, respectively,
\beqa
\lambda_\pm = \pm \sqrt{(k_x \mp b_y)^2 +(k_y \mp b_x)^2},
\eeqa
and it is worth noticing that $\lambda_- \to \lambda_+(b_i \to -b_i)$, $i=x,y$.

One obtains the solution of the wave functions
\beqa
\ii\ii\ii \left(
             \begin{array}{c}
               \Psi_+ \\
               \Psi_- \\
             \end{array}
           \right) = U_1^{-1}\left(      \begin{array}{c}
               \tilde\Psi_+ \\
               \tilde\Psi_- \\
             \end{array}
           \right)
           = \frac{1}{\sqrt{2}}\left(      \begin{array}{c}
               \tilde\Psi_+ + i\tilde\Psi_- \\
               \tilde\Psi_- + i\tilde\Psi_+ \\
             \end{array}
           \right) ,
\eeqa
and according to Eq.(\ref{Eq:transU2_tPsi_chi}) and Eq.(\ref{Eq:ttPsi-chi_pm-y}), one has
\beqa
\ii \tilde\Psi_{\beta} = U_2 \tilde{\tilde\Psi}_{\beta}  = \tilde{R}_\beta \otimes \left(\begin{array}{c}
                                                                                             -i\frac{1}{\sqrt{2}}( \chi_+ - \chi_- ) \\
                                                                                             \frac{1}{\sqrt{2}}(\chi_+ + \chi_-)  \\
                                                                                           \end{array}
                                                                                         \right),
\eeqa
where $\beta=\pm$ and $\chi_\pm=\chi_\pm(y)$.
By comparing with those defined in Eq.(\ref{Eq:tPsi-chi_12-y}), we obtain
\beqa
\ii\ii \chi_1 = -i\frac{1}{\sqrt{2}}( \chi_+ - \chi_- ) , ~\chi_2 = \frac{1}{\sqrt{2}}(\chi_+ + \chi_-), \label{Eq:chi_12-chi_pm}
\eeqa
where $\chi_{1,2}=\chi_{1,2}(y)$.

It is worth emphasizing that the form of $\chi_\pm(y)$ doesn't affect the radial sector, since they do not depend on the radial coordinates, meanwhile, the transformation matrix between $\tilde\Psi_\beta$ and $\tilde{\tilde\Psi}_\beta$, i.e., $U_2$, does not affect the radial sector of $\Psi_\beta$($\beta=\pm$), since $\tilde\Psi_\alpha \propto R_\beta$ thus in fact $(\chi_1,\chi_2)^T=U_2(\chi_+,\chi_-)^T$. The transformation that does affect the radial sector is $U_1$, since $(\Psi_+,\Psi_-)^T=U_1^{-1}(\tilde\Psi_+,\tilde\Psi_-)^T\sim U_1^{-1}(\tilde{R}_+,\tilde{R}_-)^T\otimes(\chi_1,\chi_2)^T$.

To be brief, we have obtained the solution corresponding to the eigenvalue $+\lambda$, namely,
\beqa
UD_{L}(r)\Psi = - UD_{T}(r)\Psi = \lambda \Psi,
\eeqa
where $UD_{T}(r)$ is given in Eq.(\ref{Eq:UD2}), which is equivalent to Eq.(\ref{Eq:M(UD2-L)Minv_Psi12}).

The other independent solutions corresponding to the eigenvalue $-\lambda$, is
\beqa
UD_{L}(r)\Psi = - UD_{T}(r)\Psi = -\lambda \Psi,
\eeqa

\subsubsection{The case with external magnetic field}

\begin{widetext}
In the case that there is an external magnetic field, e.g., by imposing an external magnetic field passing through the $x$-$y$ plane. Then there will be a gauge vector along the direction of partial derivative with respect to $x$ (or $y$), namely $\partial_x \to \partial_x \!-\! i q A_x(y)$ (or $\partial_y \to \partial_y \!-\! i q A_y(x)$), then Eq.(\ref{Eq:DiracEOMs_D2Psi}) becomes
\beqa
U D_{T}(r)\Psi = [U\Gamma^{\underline{1}} (\partial_x \!-\! i q A_x(y) \!-\! i b_x \Gamma^{\underline{5}}  ) + U\Gamma^{\underline{2}} (\partial_y \!-\! i b_y \Gamma^{\underline{5}} )] \Psi =  - L\Psi,
\eeqa
where the spatial component of $U(1)$ gauge field is chosen as in Eq.(\ref{Eq:A_M}). Thus, Eq.(\ref{Eq:UD2}) becomes
\beqa
U D_{T}(r)
= \left(
     \begin{array}{cc}
     i \textbf{1}_2 (\partial_x  + i q h y ) + i b_y \sigma^2 & i b_x \sigma^2 - i\textbf{1}_2 \partial_y \\
     i b_x \sigma^2 - i\textbf{1}_2 \partial_y & - i \textbf{1}_2 (\partial_x  + i q h y ) - i b_y \sigma^2   \\
     \end{array}
   \right). \label{Eq:UD2_b}
\eeqa
By doing Fourier transformation as shown in Eq.(\ref{Eq:psi-Psi-Fourier}), then in the momentum space, by making the replacement $\partial_x \to i k_x$, the Dirac equation in Eq.(\ref{Eq:DiracEOMs_D2Psi}) becomes
\beqa
\left(
     \begin{array}{cc}
     - (k_x + qh y) \textbf{1}_2 + i b_y \sigma^2 & i b_x \sigma^2 - i\textbf{1}_2 \partial_y \\
     i b_x \sigma^2 - i\textbf{1}_2 \partial_y &  (k_x+q h y) \textbf{1}_2 - i b_y \sigma^2   \\
     \end{array}
   \right) \left(
             \begin{array}{c}
               \Psi_+ \\
               \Psi_- \\
             \end{array}
           \right)  = - \lambda \textbf{1}_4 \left(
             \begin{array}{c}
               \Psi_+ \\
               \Psi_- \\
             \end{array}
           \right) ,
\eeqa
one can make a transformation as shown in Eq.(\ref{Eq:transU1_tPsi-Psi}), namely $(\tilde\Psi_+,\tilde\Psi_-)^T=U_1(\Psi_+,\Psi_-)^T$, which gives
\beqa
U_1\left(
     \begin{array}{cc}
     [\lambda - (k_x+qhy)] \textbf{1}_2 + i b_y \sigma^2 & i b_x \sigma^2 - i\textbf{1}_2 \partial_y \\
     i b_x \sigma^2 - i\textbf{1}_2 \partial_y &  [\lambda + (k_x+qhy)] \textbf{1}_2 - i b_y \sigma^2   \\
     \end{array}
   \right) U_1^{-1}  \left(      \begin{array}{c}
               \tilde\Psi_+ \\
               \tilde\Psi_- \\
             \end{array}
           \right)  = 0,
\eeqa
by using Eq.(\ref{Eq:MAMinv}), it is more explicit
\beqa
\left(
     \begin{array}{cc}
     \lambda \textbf{1}_2  & i b_x \sigma^2 - i\textbf{1}_2 \partial_y - i [  (k_x+qh y) \textbf{1}_2 - i b_y \sigma^2 ] \\
     i b_x \sigma^2 - i\textbf{1}_2 \partial_y + i [  (k_x+qhy) \textbf{1}_2 - i b_y \sigma^2 ] &  \lambda  \textbf{1}_2    \\
     \end{array}
   \right)  \left(      \begin{array}{c}
               \tilde\Psi_+ \\
               \tilde\Psi_- \\
             \end{array}
           \right)  = 0,
\eeqa
from which, we obtain
\beqa
\ii \lambda \textbf{1}_2 \Psi_+ + [i (b_x+i b_y) \sigma^2 - i \textbf{1}_2 (\partial_y + k_x + qh y) ]\tilde\Psi_- = 0, \quad \lambda \textbf{1}_2 \Psi_- + [i (b_x-i b_y) \sigma^2 - i \textbf{1}_2 (\partial_y - k_x - qh y) ]\tilde\Psi_+ = 0, \nn
\eeqa
which can be reduced in the absence of chiral gauge field to
\beqa
 \lambda \textbf{1}_2 \Psi_+ - i \textbf{1}_2 [\partial_y + (k_x + qh y) ]\tilde\Psi_- = 0, \quad \lambda \textbf{1}_2 \Psi_- - i \textbf{1}_2 [\partial_y - (k_x + qh y) ]\tilde\Psi_+ = 0. \nn
\eeqa
Thus one obtains two decoupled 2nd order ODEs,
\beqa
&&   [ -\lambda^2 \textbf{1}_2 -(b_x^2+b_y^2)\textbf{1}_2+ 2\sigma^2 [b_x \partial_y - i b_y (k_x + qhy)] + \partial_y(b_x - i b_y)\sigma^2  -(\partial_y^2 - (k_x+qhy)^2)\textbf{1}_2  ] \tilde\Psi_+ = 0, \nn\\
&&   [ -\lambda^2 \textbf{1}_2 -(b_x^2+b_y^2)\textbf{1}_2+ 2\sigma^2 [b_x \partial_y - i b_y (k_x + qhy)] + \partial_y(b_x + i b_y)\sigma^2  -(\partial_y^2 - (k_x+qhy)^2)\textbf{1}_2  ] \tilde\Psi_- = 0. \nn
\eeqa
Assuming that $b_x, b_y$ are two constants, thus $\partial_y b_x =0,  \partial_y b_y=0$. Thus we have
\beqa
   [ \lambda^2 \textbf{1}_2 + (b_x^2+b_y^2)\textbf{1}_2 - 2\sigma^2 (b_x \partial_y - i b_y k) + (\partial_y^2 - (k_x+qhy)^2)\textbf{1}_2  ] \tilde\Psi_\beta = 0, \nn
\eeqa
where $\beta=\pm$. In the absence of chiral gauge field, the equations reduce to $  [ \lambda^2 \textbf{1}_2 + (\partial_y^2 - (k_x+qhy)^2)\textbf{1}_2  ] \tilde\Psi_\beta = 0$. One can make a transformation as shown in Eq.(\ref{Eq:transU2_tPsi_chi}), i.e., $\tilde\Psi_\beta=U_2 \tilde{\tilde\Psi}_\beta$($\beta=1,2$),  which leads to
\beqa
& [ \lambda^2 \textbf{1}_2 + (b_x^2+b_y^2)\textbf{1}_2 - 2\sigma^3 (b_x \partial_y - i b_y k_x) + (\partial_y^2 - (k_x+qhy)^2)\textbf{1}_2  ] \tilde{\tilde\Psi}_\beta = 0.
\eeqa
By assuming the wave function as that in Eq.(\ref{Eq:ttPsi-chi_pm-y}), one obtains EOMs for the wave functions $\chi_\pm(y)$,
\beqa
 && [ \lambda^2  + (b_x^2+b_y^2) - 2(b_x \partial_y - i b_y k_x) + (\partial_y^2 - (k_x+qhy)^2)  ] \chi_+ = 0, \nn\\
 && [ \lambda^2  + (b_x^2+b_y^2) + 2(b_x \partial_y - i b_y k_x) + (\partial_y^2 - (k_x+qhy)^2)  ] \chi_- = 0.
\eeqa
which are typical Strum-Liouville problems: $a(y)\chi_\pm^{\prime\prime}(y) + b(y)\chi_\pm^\prime(y) + c_\pm(y)\chi_\pm(y) =0$ with $a(y)=1$, $b(y) = -2 b_x$, and $c_\pm(y) = \lambda^2 +(b_x^2+b_y^2)\textbf{1}_2 \pm 2i b_y k_x - (k_x+ qhy)^2$.
The ODE can be expressed as the familiar form $-\frac{d}{dy}[ p(x) \frac{d}{dy}\chi_\pm(y) ] + q(y)\chi_\pm(y) = \tilde\lambda w(y) \chi_\pm(y)$, with $p(y) = e^{-2b_x y}$, and $\tilde\lambda w(y) - q(y) = [\lambda^2 +(b_x^2+b_y^2)\textbf{1}_2 \pm 2i b_y k_x - (k_x+ qhy)^2] e^{-2b_x y}$.
It turns out that the wave function can be expressed as
\beqa
\chi_\pm(\bar{y}) = e^{-\frac{1}{2}(\bar{y}\mp\frac{b_x}{\sqrt{qh}})^2+\frac{1}{2}(k_x\mp b_x)^2} \bigg[ C_{\pm} H_{n_\pm}(\bar{y})
+ D_{\pm} \, {}_1F_1\big[-\frac{n_\pm}{2}, \frac{1}{2}, \bar{y}^2 \big] \bigg], \quad n_\pm \equiv \frac{\lambda^2+b_y^2 \pm 2 i k_x b_y}{2qh}-\frac{1}{2},
\eeqa
\end{widetext}
where we have introduced $\bar{y} \equiv \sqrt{qh}y + \frac{k_x}{\sqrt{qh}}$ and $c_\pm$ and $D_\pm$ are all constants. Since ${}_1F_1[-m/2,1/2,z]$ will be exponentially divergent when $m\in$ odd numbers, consequently, the physical wave functions with the quantum numbers $n_\pm\in {\mathbb Z}$ is chosen by setting $D_{\pm}=0$.

The first case that with the number $n_\pm$ being a real number can be realized by making the replacement from the beginning that $b_y \to -i b_y$, then the correspond quantum numbers become real
\beqa
n_\pm \equiv \frac{\lambda^2+b_y^2 \pm 2  k_x b_y}{2qh}-\frac{1}{2}.
\eeqa
which leads to two nondegenerate eigenvalue as
\beqa
\ii\ii\ii \lambda_{n\pm}^2 = 2qh\big(n_\pm + \frac{1}{2}\big)+k_x^2 - (k_x\pm b_y)^2, ~ n_\pm \in {\mathbb Z}. \label{Eq:lambda_n_pm}
\eeqa
The other case with a real number $n_\pm$ is realized by assuming that the chiral gauge field along the $y$ direction is absent. In that case,
\beqa
\ii\ii n_\pm  \equiv  n = \frac{\lambda^2}{2 q h} - \frac{1}{2},  \quad \Rightarrow \quad \lambda_n^2 = 2q h \big( n + \frac{1}{2} \big), \label{Eq:lambda_n-h}
\eeqa
with $n =0,1,2\ldots,  \in {\mathbb Z}$. The corresponding wave functions are
\beqa
\chi_\pm (\bar{y}) &=& e^{-\frac{1}{2}(\bar{y}\mp\frac{b_x}{\sqrt{qh}})^2+\frac{1}{2}(k_x\mp b_x)^2} \bigg[ C_{\pm} H_{n}(\bar{y}) \nn\\
&+& D_{\mp} \, {}_1F_1\big[-\frac{n}{2}, \frac{1}{2}, \bar{y}^2 \big] \bigg], \quad n \equiv \frac{\lambda^2}{2qh}-\frac{1}{2}. \nn
\eeqa
It is worth noticing that the wave function with the quantum number $n=0$ becomes
\beqa
\chi_\pm (\bar{y}) = c_{\pm} e^{-\frac{1}{2}(\bar{y}\mp\frac{b_x}{\sqrt{qh}})^2+\frac{1}{2}(k_x\mp b_x)^2} ,
\eeqa
where $\quad c_{\pm} = (C_{\pm}+D_{\pm})$, since $H_{0}(z)=1$ and ${}_1 F_1[0,1/2,z]=1$. The corresponding eigenvalue is
\beqa
\lambda_0^2 = qh, \quad \lambda_0 = \pm \sqrt{|qh|}.
\eeqa
By redefinition of the eigenvalue, one can obtain
\beqa
 \tilde\lambda_n \equiv \frac{\lambda}{\sqrt{q h}}, \quad \tilde\lambda_n^2 = 2n + 1 , \quad n \in {\mathbb Z}.
\eeqa
Alternatively, the number $n_\pm$ could be real number, in the situation that the wave function is independent of the $x$-direction, namely, $k_x=0$, we obtain the degenerate wave functions as:
\beqa
\chi_\pm(\bar{y}) &=&  C_{\pm} e^{-\frac{1}{2}\bar{y}^2} H_{n}(\bar{y})
+ D_{\pm} e^{\frac{1}{2}\bar{y}^2} H_{-n -1}(i \bar{y})  , \nn\\
 n & \equiv & \frac{\lambda^2+b_y^2}{2qh}-\frac{1}{2},
\eeqa
where $\bar{y}\equiv \sqrt{qh}y$. This leads to the eigenvalues
\beqa
\lambda_n^2 = 2qh\big(n+\frac{1}{2}\big)  - b_y^2. \label{Eq:lambda_n-h-b}
\eeqa
In the absence of chiral gauge field, i.e., $b_y=0$, the results above just recover those in ref.~\cite{Basu:2009qz}.

\subsection{Time and radial coordinates dependence}

In the following, we will try to solve the time and radial coordinate dependent part of the equation, namely, the first sector in Eq.(\ref{Eq:UD1-UD2-L}),
\begin{widetext}
\beqa
U D_{L}(r)\Psi = \frac{\sqrt{g_{xx}}}{\sqrt{g_{tt}}} U\Gamma^{\underline{t}} ( \partial_t \!-\! i q A_t \!+\! i b_0 \Gamma^{\underline{5}} )
\!+\!  \frac{\sqrt{g_{xx}}}{\sqrt{g_{rr}}} U\Gamma^{\underline{r}} (\partial_r - i b_r \Gamma^{\underline{5}}) \!-\! m_D  \sqrt{g_{xx}}U =  L \Psi, \label{Eq:DiracEOMs_D1Psi}
\eeqa
where by using Eq.(\ref{Eq:UGamma_trx}) in the representation as shown in Eq.(\ref{Eq:Gamma-rep-1}), $UD_{L}(r)$ is given as
\beqa
U D_{L}(r) = \sqrt{g_{xx}}  \left(
     \begin{array}{cc}
     - i \sigma^3 \frac{(\partial_t \!-\! i q A_t)}{\sqrt{g_{tt}}}- \sigma^1 \frac{\partial_r}{\sqrt{g_{rr}}} + i\sigma^2 m_D & \sigma^1 \frac{b_0}{\sqrt{g_{tt}}} -i \sigma^3  \frac{b_r}{\sqrt{g_{rr}}} \\
     - \sigma^1 \frac{b_0}{\sqrt{g_{tt}}} + i \sigma^3 \frac{b_r}{\sqrt{g_{rr}}}  & - i \sigma^3 \frac{(\partial_t \!-\! i q A_t)}{\sqrt{g_{tt}}} - \sigma^1  \frac{\partial_r}{\sqrt{g_{rr}}}  + i\sigma^2 m_D   \\
     \end{array}
   \right) .   \label{Eq:UD1}
\eeqa

Thus in the momentum space time as shown in Eq.(\ref{Eq:psi-Psi-Fourier}), by making the replacement $\partial_t \to -i\omega $, and expressing the bulk Dirac spinor $\Psi$ in terms of two Weyl fermions $\Psi_\pm$ through spin projection operators, as shown in Eq.(\ref{Eq:Psi}), the field equation becomes
\beqa
 \sqrt{g_{xx}}  \left(
     \begin{array}{cc}
      - \sigma^3 \frac{(\omega \!+\! q A_t)}{\sqrt{g_{tt}}}- \sigma^1 \frac{\partial_r}{\sqrt{g_{rr}}} + i\sigma^2 m_D & \sigma^1 \frac{b_0}{\sqrt{g_{tt}}} -i \sigma^3  \frac{b_r}{\sqrt{g_{rr}}} \\
     - \sigma^1 \frac{b_0}{\sqrt{g_{tt}}} + i \sigma^3 \frac{b_r}{\sqrt{g_{rr}}}  & - \sigma^3 \frac{(\omega \!+\! q A_t)}{\sqrt{g_{tt}}} - \sigma^1  \frac{\partial_r}{\sqrt{g_{rr}}}  + i\sigma^2 m_D   \\
     \end{array}
   \right) \left(
             \begin{array}{c}
               \Psi_+ \\
               \Psi_- \\
             \end{array}
           \right)   =  \lambda \left(
                  \begin{array}{cc}
                    \textbf{1}_2 & 0 \\
                    0 & \textbf{1}_2 \\
                  \end{array}
                \right)
 \left(
             \begin{array}{c}
               \Psi_+ \\
               \Psi_- \\
             \end{array}
           \right). \quad \nn
\eeqa
We have assumed that both $\Psi_+$ and $\Psi_-$ correspond to the eigenvalue $\lambda \textbf{1}_4$, alternatively, one can also assume the eigenvalue to be $-\lambda \textbf{1}_4$. This ambiguity is due to the energy degeneracy existing in the square of the operator $L^2\sim \lambda^2$ when solving spatial sector of the bulk Dirac equation. This reflects the fact that there are other two independent solutions $\Psi_{-\lambda}=(\Psi_{-\lambda,+},\Psi_{-\lambda,-})^T$, which correspond to the eigenvalue $-\lambda \textbf{1}_4$. In the following, we introduce the following notation to stand for the wavefunctions with eigenvalue $\pm \lambda$, respectively, as
\beqa
\Psi_{\pm \lambda,\beta} \equiv \Psi_{\alpha,\beta}, \label{Eq:Psi_alpha_beta}
\eeqa
where $\alpha=1,2$ and $\beta=\pm$. e.g., $\Psi_{+\lambda,\pm} \equiv \Psi_{1,\pm}$, and $\Psi_{-\lambda,\pm} \equiv \Psi_{2,\pm}$. Therefore, we have the equations of motion for the bulk Dirac fermion in $(3+1)$-dimensions as
\beqa
\sqrt{g_{xx}}  \left(
     \begin{array}{cc}
     -  \sigma^3 \frac{(\omega \!+\!  q A_t)}{\sqrt{g_{tt}}}- \sigma^1 \frac{\partial_r}{\sqrt{g_{rr}}} + i\sigma^2 m_D - \frac{\lambda}{\sqrt{g_{xx}}} \textbf{1}_2 & \sigma^1 \frac{b_0}{\sqrt{g_{tt}}} -i \sigma^3  \frac{b_r}{\sqrt{g_{rr}}} \\
     - \sigma^1 \frac{b_0}{\sqrt{g_{tt}}} + i \sigma^3 \frac{b_r}{\sqrt{g_{rr}}}  & - \sigma^3 \frac{(\omega \!+\!  q A_t)}{\sqrt{g_{tt}}} - \sigma^1  \frac{\partial_r}{\sqrt{g_{rr}}}  + i\sigma^2 m_D - \frac{\lambda}{\sqrt{g_{xx}}}  \textbf{1}_2   \\
     \end{array}
   \right) \left(
             \begin{array}{c}
               \Psi_+ \\
               \Psi_- \\
             \end{array}
           \right) = 0. \nn
\eeqa
According to discussion in Eqs.(\ref{Eq:Phipm}) and (\ref{Eq:Psi}), one has $\left(\Psi_+,\Psi_- \right)^T = \Phi_+ + \Phi_-$, where $\Phi_\beta \equiv P_\beta \Psi$ with $\beta=\pm$.

\subsubsection{The case without chiral gauge fields $b_0,b_r$}

At this step, it is worth noticing that the chiral gauge fields $b_0,b_r$ are present on the off-diagonal block of the matrix, which mixes $\Psi_+$ and $\Psi_-$. Consequently, if the chiral gauge fields are absent, i.e., $b_0=b_r=0$, then the field equations of motion can be decoupled,
\beqa
\sqrt{g_{xx}}  \left(
     \begin{array}{cc}
     -  \sigma^3 \frac{(\omega \!+\!  q A_t)}{\sqrt{g_{tt}}}- \sigma^1 \frac{\partial_r}{\sqrt{g_{rr}}} + i\sigma^2 m_D - \frac{\lambda}{\sqrt{g_{xx}}} \textbf{1}_2 & 0 \\
     0  & - \sigma^3 \frac{(\omega \!+\!  q A_t)}{\sqrt{g_{tt}}} - \sigma^1  \frac{\partial_r}{\sqrt{g_{rr}}}  + i\sigma^2 m_D - \frac{\lambda}{\sqrt{g_{xx}}}  \textbf{1}_2   \\
     \end{array}
   \right) \left(
             \begin{array}{c}
               \Psi_+ \\
               \Psi_- \\
             \end{array}
           \right) = 0 , \nn
\eeqa
which implies that $\Psi_{\pm}$ satisfy the same equations of motion,
\beqa
\bigg( -  \sigma^3 \frac{(\omega \!+\!  q A_t)}{\sqrt{g_{tt}}}- \sigma^1 \frac{\partial_r}{\sqrt{g_{rr}}} + i\sigma^2 m_D - \frac{\lambda}{\sqrt{g_{xx}}} \textbf{1}_2 \bigg)\Psi_{1,\pm} =0,
\eeqa
where $\Psi_{1,\pm}\equiv \Psi_{+\lambda,\pm}$, i.e., we have introduced the subscript $1$ indicating the corresponding eigenvalue $+\lambda$. It is obvious that other independent wave functions satisfying the equations of motion as,
\beqa
\bigg( -  \sigma^3 \frac{(\omega \!+\!  q A_t)}{\sqrt{g_{tt}}}- \sigma^1 \frac{\partial_r}{\sqrt{g_{rr}}} + i\sigma^2 m_D + \frac{\lambda}{\sqrt{g_{xx}}} \textbf{1}_2 \bigg)\Psi_{2,\pm} =0,
\eeqa
where $\Psi_{2,\pm}\equiv \Psi_{-\lambda,\pm}$, and the subscript $2$ indicating the corresponding eigenvalue $-\lambda$. Since $\Psi_{1,\pm}$ satisfy the same equations of motion, and so do $\Psi_{2,\pm}$. Without loss of generality, one can select $\Psi_{1,+}$ and $\Psi_{2,-}$ (Equivalently, one may select $\Psi_{1,-}$ and $\Psi_{2,-}$, or $\Psi_{1,+}$ and $\Psi_{2,+}$, or $\Psi_{1,-}$ and $\Psi_{2,-}$) as two independent wave functions and denote them as $\Phi_+ \equiv P_+ \Psi $ and $\Phi_2 \equiv P_- \Psi$.  Among the four, only two of them are independent, e.g., $\Phi_+ \equiv (\Psi_{1,+},0)^T \sim (\Psi_{2,+},0)^T$ and $\Phi_- \equiv (0,\Psi_{1,-})^T\sim (0,\Psi_{2,-})^T$.

In fact, since it is obvious that the second subscript of the wavefunctions, i.e., $\beta=\pm$, which indicate the eigenstates of spin projection, do not affect the EOMs, thus the subscript $\beta=\pm$ can be dropped. In this case, one keeps only the subscript $\alpha=1,2$ to indicate wavefunctions with the eigenvalue $\pm\lambda$, respectively, i.e., $\Psi_{\alpha,\beta} \equiv \Psi_{\alpha}$ with $\alpha=1,2$, respectively.

By using the notation introduced in Eq.(\ref{Eq:Psi}), the two independent Dirac equations in the absence of chiral gauge field can be expressed in a more tight form as
\beqa
\bigg( -  \sigma^3 \frac{(\omega \!+\!  q A_t)}{\sqrt{g_{tt}}}- \sigma^1 \frac{\partial_r}{\sqrt{g_{rr}}} + i\sigma^2 m_D + (-1)^\alpha \frac{\lambda}{\sqrt{g_{xx}}} \textbf{1}_2 \bigg)\Psi_\alpha = 0, \quad \alpha=1,2. \label{Eq:Marster_Dirac-EOMs}
\eeqa
For the case with only $x$ direction dependence, $\lambda=k_x \equiv k$, by multiplying a $-\sigma^1$ on the left hand side of the equations of motion, one has
\beqa
\bigg( -  i\sigma^2 \frac{(\omega \!+\!  q A_t)}{\sqrt{g_{tt}}} + \textbf{1}_2 \frac{\partial_r}{\sqrt{g_{rr}}} + \sigma^3 m_D - \sigma^1 (-1)^\alpha \frac{k_x}{\sqrt{g_{xx}}}  \bigg) \Phi_\alpha = 0, \quad \alpha=1,2. \label{Eq:Marster_Dirac-EOMs-Phi_alpha}
\eeqa
where $\Phi_+ \equiv (\Psi_{+},0)^T$ and $\Phi_- \equiv (0,\Psi_{-})^T$. These equations reduce to the bulk Dirac equation without chiral gauge field, e.g., Eq.(A14) in ref.~\cite{Faulkner:2009wj}.

\subsubsection{The case with chiral gauge fields $b_0,b_r$}

In the following, we will consider the most general case with chiral gauge fields $b_{0,r}$. By using the similar transformation as shown in Eq.(\ref{Eq:transU1_tPsi-Psi}), $U_1$, which also acts on the radial and time coordinate sector, namely, $r,t$,
\beqa
\left(
  \begin{array}{c}
    \tilde\Psi_+ \\
    \tilde\Psi_- \\
  \end{array}
\right)
= U_1 \left(
      \begin{array}{c}
        \Psi_+ \\
        \Psi_- \\
      \end{array}
    \right),  \label{Eq:tPsi_alpha-Psi_alpha}
\eeqa
and by using Eq.(\ref{Eq:MAMinv}), one obtains
\beqa
\left(
             \begin{array}{cc}
              -  \sigma^3 \big( \frac{(\omega \!+\!  q A_t)}{\sqrt{g_{tt}}} \!-\! \frac{b_r}{\sqrt{g_{rr}}} \big)  \!-\! \sigma^1 \big( \frac{\partial_r}{\sqrt{g_{rr}}} \!-\! i \frac{b_0}{\sqrt{g_{tt}}} \big) \!+\! i\sigma^2 m_D \!-\! \frac{\lambda}{\sqrt{g_{xx}}} \textbf{1}_2 \ii\ii\ii\ii\ii & \ii\ii\ii\ii\ii 0 \\
                0 \ii\ii\ii\ii\ii & \ii\ii\ii\ii\ii -  \sigma^3 \big( \frac{(\omega \!+\!  q A_t)}{\sqrt{g_{tt}}} \!+\! \frac{b_r}{\sqrt{g_{rr}}} \big) \!-\! \sigma^1 \big( \frac{\partial_r}{\sqrt{g_{rr}}} \!+\! i \frac{b_0}{\sqrt{g_{tt}}} \big) \!+\! i\sigma^2 m_D  \!-\! \frac{\lambda}{\sqrt{g_{xx}}} \textbf{1}_2 \\
             \end{array}
           \right) \left(
             \begin{array}{c}
               \tilde\Psi_+ \\
               \tilde\Psi_- \\
             \end{array}
           \right) = 0. \nn
\eeqa
It is obvious that associated with the eigenvalue transformation $\lambda \textbf{1}_4 \to -\lambda \textbf{1}_4$, there are another two independent wave functions.

One may relabel them as $\Psi_{1,\beta}$ and $\Psi_{2,\beta}$, respectively, so that the equations of motion can be expressed as an unified one as,
\beqa
\bigg[ -  \sigma^3 \bigg( \frac{(\omega \!+\!  q A_t)}{\sqrt{g_{tt}}} \!-\! \beta \frac{b_r}{\sqrt{g_{rr}}} \bigg) \!-\! \sigma^1 \bigg( \frac{\partial_r}{\sqrt{g_{rr}}} \!-\! i \beta  \frac{b_0}{\sqrt{g_{tt}}} \bigg) \!+\! i\sigma^2 m_D \!+\! (-1)^\alpha \frac{\lambda}{\sqrt{g_{xx}}} \textbf{1}_2 \bigg] \tilde\Psi_{\alpha,\beta} = 0 ,  \label{Eq:Phi-Marster_Dirac-EOMs_b0_br}
\label{Eq:Marster_Dirac-EOMs_b0_br}
\eeqa
where $\tilde\Psi_{1,\beta}\equiv \tilde\Psi_{+\lambda,\beta}$, $\tilde\Psi_{2,\beta}\equiv \tilde\Psi_{-\lambda,\beta}$ with $\beta=\pm$. Therefore in the presence of chiral gauge fields, or, equivalently, in the presence of chiral anomaly, there are four sets of independent EOMs for bulk Dirac fermion wavefunctions, namely $\tilde\Psi_{\alpha,\beta}\equiv \{\tilde\Psi_{1,+},\tilde\Psi_{1,-},\tilde\Psi_{2,+},\tilde\Psi_{2,-} \}$, instead of two sets, namely $\tilde\Psi_{\alpha}\equiv \{\tilde\Psi_{1},\tilde\Psi_{2}\}$, of independent EOMs for bulk Dirac fermion wavefunctions when the chiral gauge fields $b_0,b_r$ are absent. In other words, the degrees of freedom (d.o.f) of the EOMs are doubled, compared to the case without chiral gauge fields. It is obvious that half of Dirac wave functions will be degenerate when the chiral gauge field are turned off, as shown in Eq.(\ref{Eq:Marster_Dirac-EOMs}) by setting $\beta=0$. As a generalization of Eq.(\ref{Eq:Marster_Dirac-EOMs-Phi_alpha}), one obtains
\beqa
 \bigg[ -  \sigma^3 \bigg( \frac{(\omega \!+\!  q A_t)}{\sqrt{g_{tt}}} \!-\! \beta \frac{b_r}{\sqrt{g_{rr}}} \bigg) \!-\! \sigma^1 \bigg( \frac{\partial_r}{\sqrt{g_{rr}}} \!-\! i \beta  \frac{b_0}{\sqrt{g_{tt}}} \bigg) \!+\! i\sigma^2 m_D \!+\! (-1)^\alpha \frac{\lambda}{\sqrt{g_{xx}}} \textbf{1}_2 \bigg] \tilde\Phi_{\alpha,\beta} = 0 ,  \label{Eq:Marster_Dirac-EOMs_b0_br-Phi_alpha_beta}
\eeqa
where $\tilde\Phi_{\alpha,\beta}\equiv P_{\beta} \tilde\Psi_\alpha$ with $\beta = \pm$, i.e., $\tilde\Phi_{\alpha,+} \equiv (\tilde\Psi_{\alpha,+},0)^T$ and $\tilde\Phi_{\alpha,-} \equiv (0,\tilde\Psi_{\alpha,-})^T$ with $\alpha=1,2$. The EOMs are general in the sense that they include several special cases with interesting physical consequence. For example,
\begin{itemize}
  \item[(a)]  Weyl fermions with same helicity and same energy eigenvalues: Same spin projection eigenstates with same energy eigenvalues, $\alpha = \alpha^\prime$, $\beta = \beta^\prime$. e.g. $\alpha  = \alpha^\prime = 1$, $\beta = \beta^\prime =+$:
\beqa
\tilde\Phi_{1,2} \equiv  \tilde\Phi_{1,+} \equiv P_+ \tilde\Psi_1 : \tilde\Phi_1 \equiv (\tilde\Psi_{1,+},\textbf{0})^T ; \quad  \tilde\Phi_2 \equiv (\tilde\Psi_{1,+},\textbf{0})^T.
\eeqa
  \item[(b)]  Weyl fermions with same helicity but inverse energy eigenvalues: Same spin projection eigenstates with different energy eigenvalues, $\alpha \ne \alpha^\prime$, $\beta = \beta^\prime$. e.g. $\alpha = 1,2$, $\beta=+$:
\beqa
\tilde\Phi_\alpha \equiv \tilde\Phi_{\alpha,+}  =  P_+ \tilde\Psi_\alpha : \tilde\Phi_1 \equiv (\tilde\Psi_{1,+},\textbf{0})^T ; \quad  \tilde\Phi_2 \equiv (\tilde\Psi_{2,+},\textbf{0})^T.
\eeqa
  \item[(c)] Weyl fermions with different helicity but same energy eigenvalues: Same energy eigenstates with different spin projection eigenvalues, $\alpha = \alpha^\prime$, $\beta\ne \beta^\prime$. e.g., $\alpha=1$, $\beta=\pm$,
\beqa
 \tilde\Phi_{\beta} \equiv \tilde\Phi_{1,\beta}  =  P_\beta \tilde\Psi_1 :   \tilde\Phi_1 \equiv (\tilde\Psi_{1,+},\textbf{0})^T , \quad \tilde\Phi_2 \equiv (\textbf{0},\tilde\Psi_{1,-})^T.
\eeqa
  \item[(d)] Weyl fermions with different helicity and inverse energy eigenvalues: Different energy eigenstates with different spin projection eigenvalues, $\alpha\ne \alpha^\prime$, $\beta\ne \beta^\prime$. e.g., $\alpha= 1, \beta = +; \alpha^\prime= 2, \beta^\prime = - $
  \beqa
   \tilde\Phi_1 \equiv \tilde\Phi_{1,+} \equiv  (\tilde\Psi_{1,+}, \textbf{0})^T, \quad  \tilde\Phi_2 \equiv \tilde\Phi_{2,-} \equiv ( \textbf{0}, \tilde\Psi_{2,-})^T  .
  \eeqa
\end{itemize}

According to Eq.(\ref{Eq:tPsi-chi_12-y}), we have
\beqa
\tilde\Phi_{\alpha,\beta} = \tilde{R}_{\alpha,\beta}\otimes \left(\begin{array}{c}
                                                                                             \chi_1(y) \\
                                                                                             \chi_2(y)  \\
                                                                                           \end{array}
                                                                                         \right),
\eeqa
where $\chi_{1,2}(y)$ are related to $\chi_\pm(y)$ through Eq.(\ref{Eq:chi_12-chi_pm}).

By multiplying $-\sigma^1$ on the left hand side of Eq.(\ref{Eq:Marster_Dirac-EOMs_b0_br-Phi_alpha_beta}), one has
\beqa
\bigg[ -  i\sigma^2 \bigg( \frac{(\omega \!+\!  q A_t)}{\sqrt{g_{tt}}} \!-\! \beta \frac{b_r}{\sqrt{g_{rr}}} \bigg) \!+\! \textbf{1}_2 \bigg( \frac{\partial_r}{\sqrt{g_{rr}}} \!-\! i \beta  \frac{b_0}{\sqrt{g_{tt}}} \bigg) \!+\! \sigma^3 m_D \!-\!  (-1)^\alpha \frac{\lambda}{\sqrt{g_{xx}}} \sigma^1 \bigg] \tilde{R}_{\alpha,\beta} = 0 . \label{Eq:Dirac-EOMs-R1-R2}
\eeqa
In the following section, we will solve the above euqations.

\subsubsection{Dirac equation}

The above equations can also be re-expressed as
\beqa
\bigg[  \textbf{1}_2 \bigg( \partial_r \!-\! i \beta  \sqrt\frac{g_{rr}}{g_{tt}} b_0\bigg) \!+\! \sigma^3 m_D \sqrt{g_{rr}} \bigg] \tilde\Phi_{\alpha,\beta} = \bigg[ i\sigma^2 \bigg( \sqrt\frac{g_{rr}}{g_{tt}}(\omega \!+\!  q A_t) \!-\! \beta b_r  \bigg) \!+\!  (-1)^\alpha \sigma^1 \sqrt\frac{g_{rr}}{g_{xx}} \lambda\bigg] \tilde\Phi_{\alpha,\beta}, \label{Eq:Dirac-Psi_alpha-beta}
\eeqa
Or, in the $\zeta$ coordinate defined in Eq.(\ref{Eq:zeta-eta-k=0}), the EOM become
\beqa
\bigg[  \textbf{1}_2 \bigg( \partial_\zeta \!+\! i \beta  \sqrt\frac{g_{\zeta\zeta}}{g_{tt}} b_0\bigg) \!-\! \sigma^3 m_D \sqrt{g_{\zeta\zeta}} \bigg] \tilde\Phi_{\alpha,\beta} = -\bigg[ i\sigma^2 \bigg( \sqrt\frac{g_{\zeta\zeta}}{g_{tt}}(\omega \!+\!  q A_t) \!+\! \beta b_\zeta  \bigg) \!+\!  (-1)^\alpha \sigma^1 \sqrt\frac{g_{\zeta\zeta}}{g_{xx}} \lambda\bigg] \tilde\Phi_{\alpha,\beta}, \label{Eq:Dirac-Psi_alpha-beta_zeta}
\eeqa
where we have introduced $b_\zeta\equiv -b_r$ as shown in Eq.(\ref{Eq:fg_zeta_2nd}) and Eq.(\ref{Eq:gamma-1234-zeta}). In the absence of chiral gauge fields $b_{0,r}$, i.e., $b_\mu=0$, $b_r=0$, and also $b_i=0$, then $\lambda=|\vec{k}|$, Eqs.(\ref{Eq:Dirac-Psi_alpha-beta}) and (\ref{Eq:Dirac-Psi_alpha-beta_zeta}) reduce to
\beqa
(\partial_r \!+\! m_D\sqrt{g_{rr}}\sigma^3 )\tilde\Psi_\alpha  \!=\! \sqrt\frac{g_{rr}}{g_{xx}}[ i\sigma^2 K_0(r) \!+\! (-1)^\alpha \sigma^1 k ]\tilde\Psi_\alpha, ~ (\partial_\zeta \!-\! m_D\sqrt{g_{\zeta\zeta}}\sigma^3 )\tilde\Psi_\alpha  \!=\! \sqrt\frac{g_{\zeta\zeta}}{g_{xx}}[ -i\sigma^2 K_0(\zeta) \!-\! (-1)^\alpha k \sigma^1 ]\tilde\Psi_\alpha. \nn
\eeqa
In the presence of chiral gauge field $b_{0,r}$, the Dirac equation in Eq.(\ref{Eq:Dirac-Psi_alpha-beta}) can be expressed more explicitly as
\beqa
\left(\begin{array}{cc}
A(-m_D) \!-\! i \beta B_0(r) & -  [K_0(r)  - \beta B_r(r) + (-1)^\alpha \lambda ]   \\
K_0(r) - \beta B_r(r) - (-1)^\alpha \lambda    &  A(m_D)\!-\! i \beta B_0(r)
\end{array}
\right)  \tilde\Phi_{\alpha,\beta} =0, \label{Eq:Dirac-Psi_alpha-beta_M1}
\eeqa
where we have introduced the differential operators defined as
\beqa
A(m_D) \equiv \sqrt\frac{g_{xx}}{g_{rr}}(\partial_r - m_D\sqrt{g_{rr}} ),  \label{Eq:AmD}
\eeqa
and
\beqa
K_0(r) \equiv \sqrt\frac{g_{xx}}{g_{tt}}(\omega \!+\!  q A_t), \quad B_0(r) \equiv \sqrt\frac{g_{xx}}{g_{tt}}b_0, \quad B_r(r) \equiv \sqrt\frac{g_{xx}}{g_{rr}} b_r. \label{Eq:K0-B0-Br}
\eeqa
According to Eq.(\ref{Eq:Rt_alpha}), one obtains
\beqa
&&[A(-m_D) - i \beta B_0(r) ] \tilde{F}_{\alpha,\beta} = [K_0(r) - \beta B_r(r)+(-1)^\alpha\lambda] \tilde{G}_{\alpha,\beta}, \nn\\
&&[A(+m_D) - i \beta B_0(r)] \tilde{G}_{\alpha,\beta} = -[K_0(r) - \beta B_r(r)-(-1)^\alpha\lambda] \tilde{F}_{\alpha,\beta}, \label{Eq:Dirac-EOMs-Psi_alpha-beta-1st}
\eeqa
for $\alpha=1,2$ and $\beta=\pm$, respectively. For the case (a), the EOMs can be simplified in the absence of chiral gauge field~\cite{Iqbal:2009fd}, as
\beqa
A(-m_D) \tilde{F}_{\alpha} = [K_0(r)+ (-1)^\alpha k] \tilde{G}_{\alpha}, \quad A(m_D) \tilde{G}_{\alpha} = -[K_0(r)- (-1)^\alpha k] \tilde{F}_{\alpha}.  \label{Eq:Dirac-EOMs-Psi_alpha-1st}
\eeqa
As it is discussed before that, the EOM for $\tilde\Phi_\beta\sim \tilde{R}_\beta =(\tilde{F}_\beta, \tilde{G}_\beta)^T$ will become $\beta$ independent, and they just reduce to $\tilde\Psi_\alpha\sim \tilde{R}_\alpha=(\tilde{F}_\alpha, \tilde{G}_\alpha)^T$ with $\alpha=1$.
By combining the two coupled equations, one obtains the equivalent $2$-nd PDEs as
\beqa
&& \bigg(A(+m_D)A(-m_D) - i \beta \sqrt\frac{g_{xx}}{g_{rr}} [2B_0(r)\partial_r + B_0^\prime(r) ] - B_0(r)^2  + [K_0 - \beta B_r(r)]^2-\lambda^2\bigg) \tilde{F}_{\alpha,\beta} \nn\\
&& = \sqrt\frac{g_{xx}}{g_{rr}}[K_0(r) - \beta B_r(r)]^\prime \tilde{G}_{\alpha,\beta} = \sqrt\frac{g_{xx}}{g_{rr}}\ln^\prime[K_0(r) - \beta B_r(r)+(-1)^\alpha \lambda ][A(-m_D) - i \beta B_0] \tilde{F}_{\alpha,\beta} , \nn \\
&& \bigg(A(-m_D)A(+m_D) - i \beta \sqrt\frac{g_{xx}}{g_{rr}} [2B_0(r)\partial_r + B_0^\prime(r) ] - B_0(r)^2  + [K_0 - \beta B_r(r)]^2-\lambda^2\bigg) \tilde{G}_{\alpha,\beta} \nn\\
&&= - \sqrt\frac{g_{xx}}{g_{rr}}[K_0(r) - \beta B_r(r)]^\prime \tilde{F}_{\alpha,\beta} =\sqrt\frac{g_{xx}}{g_{rr}}\ln^\prime[K_0(r) - \beta B_r(r)-(-1)^\alpha \lambda] [A(+m_D) - i \beta B_0] \tilde{G}_{\alpha,\beta} ,  \qquad \label{Eq:Dirac-EOMs-Psi_alpha_beta}
\eeqa
where $K_0(r)$, $B_0(r)$ and $B_r(r)$ depend on the radial coordinate $r$, as shown in Eq.(\ref{Eq:K0-B0-Br}). Assuming that $B_r$ is absorbed into the redefinition of wavefunction $\tilde\Psi$, e.g., a phase factor which depends on the radial coordinate $r$, then one has
\beqa
&& \bigg(A(+m_D)A(-m_D) - i \beta \sqrt\frac{g_{xx}}{g_{rr}} [2B_0(r)\partial_r + B_0^\prime(r) ] - B_0(r)^2  + K_0^2-\lambda^2\bigg) \tilde{F}_{\alpha,\beta} \nn\\
&& = \sqrt\frac{g_{xx}}{g_{rr}}K_0(r)^\prime \tilde{G}_{\alpha,\beta} = \sqrt\frac{g_{xx}}{g_{rr}}\ln^\prime[K_0(r)+(-1)^\alpha \lambda ][A(-m_D) - i \beta B_0] \tilde{F}_{\alpha,\beta} , \nn \\
&& \bigg(A(-m_D)A(+m_D) - i \beta \sqrt\frac{g_{xx}}{g_{rr}} [2B_0(r)\partial_r + B_0^\prime(r) ] - B_0(r)^2  + K_0^2-\lambda^2\bigg) \tilde{G}_{\alpha,\beta} \nn\\
&&= - \sqrt\frac{g_{xx}}{g_{rr}}K_0(r)^\prime \tilde{F}_{\alpha,\beta} =\sqrt\frac{g_{xx}}{g_{rr}}\ln^\prime[K_0(r)-(-1)^\alpha \lambda] [A(+m_D) - i \beta B_0] \tilde{G}_{\alpha,\beta} .  \qquad \label{Eq:Dirac-EOMs-Psi_alpha_beta-Br=0}
\eeqa
At the infinite boundary $r\to\infty$, $K_0$, $B_0$ are independent of radial coordinate. Thus, the right-hand side of the EOMs vanishes. Then the EOMs can also be transformed into two $2$-nd order PDEs as
\beqa
&& \bigg( A(+m_D)A(-m_D) - i \beta\sqrt\frac{g_{xx}}{g_{rr}} 2B_0 \partial_r  - B_0^2  + K_0^2-\lambda^2\bigg) \tilde{F}_\beta =0, \nn \\
&& \bigg( A(-m_D)A(+m_D) - i \beta\sqrt\frac{g_{xx}}{g_{rr}} 2B_0 \partial_r  - B_0^2  + K_0^2-\lambda^2\bigg) \tilde{G}_\beta =0,
\eeqa
where we have neglected the subscript $\alpha$ since the EOMs are independent of $\alpha$.

In the absence of chiral gauge fields, i.e., $B_0(r)=B_r(r)=0$ and $\lambda=|\vec{k}|\equiv k$, the equations of motion reduce to
\beqa
&& \bigg( A(+m_D)A(-m_D) - \sqrt\frac{g_{xx}}{g_{rr}} \ln^\prime[K_0(r) + (-1)^\alpha k] A(-m_D) + K_0(r)^2-k^2 \bigg) \tilde{F}_{\alpha} = 0 \nn, \\
&& \bigg( A(-m_D)A(+m_D) - \sqrt\frac{g_{xx}}{g_{rr}} \ln^\prime[K_0(r) - (-1)^\alpha k] A(+m_D) + K_0(r)^2-k^2 \bigg) \tilde{G}_{\alpha} = 0. \label{Eq:Dirac-EOMs-Psi} \qquad
\eeqa
At the infinite boundary $r\to\infty$, $K_0$ is independent of radial coordinate, then the EOM can also be transformed into two $2$-nd order PDEs as
\beqa
[A(m_D)A(-m_D) + (K_0^2-k^2)] \tilde{F}_{\alpha} = 0, \quad [A(-m_D)A(m_D) + (K_0^2-k^2)] \tilde{G}_{\alpha} = 0, \label{Eq:Dirac-EOMs-Psi-infty}
\eeqa
which just are those equations in ref.~\cite{Iqbal:2009fd} in the case without the chiral gauge field.

\subsubsection{Flow equation}

From the two sets of coupled $1$-st order ODEs in Eq.(\ref{Eq:Dirac-EOMs-Psi_alpha-beta-1st}), which can also be expressed more explicitly, according to Eq.(\ref{Eq:Dirac-Psi_alpha-beta_M1}), as
\beqa
\left(\begin{array}{cc}
\partial_r \!-\! i \beta \sqrt\frac{g_{rr}}{g_{xx}} B_0(r) + m_D \sqrt{g_{rr}} & - \sqrt\frac{g_{rr}}{g_{xx}} [K_0(r)  - \beta B_r(r) + (-1)^\alpha \lambda ]   \\
\sqrt\frac{g_{rr}}{g_{xx}}[K_0(r) - \beta B_r(r) - (-1)^\alpha \lambda ]   &  \partial_r \!-\! i \beta \sqrt\frac{g_{rr}}{g_{xx}}B_0(r) - m_D \sqrt{g_{rr}}
\end{array}
\right)  \left(\begin{array}{c}
                                                                                            \tilde{F}_{\alpha,\beta}\\
                                                                                            \tilde{G}_{\alpha,\beta}  \\
                                                                                           \end{array}
                                                                                         \right) =0, \label{Eq:Dirac-Psi_alpha-beta_M2}
\eeqa
one can obtain the flow equation as
\beqa
\bigg( \frac{\tilde{F}_{\alpha,\beta}}{\tilde{G}_{\alpha^\prime,\beta^\prime}}\bigg)^\prime &=&
\sqrt\frac{g_{rr}}{g_{xx}}\bigg( [K_0(r) - \beta B_r(r)+(-1)^\alpha\lambda] \big( \frac{\tilde{F}_{\alpha,\beta}}{\tilde{G}_{\alpha,\beta}} \big)^{-1} \! + \! [K_0(r) -\beta^\prime B_r(r)-(-1)^{\alpha^\prime}\lambda] \frac{\tilde{F}_{\alpha^\prime,\beta^\prime}}{\tilde{G}_{\alpha^\prime,\beta^\prime}}  \nn\\
&+&  i  (\beta - \beta^\prime ) B_0(r) - 2m_D\sqrt{g_{xx}}\bigg) \frac{\tilde{F}_{\alpha,\beta}}{\tilde{G}_{\alpha^\prime,\beta^\prime}}, \label{Eq:flow-xi_beta-alpha_betap-alphap}
\eeqa
If $B_r(r)$ is absorbed to the phase of the wave function, through its redefinition in Eq.(\ref{Eq:psi-Psi-x}), and by defining
\beqa
\xi_\alpha \equiv  \frac{\tilde{F}_{\alpha}}{\tilde{G}_{\alpha}} \equiv \frac{\tilde{F}_{\alpha,\beta}}{\tilde{G}_{\alpha,\beta}} \equiv \frac{\tilde{F}_{\alpha,\beta^\prime}}{\tilde{G}_{\alpha,\beta^\prime}}, \quad \xi_{\alpha\alpha^\prime} \equiv \frac{\tilde{F}_{\alpha}}{\tilde{G}_{\alpha^\prime}}  \equiv \frac{\tilde{F}_{\alpha,\beta}}{\tilde{G}_{\alpha^\prime,\beta}} \equiv \frac{\tilde{F}_{\alpha,\beta^\prime}}{\tilde{G}_{\alpha^\prime,\beta^\prime}}, \quad \xi_{\alpha,\beta\beta^\prime} \equiv \frac{\tilde{F}_{\alpha,\beta}}{\tilde{G}_{\alpha,\beta^\prime}}, \quad \xi_{\alpha\alpha^\prime,\beta\beta^\prime} \equiv \frac{\tilde{F}_{\alpha,\beta}}{\tilde{G}_{\alpha^\prime,\beta^\prime}}, \label{Eq:xi-F/G}
\eeqa
we have three special cases as
\begin{itemize}
  \item[(a)] $\alpha = \alpha^\prime$, $\beta = \beta^\prime$ :
  \beqa
&& \xi_\alpha^\prime = \sqrt\frac{g_{rr}}{g_{xx}}\bigg( [K_0(r) + (-1)^\alpha\lambda]  \! + \! [K_0(r) -(-1)^{\alpha}\lambda] \xi_\alpha^2 - 2m_D\sqrt{g_{xx}} \xi_\alpha\bigg) , \label{Eq:xi_alpha}
\eeqa
  \item[(b)]  $\alpha \ne \alpha^\prime$, $\beta=\beta^\prime$:
  \beqa
\xi_{\alpha\alpha^\prime}^\prime = \sqrt\frac{g_{rr}}{g_{xx}}\bigg( [K_0(r)+(-1)^\alpha\lambda] \xi_\alpha^{-1} \! + \! [K_0(r)-(-1)^{\alpha^\prime}\lambda] \xi_{\alpha^\prime} - 2m_D\sqrt{g_{xx}}\bigg) \xi_{\alpha\alpha^\prime}, \label{Eq:xi_alpha-alphap}
\eeqa
  \item[(c)] $\alpha = \alpha^\prime$, $\beta \ne \beta^\prime$:
  \beqa
  \xi_{\alpha,\beta\beta^\prime}^\prime =
\sqrt\frac{g_{rr}}{g_{xx}}\bigg( [K_0(r)+(-1)^\alpha\lambda] \xi_{\alpha}^{-1} \! + \! [K_0(r) -(-1)^{\alpha}\lambda] \xi_{\alpha} - 2m_D\sqrt{g_{xx}} +  i  (\beta - \beta^\prime ) B_0(r) \bigg) \xi_{\alpha,\beta\beta^\prime}, \label{Eq:xi_alpha_beta-betap}
  \eeqa
Note that $\xi_{\alpha,\beta\beta}=\xi_\alpha$ in case $(a)$.
    \item[(d)] $\alpha \ne \alpha^\prime$, $\beta \ne \beta^\prime$:
  \beqa
\xi_{\alpha\alpha^\prime,\beta\beta^\prime}^\prime &=&
\sqrt\frac{g_{rr}}{g_{xx}}\bigg( [K_0(r)+(-1)^\alpha\lambda] \xi_\alpha^{-1} \! + \! [K_0(r) -(-1)^{\alpha^\prime}\lambda] \xi_{\alpha^\prime} - 2m_D\sqrt{g_{xx}} +  i  (\beta - \beta^\prime ) B_0(r) \bigg) \xi_{\alpha\alpha^\prime,\beta\beta^\prime}. \label{Eq:xi_alpha-alphap_beta-betap}
\eeqa
Note that $\xi_{\alpha\alpha^\prime,\beta\beta}=\xi_{\alpha\alpha^\prime}$ in case $(b)$, $\xi_{\alpha\alpha,\beta\beta^\prime}=\xi_{\alpha,\beta\beta^\prime}$ in case $(c)$.
\end{itemize}
In the following, we will firstly solve flow equations in case $(a)$ to obtain $\xi_\alpha$. Secondly, by substituting $\xi_\alpha$ back into the flow equations in cases $(b)$,$(c)$ and $(d)$, then $\xi_{\alpha\alpha^\prime}$, $\xi_{\alpha,\beta\beta^\prime}$, and $\xi_{\alpha\alpha^\prime,\beta\beta^\prime}$ can be solved.

To be more explicit,
\beqa
(a)&& \xi_\alpha^\prime = \sqrt\frac{g_{rr}}{g_{xx}}\bigg( [K_0(r) + (-1)^\alpha\lambda]  \! + \! [K_0(r) -(-1)^{\alpha}\lambda] \xi_\alpha^2 - 2m_D\sqrt{g_{xx}} \xi_\alpha\bigg) , \nn \\
(b)&& \xi_{12}^\prime = \sqrt\frac{g_{rr}}{g_{xx}}\bigg( [K_0(r) - \lambda] (\xi_{1}^{-1} \! + \! \xi_{2}) - 2m_D\sqrt{g_{xx}}\bigg) \xi_{12}, \nn \\
&& \xi_{21}^\prime = \sqrt\frac{g_{rr}}{g_{xx}}\bigg( [K_0(r)+\lambda] (\xi_{2}^{-1} \! + \! \xi_{1}) - 2m_D\sqrt{g_{xx}}\bigg) \xi_{21}, \nn \\
(c)&& \xi_{\alpha,+-}^\prime =
\sqrt\frac{g_{rr}}{g_{xx}}\bigg( [K_0(r)+(-1)^\alpha\lambda] \xi_{\alpha}^{-1} \! + \! [K_0(r) -(-1)^{\alpha}\lambda] \xi_{\alpha} - 2m_D\sqrt{g_{xx}} +  2 i B_0(r) \bigg) \xi_{\alpha,+-}, \nn \\
&& \xi_{\alpha,-+}^\prime =
\sqrt\frac{g_{rr}}{g_{xx}}\bigg( [K_0(r)+(-1)^\alpha\lambda] \xi_{\alpha}^{-1} \! + \! [K_0(r) -(-1)^{\alpha}\lambda] \xi_{\alpha} - 2m_D\sqrt{g_{xx}} -  2 iB_0(r) \bigg) \xi_{\alpha,-+}, \nn \\
(d)&& \xi_{12,\beta\beta^\prime}^\prime =
\sqrt\frac{g_{rr}}{g_{xx}}\bigg( [K_0(r)-\lambda] (\xi_{1}^{-1} \! + \! \xi_{2}) - 2m_D\sqrt{g_{xx}} +  i  (\beta - \beta^\prime ) B_0(r) \bigg) \xi_{12,\beta\beta^\prime}, \nn \\
&& \xi_{21,\beta\beta^\prime}^\prime =
\sqrt\frac{g_{rr}}{g_{xx}}\bigg( [K_0(r)+\lambda] (\xi_{2}^{-1} \! + \! \xi_{1}) - 2m_D\sqrt{g_{xx}} +  i  (\beta - \beta^\prime ) B_0(r) \bigg) \xi_{21,\beta\beta^\prime}, \label{Eq:flow-xi_alpha}
\eeqa
where the flow functions $\xi_{\alpha\alpha^\prime,\beta\beta^\prime}$ are defined in Eq.(\ref{Eq:xi-F/G}), and explicit forms are as follows.
\beqa
&& \xi_\alpha \equiv  \frac{\tilde{F}_{\alpha}}{\tilde{G}_{\alpha}} \equiv \frac{\tilde{F}_{\alpha,\pm}}{\tilde{G}_{\alpha,\pm}}, \quad \xi_{12} \equiv \frac{\tilde{F}_{1}}{\tilde{G}_{2}}  \equiv \frac{\tilde{F}_{1,\pm}}{\tilde{G}_{2,\pm}} , \quad \xi_{21} \equiv \frac{\tilde{F}_{2}}{\tilde{G}_{1}}  \equiv \frac{\tilde{F}_{2,\pm}}{\tilde{G}_{1,\pm}} , \nn\\
&& \xi_{\alpha,+-} \equiv \frac{\tilde{F}_{\alpha,+}}{\tilde{G}_{\alpha,-}}, \quad \xi_{\alpha,-+} \equiv \frac{\tilde{F}_{\alpha,-}}{\tilde{G}_{\alpha,+}},  \quad \xi_{12,\beta\beta^\prime} \equiv \frac{\tilde{F}_{1,\beta}}{\tilde{G}_{2,\beta^\prime}}, \quad \quad \xi_{21,\beta\beta^\prime} \equiv \frac{\tilde{F}_{2,\beta}}{\tilde{G}_{1,\beta^\prime}}, \label{Eq:xi-F/G-2}
\eeqa
where $\alpha=1,2$, $\beta,\beta^\prime =\pm$. It is obvious that the flow equations in cases (c) and (d) reduce to those in cases (a) and (b), respectively, in the absence of chiral gauge fields $b_r,b_0$. In addition, if $b_i$ vanishes, i.e., $\lambda=|\vec{k}|\equiv k$, then one obtains the flow equations without chiral gauge field,
\beqa
(a)&& \xi_\alpha^\prime = \sqrt\frac{g_{rr}}{g_{xx}}\bigg( [K_0(r) + (-1)^\alpha k ]  \! + \! [K_0(r) -(-1)^{\alpha} k] \xi_\alpha^2 - 2m_D\sqrt{g_{xx}} \xi_\alpha\bigg) , \nn \\
(b)&& \xi_{12}^\prime = \sqrt\frac{g_{rr}}{g_{xx}}\bigg( [K_0(r) - k] (\xi_{1}^{-1} \! + \! \xi_{2}) - 2m_D\sqrt{g_{xx}}\bigg) \xi_{12}, \nn \\
&& \xi_{21}^\prime = \sqrt\frac{g_{rr}}{g_{xx}}\bigg( [K_0(r)+k] (\xi_{2}^{-1} \! + \! \xi_{1}) - 2m_D\sqrt{g_{xx}}\bigg) \xi_{21}, \label{Eq:flow-xi_alpha-B=0}
\eeqa
where (a) just recovers the result, Eq.(A22), in ref.~\cite{Faulkner:2009wj}.

In summary, we have obtained the flow equations associated with Dirac equation as shown in Eq.(\ref{Eq:flow-xi_alpha}), in the near horizon geometry of black brane as discussed in Sec.(\ref{sec:Horizon_BCs}).

\section{UV Retarded Green's Function from Bulk Gravity with Spatial Rotational Invariance}
\label{sec:UVGR-AdS(d+1)-SpatialRotationalInvariance}

\subsection{Retarded Green's function of Dirac fermion from bulk gravity with Ricci flat hypersurface}

For AdS$_4$, the generic bulk metric with flat Ricci horizon $k=0$ as shown in Eq.(\ref{Eq:ds2-fr}) and Eq.(\ref{Eq:fr-gr}) is given by
\beqa
\ii
ds^2 =  -    \frac{ r^2}{\ell^2} g(r) dt^2 +   \frac{\ell^2}{ r^2} \frac{1}{g(r)}{dr^2} + \frac{r^2}{\ell^2}dx_{d-1}^2 .
\eeqa
From the above one can read off
\beqa
\ii
g_{tt}(r) \!=\!     \frac{r^2}{\ell^2} g(r), ~ g_{rr}(r) \!=\! g_{tt}^{-1}, ~  g_{xx}(r) \!=\! \frac{r^2}{\ell^2}. \label{Eq:gtt-grr-gxx_AdS}
\eeqa
The differential operators defined in Eq.(\ref{Eq:AmD}) becomes
\beqa
A(m_D) \equiv  \frac{r^2}{\ell^2} \sqrt{g(r)} \bigg(\partial_r - m_D  \frac{\ell}{r} \frac{1}{\sqrt{g(r)}} \bigg) . \label{Eq:AmD-AdS}
\eeqa
$K_0(r)$, $B_0(r)$ and $B_r(r)$, defined in Eq.(\ref{Eq:K0-B0-Br}), in this case, are given by
\beqa
K_0(r) \equiv
 \frac{\omega + q A_t(r)}{\sqrt{g(r)}} , \quad B_0(r) \equiv  \frac{1}{\sqrt{g(r)}} b_0, \quad B_r(r) \equiv \frac{r^2}{\ell^2} \sqrt{g(r)} b_r.
\label{Eq:K0-AdS}
\eeqa

Assuming that $b_r$ is absorbed into the phase of the wavefunction $\Psi$ through its redefinition. Then in a given bulk gravity background, one can obtain Dirac equation in Eq.(\ref{Eq:Dirac-Psi_alpha-beta}) with $b_r$, and flow equation in Eq.(\ref{Eq:flow-xi_alpha}) as
\beqa
&& \bigg[  \textbf{1}_2 \bigg( \partial_r \!-\! i \beta    \frac{\ell^2}{r^2} \frac{1}{g(r)} b_0\bigg) \!+\! \sigma^3 m_D  \frac{\ell}{r} \frac{1}{\sqrt{g(r)}} \bigg] \tilde\Phi_{\alpha,\beta} \!=\! \bigg( i\sigma^2   \frac{\ell^2}{r^2} \frac{1}{g(r)}[\omega \!+\!  q A_t(r)]  \!+\!  (-1)^\alpha \sigma^1  \frac{\ell^2}{r^2} \frac{1}{\sqrt{g(r)}} \lambda\bigg) \tilde\Phi_{\alpha,\beta}, \nn\\
(a)&& \xi_\alpha^\prime =  \frac{\ell^2}{r^2} \frac{1}{\sqrt{g(r)}} \bigg[ \bigg(  \frac{\omega + q A_t(r)}{\sqrt{g(r)}} + (-1)^\alpha\lambda \bigg)  \! + \! \bigg( \frac{\omega + q A_t(r)}{\sqrt{g(r)}} -(-1)^{\alpha}\lambda \bigg) \xi_\alpha^2 - 2m_D \frac{r}{\ell} \xi_\alpha\bigg] , \nn \\
(b)&& \xi_{12}^\prime =  \frac{\ell^2}{r^2} \frac{1}{\sqrt{g(r)}} \bigg[ \bigg(  \frac{\omega + q A_t(r)}{\sqrt{g(r)}} - \lambda \bigg) (\xi_{1}^{-1} \! + \! \xi_{2}) - 2m_D \frac{r}{\ell}\bigg] \xi_{12}, \nn \\
&& \xi_{21}^\prime =  \frac{\ell^2}{r^2} \frac{1}{\sqrt{g(r)}} \bigg[ \bigg(  \frac{\omega + q A_t(r)}{\sqrt{g(r)}} +\lambda \bigg) (\xi_{2}^{-1} \! + \! \xi_{1}) - 2m_D \frac{r}{\ell}\bigg] \xi_{21}, \nn \\
(c)&& \xi_{\alpha,+-}^\prime =
 \frac{\ell^2}{r^2} \frac{1}{\sqrt{g(r)}} \bigg[ \bigg(  \frac{\omega + q A_t(r)}{\sqrt{g(r)}} +(-1)^\alpha\lambda \bigg) \xi_{\alpha}^{-1} \! + \! \bigg( \frac{\omega + q A_t(r)}{\sqrt{g(r)}} -(-1)^{\alpha}\lambda \bigg) \xi_{\alpha} - 2m_D\frac{r}{\ell} +  2 i  \frac{1}{\sqrt{g(r)}} b_0 \bigg] \xi_{\alpha,+-}, \nn \\
&& \xi_{\alpha,-+}^\prime =
 \frac{\ell^2}{r^2} \frac{1}{\sqrt{g(r)}} \bigg[ \bigg(  \frac{\omega + q A_t(r)}{\sqrt{g(r)}} +(-1)^\alpha\lambda \bigg) \xi_{\alpha}^{-1} \! + \! \bigg( \frac{\omega + q A_t(r)}{\sqrt{g(r)}} -(-1)^{\alpha}\lambda \bigg) \xi_{\alpha} - 2m_D\frac{r}{\ell} -  2 i  \frac{1}{\sqrt{g(r)}} b_0 \bigg] \xi_{\alpha,-+}, \nn \\
(d)&& \xi_{12,\beta\beta^\prime}^\prime =
 \frac{\ell^2}{r^2} \frac{1}{\sqrt{g(r)}} \bigg[ \bigg(  \frac{\omega + q A_t(r)}{\sqrt{g(r)}} -\lambda \bigg) (\xi_{1}^{-1} \! + \! \xi_{2}) - 2m_D \frac{r}{\ell} +  i  (\beta - \beta^\prime )  \frac{1}{\sqrt{g(r)}} b_0 \bigg] \xi_{12,\beta\beta^\prime}, \nn \\
&& \xi_{21,\beta\beta^\prime}^\prime =
 \frac{\ell^2}{r^2} \frac{1}{\sqrt{g(r)}} \bigg[ \bigg(  \frac{\omega + q A_t(r)}{\sqrt{g(r)}} +\lambda \bigg) (\xi_{2}^{-1} \! + \! \xi_{1}) - 2m_D \frac{r}{\ell} +  i  (\beta - \beta^\prime )  \frac{1}{\sqrt{g(r)}} b_0 \bigg] \xi_{21,\beta\beta^\prime}, \label{Eq:DiracEOM-flow-xi_alpha-UV-AdS(d+1)-omega!=0}
\eeqa
where $\tilde\Phi_{\alpha,\beta}$ are bulk Dirac fermions with $\beta=\pm$ indicating left and right helicity states.

In the absence of chiral gauge fields $b_r$ and $b_0$, it is not necessary to make the rotation between $\Psi$ and $\tilde\Psi$, the Dirac equation and flow equations reduce to
\beqa
&& \bigg(  \textbf{1}_2 \partial_r \!+\! \sigma^3 m_D  \frac{\ell}{r} \frac{1}{\sqrt{g(r)}} \bigg) \tilde\Phi_{\alpha} \!=\! \bigg( i\sigma^2   \frac{\ell^2}{r^2} \frac{1}{g(r)}[\omega \!+\!  q A_t(r)]  \!+\!  (-1)^\alpha \sigma^1  \frac{\ell^2}{r^2} \frac{1}{\sqrt{g(r)}} \lambda\bigg) \tilde\Phi_{\alpha}, \label{Eq:EOM-Diracfermion-UV-AdS(d+1)-omega!=0}\\
(a)(c)&& \xi_\alpha^\prime =  \frac{\ell^2}{r^2} \frac{1}{\sqrt{g(r)}} \bigg[ \bigg(  \frac{\omega + q A_t(r)}{\sqrt{g(r)}} + (-1)^\alpha\lambda \bigg)  \! + \! \bigg( \frac{\omega + q A_t(r)}{\sqrt{g(r)}} -(-1)^{\alpha}\lambda \bigg) \xi_\alpha^2 - 2m_D \frac{r}{\ell} \xi_\alpha\bigg] , \nn \\
(b)(d)&& \xi_{12}^\prime =  \frac{\ell^2}{r^2} \frac{1}{\sqrt{g(r)}} \bigg[ \bigg(  \frac{\omega + q A_t(r)}{\sqrt{g(r)}} - \lambda \bigg) (\xi_{1}^{-1} \! + \! \xi_{2}) - 2m_D \frac{r}{\ell}\bigg] \xi_{12}, \nn \\
&& \xi_{21}^\prime =  \frac{\ell^2}{r^2} \frac{1}{\sqrt{g(r)}} \bigg[ \bigg(  \frac{\omega + q A_t(r)}{\sqrt{g(r)}} +\lambda \bigg) (\xi_{2}^{-1} \! + \! \xi_{1}) - 2m_D \frac{r}{\ell}\bigg] \xi_{21}, \label{Eq:flow-xi_alpha-AdS(d+1)-omega!=0}
\eeqa
where in the Dirac equations and flow equations for $\xi_{\alpha,\beta\beta^\prime}$ and $\xi_{\alpha\alpha^\prime,\beta\beta^\prime}$, we have dropped their spin projection subscript $\beta$. The eigenvalue is $\lambda=\sqrt{|\vec{k}|^2-|\vec{b}|^2}$, and $\tilde\Phi_\alpha = \Phi_\alpha$ since in the absence of chiral gauge field, one does not need introduce the unitary transformation between $\tilde\Psi$ and $\Psi$ to diagonalize the eigenvalue matrix as shown in Eq.(\ref{Eq:tPsi_alpha-Psi_alpha}).

\subsubsection{Dirac equation and flow equation at infinite boundary}

At the infinite boundary $r\to \infty$, $g(r) \overset{r\to\infty}{=} 1$, according to Eq.(\ref{Eq:gtt-grr-gxx_AdS}), the metric and gauge field become
\beqa
ds^2 \overset{r\to\infty}{=} \frac{r^2}{\ell^2} (-   dt^2 + dx_{d-1}^2) +  \frac{\ell^2}{r^2}dr^2, \quad A_t(r) \overset{r\to\infty}{=} \mu.
\label{Eq:infinite-UV_AdS}
\eeqa
In this case, the differential operator in Eq.(\ref{Eq:AmD-AdS}) and $K_0(r)$ given in Eq.(\ref{Eq:K0-AdS}) become, respectively,
\beqa
A(m_D) \equiv  \frac{r^2}{\ell^2}  \bigg(\partial_r - m_D  \frac{\ell}{r} \bigg), \quad K_0(r) \overset{r\to\infty}{=}  (\omega + q\mu).
\eeqa
In the background of AdS metric in Eq.(\ref{Eq:infinite-UV_AdS}), the Dirac equation and flow equations in Eq.(\ref{Eq:DiracEOM-flow-xi_alpha-UV-AdS(d+1)-omega!=0}) are obtained.

In the absence of chiral gauge fields $b_r$ (or $b_r$ is absorbed into the radial sector of wave function) and $b_0$, the Dirac equation and flow equation in Eqs.(\ref{Eq:EOM-Diracfermion-UV-AdS(d+1)-omega!=0}) and (\ref{Eq:flow-xi_alpha-AdS(d+1)-omega!=0}) reduce to
\beqa
&& \bigg(  \textbf{1}_2 \partial_r \!+\! \sigma^3 m_D  \frac{\ell}{r}  \bigg) \tilde\Phi_{\alpha} \!=\! \bigg( i\sigma^2   \frac{\ell^2}{r^2} ( \omega + q \mu )  \!+\!  (-1)^\alpha \sigma^1  \frac{\ell^2}{r^2}  \lambda\bigg) \tilde\Phi_{\alpha}, \nn\\
(a)(c)&& \xi_\alpha^\prime =  \frac{\ell^2}{r^2}  \bigg( [ ( \omega + q \mu ) + (-1)^\alpha\lambda]  \! + \! [ ( \omega + q \mu ) -(-1)^{\alpha}\lambda] \xi_\alpha^2 - 2m_D \frac{r}{\ell} \xi_\alpha\bigg) , \nn \\
(b)(d)&& \xi_{12}^\prime =  \frac{\ell^2}{r^2}  \bigg( [ ( \omega + q \mu ) - \lambda] (\xi_{1}^{-1} \! + \! \xi_{2}) - 2m_D \frac{r}{\ell}\bigg) \xi_{12}, \nn \\
&& \xi_{21}^\prime =  \frac{\ell^2}{r^2}  \bigg( [ ( \omega + q \mu )+\lambda] (\xi_{2}^{-1} \! + \! \xi_{1}) - 2m_D \frac{r}{\ell}\bigg) \xi_{21}, \label{Eq:flow-xi_alpha-AdS(d+1)}
\eeqa
where $\lambda=\sqrt{|\vec{k}|^2-|\vec{b}|^2}$, $\tilde\Phi_\alpha=\Phi_\alpha$ since there is no need for rotation to diagonalize the matrix.

In the infinite boundary $r\to \infty$, Eqs.(\ref{Eq:flow-xi_alpha-AdS(d+1)}) become
\beqa
&& \bigg(  \textbf{1}_2 \partial_r  \!+\! \sigma^3 m_D  \frac{\ell}{r} \bigg) \tilde\Phi_{\alpha,\beta} \!\overset{r\to \infty}{=}\!  0, \nn \\
(a)(c)&& \xi_{\alpha}^\prime \overset{r\to \infty}{=}   - 2m_D  \frac{\ell}{r} \xi_{\alpha} , \nn \\
(b)(d)&& \xi_{12}^\prime \overset{r\to \infty}{=}   - 2m_D  \frac{\ell}{r}  \xi_{12}, \quad \xi_{21}^\prime \overset{r\to \infty}{=}  - 2m_D  \frac{\ell}{r}  \xi_{21},  \label{Eq:flow-xi_alpha-AdS(d+1)_infty}
\eeqa
which means that both $b_0$ and $b_i$ do not affect the Dirac equation and the flow equations, so do the energy eigen-values $\lambda$, thus independent of $\alpha$. In other words, they have no relation to the UV limit, but they are relevant in the IR limit, which means their physical consequence can not be neglected in the IR limit.

\subsubsection{UV retarded Green's functions with spin helicity}

From the above equations, we have two linearly independent solutions
\beqa
&& \tilde\Phi_{\alpha,\beta\beta^\prime}
\overset{r\to\infty}{=}  \tilde{B}_{\alpha,\beta} r^{-   m_D\ell} \left(\begin{array}{c}
1\\
0
\end{array}
\right) + \tilde{A}_{\alpha,\beta^\prime} r^{  m_D \ell} \left(\begin{array}{c}
0\\
1
\end{array}
\right) , \nn\\
(a)&& \xi_\alpha \equiv \frac{\tilde{F}_\alpha}{\tilde{G}_\alpha} = \xi_{\alpha,\beta\beta}= \frac{\tilde{F}_{\alpha,\beta}}{\tilde{G}_{\alpha,\beta}} \overset{r\to \infty}{=} r^{-2 m_D \ell} \tilde{B}_\alpha\tilde{A}_\alpha^{-1}
, \nn\\
(b)&& \xi_{\alpha\alpha^\prime} \equiv \frac{\tilde{F}_\alpha}{\tilde{G}_{\alpha^\prime}} = \xi_{\alpha\alpha^\prime,\beta\beta} = \frac{\tilde{F}_{\alpha,\beta}}{\tilde{G}_{\alpha^\prime,\beta}} \overset{r\to \infty}{=} r^{-2 m_D \ell} \tilde{B}_\alpha \tilde{A}_{\alpha^\prime}^{-1}  \nn\\
(c)&& \xi_{\alpha,\beta\beta^\prime} \equiv \frac{\tilde{F}_{\alpha,\beta}}{\tilde{G}_{\alpha,\beta^\prime}}  \overset{r\to \infty}{=} r^{-2 m_D \ell} \tilde{B}_{\alpha,\beta} \tilde{A}_{\alpha,\beta^\prime}^{-1}, \nn \\
(d)&& \xi_{\alpha\alpha^\prime,\beta\beta^\prime} \equiv \frac{\tilde{F}_{\alpha,\beta}}{\tilde{G}_{\alpha^\prime,\beta^\prime}} = \frac{\tilde{F}_{\alpha,\beta}}{\tilde{G}_{\alpha^\prime,\beta^\prime}} \overset{r\to \infty}{=} r^{-2 m_D \ell} \tilde{B}_{\alpha,\beta} \tilde{A}_{\alpha^\prime,\beta^\prime}^{-1},\label{Eq:Psi_alpha-UV}
\eeqa
where $\tilde\Phi_{\alpha,\beta\beta^\prime} \equiv (\tilde{F}_{\alpha,\beta},\tilde{G}_{\alpha,\beta^\prime})$ are defined in Eq.(\ref{Eq:Rt_alpha}). In the massless limit, the flow functions $\xi_{\alpha\alpha^\prime,\beta\beta^\prime}$ all approach to constants,
\beqa
\xi_\alpha \overset{m_D\to 0}{=}  \tilde{B}_\alpha \tilde{A}_\alpha^{-1}, \quad \xi_{\alpha\alpha^\prime} \overset{m_D\to 0}{=}  \tilde{B}_\alpha \tilde{A}_{\alpha^\prime}^{-1} , \quad \xi_{\alpha,\beta\beta^\prime} \overset{m_D\to 0}{=}  \tilde{B}_{\alpha,\beta} \tilde{A}_{\alpha,\beta^\prime}^{-1} , \quad \xi_{\alpha\alpha^\prime,\beta\beta^\prime} \overset{m_D\to 0}{=}  \tilde{B}_{\alpha,\beta} \tilde{A}_{\alpha^\prime,\beta^\prime}^{-1},
\eeqa
where $\tilde{B}_{\alpha,\beta}$ and $\tilde{A}_{\alpha^\prime,\beta^\prime}$ are viewed as response and source, respectively. Suppose the two coefficients above are related by
\beqa
&& \tilde{B}_{\alpha,\beta} \left(\begin{array}{cc}
1\\
0
\end{array}
\right) = {\mathcal S} \tilde{A}_{\alpha^\prime,\beta^\prime} \left(\begin{array}{cc}
0\\
1
\end{array}
\right), \nn
\eeqa
then the UV retarded Green's function in the infinite boundary can be read off as
\beqa
&& \tilde{G}_\alpha^R = -i {\mathcal S}\gamma^t =\lim_{r\to\infty}r^{2  m_D\ell} \left(\begin{array}{cc}
\xi_1 & \\
 & \xi_2
\end{array}
\right) \equiv \left(\begin{array}{cc}
\tilde{G}_1^R & \\
 & \tilde{G}_2^R
\end{array}
\right),
\eeqa
where $\gamma^t = i\sigma^1$ and we have defined the generalized boundary spinor Green's function, which has two sets of eigenvalues given by the ratio
\beqa
\tilde{G}_{\alpha\alpha^\prime,\beta\beta^\prime}^R(\omega,\lambda) \equiv \tilde{B}_{\alpha,\beta} \tilde{A}_{\alpha^\prime,\beta^\prime}^{-1} \overset{r\to\infty}{=} \frac{r^{  m_D\ell}\tilde{F}_{\alpha,\beta}}{r^{-  m_D\ell}\tilde{G}_{\alpha^\prime,\beta^\prime}}  = \lim_{r\to \infty} r^{{2 }m_D\ell} \xi_{\alpha\alpha^\prime,\beta\beta^\prime} ,\label{Eq:G_alpha-alphap-beta-betap-AdS}
\eeqa
where $\xi_{\alpha \alpha^\prime, \beta\beta^\prime} \equiv {\tilde{F}_{\alpha,\beta}}/{\tilde{G}_{\alpha^\prime,\beta^\prime}}$ with $\alpha,\alpha^\prime=1,2$ and $\beta=\pm$. The elements of the diagonal term are
\beqa
(a) && \tilde{G}_\alpha^R(\omega,\lambda) \equiv \tilde{G}_{\alpha,\beta\beta}^R(\omega,\lambda) \equiv \tilde{G}_{\alpha\alpha,\beta\beta}^R(\omega,\lambda) \equiv \tilde{B}_{\alpha,\beta} \tilde{A}_{\alpha,\beta}^{-1} = \tilde{B}_{\alpha} \tilde{A}_{\alpha}^{-1} = \lim_{r\to \infty} r^{{2 }m_D\ell} \xi_{\alpha,\beta\beta}= \lim_{r\to \infty} r^{{2 }m_D\ell} \xi_{\alpha}, \nn\\
(b) && \tilde{G}_{\alpha\alpha^\prime}^R(\omega,\lambda) \equiv \tilde{G}_{\alpha\alpha^\prime,\beta\beta}^R(\omega,\lambda) \equiv \tilde{B}_{\alpha,\beta} \tilde{A}_{\alpha^\prime,\beta}^{-1} \equiv \tilde{B}_{\alpha^\prime} \tilde{A}_{\alpha}^{-1} = \lim_{r\to \infty} r^{{2 }m_D\ell} \xi_{\alpha\alpha^\prime,\beta\beta}= \lim_{r\to \infty} r^{{2 }m_D\ell} \xi_{\alpha\alpha^\prime}, \nn\\
(c) && \tilde{G}_{\alpha,\beta\beta^\prime}^R(\omega,\lambda) \equiv \tilde{G}_{\alpha\alpha,\beta\beta^\prime}^R(\omega,\lambda) \equiv  \tilde{B}_{\alpha,\beta} \tilde{A}_{\alpha,\beta^\prime}^{-1}  = \lim_{r\to \infty} r^{{2 }m_D\ell} \xi_{\alpha,\beta\beta^\prime}, \label{Eq:G_alpha-alphap-beta-betap-AdS-2}
\eeqa
where $ \xi_\alpha \equiv \xi_{\alpha\alpha} = {\tilde{F}_{\alpha}}/{\tilde{G}_{\alpha}} $, with $\alpha=1,2$. To be more explicit, take the case (b) as an example, the non-diagonal terms are denoted by
\beqa
\tilde{G}_{12}^R = \lim_{r\to \infty} r^{{2 }m_D\ell} \xi_{12}, \quad \tilde{G}_{21}^R = \lim_{r\to \infty} r^{{2 }m_D\ell} \xi_{21}, \quad \xi_{12} = \frac{\tilde{F}_{1}}{\tilde{G}_{2}}, \quad \xi_{21} = \frac{\tilde{F}_{2}}{\tilde{G}_{1}},
\eeqa
By imposing the in-going boundary condition for $\Phi_{\alpha,\beta}$ at the horizon $r\to r_\star $, as will be discussed in the following section, the ratio $\xi_{\alpha\alpha^\prime}$ can be determined through solving flow equations in Eqs.(\ref{Eq:DiracEOM-flow-xi_alpha-UV-AdS(d+1)-omega!=0}). Therefore, the problem of solving Dirac equation to obtain the retarded Green's function is transformed into that of solving the flow equation. In the following, we will mainly focus on the cases (a) and (c), which correspond to the wavefunctions with same energy eigenvalues and the same or different helicity eigenvalues of the projection operators $P_\beta$ with $\beta=\pm$.

According to Eq.(\ref{Eq:tPsi_alpha-Psi_alpha}), the original Weyl fermion wavefunctions $\Psi_\beta$ are obtained through a transformation acting upon energy eigenvalue wavefunctions $\tilde\Psi_\beta$, $ ( \tilde\Psi_+, \tilde\Psi_- )^T = U_1 ( \Psi_+, \Psi_- )^T $, where $U_1=(1-i\sigma^1)/\sqrt{2}$ is introduced in Eq.(\ref{Eq:transU1_tPsi-Psi}). Since $\tilde\Psi_\beta$ have two different energy eivenvalues, namely $+\lambda$ and $-\lambda$. The transformations between the two eigen wave functions are of the same form, i.e., $(\tilde\Psi_{\alpha,+},\tilde\Psi_{\alpha,-})^T=U_1(\Psi_{\alpha,+},\Psi_{\alpha,-})^T$, where the subscript $\alpha$ of eigen wavefunctions indicates that they correspond to the eigenvalues $\lambda$ and $-\lambda$, respectively. The explicit notations have been introduced in Eq.(\ref{Eq:Marster_Dirac-EOMs_b0_br}), $\tilde\Psi_{\alpha,\beta}$ with $\alpha=1,2$ and $\beta=\pm$. Therefore, one has
\beqa
\Psi_\alpha \equiv \left(
                 \begin{array}{c}
                   \Psi_{\alpha,+} \\
                   \Psi_{\alpha,-} \\
                 \end{array}
               \right)
 = U_1^{-1} \left(
                 \begin{array}{c}
                   \tilde\Psi_{\alpha,+} \\
                   \tilde\Psi_{\alpha,-} \\
                 \end{array}
               \right) = \frac{1}{\sqrt{2}} \left(
                                              \begin{array}{c}
                                                \tilde\Psi_{\alpha,+} + i \tilde\Psi_{\alpha,-} \\
                                                \tilde\Psi_{\alpha,-} + i \tilde\Psi_{\alpha,+} \\
                                              \end{array}
                                            \right) \equiv ( \tilde\Phi_{1} + \tilde\Phi_{2} ), \nn
\eeqa
where $\tilde\Phi_{\alpha,\beta} \equiv P_\beta \tilde\Psi_\alpha$, and the subscript $\beta=\pm$ stand for the left and right helicity sector of the bulk Dirac fermion, respectively. Therefore,
\beqa
&& \Psi_{\alpha,\beta} \equiv \frac{1}{\sqrt{2}}(\tilde\Psi_{\alpha,\beta} + i \tilde\Psi_{\alpha,-\beta}) \overset{r\to\infty}{=}  r^{-   m_D\ell}  B_{\alpha,\beta} \left(\begin{array}{c}
1\\
0
\end{array}
\right) \!+\! A_{\alpha,\beta}  r^{   m_D\ell} \left(\begin{array}{c}
0\\
1
\end{array}
\right), \nn\\
&& B_{\alpha,\beta} = \frac{\tilde{B}_{\alpha,\beta}+i\tilde{B}_{\alpha,-\beta}}{\sqrt{2}}, \quad A_{\alpha,\beta} = \frac{\tilde{A}_{\alpha,\beta}+i\tilde{A}_{\alpha,-\beta}}{\sqrt{2}}, \label{Eq:Psi_alpha-beta_Bt-At}
\eeqa
where $\alpha=1,2$ and $\beta=\pm$.

By the definition of boundary retarded Green's functions, one obtains
\beqa
G_{\alpha,\beta\beta^\prime}^R = \frac{\tilde{B}_{\alpha,\beta} \!+\! i \tilde{B}_{\alpha,-\beta}}{\tilde{A}_{\alpha,\beta^\prime} \!+\! i \tilde{A}_{\alpha,-\beta^\prime}}, \label{Eq:GR_alpha_beta-betap-B/A}
\eeqa
with $\beta,\beta^\prime=\pm$. To be more explicit,
\beqa
G_{\alpha,++}^R = \frac{\tilde{B}_{\alpha,+} \!+\! i \tilde{B}_{\alpha,-}}{\tilde{A}_{\alpha,+} \!+\! i \tilde{A}_{\alpha,-}}, \quad G_{\alpha,--}^R = \frac{\tilde{B}_{\alpha,-} \!+\! i \tilde{B}_{\alpha,+}}{\tilde{A}_{\alpha,-} \!+\! i \tilde{A}_{\alpha,+}}, \quad G_{\alpha,+-}^R = \frac{\tilde{B}_{\alpha,+} \!+\! i \tilde{B}_{\alpha,-}}{\tilde{A}_{\alpha,-} \!+\! i \tilde{A}_{\alpha,+}}, \quad G_{\alpha,-+}^R = \frac{\tilde{B}_{\alpha,-} \!+\! i \tilde{B}_{\alpha,+}}{\tilde{A}_{\alpha,+} \!+\! i \tilde{A}_{\alpha,-}},
\eeqa
Using Eq.(\ref{Eq:G_alpha-alphap-beta-betap-AdS-2}), Eq.(\ref{Eq:GR_alpha_beta-betap-B/A}) can be rewritten as
\beqa
G_{\alpha,\beta\beta^\prime}^R  = \frac{ \tilde{G}_{\alpha,\beta\beta^\prime}^R \!+\! i \tilde{G}_{\alpha,-\beta\beta^\prime}^R  }{1 \!+\! i (\tilde{G}_{\alpha,-\beta-\beta^\prime}^R)^{-1} \tilde{G}_{\alpha,-\beta\beta^\prime}^R } = \frac{ \tilde{G}_{\alpha,\beta\beta^\prime}^R \!+\! i \tilde{G}_{\alpha,-\beta\beta^\prime}^R  }{ \tilde{G}_{\alpha,-\beta-\beta^\prime}^R \!+\! i  \tilde{G}_{\alpha,-\beta\beta^\prime}^R } \tilde{G}_{\alpha,-\beta-\beta^\prime}^R.
\eeqa
It is intuitive to observe that $G_{\alpha,\beta\beta^\prime}^R$ recover $G_{\alpha}^R = \tilde{G}_{\alpha}^R$, since the subscript $\beta$ will be unnecessary in the absence of chiral gauge field $b_0$. In this case, it is also obvious through definition in Eq.(\ref{Eq:GR_alpha_beta-betap-B/A}), since $G_{\alpha}^R = ({\tilde{B}_{\alpha} \!+\! i \tilde{B}_{\alpha}})/({\tilde{A}_{\alpha} \!+\! i \tilde{A}_{\alpha}})=\tilde{B}_{\alpha}\tilde{A}_{\alpha}^{-1}=\tilde{G}_\alpha^R$.

\begin{itemize}
  \item
If $\beta=\beta^\prime$, then one obtains
\beqa
 G_{\alpha}^R \equiv G_{\alpha,\beta\beta}^R  = \frac{ \tilde{G}_{\alpha,\beta\beta} \!+\! i \tilde{G}_{\alpha,-\beta\beta}  }{ \tilde{G}_{\alpha,-\beta-\beta} \!+\! i  \tilde{G}_{\alpha,-\beta\beta} } \tilde{G}_{\alpha,-\beta-\beta}  = \frac{ \tilde{G}_{\alpha} \!+\! i \tilde{G}_{\alpha,-\beta\beta}  }{ \tilde{G}_{\alpha} \!+\! i  \tilde{G}_{\alpha,-\beta\beta} } \tilde{G}_{\alpha} = \tilde{G}_{\alpha}^R ,
\eeqa
where we have made the abbreviation that $\tilde{G}_{\alpha,-\beta-\beta}=\tilde{G}_{\alpha,\beta\beta}\equiv \tilde{G}_{\alpha}$, since $\xi_{\alpha,\beta\beta}$ are independent of helicity $\beta$, which is obvious by observing the flow equations in Eq.(\ref{Eq:xi_alpha_beta-betap}). This means that the boundary retarded Green's functions for the original wave function $\Psi_{\alpha}$ with same helicity, i.e., $G^R_{\alpha}$ is completely determined by that of $\tilde\Psi_{\alpha}$, i.e., $\tilde{G}_\alpha$.

  \item If $\beta=-\beta^\prime$, then one has
\beqa
G_{\alpha,\beta-\beta}^R  = \frac{ \tilde{G}_{\alpha,\beta-\beta} \!+\! i \tilde{G}_{\alpha,-\beta-\beta}  }{ \tilde{G}_{\alpha,-\beta\beta} \!+\! i  \tilde{G}_{\alpha,-\beta-\beta} } \tilde{G}_{\alpha,-\beta\beta} = \frac{ \tilde{G}_{\alpha,\beta-\beta} \!+\! i \tilde{G}_{\alpha}  }{ \tilde{G}_{\alpha,-\beta\beta} \!+\! i  \tilde{G}_{\alpha} } \tilde{G}_{\alpha,-\beta\beta} = \frac{ \tilde{G}_{\alpha} \!+\! i \tilde{G}_{\alpha,-\beta\beta}  }{ \tilde{G}_{\alpha,-\beta\beta} \!+\! i  \tilde{G}_{\alpha} } \tilde{G}_{\alpha} = \frac{ \tilde{G}_{\alpha,\beta-\beta} \!+\! i \tilde{G}_{\alpha}  }{ \tilde{G}_{\alpha} \!+\! i  \tilde{G}_{\alpha,\beta-\beta} } \tilde{G}_{\alpha},
\eeqa
where in the last two equalities, we have used the identity that $\tilde{G}_{\alpha,\beta-\beta}\tilde{G}_{\alpha,-\beta\beta}=\tilde{G}_{\alpha,\beta\beta}\tilde{G}_{\alpha,-\beta-\beta}=\tilde{G}_{\alpha}^2$. This means that the boundary retarded Green's functions for the original wave function $\Psi_{\alpha}$ with opposite helicity, i.e., $G^R_{\alpha,\beta-\beta}$ is completely determined by $\tilde{G}_{\alpha,\beta-\beta}$ and $\tilde{G}_{\alpha}$.
\end{itemize}
In summary, we have
\beqa
&& G_{\alpha,++}^R  = G_{\alpha,--}^R  = G_{\alpha}^R = \tilde{G}_{\alpha}^R, \nn\\
&& G_{\alpha,+-}^R = \frac{ \tilde{G}_{\alpha} \!+\! i \tilde{G}_{\alpha,-+}  }{ \tilde{G}_{\alpha,-+} \!+\! i  \tilde{G}_{\alpha} } \tilde{G}_{\alpha} = \frac{ \tilde{G}_{\alpha,+-} \!+\! i \tilde{G}_{\alpha}  }{ \tilde{G}_{\alpha} \!+\! i  \tilde{G}_{\alpha,+-} } \tilde{G}_{\alpha} , \quad G_{\alpha,-+}^R = \frac{ \tilde{G}_{\alpha} \!+\! i \tilde{G}_{\alpha,+-}  }{ \tilde{G}_{\alpha,+-} \!+\! i  \tilde{G}_{\alpha} } \tilde{G}_{\alpha} = \frac{ \tilde{G}_{\alpha,-+} \!+\! i \tilde{G}_{\alpha}  }{ \tilde{G}_{\alpha} \!+\! i  \tilde{G}_{\alpha,-+} } \tilde{G}_{\alpha}, \label{Eq:GR-tGR}
\eeqa
where $\beta=\pm$ indicate the left and right handed helicity states of the original bulk Dirac fermions $\Psi_\alpha$ with eigen energy $\pm \lambda$, respectively.
It is also obvious that there is an identity $G_{\alpha,+-}^R G_{\alpha,+-}^R = (\tilde{G}_{\alpha}^R)^2 = G_{\alpha,++}^R G_{\alpha,--}^R= (G_{\alpha}^R)^2$. In a word, we have obtained the boundary retarded Green's functions from the original bulk wavefunctions $\Psi_{\alpha}$ with the same or opposite helicity states. They can be calculated through Eqs.(\ref{Eq:G_alpha-alphap-beta-betap-AdS}) and (\ref{Eq:G_alpha-alphap-beta-betap-AdS-2}).

\subsubsection{Properties of UV retarded Green's function at zero temperature}
\label{sec:GR-prop}

In this section, we briefly summarize the physical property of retarded Green's function for fermions. According to Eq.(\ref{Eq:Dirac-Psi_alpha-beta}), the equation for $\tilde\Psi_2$ is related to that of $\tilde\Psi_1$ by inversion of the momentum $\lambda \to -\lambda$,
e.g., for the case with Ricci flat hypersurface without chiral gauge field, $\lambda \to - \lambda$, thus it is obvious that
\beqa
G_2(\omega,\lambda) = G_1(\omega,-\lambda).  \label{Eq:GI-property-1}
\eeqa
As a result, the trace $\sim G_1+G_2$ and determinant $\sim G_1 G_2$ of $G^R$ are invariant under the inversion symmetry of momentum $\lambda \to -\lambda$.
Therefore, one can focus only on one of them, e.g., $G_1(\omega,\lambda)$, without loss of generality,
\beqa
 \sqrt\frac{g_{xx}(r)}{g_{rr}(r)}\partial_r \xi_\alpha = [K_0(r) -B_r(r)+ (-1)^\alpha \lambda] + [ K_0(r) - B_r(r) - (-1)^\alpha \lambda ]\xi_\alpha^2  - 2m_D\sqrt{g_{xx}(r)}\xi_\alpha. \qquad \label{Eq:flow-xi_alpha}
\eeqa
For $\alpha=1$, by neglecting the subscript, the flow equation in Eq.(\ref{Eq:flow-xi_alpha}) becomes
\beqa
 \sqrt\frac{g_{xx}(r)}{g_{rr}(r)}\partial_r \xi = [K_0(r) - B_r(r) - \lambda ] + [ K_0(r) - B_r(r) + \lambda ]\xi^2 - 2m_D\sqrt{g_{xx}(r)}\xi, \nn
\eeqa
where $\xi \equiv {F_1}/{G_1}$. With the in-falling boundary condition at the horizon for the $\omega\ne 0 $ case, i.e., $\xi |_{r=r_0} = i$, as given in Eq.(\ref{Eq:xi_alpha-horizon-omega!=0}), and according to Eq.(\ref{Eq:K0-B0-Br}), by taking
\beqa
 q \to -q, ~ \omega \to - \omega, ~ \lambda \to -\lambda, ~  b_r \to - b_r,  \Rightarrow   K_0(r) \to -K_0(r), ~ B_r(r) \to - B_r(r),
\eeqa
the flow equation for $ \xi \to -\xi $ doesn't change its form, which implies that
\beqa
G_\alpha(\omega,\lambda, q, b_r ) =- G_\alpha^\star(-\omega,-\lambda, -q, -b_r ),  \label{Eq:GI-property-2}
\eeqa
where the complex conjugation is due to that by flipping sign of frequency $\omega\to -\omega$, the in-falling horizon boundary condition turns out to be the out-going one. By dividing the flow equation by $\xi^2$, one has
\beqa
 -\sqrt\frac{g_{xx}(r)}{g_{rr}(r)}\partial_r  \xi^{-1}  =  [K_0(r) - B_r(r) - \lambda]\xi^{-2} + [ K_0(r) - B_r(r) + \lambda ] -  2m_D\sqrt{g_{xx}(r)}\xi^{-1}, \nn
\eeqa
where $\xi^{-1} \! \equiv \! {G}/{F}$ with subscripts omitted.
If one also takes
\beqa
 \lambda \to -\lambda, \quad m_D \to - m_D,
\eeqa
one can obtain an identical equation for $-1/\xi$,
\beqa
 \sqrt\frac{g_{xx}(r)}{g_{rr}(r)}\partial_r  (-\xi^{-1}) =  [K_0(r) - B_r(r) + \lambda]\xi^{-2} + [ K_0(r) - B_r(r) - \lambda ] -  2m_D\sqrt{g_{xx}(r)}(-\xi^{-1}),  \nn
\eeqa
which implies that the retarded Green's function for the standard quantization has the property
\beqa
G_\alpha(\omega,\lambda, m_D) = - G^{-1}_\alpha(\omega,-\lambda, -m_D). \label{Eq:GI-property-3}
\eeqa
Thus for the standard quantization, the quantity $\tilde{m}_\alpha \equiv -(-1)^\alpha \lambda \ell/r_\star $ carries with subscript $\alpha$, which flips the sign of eigenvalue $\lambda$, e.g., for the Ricci flat hypersurface, flips the momentum($|\vec{k}| $). In this case, by combining Eq.(\ref{Eq:GI-property-1}) and Eq.(\ref{Eq:GI-property-3}), one can obtain
\beqa
G_2(\omega,\lambda,m_D) = G_1(\omega,-\lambda,m_D) = - G_1^{-1}(\omega,\lambda, - m_D). \label{Eq:GI-property-4}
\eeqa

For the alternative quantization, by using Eq.(\ref{Eq:GI-property-3}), one has
\beqa
\3i
G_{a\alpha}(\omega,\lambda,m_D) & \equiv & - G_{s\alpha}^{-1}(\omega,\lambda,m_D) = G_{s\alpha}(\omega,-\lambda, -m_D), \label{Eq:GIt-property-1}
\eeqa
where subscripts $a$ and $s$ stand for alternative and standard quantization, respectively.
Thus the alternative quantization can be included by extending the mass range for the standard quantization for $G(\omega,\lambda,m_D)$ from $m_D\ell \ge 0$ to $m_D \ell>-1/2$. This implies that the alternative quantization is related to the negative mass range of the standard quantization. On the other hand, for $m_D=0$,
\beqa
\ii
G_2(\omega,\lambda,0) = G_1(\omega,-\lambda ,0)= - G_1^{-1}(\omega,\lambda,0),  \label{Eq:GI-property-4-mD=0}
\eeqa
which implies that
\beqa
\ii
\det G^R(\omega,\lambda) = G_1(\omega,\lambda,0) G_2(\omega,\lambda,0) = -1,
\eeqa
and for $m_D=0$, the alternative quantization is equivalently to the original one, since
\beqa
\ii
G_{a\alpha}(\omega,\lambda,0) \equiv - G_{s\alpha}^{-1}(\omega,\lambda,0) = G_{\alpha}(\omega,-\lambda,0), \label{Eq:GIt-property-1-mD=0}
\eeqa
Since neither basis change nor Lorentz rotation change the determinant of $G^R$, the equation above applies to any basis of Gamma matrices and any momentum. For the zero eigenvalue case $\lambda=0$, e.g., $\sqrt{|\vec{k}|^2-|\vec{b}|^2}=0$, with massless Dirac femion $m_D=0$,
\beqa
\ii\ii
G_2(\omega,0,0) = G_1(\omega,0,0) = - G_1^{-1}(\omega,0,0), ~ \lambda =0, \label{Eq:GI-property-5}
\eeqa
which gives
\beqa
G_1(\omega,0,0)=G_2(\omega,0,0)=i, \quad \lambda=0. \label{Eq:GI-property-6}
\eeqa

\subsubsection{IR boundary conditions of near horizon in-falling wave function}
\label{sec:Horizon_BCs}

With the metric in Eq.(\ref{Eq:gr-At-star-AdS(d+1)-T=0}) and gauge field in Eq.(\ref{Eq:At-AdS(3+1)}) in the Ricci flat case, one obtains the near horizon behavior of metric and gauge field as
\beqa
 g(r)  \overset{r\to r_\star}{=}     \frac{6(r-r_\star)^2}{r_\star^2}, \quad A_t(r)  \overset{r\to r_\star}{=}  \mu\frac{r-r_\star}{r_\star}, \label{Eq:g(r)-At(r)-rs}
\eeqa
where we have defined $q_0=\mu r_\star$.

Note that $g(r_\star)=0$ and $A_t(r_\star)=0$. It is worth emphasizing that the associated $K_0(r)$ defined in Eq.(\ref{Eq:K0-AdS}) near the horizon becomes,
\beqa
\ii
K_0(r)  \!\overset{r \to r_\star}{=} \!
\left\{ \begin{aligned}
        &\frac{1}{\sqrt{6}}\frac{r_\star}{r-r_\star}  \omega, \!&\! \omega\ne 0; \\
                  & \frac{1}{\sqrt{6}} q \mu, \!&\! \omega = 0.
                          \end{aligned} \right. \quad B_0(r)  \!\overset{r \to r_\star}{=}\!  \frac{1}{\sqrt{6}}\frac{r_\star}{r-r_\star}b_0, \quad
B_r(r)  \!\overset{r \to r_\star}{=}\!  \sqrt{6}\frac{r_\star}{\ell^2}(r-r_\star) b_r, \label{Eq:K0-AdS_rs}
\eeqa
which has singularity for the finite frequency case with $\omega\ne 0$, while it is regular for the zero frequency case with $\omega=0$.

The EOM can be approximated by using Eq.(\ref{Eq:DiracEOM-flow-xi_alpha-UV-AdS(d+1)-omega!=0}).
\begin{enumerate}
\item Finite frequency case ($\omega\ne 0$): In the presence of chiral gauge field, according to Eq.(\ref{Eq:DiracEOM-flow-xi_alpha-UV-AdS(d+1)-omega!=0}), the term with $\omega$ $ \sim (r-r_\star)^{-2} $ dominates over the term with $k$ and Dirac mass term with $m_D$ (both $ \sim (r-r_\star)^{-1} $), since $g(r)\sim (r-r_\star)^2$. Then we have
\beqa
&&  \textbf{1}_2  \partial_r   \tilde\Phi_{\alpha,\beta} \!=\!    \frac{\ell^2}{r^2} \frac{1}{g(r)} i(\omega \sigma^2 + \beta b_0 \textbf{1}_2)  \tilde\Phi_{\alpha,\beta}, \quad \frac{\ell^2}{r^2} \frac{1}{g(r)} \overset{r\to r_\star}{=} \frac{\ell^2}{r_\star^2}\frac{1}{d(d-1)}\frac{r_\star^2}{(r-r_\star)^2} = \frac{\ell_2^2}{(r-r_\star)^2} \\
(a)&& \xi_\alpha^\prime =   \frac{\ell^2}{r^2} \frac{1}{g(r)}\omega ( 1 \! + \! \xi_\alpha^2 ) , \nn \\
(b)&& \xi_{\alpha\alpha^\prime}^\prime =   \frac{\ell^2}{r^2} \frac{1}{g(r)}    \omega   (\xi_{\alpha}^{-1} \! + \! \xi_{\alpha^\prime})  \xi_{\alpha\alpha^\prime}, \quad \xi_{21}^\prime =   \frac{\ell^2}{r^2} \frac{1}{g(r)}    \omega  (\xi_{2}^{-1} \! + \! \xi_{1})  \xi_{21}, \nn \\
(c)&& \xi_{\alpha,+-}^\prime =
  \frac{\ell^2}{r^2} \frac{1}{g(r)} [  \omega (\xi_{\alpha}^{-1} \! + \! \xi_{\alpha})   +  2 i  b_0 ] \xi_{\alpha,+-}, \quad \xi_{\alpha,-+}^\prime =
  \frac{\ell^2}{r^2} \frac{1}{g(r)} [  \omega (\xi_{\alpha}^{-1} \! + \! \xi_{\alpha})   -  2 i  b_0 ] \xi_{\alpha,-+}, \nn \\
(d)&& \xi_{\alpha\alpha^\prime,\beta\beta^\prime}^\prime =
  \frac{\ell^2}{r^2} \frac{1}{g(r)} [  \omega (\xi_{\alpha}^{-1} \! + \! \xi_{\alpha^\prime})  +  i  (\beta - \beta^\prime )  b_0 ] \xi_{\alpha\alpha^\prime,\beta\beta^\prime}, \label{Eq:flow-xi_alpha-alphap-beta-betap}
\eeqa
where $\ell_2$ is defined in Eq.(\ref{Eq:ell_2}).

Note that here we have neglected the mass $m_D$, since the mass term is not important at the horizon with a non-vanishing momentum $\omega$. Compared to the frequency term, the momentum $\lambda$ term can also be neglected. By using the transformation that $U_2^{-1} \sigma^2 U_2 = \sigma^3$, the EOM can be transformed into
\beqa
\textbf{1}_2  \partial_r  ( U_2^{-1} \tilde\Phi_{\alpha,\beta} ) \!=\!    \frac{\ell^2}{r^2} \frac{1}{g(r)} i(\omega \sigma^3 + \beta b_0 \textbf{1}_2) ( U_2^{-1} \tilde\Phi_{\alpha,\beta} ) \overset{r\to r_\star}{=} \frac{\ell_2^2}{(r-r_\star)^2} i(\omega \sigma^3 + \beta b_0 \textbf{1}_2) ( U_2^{-1} \tilde\Phi_{\alpha,\beta} ),
\eeqa
where $U_2$ is given in Eq.(\ref{Eq:transU2_tPsi_chi}).
For the above EOM, one has the solution,
\beqa
\ii U_2^{-1} \tilde\Phi_{\alpha,\beta} \!=\! e^{i \beta \theta_b(r)}\bigg( C_{\alpha,+} e^{i \theta(r) }\left(
 \begin{array}{c}
 1 \\
 0 \\
 \end{array}
 \right)
 \!+\! C_{\alpha,-} e^{-i \theta(r) }\left(
 \begin{array}{c}
 0 \\
 1 \\
 \end{array}
 \right) \bigg), \label{Eq:Pinv-tPsi_alpha-tPhi_alpha-beta-1}
\eeqa
where the phase factors are given by
\beqa
\theta(r) \equiv \int dr    \frac{\ell^2}{r^2} \frac{1}{g(r)} \omega , \quad \theta_b(r) \equiv \int dr   \frac{\ell^2}{r^2}  \frac{1}{g(r)} b_0 .  \label{Eq:theta-theta_b}
\eeqa
It is worthy to notice that $\theta(r)$ and $\theta_b(r)$ are vanishing at the infinite boundary, i.e., $\theta(r),\theta_b(r)\overset{r\to\infty}{=}0$.
In the near horizon limit, the phase factors are integrable and can be expressed explicitly as
\beqa
 \theta(r)  \overset{r\to r_\star}{=}  \int  dr \frac{\ell_2^2}{(r-r_\star)^2}\omega = -\frac{\ell_2^2}{r-r_\star}\omega, \quad \theta_b(r) \overset{r\to r_\star}{=} -\frac{\ell_2^2}{r-r_\star}  b_0. \label{Eq:theta-theta_b_2}
\eeqa

Therefore, one obtains the solutions to EOMs from Eq.(\ref{Eq:Pinv-tPsi_alpha-tPhi_alpha-beta-1}) and to the flow equations from Eqs.(\ref{Eq:flow-xi_alpha-alphap-beta-betap}),
\beqa
&& \ii \tilde\Phi_{\alpha,\beta} \sim \left(
                         \begin{array}{c}
                           \tilde{F}_{\alpha,\beta} \\
                           \tilde{G}_{\alpha,\beta} \\
                         \end{array}
                       \right)
 \!=\! \frac{e^{i \beta \theta_b(r)}}{\sqrt{2}} \left(
                                         \begin{array}{cc}
                                           \cos{\theta(r)} & \sin{\theta(r)} \\
                                           -\sin{\theta(r)} & \cos{\theta(r)} \\
                                         \end{array}
                                       \right) \left(
                                                 \begin{array}{c}
                                                   C_{\alpha,1} \\
                                                   C_{\alpha,2} \\
                                                 \end{array}
                                               \right),  \nn \\
(a)&& \xi_\alpha(r) \overset{r\to r_\star}{=} \tan{[C_{\alpha,3} + \theta(r)]} = \frac{\cos\theta(r)C_{\alpha,1} C_{\alpha,2}^{-1} + \sin\theta(r) }{-\sin\theta(r)C_{\alpha,1} C_{\alpha,2}^{-1} + \cos\theta(r) } , \nn \\
(b)&& \xi_{\alpha\alpha^\prime}(r) \overset{r\to r_\star}{=} C_{\alpha\alpha^\prime} \frac{\sin{[C_{\alpha,3}+\theta(r)]}}{\cos{[C_{\alpha^\prime,3}+\theta(r)]}} = \frac{\cos\theta(r)C_{\alpha,1} C_{\alpha,2}^{-1} + \sin\theta(r) }{-\sin\theta(r)C_{\alpha^\prime,1} C_{\alpha^\prime,2}^{-1} + \cos\theta(r) }, \quad C_{\alpha\alpha^\prime} = \frac{C_{\alpha,3}}{C_{\alpha^\prime,3}} \frac{\cos{C_{\alpha^\prime,3}}}{\cos{C_{\alpha,3}}}, \nn\\
(c)&& \xi_{\alpha,\beta\beta^\prime} \overset{r\to r_\star}{=} C_{\alpha,\beta\beta^\prime}e^{i(\beta-\beta^\prime)\theta_b(r)}\tan{[C_{\alpha,3}+\theta(r)]}, \quad C_{\alpha,\beta\beta^\prime} = C_{\alpha\alpha} = 1, \nn\\
(d)&& \xi_{\alpha\alpha^\prime,\beta\beta^\prime} \overset{r\to r_\star}{=} C_{\alpha\alpha^\prime,\beta\beta^\prime} e^{i(\beta-\beta^\prime)\theta_b(r)}\frac{\sin{[C_{\alpha,3}+\theta(r)]}}{\cos{[C_{\alpha^\prime,3}+\theta(r)]}}, \quad C_{\alpha\alpha^\prime,\beta\beta^\prime} = C_{\alpha\alpha^\prime}, \label{Eq:xi_rs}
\eeqa
where $C_{\alpha,\pm} =(C_{\alpha,2} \pm i C_{\alpha,1})/2 $ and $C_{\alpha,3} \equiv \arctan{C_{\alpha,1} C_{\alpha,2}^{-1}}$. Notice that the time component of chiral gauge field $b_0$ will affect the Green's function through its presence as a phase factor of the wave functions. In the following, let us give the near horizon boundary conditions (BCs) for the flow equations.

\begin{enumerate}

\item[(1)] If $b_0 = 0$, according to Eq.(\ref{Eq:theta-theta_b}) and Eq.(\ref{Eq:theta-theta_b_2}), it is direct to observe that the in-falling wave solution of Eq.(\ref{Eq:Pinv-tPsi_alpha-tPhi_alpha-beta-1}) near the horizon is $e^{-i\omega t } e^{-i \theta(r)}  = e^{-i \omega [t + \omega^{-1}\theta(r) ]} \overset{r\to r_\star}{=} e^{ - i \omega[t - \ell_2^2/(r-r_\star) ]} $, since $\theta(r)$ is an increasing function of $r$. Therefore, imposing in-falling wave boundary condition, is equivalent to turning off the out-going wave solution, namely, by setting $C_{\alpha,+}=0$, which implies that $C_{\alpha,2}=-i C_{\alpha,1}$, or $C_{\alpha,1}=iC_{\alpha,2}$, or equivalently, $C_{\alpha,3} \equiv \arctan{(C_{\alpha,1}C_{\alpha,2}^{-1})} = \arctan(i)=i\infty$. The in-falling wave boundary conditions are
\beqa
(a)(c) ~~ \xi_\alpha|_{r\to r_\star} = i ; \quad (b)(d) ~~ \xi_{\alpha\alpha^\prime}|_{r\to r_\star} = i. \label{Eq:xi_alpha-horizon-omega!=0_b0=0}
\eeqa

\item[(2)] If $b_0 \ne 0$, near the horizon, $e^{-i\omega t } e^{-i \theta(r)} e^{i\beta\theta_b(r)} = e^{ - i \omega[t - \ell_2^2/(r-r_\star) ]} e^{ - i \beta b_0 \ell_2^2/(r-r_\star) } $ is in-falling wave, while $e^{-i\omega t } e^{i \theta(r)} e^{i \beta \theta_b(r)} = e^{ - i \omega[t + \ell_2^2/(r-r_\star) ]} e^{ - i  \beta b_0\ell_2^2/(r-r_\star) } $ is out-going wave. As the case with $b_0=0$, the in-falling wave boundary condition is $C_{\alpha,+}=0$, which implies that $C_{\alpha,1}=iC_{\alpha,2}$, or equivalently, $C_{\alpha,3} = \arctan(i)$.

By imposing the in-falling wave boundary conditions to the wavefunctions in Eq.(\ref{Eq:Pinv-tPsi_alpha-tPhi_alpha-beta-1}), one obtains
\beqa
\tilde\Phi_{\alpha,\beta} \sim \left(
                         \begin{array}{c}
                           \tilde{F}_{\alpha,\beta} \\
                           \tilde{G}_{\alpha,\beta} \\
                         \end{array}
                       \right) = e^{i \beta \theta_b(r)}  C_{\alpha,-} e^{-i \theta(r) } P \left(
 \begin{array}{c}
 0 \\
 1 \\
 \end{array}
 \right) \overset{r\to r_\star}{=}  C_{\alpha,-} e^{i \frac{\ell_2^2}{r-r_\star}(\omega - \beta b_0)  }  \left(
 \begin{array}{c}
 i \\
 1 \\
 \end{array}
 \right).
\eeqa
According to Eq.(\ref{Eq:xi_rs}), the in-falling wave boundary conditions for the flow equation functions $\xi$ are
\beqa
(a)&& \xi_\alpha(r)|_{r\to r_\star} = \tan{[C_{\alpha,3}+\theta(r)]} = i, \nn\\
(b)&& \xi_{\alpha\alpha^\prime}(r)|_{r\to r_\star} = \frac{C_{\alpha,3}}{C_{\alpha^\prime,3}} \frac{\cos{C_{\alpha^\prime,3}}}{\cos{C_{\alpha,3}}} \frac{\sin{[C_{\alpha,3}+\theta(r)]}}{\cos{[C_{\alpha^\prime,3}+\theta(r)]}} = i, \nn\\
(c)&& \xi_{\alpha,\beta\beta^\prime}(r)|_{r\to r_\star} = e^{i(\beta-\beta^\prime)\theta_b(r)}\tan{[C_{\alpha,3}+\theta(r)]} = i e^{-i(\beta-\beta^\prime) \frac{\ell_2^2}{r-r_\star}b_0} , \nn\\
(d)&& \xi_{\alpha\alpha^\prime,\beta\beta^\prime}(r)|_{r\to r_\star}  = e^{-i(\beta-\beta^\prime) \frac{\ell_2^2}{r-r_\star}b_0} \frac{C_{\alpha,3}}{C_{\alpha^\prime,3}} \frac{\cos{C_{\alpha^\prime,3}}}{\cos{C_{\alpha,3}}} \frac{\sin{[C_{\alpha,3}+\theta(r)]}}{\cos{[C_{\alpha^\prime,3}+\theta(r)]}} = i e^{-i(\beta-\beta^\prime) \frac{\ell_2^2}{r-r_\star}b_0}.
\eeqa
Therefore, in the near horizon limit, one obtains the complete in-falling wave boundary conditions for the flow equations,
\beqa
(a)(b) && \xi_{\alpha}(r)|_{r\to r_\star} = \xi_{\alpha\alpha^\prime}(r)|_{r\to r_\star} = i ;\nn\\
(c) && \xi_{\alpha,+-}(r)|_{r\to r_\star}  = i e^{-2i \frac{\ell_2^2 b_0}{r-r_\star}} , \quad \xi_{\alpha,-+}(r)|_{r\to r_\star}  = i e^{+2i \frac{\ell_2^2 b_0}{r-r_\star}} ,\nn\\
(d) && \xi_{\alpha\alpha^\prime,\beta\beta^\prime}(r)|_{r\to r_\star} = i e^{-i(\beta-\beta^\prime) \frac{\ell_2^2 b_0}{r-r_\star}}. \label{Eq:xi_alpha-horizon-omega!=0}
\eeqa
One can see that the IR BCs in Eq.(\ref{Eq:xi_alpha-horizon-omega!=0}) just recover Eq.(\ref{Eq:xi_alpha-horizon-omega!=0_b0=0}), if $b_0=0$. It is also worth noticing that $b_0$ can be viewed as an IR observable that interpolates to the UV physics, since the phase factor controls the influence of $b_0$ in the near horizon limit, but vanishes completely in the infinite boundary, i.e., $\exp{(-i(\beta-\beta^\prime){\ell_2^2 b_0}{(r-r_\star)^{-1}})}\overset{r\to \infty}{\to} 0 $. Consequently, one would expect that the chiral anomaly will only modify the IR behavior of UV retarded Green's function $G^R$, but not affect the UV behavior of $G^R$ in the infinite boundary.

\end{enumerate}

In summary, in the case of $b_0=0$, it is unnecessary to introduce the unitary matrix $U_1$ in Eq.(\ref{Eq:transU1_tPsi-Psi}) for transformation, thus $G^R_\alpha$ can be obtained from $\tilde{G}_\alpha^R$, since the subscript $\beta$ can be dropped. In this case, the near horizon BCs for $\xi_{\alpha}|_{r\to r_\star}=i$. These results are consistent with Eq.(25) in ref.~\cite{Liu:2009dm}. While it is worthy to notice that, in the presence of $b_0\ne 0$, the helicity states are non-degenerate, namely the subscript $\beta$ are physical distinguishable, so that they can not be dropped. Consequently, it is required to entail the near horizon BCs, not only for $\xi_\alpha$, but also for both $\xi_{\alpha,+-}$ and $\xi_{\alpha,-+}$. When $b_0$ is non-zero, it turns out that the near horizon BCs for the flow equations are associated with a phase factor $e^{-i(\beta-\beta^\prime)\ell_2^2b_0(r-r_\star)}$.

\item Zero frequency case ($\omega=0$): In this case, according to Eq.(\ref{Eq:DiracEOM-flow-xi_alpha-UV-AdS(d+1)-omega!=0}), the Dirac equations and flow equations become,
\beqa
&& \bigg[  \textbf{1}_2 \bigg( \partial_r \!-\! i \beta    \frac{\ell^2}{r^2} \frac{1}{g(r)} b_0\bigg) \!+\! \sigma^3 m_D  \frac{\ell}{r} \frac{1}{\sqrt{g(r)}} \bigg] \tilde\Phi_{\alpha,\beta} \!=\! \bigg( i\sigma^2   \frac{\ell^2}{r^2} \frac{1}{g(r)} q A_t(r) \!+\!  (-1)^\alpha \sigma^1  \frac{\ell^2}{r^2} \frac{1}{\sqrt{g(r)}} \lambda\bigg) \tilde\Phi_{\alpha,\beta}, \nn\\
(a)&& \xi_\alpha^\prime =  \frac{\ell^2}{r^2} \frac{1}{\sqrt{g(r)}} \bigg[ \bigg(  \frac{1}{\sqrt{g(r)}} q A_t(r)  + (-1)^\alpha\lambda \bigg)  \! + \! \bigg( \frac{1}{\sqrt{g(r)}} q A_t(r) -(-1)^{\alpha}\lambda \bigg) \xi_\alpha^2 - 2m_D \frac{r}{\ell} \xi_\alpha\bigg] , \nn \\
(b)&& \xi_{12}^\prime =  \frac{\ell^2}{r^2} \frac{1}{\sqrt{g(r)}} \bigg[ \bigg(  \frac{1}{\sqrt{g(r)}} q A_t(r) - \lambda \bigg) (\xi_{1}^{-1} \! + \! \xi_{2}) - 2m_D \frac{r}{\ell} \bigg] \xi_{12}, \nn \\
&& \xi_{21}^\prime =  \frac{\ell^2}{r^2} \frac{1}{\sqrt{g(r)}} \bigg[ \bigg(  \frac{1}{\sqrt{g(r)}} q A_t(r) + \lambda \bigg) (\xi_{2}^{-1} \! + \! \xi_{1}) - 2m_D \frac{r}{\ell}\bigg] \xi_{21}, \nn \\
(c)&& \xi_{\alpha,+-}^\prime \!=\!
 \frac{\ell^2}{r^2} \frac{1}{\sqrt{g(r)}} \bigg[ \bigg(  \frac{1}{\sqrt{g(r)}} q A_t(r) + (-1)^\alpha\lambda \bigg) \xi_{\alpha}^{-1} \! + \! \bigg(  \frac{1}{\sqrt{g(r)}}  q A_t(r)  \!-\! (-1)^{\alpha}\lambda \bigg) \xi_{\alpha} \!-\! 2m_D\frac{r}{\ell} \!+\! 2 i  \frac{1}{\sqrt{g(r)}} b_0 \bigg] \xi_{\alpha,+-}, \nn \\
&& \xi_{\alpha,-+}^\prime \!=\!
 \frac{\ell^2}{r^2} \frac{1}{\sqrt{g(r)}} \bigg[ \bigg(  \frac{1}{\sqrt{g(r)}}  q A_t(r) +(-1)^\alpha\lambda \bigg) \xi_{\alpha}^{-1} \! + \! \bigg(  \frac{1}{\sqrt{g(r)}}  q A_t(r)  \!-\! (-1)^{\alpha}\lambda \bigg) \xi_{\alpha} \!-\! 2m_D\frac{r}{\ell} \!-\! 2 i  \frac{1}{\sqrt{g(r)}} b_0 \bigg] \xi_{\alpha,-+}, \nn \\
(d)&& \xi_{12,\beta\beta^\prime}^\prime =
 \frac{\ell^2}{r^2} \frac{1}{\sqrt{g(r)}} \bigg[ \bigg(  \frac{1}{\sqrt{g(r)}}  q A_t(r) - \lambda \bigg) (\xi_{1}^{-1} \! + \! \xi_{2}) - 2m_D \frac{r}{\ell} +  i  (\beta - \beta^\prime )  \frac{1}{\sqrt{g(r)}} b_0 \bigg] \xi_{12,\beta\beta^\prime}, \nn \\
&& \xi_{21,\beta\beta^\prime}^\prime =
 \frac{\ell^2}{r^2} \frac{1}{\sqrt{g(r)}} \bigg[ \bigg(  \frac{1}{\sqrt{g(r)}}  q A_t(r) + \lambda \bigg) (\xi_{2}^{-1} \! + \! \xi_{1}) - 2m_D \frac{r}{\ell} +  i  (\beta - \beta^\prime )  \frac{1}{\sqrt{g(r)}} b_0 \bigg] \xi_{21,\beta\beta^\prime}. \label{Eq:DiracEOM-flow-xi-alpha-UV-AdS(d+1)-omega=0}
\eeqa
It is worth observing that in both the field equations and flow equations, in the near horizon region, $b_0$ term is more singular than other terms, i.e., there is an additional $(r-r_\star)^{-1}$ factor, with respect to the other terms in the $r\to r_\star$ limit. Therefore, the $b_0$ term is a relevant term dominant at the IR boundary if $b_0\ne 0$.

In the near horizon limit, i.e., $r\to r_\star$, according to Eqs.(\ref{Eq:g(r)-At(r)-rs}), we obtain
\beqa
  \frac{\ell^2}{r^2} \frac{1}{g(r)} \overset{r\to r_\star}{=}  \frac{\ell_2^2}{(r-r_\star)^2}, \quad  \frac{\ell^2}{r^2} \frac{1}{\sqrt{g(r)}} \overset{r\to r_\star}{=} \frac{\ell}{r_\star} \frac{\ell_2}{r-r_\star}, \quad  \frac{\ell}{r} \frac{1}{\sqrt{g(r)}} A_t(r) \overset{r\to r_\star}{=} \mu\frac{\ell_2}{r_\star} = \frac{e_3}{\ell_2},
\eeqa
where $\ell_2\equiv {\ell}/{\sqrt{6}}$ defined through Eq.(\ref{Eq:ell_2}), and the definition of effective IR gauge coupling through chemical potential, $e_3 \equiv q_0 \ell_2^2/r_\star^2 = \mu \ell_2^2/r_\star$ defined in Eq.(\ref{Eq:ed-r_star}). It is clear that the term $ \sim A_t(r)/\sqrt{g(r)} $ is at the same order as the Dirac mass term $m_D$ and $\lambda$ term, since $A_t(r)\sim (r-r_\star)$ and $g(r)\sim (r-r_\star)$ in the near horizon region.

In the near horizon region $r\sim r_\star$, according to Eq.(\ref{Eq:DiracEOM-flow-xi-alpha-UV-AdS(d+1)-omega=0}), the Dirac equation is dominated by
\beqa
\bigg[  \textbf{1}_2 \bigg( \partial_r \!-\! i \beta   \frac{\ell_2^2}{(r-r_\star)^2} b_0\bigg) \!+\! \sigma^3 m_D \frac{\ell_2}{r-r_\star} \bigg] \tilde\Phi_{\alpha,\beta} \!=\! \bigg( i\sigma^2 \frac{q e_3}{r-r_\star} \!+\!  (-1)^\alpha \sigma^1 \frac{\ell_2}{r-r_\star} \frac{\ell}{r_\star}  \lambda\bigg) \tilde\Phi_{\alpha,\beta}, \nn
\eeqa
where the term associated with $b_0$, has a double pole singularity since $g(r) \sim (r-r_\star)^2$, while all other terms have only single pole singularities, i.e., $\sqrt{g(r)}\sim (r-r_\star)$. This implies that temporal component of chiral gauge field, i.e., the term associated with $b_0$, will be the dominant relevant term near the horizon, unless $b_0(r)$ has the near horizon behavior as that for the pole of $\sqrt{g(r)}$. In this paper, we consider that $b_0$ is a constant as the simplest example.

In the near horizon region, $r\sim r_\star$, the Dirac equations and flow equations in Eq.(\ref{Eq:DiracEOM-flow-xi-alpha-UV-AdS(d+1)-omega=0}), can be simplified as
\beqa
&& \bigg[  \textbf{1}_2 \bigg( \partial_r \!-\! i \beta   \frac{\ell_2^2}{(r-r_\star)^2} b_0\bigg) \!+\!  \frac{\sigma^3 m_D\ell_2 - i \sigma^2 q e_3 + \sigma^1 \tilde{m}_\alpha \ell_2 }{r-r_\star} \bigg] \tilde\Phi_{\alpha,\beta} \!=\! 0, \label{Eq:EOM-Diracfermion-IR-AdS(d+1)-omega=0-b0!=0}  \\
(a)&& \xi_\alpha^\prime =  \frac{q e_3  (1+\xi_\alpha^2)  \! - \! \tilde{m}_\alpha \ell_2 (1-\xi_\alpha^2) - 2m_D \ell_2 \xi_\alpha}{r-r_\star} , \nn \\
(b)&& \xi_{12}^\prime =  \frac{( q e_3  - \tilde{m} \ell_2 ) (\xi_{1}^{-1} \! + \! \xi_{2}) - 2m_D \ell_2}{r-r_\star}  \xi_{12}, \quad \xi_{21}^\prime =  \frac{( q e_3  + \tilde{m} \ell_2 ) (\xi_{2}^{-1} \! + \! \xi_{1}) - 2m_D \ell_2}{r-r_\star}  \xi_{21}, \nn \\
(c)&& \xi_{\alpha,+-}^\prime \!=\!
\bigg( \frac{q e_3(\xi_{\alpha}^{-1}+\xi_{\alpha}) - \tilde{m}_\alpha \ell_2 (\xi_{\alpha}^{-1}-\xi_{\alpha}) \!-\! 2m_D \ell_2}{r-r_\star} \!+\! 2 i \frac{\ell_2^2}{(r-r_\star)^2} b_0  \bigg) \xi_{\alpha,+-}, \nn \\
&& \xi_{\alpha,-+}^\prime \!=\!
\bigg( \frac{q e_3(\xi_{\alpha}^{-1}+\xi_{\alpha}) - \tilde{m}_\alpha \ell_2 (\xi_{\alpha}^{-1}-\xi_{\alpha}) \!-\! 2m_D \ell_2}{r-r_\star} \!-\! 2 i \frac{\ell_2^2}{(r-r_\star)^2} b_0 \bigg) \xi_{\alpha,-+}, \nn \\
(d)&& \xi_{12,\beta\beta^\prime}^\prime =
\bigg( \frac{( q e_3  - \tilde{m} \ell_2 ) (\xi_{1}^{-1} \! + \! \xi_{2}) - 2m_D \ell_2}{r-r_\star}  +  i  (\beta - \beta^\prime ) \frac{\ell_2^2}{(r-r_\star)^2} b_0 \bigg) \xi_{12,\beta\beta^\prime}, \nn \\
&& \xi_{21,\beta\beta^\prime}^\prime =
\bigg( \frac{( q e_3  + \tilde{m} \ell_2 ) (\xi_{2}^{-1} \! + \! \xi_{1}) - 2m_D \ell_2}{r-r_\star}  +  i  (\beta - \beta^\prime ) \frac{\ell_2^2}{(r-r_\star)^2} b_0 \bigg) \xi_{21,\beta\beta^\prime}. \label{Eq:xi_alpha-horizon-omega=0-b0!=0}
\eeqa
where
\beqa
\tilde{m}_\alpha \equiv -(-1)^{\alpha} \lambda \frac{\ell}{r_\star}. \label{Eq:mt_alpha}
\eeqa
and $\tilde{m}\equiv \tilde{m}_1=\lambda \ell/r_\star$.

In the following, we will discuss the solution in the cases with/without $b_0$.
\begin{enumerate}
\item[(1)] If $b_0=0$, the Dirac EOM and flow equation can be simplified as
\beqa
&&   \textbf{1}_2  \tilde\Phi_{\alpha}^\prime  \!=\!  \frac{i \sigma^2 q e_3 - \sigma^1 \tilde{m}_\alpha \ell_2 - \sigma^3 m_D\ell_2 }{r-r_\star}  \tilde\Phi_{\alpha} , \label{Eq:EOM-Diracfermion-IR-AdS(d+1)-omega=0-b0=0} \\
(a)(c)&& \xi_\alpha^\prime =  \frac{q e_3  (1+\xi_\alpha^2)  \! - \! \tilde{m}_\alpha \ell_2 (1-\xi_\alpha^2) - 2m_D \ell_2 \xi_\alpha}{r-r_\star} , \nn \\
(b)(d)&& \xi_{12}^\prime =  \frac{( q e_3  - \tilde{m} \ell_2 ) (\xi_{1}^{-1} \! + \! \xi_{2}) - 2m_D \ell_2}{r-r_\star}  \xi_{12}, \quad \xi_{21}^\prime =  \frac{( q e_3  + \tilde{m} \ell_2 ) (\xi_{2}^{-1} \! + \! \xi_{1}) - 2m_D \ell_2}{r-r_\star}  \xi_{21}. \nn
\eeqa
where the helicity $\beta=\pm$ can be dropped, due to the absence of $b_0$.

The eigenvalues and eigenvectors of wavefunctions to the Dirac EOM can be solved through a transformation matrix $S$, which is independent of radial coordinate $r$,
\beqa
S^{-1} (i \sigma^2 q e_3 - \sigma^1 \tilde{m}_\alpha \ell_2 - \sigma^3 m_D\ell_2) S = \nu_\lambda \sigma^3, \quad S \equiv \left(
            \begin{array}{cc}
              \frac{m_D\ell_2 - \nu_\lambda}{q e_3 + \tilde{m}_\alpha \ell_2} & \frac{m_D\ell_2 + \nu_\lambda}{q e_3 + \tilde{m}_\alpha \ell_2} \\
              1 & 1 \\
            \end{array}
          \right),
\eeqa
where $\nu_\lambda \equiv \sqrt{(m_D^2+\tilde{m}^2)\ell_2^2- q^2 e_3^2}$. Then the original Dirac EOM becomes an eigen-value equation with diagonal eigen-values, namely,
\beqa
(S^{-1} \tilde\Phi_\alpha)^\prime  = \frac{1}{r-r_\star}\nu_\lambda \sigma^3 (S^{-1} \tilde\Phi_\alpha) ,
\eeqa
where the solution to the EOM turns out to be
\beqa
S^{-1}\tilde\Phi_\alpha = C_{\alpha,+} (r-r_\star)^{+\nu_\lambda} \left(
                                                    \begin{array}{c}
                                                      1 \\
                                                      0 \\
                                                    \end{array}
                                                  \right) +  C_{\alpha,-} (r-r_\star)^{-\nu_\lambda} \left(
                                                    \begin{array}{c}
                                                      0 \\
                                                      1 \\
                                                    \end{array}
                                                  \right), \nn
\eeqa
where $C_{\pm}$ are two integration constant.
By using the definition of $\tilde{R}_\alpha$ as below, the solution to the wave function $\tilde\Phi_\alpha  \sim   \tilde{R}_\alpha$ is obtained,
\beqa
 \tilde{R}_\alpha  \equiv  \left(
                                          \begin{array}{c}
                                            \tilde{F}_\alpha \\
                                            \tilde{G}_\alpha \\
                                          \end{array}
                                        \right) & \sim & C_{\alpha,+} (r-r_\star)^{+\nu_\lambda} \left(
            \begin{array}{cc}
              \frac{m_D\ell_2 - \nu_\lambda}{q e_3 + \tilde{m}_\alpha \ell_2}  \\
              1 \\
            \end{array}
          \right)  + C_{\alpha,-} (r-r_\star)^{-\nu_\lambda}\left(
            \begin{array}{cc}
               \frac{m_D\ell_2 + \nu_\lambda}{q e_3 + \tilde{m}_\alpha \ell_2} \\
              1 \\
            \end{array}
          \right)  \nn\\
&=& c_{\alpha,+} v_{\alpha,-}  (r-r_\star)^{+\nu_\lambda}   + c_{\alpha,-}  v_{\alpha,+} (r-r_\star)^{-\nu_\lambda}, \nn
\eeqa
where we have chosen the conversion, e.g., (A33) in ref.~\cite{Faulkner:2009wj}, in order to compare with the results without chiral gauge field,
\beqa
v_{\alpha,\mp} \equiv \left(
            \begin{array}{cc}
               m_D\ell_2 \mp \nu_\lambda \\
              q e_3 + \tilde{m}_\alpha \ell_2 \\
            \end{array}
          \right), \quad c_{\alpha,\pm} = \frac{C_{\alpha,\pm}}{q e_3 + \tilde{m}_\alpha \ell_2}, \label{Eq:vpm}
\eeqa
from which, one obtains
\beqa
\tilde{R}_\alpha \sim (c_{\alpha,+} v_{\alpha,-} + c_{\alpha,-} v_{\alpha,+})   \cosh[\nu_\lambda \log(r-r_\star) ] + (c_{\alpha,+} v_{\alpha,-} - c_{\alpha,-} v_{\alpha,+})\sinh[\nu_\lambda \log(r-r_\star) ].
\eeqa
Thus the flow function can be analytically expressed as
\beqa \xi_\alpha(r) \equiv \frac{\tilde{F}_\alpha}{\tilde{G}_\alpha} = \frac{m_D \ell_2 - \nu_\lambda \tau_\alpha(r)}{q e_3 + \tilde{m}_\alpha \ell_2},\label{Eq:F_alpha-G_alpha-omega=0}
\eeqa
where
\beqa
\tau_\alpha(r) \equiv 1 - \frac{2c_{\alpha,-}}{c_{\alpha,-} + c_{\alpha,+} e^{2\nu_\lambda \ln{(r-r_\star)}}} = 1 - \frac{2C_{\alpha,-}}{C_{\alpha,-} + C_{\alpha,+} e^{2\nu_\lambda \ln{(r-r_\star)}}}. \label{Eq:tau(r)}
\eeqa

In the oscillatory region the IR CFT scaling weight becomes imaginary, i.e., $\nu_\lambda = -i \rho_\lambda$, with $\rho_\lambda>0$ then one obtains a general wave solution with both the in-falling and out-going modes, \beqa
\tilde{R}_\alpha \sim  c_{\alpha,+} v_{\alpha,-} (r-r_\star)^{-i\rho_\lambda} + c_{\alpha,-} v_{\alpha,+} (r-r_\star)^{+i \rho_\lambda}.
\eeqa

The in-falling mode is
$e^{-i\omega t}  (r-r_\star)^{-i\rho_\lambda} = e^{-i \omega [t + \omega^{-1}\rho_\lambda\ln{(r-r_\star)}  ] }$, thus one needs to impose the in-falling boundary condition at the horizon, which is equivalent to turning off the out-going mode, i.e., $c_{\alpha,-}=0 $, or equivalently $C_{\alpha,-}=0$, to make the out-going wave $(r-r_\star)^{+i \rho_\lambda} $ vanishes. In this case, $\tau(r)=1$, the solution can be re-expressed as $\tilde\Psi_\alpha \sim \tilde{R}_\alpha$,
\beqa
 \tilde{R}_\alpha  \sim c_{\alpha,+} v_{\alpha,-} (r-r_\star)^{\nu_\lambda} = c_{\alpha,+} v_{\alpha,-} e^{\nu_\lambda \ln{(r-r_\star)}}, \quad \frac{\tilde{F}_\alpha}{\tilde{G}_\alpha } = \frac{v_{\alpha,-\uparrow}}{v_{\alpha,-\downarrow}} = \frac{m_D \ell_2 - \nu_\lambda}{q e_3 + \tilde{m}_\alpha \ell_2}.
\eeqa
where we have introduced $v_{\alpha,-\uparrow}$ and $v_{\alpha,-\downarrow}$ to label the up and down components of the vector $v_{\alpha,-}$.  The general solutions to the flow equations in Eqs.(\ref{Eq:EOM-Diracfermion-IR-AdS(d+1)-omega=0-b0=0}), at $\omega=0$, turn out to be
\beqa
(a)(c)&& \xi_{\alpha}(r) = \frac{m_D \ell_2 - \nu_\lambda \tanh{(\nu_\lambda(C_{\alpha,3}+\log{(r-r_\star)}))}}{q e_3 + \tilde{m}_\alpha \ell_2},  \nn\\
(b)(d)&& \xi_{\alpha\alpha^\prime}(r) = C_{\alpha}\bigg( m_D \ell_2 \frac{\cosh{(\nu_\lambda(C_{\alpha,3}+\log{(r-r_\star)}))}}{\cosh{(\nu_\lambda(C_{\alpha^\prime,3}+\log{(r-r_\star)}))}} - \nu_\lambda \frac{\sinh{(\nu_\lambda(C_{\alpha,3}+\log{(r-r_\star)}))}}{\cosh{(\nu_\lambda(C_{\alpha^\prime,3}+\log{(r-r_\star)}))}} \bigg),
\eeqa
where
\beqa
C_{\alpha,3} = \nu_\lambda^{-1} \tanh^{-1}{\bigg(1 - \frac{2C_{\alpha,-}}{C_{\alpha,-}+C_{\alpha,+}(r-r_\star)^{2\nu_\lambda}} \bigg)} - \log{(r-r_\star)} = \nu_\lambda^{-1}\frac{1}{2}\ln\bigg(\frac{C_{\alpha,+}}{C_{\alpha,-}}\bigg), \quad C_{\alpha} = \frac{1}{q e_3 + \tilde{m}_{\alpha}\ell_2}. \nn
\eeqa
The flow equation for $\xi_{\alpha\alpha^\prime}(r)$ can be solved, by substituting $\xi_\alpha(r)$ back into its flow equation. Namely, we have
\beqa
 \xi_{\alpha}^{-1}(r) = \frac{qe_3 + \tilde{m}_\alpha\ell_2}{m_D \ell_2 -\nu_\lambda\tau_\alpha(r)} \overset{\tau_\alpha(r)\to 1}{=} \frac{qe_3 + \tilde{m}_\alpha\ell_2}{m_D \ell_2 -\nu_\lambda} = \frac{m_D \ell_2 + \nu_\lambda}{qe_3 - \tilde{m}_\alpha\ell_2}, \quad \tau_\alpha(r) \equiv \tanh(\nu_\lambda(C_{\alpha,3}+\ln{(r-r_\star)})) \3i \overset{C_{\alpha,3} \to \infty}{=} \3i 1. \nn
\eeqa

To be brief, by imposing the in-falling wave condition, one obtains the near horizon boundary conditions for $\xi_{\alpha}$ and $\xi_{\alpha\alpha^\prime}$ as
\beqa
(a)(c)~~\xi_{\alpha}|_{r\to r_\star} = \frac{m_D \ell_2 - \nu_\lambda}{q e_3 + \tilde{m}_\alpha \ell_2}, \quad  (b)(d)~~\xi_{\alpha\alpha^\prime}|_{r\to r_\star} = \frac{m_D \ell_2 - \nu_\lambda}{q e_3 + \tilde{m}_{\alpha^\prime} \ell_2}.
\eeqa

\item[(2)] If $b_0 \ne 0$, then the $b_0$ term will be the dominant one in the field equation in Eq.(\ref{Eq:DiracEOM-flow-xi-alpha-UV-AdS(d+1)-omega=0}), since all other terms will be less singular than the $b_0$ term.  The $b_0$ term is relevant in Dirac equation and flow equation as shown in Eqs.(\ref{Eq:EOM-Diracfermion-IR-AdS(d+1)-omega=0-b0!=0}) and (\ref{Eq:xi_alpha-horizon-omega=0-b0!=0}), respectively. From these equations, one can obtain
\beqa
S^{-1}\tilde\Phi_\alpha = e^{-i \beta \frac{\ell_2^2}{r-r_\star} b_0 } \bigg( C_{\alpha,+} (r-r_\star)^{+\nu_\lambda} \left(
                                                    \begin{array}{c}
                                                      1 \\
                                                      0 \\
                                                    \end{array}
                                                  \right) +  C_{\alpha,-} (r-r_\star)^{-\nu_\lambda} \left(
                                                    \begin{array}{c}
                                                      0 \\
                                                      1 \\
                                                    \end{array}
                                                  \right) \bigg). \nn
\eeqa
Thus,  $\tilde\Phi_\alpha \sim  \tilde{R}_\alpha$ with
\beqa
 \tilde{R}_{\alpha,\beta}  \equiv  \left(
                                          \begin{array}{c}
                                            \tilde{F}_{\alpha,\beta} \\
                                            \tilde{G}_{\alpha,\beta} \\
                                          \end{array}
                                        \right)
= e^{i \beta \theta_b(r) } [ c_{\alpha,+} v_{\alpha,-}  (r-r_\star)^{+\nu_\lambda}   + c_{\alpha,-}  v_{\alpha,+} (r-r_\star)^{-\nu_\lambda} ] . \nn
\eeqa

In the oscillatory region the IR CFT scaling weight becomes imaginary, i.e., $\nu_\lambda = -i \rho_\lambda$, with $\rho_\lambda>0$ then one obtains a general wave solution with both the in-falling and out-going modes, \beqa
\tilde{R}_{\alpha,\beta} \sim  e^{-i \beta \frac{\ell_2^2}{r-r_\star} b_0 } [v_{\alpha,-} (r-r_\star)^{-i\rho_\lambda} + c_{\alpha,-} v_{\alpha,+} (r-r_\star)^{+i \rho_\lambda}],
\eeqa
where $\theta_b(r)$ is given in Eq.(\ref{Eq:theta-theta_b}) and Eq.(\ref{Eq:theta-theta_b_2}).

By imposing the in-falling wave condition, $C_{\alpha,-}=0$, i.e., $C_{\alpha,3} = \ln{(+\infty)} = +\infty$, then one obtains the IR BCs for the flow equations as
\beqa
(a)(b)&& \xi_{\alpha}(r)|_{r\to r_\star}  = \xi_{\alpha\alpha^\prime}(r)|_{r\to r_\star}  = \frac{m_D \ell_2 - \nu_\lambda }{q e_3 + \tilde{m}_\alpha \ell_2}; \nn\\
(c)&&\xi_{\alpha,+-}(r)|_{r\to r_\star}  = e^{-2i  \frac{\ell_2^2}{r-r_\star} b_0} \frac{m_D \ell_2 - \nu_\lambda}{q e_3 + \tilde{m}_\alpha \ell_2}, \quad \xi_{\alpha,-+}|_{r\to r_\star}  = e^{+2i  \frac{\ell_2^2}{r-r_\star} b_0} \frac{m_D \ell_2 - \nu_\lambda}{q e_3 + \tilde{m}_\alpha \ell_2}; \nn\\
(d)&&\xi_{\alpha\alpha^\prime,\beta\beta^\prime}(r)|_{r\to r_\star} = e^{-i (\beta - \beta^\prime) \frac{\ell_2^2}{r-r_\star} b_0} \frac{m_D \ell_2 - \nu_\lambda}{q e_3 + \tilde{m}_{\alpha} \ell_2}. \label{Eq:xi_alpha-horizon-omega=0}
\eeqa
Note that according to Eq.(\ref{Eq:mt_alpha}), $\tilde{m}_\alpha \equiv -(-1)^{\alpha}\lambda {\ell}/{r_\star}$ carries the subscript $\alpha=1,2$, thus they own a relative minus sign for upper and down components of bulk Dirac fermions. The result is consistent with Eq.(26) in ref.~\cite{Liu:2009dm}. Therefore at $\omega=0$, the in-falling condition $\xi_{\alpha}|_{r_0}=i$ does not apply, thus Eq.(\ref{Eq:xi_alpha-horizon-omega!=0}) should be replaced by one with a function depending on $m_D$ as shown in Eq.(\ref{Eq:xi_alpha-horizon-omega=0}).

\end{enumerate}

\end{enumerate}

In the end, for the standard quantization, the Green's function can be obtained, by using Eq.(\ref{Eq:G_alpha-alphap-beta-betap-AdS}) and (\ref{Eq:G_alpha-alphap-beta-betap-AdS-2}), as
\beqa
&& (a)~~\tilde{G}^R_{s\alpha}(\omega,\lambda) = \lim_{r\to \infty} r^{2 m_D\ell} \xi_\alpha(r) , \quad (b)~~\tilde{G}^R_{s\alpha\alpha^\prime}(\omega,\lambda) =  \lim_{r\to \infty} r^{2 m_D\ell} \xi_{\alpha\alpha^\prime}(r) , \nn\\
&& (c)~~\tilde{G}^R_{s\alpha,\beta\beta^\prime}(\omega,\lambda) = \lim_{r\to \infty} r^{2 m_D\ell} \xi_{\alpha,\beta\beta^\prime}(r) ,  \quad (d)~~\tilde{G}^R_{s\alpha\alpha^\prime,\beta\beta^\prime}(\omega,\lambda) = \lim_{r\to \infty} r^{2 m_D\ell} \xi_{\alpha\alpha^\prime,\beta\beta^\prime}(r) ,  \label{Eq:G-xi_alpha-alphap-beta-betap}
\eeqa
where $\alpha=1,2$, and $\beta =\pm$. The general flow equations are listed in Eq.(\ref{Eq:DiracEOM-flow-xi_alpha-UV-AdS(d+1)-omega!=0}), and the near horizon boundary conditions are listed in Eqs.(\ref{Eq:xi_alpha-horizon-omega!=0}) and (\ref{Eq:xi_alpha-horizon-omega=0}), for $\omega\ne 0$ and $\omega=0$ cases, respectively. The general properties on the UV retarded Green's function of spinors at zero temperature are summarized in Sec.~\ref{sec:GR-prop}.

\subsection{Wave functions of Dirac fermion in the bulk}

For the background metric and gauge field in Eq.(\ref{Eq:gtt-grr-gxx_AdS}), according to Eq.(\ref{Eq:lambda_1^pm-lambda_2-r}), one has
\beqa
\lambda_1^\pm(r) =   \frac{\ell^2}{r^2} \frac{1}{g(r)} [ (\omega\pm \beta b_0)+ q A_t(r) ] - \beta b_r, \quad \lambda_2(r) \equiv  \frac{\ell}{r} \frac{1}{\sqrt{g(r)}}  m_f(r) ,
\eeqa
where
\beqa
m_f(r) \equiv m_D + i m_\alpha(r),  \quad m_\alpha(r) \equiv -(-1)^\alpha \lambda  \frac{\ell}{r} = \tilde{m}_\alpha\frac{r_\star}{r}, \label{Eq:m_f(r)} \label{Eq:m_alpha(r)-mt_alpha}
\eeqa
and $\tilde{m}_\alpha$ is defined in Eq.(\ref{Eq:mt_alpha}) with $\alpha=1,2$ and for $d\ge 3$, $g(r)$ and gauge field are defined in Eq.(\ref{Eq:gr-At-star-AdS(d+1)-T=0}) and Eq.(\ref{Eq:At-AdS(3+1)}), respectively, with $r_0 =r_\star$ in the zero temperature case.

By using Eq.(\ref{Eq:gamma-lambda-r}), Eq.(\ref{Eq:fg_r_2nd}) can be expressed as
\beqa
&& \tilde{f}_{\alpha,\beta}^{\prime\prime}(r) + \bigg( -2i \beta   \frac{\ell^2}{r^2}\frac{1}{g(r)} b_0 + \frac{1}{2}\frac{g^\prime(r)}{g(r)} + \frac{2}{r} - \frac{m_D}{m_D r + i \tilde{m}_\alpha r_\star} \bigg) \tilde{f}_{\alpha,\beta}^{\prime}(r) \nn\\
&& + \bigg[ \Big(   \frac{\ell^2}{r^2} \frac{1}{g(r)}[\omega \!+\!  q A_t(r)] \!-\! \beta b_r \Big)^2 -   \frac{\ell^4}{r^4} \frac{1}{g(r)^2} b_0^2 -   \frac{\ell^2}{r^2} \frac{1}{g(r)}\Big( |m_f(r)|^2  -i q A_t^\prime(r) \Big)  \nn\\
             && + i \bigg(   -   \frac{\ell^2}{r^2} \frac{1}{g(r)}[(\omega-\beta b_0)+q A_t(r)]\Big( \frac{1}{2}\frac{g^\prime(r)}{g(r)} +  \frac{m_D}{m_D r + i \tilde{m}_\alpha r_\star} \Big) - \beta b_r \Big( \frac{1}{2}\frac{g^\prime(r)}{g(r)} + \frac{2}{r} - \frac{m_D}{m_D r + i \tilde{m}_\alpha r_\star}\Big)  \bigg) \bigg]\tilde{f}_{\alpha,\beta}(r) =0, \nn\\
&& \tilde{g}_{\alpha,\beta}^{\prime\prime}(r) + \bigg( -2i \beta   \frac{\ell^2}{r^2}\frac{1}{g(r)} b_0 + \frac{1}{2}\frac{g^\prime(r)}{g(r)} + \frac{2}{r} - \frac{m_D}{m_D r - i \tilde{m}_\alpha r_\star} \bigg) \tilde{g}_{\alpha,\beta}^{\prime}(r) \nn\\
&& + \bigg[ \Big(   \frac{\ell^2}{r^2} \frac{1}{g(r)}[\omega \!+\!  q A_t(r)] \!-\! \beta b_r \Big)^2 -   \frac{\ell^4}{r^4} \frac{1}{g(r)^2} b_0^2 -   \frac{\ell^2}{r^2} \frac{1}{g(r)} \Big( |m_f(r)|^2 + i q A_t^\prime(r)\Big)  \nn\\
             && - i \bigg(   -   \frac{\ell^2}{r^2} \frac{1}{g(r)}[(\omega+\beta b_0)+q A_t(r)]\Big( \frac{1}{2}\frac{g^\prime(r)}{g(r)} + \frac{m_D}{m_D r - i \tilde{m}_\alpha r_\star} \Big) - \beta b_r \Big(  \frac{1}{2}\frac{g^\prime(r)}{g(r)} + \frac{2}{r} - \frac{m_D}{m_D r - i \tilde{m}_\alpha r_\star} \Big)  \bigg) \bigg]\tilde{g}_{\alpha,\beta}(r) =0, \nn\\
&& \label{Eq:EOM-wpm-r}
\eeqa
where we have made the replacement $\partial_t \to -i \omega$, and
\beqa
|m_f(r)|^2  =  m_D^2 + m_\alpha(r)^2 = m_D^2 + \tilde{m}_\alpha^2\frac{r_\star^2}{r^2},  \quad \frac{m_f^\prime(r)}{m_f(r)}  =  -\frac{ i\tilde{m}_\alpha r_\star}{r^2(m_D  + i  \tilde{m}_\alpha r_\star  /r)}
= -\frac{1}{r}+ \frac{m_D}{m_D r + i\tilde{m}_\alpha r_\star},\nn
\eeqa
and $\tilde{m}_\alpha$ is defined in Eq.(\ref{Eq:mt_alpha}) with $\alpha=1,2$.

When both the $b_0$ and $b_r$ vanish, the EOMs in Eq.(\ref{Eq:fg_r_2nd}) become
\beqa
&& \tilde{f}_\alpha^{\prime\prime}(r) + \bigg(  \frac{1}{2}\frac{g^\prime(r)}{g(r)} + \frac{2}{r} - \frac{m_D}{m_D r + i \tilde{m}_\alpha r_\star} \bigg) \tilde{f}_\alpha^{\prime}(r) +   \frac{\ell^2}{r^2} \frac{1}{g(r)} \bigg[   \frac{\ell^2}{r^2} \frac{1}{g(r)}[\omega \!+\!  q A_t(r)]^2 \nn\\
&& - \Big( |m_f(r)|^2  -i q A_t^\prime(r) \Big) - i [\omega+q A_t(r)]\bigg( \frac{1}{2}\frac{g^\prime(r)}{g(r)} +  \frac{m_D}{m_D r + i \tilde{m}_\alpha r_\star} \bigg) \bigg]\tilde{f}_\alpha(r) =0, \nn\\
&& \tilde{g}_\alpha^{\prime\prime}(r) + \bigg(  \frac{1}{2}\frac{g^\prime(r)}{g(r)} + \frac{2}{r} - \frac{m_D}{m_D r - i \tilde{m}_\alpha r_\star} \bigg) \tilde{g}_\alpha^{\prime}(r) +   \frac{\ell^2}{r^2} \frac{1}{g(r)} \bigg[   \frac{\ell^2}{r^2} \frac{1}{g(r)}[\omega \!+\!  q A_t(r)]^2 \nn\\
&& -  \Big( |m_f(r)|^2 + i q A_t^\prime(r)\Big) + i [\omega+q A_t(r)]\Big( \frac{1}{2}\frac{g^\prime(r)}{g(r)} + \frac{m_D}{m_D r - i \tilde{m}_\alpha r_\star} \Big) \bigg]\tilde{g}_\alpha(r) =0,
\eeqa
where the subscript $\beta$ has been dropped.

\subsubsection{Infinite boundary}

At the infinite boundary, one has $g(r) \overset{r\to \infty}{=} 1 $, $A_t(r) \overset{r\to \infty}{=} \mu $ for $d = 3$, in the $m_D\ne 0$ case. Furthermore,
\beqa
\3i
\frac{g^\prime(r)}{g(r)} \overset{r \to \infty}{=} C \frac{ 3\ell^2 }{r^{4}} , \quad \frac{m_f^\prime(r)}{m_f(r)} \overset{r\to \infty}{=} i \frac{\tilde{m}r_\star}{m_D r^2} \overset{m_D = 0}{=}  - \frac{1}{r}, \quad \frac{A_t^\prime(r)}{A_t(r)} \overset{r \to \infty}{=} \frac{r_\star}{r^{2}}.
\eeqa
where $C$ is constant. Thus, for $d = 3$, the ratio $g^\prime(r)/g(r)$ always vanishes in the infinite boundary limit $r\to \infty$.

Thus Eqs.(\ref{Eq:EOM-wpm-r}) become
\beqa
&& \tilde{f}_{\alpha,\beta}^{\prime\prime}(r) + \bigg( -2i \beta   \frac{\ell^2}{r^2} b_0 + \frac{2}{r} - \frac{m_D}{m_D r + i \tilde{m}_\alpha r_\star} \bigg) \tilde{f}_{\alpha,\beta}^{\prime}(r) + \bigg[ + i \bigg(   -   \frac{\ell^2}{r^2} [(\omega-\beta b_0)+q \mu ]  \frac{m_D}{m_D r + i \tilde{m}_\alpha r_\star} \nn\\
&& - \beta b_r \Big(  \frac{2}{r} - \frac{m_D}{m_D r + i \tilde{m}_\alpha r_\star}\Big)  \bigg) + \Big(   \frac{\ell^2}{r^2} [\omega \!+\!  q \mu ] \!-\! \beta b_r \Big)^2 - \frac{\ell^4}{r^4}  b_0^2 -   \frac{\ell^2}{r^2} \Big( m_D^2 + \tilde{m}^2\frac{r_\star^2}{r^2}  -i q \mu \frac{r_\star}{r^{2}} \Big)  \bigg]\tilde{f}_{\alpha,\beta}(r) =0, \nn\\
&& \tilde{g}_{\alpha,\beta}^{\prime\prime}(r) + \bigg( -2i \beta   \frac{\ell^2}{r^2} b_0  + \frac{2}{r} - \frac{m_D}{m_D r - i \tilde{m}_\alpha r_\star} \bigg) \tilde{g}_{\alpha,\beta}^{\prime}(r)  + \bigg[ - i \bigg(   -   \frac{\ell^2}{r^2} [(\omega+\beta b_0)+q \mu ] \frac{m_D}{m_D r - i \tilde{m}_\alpha r_\star} \nn\\
&& - \beta b_r \Big(  \frac{2}{r} - \frac{m_D}{m_D r - i \tilde{m}_\alpha r_\star} \Big)  \bigg) + \Big(   \frac{\ell^2}{r^2} [\omega \!+\!  q \mu ] \!-\! \beta b_r \Big)^2 - \frac{\ell^4}{r^4}  b_0^2 -   \frac{\ell^2}{r^2} \Big( m_D^2 + \tilde{m}^2\frac{r_\star^2}{r^2} + i q \mu \frac{r_\star}{r^{2}} \Big)   \bigg]\tilde{g}_{\alpha,\beta}(r) =0.  \label{Ew:wpm-infity}
\eeqa

In this case, the term with frequency and charge can be neglected as in the pure vacuum case, the two decoupled EOMs in Eq.(\ref{Ew:wpm-infity}) have the same form,
\beqa
&& \tilde{f}_{\alpha,\beta}^{\prime\prime}(r) + \frac{1}{r} \tilde{f}_{\alpha,\beta}^{\prime}(r) + \bigg[ b_r^2 - 2 \beta b_r   \frac{\ell^2}{r^2} (\omega \!+\!  q \mu ) -   \frac{\ell^2}{r^2} m_D^2   - i \beta b_r \bigg( \frac{1}{r} + i \frac{\tilde{m}_\alpha r_\star}{m_D r^2} \bigg)   \bigg]\tilde{f}_{\alpha,\beta}(r) =0, \nn\\
&& \tilde{g}_{\alpha,\beta}^{\prime\prime}(r) + \frac{1}{r} \tilde{g}_{\alpha,\beta}^{\prime}(r) + \bigg[ b_r^2 - 2 \beta b_r   \frac{\ell^2}{r^2} (\omega \!+\!  q \mu ) -   \frac{\ell^2}{r^2} m_D^2
+ i \beta b_r  \bigg( \frac{1}{r} - i \frac{\tilde{m}_\alpha r_\star}{m_D r^2} \bigg)    \bigg]\tilde{g}_{\alpha,\beta}(r) =0,
\eeqa
which give the solutions as
\beqa
&& \tilde{f}_{\alpha,\beta}(r) = e^{-i \beta b_r r} r^{\nu_\beta} \big[ C_1 U[1+\nu_\beta,1+2\nu_\beta,2i \beta b_r r] + C_2 L_{-1-\nu_\beta}^{2\nu_\beta}(2i \beta b_r r) \big], \nn\\
&& \tilde{g}_{\alpha,\beta}(r) = e^{-i \beta b_r r} r^{\nu_\beta} \big[ C_3 U[\nu_\beta,1+2\nu_\beta,2i \beta b_r r] + C_4 L_{-\nu_\beta}^{2\nu_\beta}(2i \beta b_r r) \big],
\eeqa
where $C_1$, $C_2$, $C_3$ and $C_4$ are all constants. $\nu_\beta \equiv \sqrt{   [m_D^2 + 2 \beta b_r(\omega+ q\mu)]\ell^2 - b_r \tilde{m}_\alpha r_\star/m_D } \overset{b_r=0}{=}  m_D \ell $ with $\beta=\pm$, $U[a,b,z]$ is Tricomi's confluent hypergeometric function, and $L_n^{a}(z)$ are generalized Laguerre polynomials, or associated Laguerre polynomials. For the massless case, i.e., $m_D=0$, the EOMs become
\beqa
&& \tilde{f}_{\alpha,\beta}^{\prime\prime}(r) + \frac{2}{r} \tilde{f}_{\alpha,\beta}^{\prime}(r) + \bigg( b_r^2 - 2 \beta b_r   \frac{\ell^2}{r^2} (\omega \!+\!  q \mu )- i \beta b_r \frac{2}{r}    \bigg)\tilde{f}_{\alpha,\beta}(r) = 0, \nn\\
&& \tilde{g}_{\alpha,\beta}^{\prime\prime}(r) + \frac{2}{r} \tilde{g}_{\alpha,\beta}^{\prime}(r) + \bigg( b_r^2 - 2 \beta b_r   \frac{\ell^2}{r^2} (\omega \!+\!  q \mu )+ i \beta b_r  \frac{2}{r}   \bigg)\tilde{g}_{\alpha,\beta}(r) = 0,
\eeqa
which lead to solution
\beqa
&& \tilde{f}_{\alpha,\beta}(r) = e^{-i \beta b_r r} r^{\nu_\beta} \big[ C_1 U[2+\nu_\beta,2+2\nu_\beta,2i \beta b_r r] + C_2 L_{-2-\nu_\beta}^{1+2\nu_\beta}(2i \beta b_r r) \big], \nn\\
&& \tilde{g}_{\alpha,\beta}(r) = e^{-i \beta b_r r} r^{\nu_\beta} \big[ C_3 U[\nu_\beta,2+2\nu_\beta,2i \beta b_r r] + C_4 L_{-\nu_\beta}^{1+2\nu_\beta}(2i \beta b_r r) \big],
\eeqa
where $\nu_\beta \equiv (\sqrt{1 + 8   \beta b_r (\omega+q\mu)  \ell^2 }-1)/2 \overset{b_r=0}{=} 0$.

In the case with $b_r\ne 0$ and $m_D\ne 0$, one has
\beqa
\tilde{f}_{\alpha,\beta}^{\prime\prime}(r) + \frac{1}{r} \tilde{f}_{\alpha,\beta}^{\prime}(r) + \bigg( b_r^2 - i \beta b_r \frac{1}{r}   \bigg)\tilde{f}_{\alpha,\beta}(r) =0, \quad \tilde{g}_{\alpha,\beta}^{\prime\prime}(r) + \frac{1}{r} \tilde{g}_{\alpha,\beta}^{\prime}(r) + \bigg( b_r^2 + i \beta b_r  \frac{1}{r}   \bigg)\tilde{g}_{\alpha,\beta}(r) =0,
\eeqa
which give
\beqa
\tilde{f}_{\alpha,\beta}(r) = e^{i \beta b_r r } [ C_1 + C_2 \text{E}_i(-2i \beta b_r r) ], \quad \tilde{g}_{\alpha,\beta}(r) = e^{-i \beta b_r r } [ C_3 + C_4 \text{E}_i(2i \beta b_r r) ],
\eeqa
where $\text{E}_i(z)$ is the exponential integral function, which has the behavior in the infinite boundary, $\text{E}_i(z) \overset{z\to \infty}{\sim} e^{z} z^{-1}$. Thus
\beqa
\tilde{f}_{\alpha,\beta}(r) \overset{r \to \infty}{\sim} C_1 e^{+i \beta b_r r }  + i \frac{C_2}{2 \beta b_r} \frac{e^{-i \beta b_r r }}{ r} , \quad \tilde{g}_{\alpha,\beta}(r) \overset{r \to \infty}{\sim} C_3 e^{-i \beta b_r r }  - i \frac{C_4}{2 \beta b_r} \frac{e^{+i \beta b_r r }}{r}.
\eeqa
In the infinite boundary,
\beqa
\ii
\left(
  \begin{array}{c}
    \tilde{f}_{\alpha,\beta} \\
    \tilde{g}_{\alpha,\beta} \\
  \end{array}
\right) \3i &=& \3i \left(\begin{array}{c}
            C_1 \\
            C_3
          \end{array} \right) J_0(\beta b_r r) + \left(\begin{array}{c}
            C_2 \\
            C_4
          \end{array}\right) Y_0(\beta b_r r) \nn\\
&\overset{r\to \infty}{\sim} & \frac{1+i}{2}\frac{1}{\sqrt{\pi \beta b_r r}}e^{-i \beta b_r r} \left(
                                                                                                                           \begin{array}{c}
                                                                                                                             C_1 + i C_2 \\
                                                                                                                             C_3 + i C_4 \\
                                                                                                                           \end{array}
                                                                                                                         \right)
           + \frac{1-i}{2}\frac{1}{\sqrt{\pi \beta b_r r}}e^{i \beta b_r r} \left(
                                                                  \begin{array}{c}
                                                                    C_1 - i C_2 \\
                                                                    C_3 - i C_4 \\
                                                                  \end{array}
                                                                \right), \nn
\eeqa
where $J_0(z)$ and $Y_0(z)$ are Bessel function of first kind $J_\nu(z)$ and that of second kind $Y_\nu(z)$ with $\nu=0$, respectively, and have the asymptotic behavior $J_0(z)\overset{z\to \infty}{\sim}\sqrt{2 (\pi z)^{-1}}\sin(z+\pi/4)$ and $Y_0(z)\overset{z\to \infty}{\sim} e^{-iz} $.

For the massless case, i.e., $m_D=0$, the EOMs become
\beqa
\tilde{f}_{\alpha,\beta}^{\prime\prime}(r) + \frac{2}{r} \tilde{f}_{\alpha,\beta}^{\prime}(r) + \bigg( b_r^2 - i \beta b_r \frac{2}{r}  \bigg)\tilde{f}_{\alpha,\beta}(r) = 0, \quad \tilde{g}_{\alpha,\beta}^{\prime\prime}(r) + \frac{2}{r} \tilde{g}_{\alpha,\beta}^{\prime}(r) + \bigg( b_r^2 + i \beta b_r \frac{2}{r}  \bigg)\tilde{g}_{\alpha,\beta}(r) = 0, \nn
\eeqa
which have solutions
\beqa
\tilde{f}_{\alpha,\beta}(r)  &=&  C_1 e^{+i \beta b_r r} - C_2 e^{-i \beta b_r r}\bigg( \frac{1}{r} + 2 i \beta b_r e^{+2i \beta b_r r} \text{E}_i(-2i \beta b_r r) \bigg) \overset{r\to \infty}{\sim}  C_1 e^{+i \beta b_r r} + C_2 {\mathcal O}(\frac{1}{\beta b_r r^2}), \nn\\
\tilde{g}_{\alpha,\beta}(r)  &=&  C_3 e^{-i \beta b_r r} + C_4 e^{+i \beta b_r r}\bigg( \frac{1}{r} - 2 i \beta b_r e^{-2i \beta b_r r} \text{E}_i(+2i \beta b_r r) \bigg) \overset{r\to \infty}{\sim}  C_3 e^{-i \beta b_r r} + C_4 {\mathcal O}(\frac{1}{\beta b_r r^2}). \ii
\eeqa
In the infinite boundary, one has
\beqa
\tilde{f}_{\alpha,\beta}(r) = \frac{1}{r}\bigg(C_1 e^{-i \beta b_r r} - iC_2 \frac{1}{2 \beta b_r}e^{i \beta b_r r} \bigg), \quad \tilde{g}_{\alpha,\beta}(r) = \frac{1}{r}\bigg(C_3 e^{-i \beta b_r r} - iC_4 \frac{1}{2 \beta b_r}e^{i \beta b_r r} \bigg).
\eeqa

For the massless case, i.e., $m_D=0$, if $b_r=0$ but $b_0\ne 0$, the EOMs in Eq.(\ref{Ew:wpm-infity}) become
\beqa
&& \tilde{f}_{\alpha,\beta}^{\prime\prime}(r) + \bigg( -2i \beta   \frac{\ell^2}{r^2} b_0 + \frac{2}{r}  \bigg) \tilde{f}_{\alpha,\beta}^{\prime}(r) + \bigg[ \Big(   \frac{\ell^2}{r^2} (\omega \!+\!  q \mu ) \Big)^2 -   \frac{\ell^4}{r^4}  b_0^2 -   \frac{\ell^2}{r^2} \Big( \tilde{m}^2\frac{r_\star^2}{r^2}  -i q \mu \frac{r_\star}{r^{2}} \Big)  \bigg]\tilde{f}_{\alpha,\beta}(r) =0, \nn\\
&& \tilde{g}_{\alpha,\beta}^{\prime\prime}(r) + \bigg( -2i \beta   \frac{\ell^2}{r^2} b_0  + \frac{2}{r} \bigg) \tilde{g}_{\alpha,\beta}^{\prime}(r) + \bigg[ \Big(   \frac{\ell^2}{r^2} (\omega \!+\!  q \mu ) \Big)^2 -   \frac{\ell^4}{r^4}  b_0^2 -   \frac{\ell^2}{r^2} \Big( \tilde{m}^2\frac{r_\star^2}{r^2}
+i q \mu \frac{r_\star}{r^{2}} \Big)  \bigg]\tilde{g}_{\alpha,\beta}(r) =0. \nn
\eeqa
For simplicity, one can assume $q=0$. In this case, one has the solution of the wave functions:
\beqa
\tilde{f}_{\alpha,\beta}, \tilde{g}_{\alpha,\beta}  \overset{m_D=0}{=}   e^{-i \beta   \ell^2 \frac{b_0}{r} } \big[  C_{1,3} \cos{\bigg( \frac{ \ell}{r}\sqrt{  \omega^2 \ell^2 - \tilde{m}^2 r_\star^2} \bigg)} +   C_{2,4} \sin{\bigg( \frac{ \ell}{r}\sqrt{   \omega^2 \ell^2 - \tilde{m}^2 r_\star^2} \bigg)} \big].
\eeqa
For the massive case, i.e., $m_D\ne 0$, one has
\beqa
\tilde{f}_\alpha^{\prime\prime}(r) + \frac{1}{r} \tilde{f}_\alpha^{\prime}(r) -   \frac{\ell^2}{r^2} m_D^2 \tilde{f}_\alpha(r) = 0, \quad \tilde{g}_\alpha^{\prime\prime}(r) + \frac{1}{r} \tilde{g}_\alpha^{\prime}(r) -   \frac{\ell^2}{r^2} m_D^2 \tilde{g}_\alpha(r) = 0,
\eeqa
where the subscript $\beta$ has been dropped.
This leads to solution
\beqa
\left(
  \begin{array}{c}
    \tilde{f}_\alpha \\
    \tilde{g}_\alpha \\
  \end{array}
\right) = \left(
            \begin{array}{c}
              C_1 \\
              C_3 \\
            \end{array}
          \right) \cosh{( m_D \ell \ln{r})} + i \left(
                                                    \begin{array}{c}
                                                      C_2 \\
                                                      C_4 \\
                                                    \end{array}
                                                  \right)
           \sinh{( m_D \ell \ln{r})}.
\eeqa
Thus, according to Eq.(\ref{Eq:fgt_alpha-transQ_FGt_alpha}), one has
\beqa
\left(
  \begin{array}{c}
    \tilde{F}_{\alpha} \\
    \tilde{G}_{\alpha} \\
  \end{array}
\right) &=& Q^{-1} \left(
  \begin{array}{c}
    \tilde{f}_\alpha \\
    \tilde{g}_\alpha \\
  \end{array}
\right) = \frac{1}{\sqrt{2}}\left(
            \begin{array}{cc}
              1 & 1 \\
             -i & i \\
            \end{array}
          \right) \left(
  \begin{array}{c}
    \tilde{f}_\alpha \\
    \tilde{g}_\alpha \\
  \end{array}
\right)  \equiv   \tilde{B}_\alpha \left(\begin{array}{c}
1  \\
0
\end{array}
\right)r^{- m_D\ell} + \tilde{A}_\alpha \left(\begin{array}{c}
0  \\
1
\end{array}
\right)r^{ m_D\ell} ,\nn\\
&=&  \frac{1}{2\sqrt{2}}\left(
      \begin{array}{c}
        C_1 \!+\! C_3 \!-\! i(C_2 \!+\! C_4) \\
        -i(C_1 \!-\! C_3) \!-\! (C_2 \!-\! C_4) \\
      \end{array}
    \right) r^{- m_D \ell}  \! + \!  \frac{1}{2\sqrt{2}}\left(
      \begin{array}{c}
        C_1 \!+\! C_3 \!+\! i(C_2 \!+\! C_4) \\
        -i(C_1 \!-\! C_3) \!+\! (C_2 \!-\! C_4) \\
      \end{array}
    \right) r^{ m_D \ell}, \nn
\eeqa
which is consistent with the solution in Eq.(\ref{Eq:Psi_alpha-UV}), if one chooses the constants $C_1=(\tilde{B}+i\tilde{A})/2$, $C_2 = (\tilde{A}+i\tilde{B})/2 $, $C_3=(\tilde{B}-i\tilde{A})/2$, and $C_4 = (-\tilde{A}+i\tilde{B})/2 $.

By keeping only the leading order up to $\omega$ and keeping the divergent term up to the lowest order for simplicity in Eq.(\ref{Ew:wpm-infity}) for the massive case $m_D\ne 0$, one has
\beqa
&& \tilde{f}_{\alpha,\beta}^{\prime\prime}(r) + \bigg( \frac{1}{r} + i \frac{\tilde{m}_\alpha r_\star}{m_D r^2} - 2i \beta   \frac{\ell^2}{r^2} b_0  \bigg) \tilde{f}_{\alpha,\beta}^{\prime}(r) + \bigg[ b_r^2 - 2 \beta b_r   \frac{\ell^2}{r^2} (\omega \!+\!  q \mu ) -   \frac{\ell^2}{r^2} m_D^2   \nn\\
             && + i \bigg(   -   \frac{\ell^2}{r^3} [(\omega-\beta b_0)+q \mu ] - \beta b_r\Big(  \frac{1}{r} + i \frac{\tilde{m}_\alpha r_\star}{m_D r^2} + \frac{\tilde{m}^2 r_\star^2}{m_D^2 r^3}   \Big)  \bigg) \bigg]\tilde{f}_{\alpha,\beta}(r) =0, \nn\\
&& \tilde{g}_{\alpha,\beta}^{\prime\prime}(r) + \bigg( \frac{1}{r} - i \frac{\tilde{m}_\alpha r_\star}{m_D r^2} - 2i \beta   \frac{\ell^2}{r^2} b_0 \bigg) \tilde{g}_{\alpha,\beta}^{\prime}(r) + \bigg[ b_r^2 - 2 \beta b_r   \frac{\ell^2}{r^2} (\omega \!+\!  q \mu ) -   \frac{\ell^2}{r^2} m_D^2 \nn\\
             && - i \bigg(   -   \frac{\ell^2}{r^3} [(\omega+\beta b_0)+q \mu ] - \beta b_r \Big( \frac{1}{r} - i \frac{\tilde{m}_\alpha r_\star}{m_D r^2} + \frac{\tilde{m}^2 r_\star^2}{m_D^2 r^3} \Big)  \bigg) \bigg]\tilde{g}_{\alpha,\beta}(r) =0,
\eeqa
by absorbing $b_r$ into the definition of wave function, but $b_0$ is present, the EOMs are solvable in the UV limit with finite chemical potential at small frequency as
\beqa
\tilde{f}_{\alpha,\beta}
&=& C_1 \bigg(-i \frac{r^+_{\star\alpha,\beta}}{r}\bigg)^{ -  m_D \ell} {}_1 F_1 \bigg[  \frac{  \ell^2(\omega-\beta b_0+q\mu)}{r_{\star\alpha,\beta}^+} -  m_D \ell, 1-2  m_D \ell, i \frac{r^+_{\star\alpha,\beta}}{r} \bigg] \nn\\
&+& C_2 \bigg(-i \frac{r^+_{\star\alpha,\beta}}{r}\bigg)^{ +  m_D \ell} {}_1 F_1 \bigg[  \frac{  \ell^2(\omega-\beta b_0+q\mu)}{r_{\star\alpha,\beta}^+} +  m_D \ell, 1+2  m_D \ell, i \frac{r^+_{\star\alpha,\beta}}{r} \bigg], \nn\\
\tilde{g}_{\alpha,\beta}
&=& C_3 \bigg(-i \frac{r^-_{\star\alpha,\beta}}{r}\bigg)^{ -  m_D \ell} {}_1 F_1 \bigg[ -\frac{  \ell^2(\omega+\beta b_0+q\mu)}{r_{\star\alpha,\beta}^-} -  m_D \ell, 1-2  m_D \ell, i \frac{r^-_{\star\alpha,\beta}}{r} \bigg] \nn\\
&+& C_4 \bigg(-i \frac{r^-_{\star\alpha,\beta}}{r}\bigg)^{ +  m_D \ell} {}_1 F_1 \bigg[ -\frac{  \ell^2(\omega+\beta b_0+q\mu)}{r_{\star\alpha,\beta}^-} +  m_D \ell, 1+2  m_D \ell, i \frac{r^-_{\star\alpha,\beta}}{r} \bigg],
\eeqa
where $C_1$, $C_2$, $C_3$ and $C_4$ are all constants. $r^\pm_{\star\alpha,\beta} \equiv (\pm \tilde{m}_\alpha r_\star - 2 \beta b_0   m_D \ell^2)/m_D$, ${}_1 F_1[a,b,z]$ is Kummer's confluent hypergeometric function, which is related to Whittaker-M function through $e^{z/2}z^{-b/2}M[-a+b/{2},-{1}/{2}+b/{2},z]={}_1 F_1[a,b,z]$.
In the absence of $b_0$, the EOMs become
\beqa
&& \tilde{f}_\alpha^{\prime\prime}(r) + \bigg( \frac{1}{r} + i\frac{\tilde{m}_\alpha r_\star}{m_D r^2}  \bigg) \tilde{f}_\alpha^{\prime}(r) +   \frac{\ell^2}{r^2} \bigg(  -  m_D^2 - i  \frac{1}{r} (\omega+q \mu )  \bigg)\tilde{f}_\alpha(r) =0, \nn\\
&& \tilde{g}_\alpha^{\prime\prime}(r) + \bigg( \frac{1}{r} - i\frac{\tilde{m}_\alpha r_\star}{m_D r^2}  \bigg) \tilde{g}_\alpha^{\prime}(r) +   \frac{\ell^2}{r^2} \bigg(  -  m_D^2 + i  \frac{1}{r} (\omega+q \mu )  \bigg)\tilde{g}_\alpha(r) =0, \label{Ew:wpm-infity-massive}
\eeqa
and the solutions turn out to be
\beqa
\tilde{f}_\alpha,\tilde{g}_\alpha \3i &\overset{m_D\ne 0}{=}& \3i C_{1,3} \bigg(\mp\frac{i\tilde{m}_\alpha}{m_D}\frac{r_\star}{r}\bigg)^{- m_D\ell} {}_1F_1 \Big[ \frac{  m_D \ell^2(q\mu+\omega)}{\tilde{m}_\alpha r_\star} -  m_D \ell, 1 - 2 m_D \ell, \pm \frac{i \tilde{m}_\alpha }{m_D}\frac{r_\star}{r} \Big]  \nn\\
\3i &+& \3i C_{2,4} \bigg(\mp\frac{i\tilde{m}_\alpha}{m_D}\frac{r_\star}{r}\bigg)^{ m_D\ell} {}_1F_1 \bigg[ \frac{  m_D \ell^2(q\mu+\omega)}{\tilde{m}_\alpha r_\star} +  m_D \ell, 1 + 2 m_D \ell, \pm \frac{i \tilde{m}_\alpha }{m_D}\frac{r_\star}{r} \bigg].
\eeqa

According to Eq.(\ref{Eq:Psi_alpha-beta_Bt-At}), one can read off the conformal scaling weight from UV CFT$_3$ as
\beqa
\psi_{\alpha,\beta\beta^\prime}  =  r^{-\frac{3}{2}}\Psi_{\alpha,\beta\beta^\prime}  \overset{r\to\infty}{=}  \left(  \begin{array}{c}
B_{\alpha,\beta} r^{-\frac{3}{2}- m_D\ell}  \\
A_{\alpha,\beta^\prime} r^{-\frac{3}{2}+ m_D\ell}
\end{array}
\right)  =  \left(\begin{array}{c}
B_{\alpha,\beta} u_\star^{-2\Delta_+} u^{\Delta_+}  \\
A_{\alpha,\beta^\prime} u_\star^{-2\Delta_-} u^{\Delta_-}
\end{array}
\right), \nn
\eeqa
where the up and down components may have same or opposite helicity, i.e., $\beta^\prime=\pm \beta$ and $\Delta_\pm \equiv 3/2 \pm  m_D \ell$. From the asymptotic behavior of the wave function in the infinite boundary, one can read off the UV retarded Green's functions by identifying $A$ and $B$ as source and response, respectively.

\subsubsection{Near horizon region}

In the near horizon region, one has
\beqa
  g(r) & \overset{r\to r_\star}{=} &   \frac{6(r-r_\star)^2}{r_\star^2}   ,  \quad \frac{g^\prime(r)}{g(r)}\overset{r\to r_\star}{=} \frac{2}{r-r_\star},  \nn \\ 
  A_t(r) & \overset{r\to r_\star}{=} & \mu \frac{r-r_\star}{r_\star}\bigg( 1 - \frac{r-r_\star}{r_\star} \bigg), \quad A_t^\prime(r)  \overset{r\to r_\star}{=}  \mu\frac{1}{r_\star}\bigg(1 - \frac{2(r-r_\star)}{r_\star} \bigg).
\eeqa
The EOMs for $\tilde{f}_{\alpha,\beta} $ and $\tilde{g}_{\alpha,\beta}$ in Eq.(\ref{Eq:EOM-wpm-r}) become
\beqa
&& \tilde{f}_{\alpha,\beta}^{\prime\prime}(r) + \bigg( -2i \beta \frac{\ell_2^2}{(r-r_\star)^2} b_0 + \frac{1}{r-r_\star} \bigg) \tilde{f}_{\alpha,\beta}^{\prime}(r) \nn\\
&& + \bigg[ \bigg( \frac{\ell_2^2}{(r-r_\star)^2}\Big(\omega \!+\!  q \mu \frac{r-r_\star}{r_\star}\big(1 - \frac{r-r_\star}{r_\star} \big) \Big) \!-\! \beta b_r \bigg)^2 - \bigg( \frac{b_0 \ell_2^2}{(r-r_\star)^2}  \bigg)^2 - \frac{\ell_2^2}{(r-r_\star)^2} \Big( m_\lambda^2  -i q \mu\frac{1}{r_\star} \Big)  \nn\\
             && + i \bigg( - \frac{\ell_2^2}{(r-r_\star)^2}\Big(\frac{\omega-\beta b_0}{r-r_\star}+q \mu \frac{1}{r_\star}\big(1 - \frac{r-r_\star}{r_\star} \big) \Big) - \frac{\beta b_r}{r-r_\star}  \bigg) \bigg]\tilde{f}_{\alpha,\beta}(r) =0, \nn\\
&& \tilde{g}_{\alpha,\beta}^{\prime\prime}(r) + \bigg( -2i \beta \frac{\ell_2^2}{(r-r_\star)^2} b_0 + \frac{1}{r-r_\star} \bigg) \tilde{g}_{\alpha,\beta}^{\prime}(r) \nn\\
&& + \bigg[ \bigg( \frac{\ell_2^2}{(r-r_\star)^2}\Big(\omega \!+\!  q \mu \frac{r-r_\star}{r_\star}\big(1 - \frac{r-r_\star}{r_\star} \big) \Big) \!-\! \beta b_r \bigg)^2 -  \bigg( \frac{b_0 \ell_2^2}{(r-r_\star)^2} \bigg)^2 - \frac{\ell_2^2}{(r-r_\star)^2} \Big( m_\lambda^2 + i q \mu\frac{1}{r_\star} \Big)  \nn\\
             && - i \bigg( - \frac{\ell_2^2}{(r-r_\star)^2}\Big(\frac{\omega+\beta b_0}{r-r_\star}+q \mu \frac{1}{r_\star}\big(1 - \frac{r-r_\star}{r_\star} \big) \Big) - \frac{\beta b_r}{r-r_\star}  \bigg) \bigg]\tilde{g}_{\alpha,\beta}(r) =0.
\eeqa
where we have considered
\beqa
 m_f & \overset{r\to r_\star}{=}& m_D + i\tilde{m}_\alpha, \quad |m_f|^2 \overset{r\to r_\star}{=} (m_D^2+\tilde{m}_\alpha^2) \equiv m_\lambda^2, \nn\\
 \frac{m_f^\prime(r)}{m_f(r)} & \overset{r\to r_\star}{=} &
\frac{1}{r_\star}\frac{m_D}{m_D+ i \tilde{m}_\alpha} - \frac{1}{r_\star} \sim \text{const.}, \quad   \frac{\ell^2}{r^2}\frac{1}{g(r)}  \overset{r\to r_\star}{=}  \frac{\ell_2^2}{(r-r_\star)^2} \bigg(1 + \frac{4}{3r_\star}(r-r_\star)\bigg),  \nn
\eeqa
and $\tilde{m}_\alpha$ is defined in Eq.(\ref{Eq:mt_alpha}) with subscript $\alpha=1,2$.


In the deep near horizon region without chiral gauge field $b_r=b_0=0$ but $b_i\ne 0$, the EOMs become
\beqa
\tilde{f}_\alpha^{\prime\prime}(r) + \frac{1}{r-r_\star} \tilde{f}_\alpha^{\prime}(r) + \frac{q^2 e_3^2 - m_\lambda^2 \ell_2^2}{(r-r_\star)^2}  \tilde{f}_\alpha(r) = 0, \quad \tilde{g}_\alpha^{\prime\prime}(r) + \frac{1}{r-r_\star} \tilde{g}_\alpha^{\prime}(r) + \frac{q^2 e_3^2 - m_\lambda^2\ell_2^2}{(r-r_\star)^2}  \tilde{g}_\alpha(r) = 0, \nn
\eeqa
where $e_3$ and $\ell_2$ defined in Eqs.(\ref{Eq:ed-r_star}) and (\ref{Eq:ell_2}). The wavefunction solutions to the equations are
\beqa
\left(
  \begin{array}{c}
    \tilde{f}_\alpha \\
    \tilde{g}_\alpha \\
  \end{array}
\right) &=& \left(
            \begin{array}{c}
              C_1 \\
              C_3 \\
            \end{array}
          \right) \cosh{[\nu_\lambda \ln{(r-r_\star)}]} + i\left(
            \begin{array}{c}
              C_2 \\
              C_4 \\
            \end{array}
          \right) \sinh{[\nu_\lambda \ln{(r-r_\star)}]} , \nn\\
&=& \frac{1}{2}\left(
            \begin{array}{c}
              C_1 + i C_2 \\
              C_3 + i C_4 \\
            \end{array}
          \right) (r-r_\star)^{-\nu_\lambda} + \frac{1}{2}\left(
            \begin{array}{c}
              C_1 - i C_2 \\
              C_3 - i C_4 \\
            \end{array}
          \right) (r-r_\star)^{+\nu_\lambda} ,
\eeqa
with the conformal weight of IR CFT given by $\nu_\lambda=\sqrt{ m_\lambda^2\ell_2^2 - q^2 e_3^2} $. Thus according Eq.(\ref{Eq:fgt_alpha-transQ_FGt_alpha}), one has
\beqa
\left(
  \begin{array}{c}
    \tilde{F}_{\alpha} \\
    \tilde{G}_{\alpha} \\
  \end{array}
\right) &=& Q^{-1} \left(
  \begin{array}{c}
    \tilde{f}_\alpha \\
    \tilde{g}_\alpha \\
  \end{array}
\right) = \frac{1}{\sqrt{2}}\left(
            \begin{array}{cc}
              1 & 1 \\
             -i & i \\
            \end{array}
          \right) \left(
  \begin{array}{c}
    \tilde{f}_\alpha \\
    \tilde{g}_\alpha \\
  \end{array}
\right)  \nn\\
&=&  \frac{1}{2\sqrt{2}}\left(
      \begin{array}{c}
        C_1 \!+\! C_3 \!-\! i(C_2 \!+\! C_4) \\
        -i(C_1 \!-\! C_3) \!-\! (C_2 \!-\! C_4) \\
      \end{array}
    \right) (r-r_\star)^{+\nu_\lambda} \! + \! \frac{1}{2\sqrt{2}}\left(
      \begin{array}{c}
        C_1 \!+\! C_3 \!+\! i(C_2 \!+\! C_4) \\
        -i(C_1 \!-\! C_3) \!+\! (C_2 \!-\! C_4) \\
      \end{array}
    \right) (r-r_\star)^{-\nu_\lambda}, \nn
\eeqa
which are consistent with Eq.(\ref{Eq:Psi-fermions-AdS2-asymptotic}),
\beqa
\tilde\Psi_{\alpha}  \sim  \tilde{R}_\alpha  =  c_+  v_- (r-r_\star)^{+\nu_\lambda} + c_-  v_+ (r-r_\star)^{-\nu_\lambda} \equiv c_+ \ell_2^{+2\nu_\lambda} v_- \zeta^{-\nu_\lambda} + c_- \ell_2^{-2\nu_\lambda} v_+ \zeta^{+\nu_\lambda},
\eeqa
where $\zeta \equiv {\ell_2^2}/{(r-r_\star)}$.
By using $e_3$ and $\ell_2$ defined in Eqs.(\ref{Eq:ed-r_star}) and (\ref{Eq:ell_2}), considering the zero frequency case, one obtains
\beqa
&& \tilde{f}_\alpha^{\prime\prime}(r) + \frac{1}{r-r_\star} \tilde{f}_\alpha^{\prime}(r) + \bigg[ \Big( \frac{q e_3}{r-r_\star}   \big(1 - \frac{r-r_\star}{r_\star} \big) \Big)^2 - \frac{m_\lambda^2 \ell_2^2}{(r-r_\star)^2}  + i   \frac{q e_3}{(r-r_\star)}  \frac{1}{r_\star}   \bigg]\tilde{f}_\alpha(r) = 0, \nn\\
&& \tilde{g}_\alpha^{\prime\prime}(r) + \frac{1}{r-r_\star} \tilde{g}_\alpha^{\prime}(r) + \bigg[ \Big( \frac{q e_3}{r-r_\star}   \big(1 - \frac{r-r_\star}{r_\star} \big) \Big)^2 - \frac{m_\lambda^2 \ell_2^2}{(r-r_\star)^2}  - i   \frac{q e_3}{(r-r_\star)}  \frac{1}{r_\star}   \bigg]\tilde{g}_\alpha(r) = 0, \label{Eq:Dirac-EOMs-fgt_alpha-IR-omega=0-b=0}
\eeqa
which have the solutions
\beqa
\tilde{f}_\alpha(r) \3i && \3i = e^{-iq e_3 \tilde{r}}(r-r_\star)^{\nu_\lambda}  \big[C_3 U [\nu_\lambda - iq e_3, 1+ 2\nu_\lambda, 2iq e_3 \tilde{r} ]  +  C_4 L_{-\nu_\lambda+i qe_3}^{2\nu_\lambda} (2 iq e_3 \tilde{r} ) \big], \label{Eq:ft-r} \\
\tilde{g}_\alpha(r) \3i && \3i = e^{-iq e_3 \tilde{r}}(r-r_\star)^{\nu_\lambda} \big[C_1 U [1+\nu_\lambda - iq e_3, 1+ 2\nu_\lambda, 2iqe_3 \tilde{r}] +  C_2 L_{-1-\nu_\lambda + i qe_3}^{2\nu_\lambda}(2iqe_3\tilde{r}) \big], \label{Eq:gt-r}
\eeqa
where $\tilde{r}\equiv {(r-r_\star)}/{r_\star}$, $U(\mu,\nu,z)$ is Tricomi's confluent hypergeometric function and $L_n^a(z) $ is the generalized Laguerre polynomial. The results are consistent with those obtained with a non-zero frequency case at zero temperature as shown in Eq.(\ref{Eq:ft_alpha-zeta}) and Eq.(\ref{Eq:gt_alpha-zeta}) with $b_r=0$ (or $b_\zeta=0$) and $b_0=0$. In this case, due to the exchange symmetry between $\tilde{m}_{1,2}(r)$ as defined in Eq.(\ref{Eq:m_alpha(r)-mt_alpha}), there exist identity relations between wavefunctions as shown in Eq.(\ref{Eq:ft-gt-m_alpha-symmetry}). Thus there exists an exchange relation between wavefunctions, i.e., $(\tilde{g}_\alpha,\tilde{f}_\alpha)(\zeta)\leftrightarrow (\tilde{f}_\alpha,\tilde{g}_\alpha)(r)$. It is worthy to notice the equivalent relation between the wavefunctions by making the replacement as
\beqa
\ii
q e_3 \tilde{r} \sim  \omega\zeta \quad \Rightarrow \quad \omega \sim q \mu \frac{(r-r_\star)^2}{r_\star^2},
\eeqa
where $\zeta$ is defined in Eq.(\ref{Eq:zeta-eta-k=0}) for Ricci flat hypersurface case with $k=0$. The last equivalent relation will be more obvious by observing the terms associated with imaginary unit $i$ in Eq.(\ref{Eq:Dirac-EOMs-fgt_alpha-IR-omega!=0-b=0}).

In the presence of chiral gauge fields of $b_r$ and $b_0$, in the zero frequency limit $\omega=0$, the EOMs can be expressed as
\beqa
&& \tilde{f}_{\alpha,\beta}^{\prime\prime}(r) + \bigg( -2i \frac{\beta b_0\ell_2^2}{(r-r_\star)^2} + \frac{1}{r-r_\star} \bigg) \tilde{f}_{\alpha,\beta}^{\prime}(r) + \bigg[ \Big( \frac{qe_3}{r-r_\star}   \big(1 - \frac{r-r_\star}{r_\star} \big) - \beta b_r \Big)^2 \nn\\
&& -  \frac{(b_0\ell_2^2)^2}{(r-r_\star)^4} - \frac{m_\lambda^2\ell_2^2}{(r-r_\star)^2} + i \bigg( + \frac{\beta b_0 \ell_2^2}{(r-r_\star)^3} + \frac{q e_3 }{(r-r_\star)}   \frac{1}{r_\star} - \frac{\beta b_r}{r-r_\star}  \bigg) \bigg]\tilde{f}_{\alpha,\beta}(r) = 0, \nn\\
&& \tilde{g}_{\alpha,\beta}^{\prime\prime}(r) + \bigg( -2i \frac{\beta b_0\ell_2^2}{(r-r_\star)^2} + \frac{1}{r-r_\star} \bigg) \tilde{g}_{\alpha,\beta}^{\prime}(r) + \bigg[ \Big( \frac{qe_3 }{r-r_\star}   \big(1 - \frac{r-r_\star}{r_\star} \big) - \beta b_r \Big)^2 \nn\\
&& -  \frac{(b_0\ell_2^2)^2}{(r-r_\star)^4} - \frac{m_\lambda^2\ell_2^2}{(r-r_\star)^2} - i \bigg( - \frac{\beta b_0 \ell_2^2}{(r-r_\star)^3} + \frac{q e_3 }{(r-r_\star)}  \frac{1}{r_\star} - \frac{\beta b_r}{r-r_\star}  \bigg) \bigg]\tilde{g}_{\alpha,\beta}(r) = 0, \label{Eq:Dirac-EOMs-fgt_alpha-IR-omega=0-b!=0}
\eeqa
which give solutions
\beqa
\tilde{f}_{\alpha,\beta}(r) \3i && \3i = e^{-i \frac{\beta b_0\ell_2^2}{r-r_\star}} e^{-i\beta b_r(r-r_\star)} e^{-iq e_3  \tilde{r}}(r-r_\star)^{\nu_\lambda}  \big[C_3 U [\nu_\lambda - iq e_3  + \beta \tilde{b}_r, 1+ 2\nu_\lambda, 2 i[q e_3  \tilde{r} + \beta b_r(r-r_\star)] ]  \nn\\
&+&  C_4 L_{-\nu_\lambda   + i q e_3  - \beta \tilde{b}_r}^{2\nu_\lambda}(2 i[q e_3  \tilde{r} + \beta b_r(r-r_\star)]) \big], \nn \label{Eq:ft-r-br-b0} \\
\tilde{g}_{\alpha,\beta}(r) \3i && \3i = e^{-i \frac{\beta b_0\ell_2^2}{r-r_\star}} e^{-i\beta b_r(r-r_\star)} e^{-iq e_3  \tilde{r}}(r-r_\star)^{\nu_\lambda} \big[C_1 U [1+\nu_\lambda - iq e_3  - \beta \tilde{b}_r, 1+ 2\nu_\lambda, 2 i[q e_3  \tilde{r} + \beta b_r(r-r_\star)]] \nn\\
&+&  C_2 L_{-1-\nu_\lambda + i q e_3  + \beta \tilde{b}_r}^{2\nu_\lambda}(2 i[q e_3  \tilde{r} + \beta b_r(r-r_\star)]) \big], \nn \label{Eq:gt-r-br-b0}
\eeqa
where $\tilde{r}\equiv {(r-r_\star)}/{r_\star}$ and $\tilde{b}_r \equiv {b_r r_\star}/( b_r r_\star + q e_3 ) $. The wave functions above can also be rearranged as
\beqa
&& \tilde{f}_\alpha(r) = e^{-i \frac{\beta b_0\ell_2^2}{r-r_\star}} \frac{1}{\sqrt{r-r_\star}}[ C_1  W_{+\frac{1}{2}+i q e_3  - \beta \tilde{b}_r,\nu_\lambda}(2i[\tilde{r} + \beta b_r (r-r_\star)]) + C_2  M_{+\frac{1}{2}+i q e_3  - \beta \tilde{b}_r,\nu_\lambda}(2i[\tilde{r} + \beta b_r (r-r_\star)]) ], \nn\\
&& \tilde{g}_\alpha(r) = e^{-i \frac{\beta b_0\ell_2^2}{r-r_\star}} \frac{1}{\sqrt{r-r_\star}}[ C_1  W_{-\frac{1}{2}+i q e_3  + \beta \tilde{b}_r,\nu_\lambda}(2i[\tilde{r} + \beta b_r (r-r_\star)]) + C_2  M_{-\frac{1}{2}+i q e_3  + \beta \tilde{b}_r,\nu_\lambda}(2i[\tilde{r} + \beta b_r (r-r_\star)]) ], \nn
\eeqa
where $W_{k,m}(z)$ and $M_{k,m}(z)$ are Whittaker's functions. The other independent wave functions can be obtained by making the replacement $(b_0,b_r)  \to - (b_0,b_r)$, as discussed after Eq.(\ref{Eq:Marster_Dirac-EOMs_b0_br}).


For the general case with $\omega\ne 0$, the EOMs are
\beqa
&& \tilde{f}_\alpha^{\prime\prime}(r) + \frac{1}{r-r_\star}  \tilde{f}_\alpha^{\prime}(r) + \bigg[ \bigg( \frac{\ell_2^2}{(r-r_\star)^2}\Big(\omega \!+\!  q \mu \frac{r-r_\star}{r_\star}\big(1 - \frac{r-r_\star}{r_\star} \big) \Big)  \bigg)^2 - \frac{m_\lambda^2\ell_2^2}{(r-r_\star)^2}      \nn\\
             && + i \bigg(   - \frac{\ell_2^2}{(r-r_\star)^2}\Big( \frac{\omega}{r-r_\star} +q \mu \frac{1}{r_\star}\big( - \frac{r-r_\star}{r_\star} \big) \Big)  \bigg) \bigg]\tilde{f}_\alpha(r) =0, \nn\\
&& \tilde{g}_\alpha^{\prime\prime}(r) + \frac{1}{r-r_\star} \tilde{g}_\alpha^{\prime}(r) + \bigg[ \bigg( \frac{\ell_2^2}{(r-r_\star)^2}\Big(\omega \!+\!  q \mu \frac{r-r_\star}{r_\star}\big(1 - \frac{r-r_\star}{r_\star} \big)\Big)\bigg)^2  - \frac{m_\lambda^2\ell_2^2}{(r-r_\star)^2}     \nn\\
             && - i \bigg(   - \frac{\ell_2^2}{(r-r_\star)^2}\Big( \frac{\omega}{r-r_\star} +q \mu \frac{1}{r_\star}\big( - \frac{r-r_\star}{r_\star} \big) \Big)   \bigg) \bigg]\tilde{g}_\alpha(r) =0,  \label{Eq:Dirac-EOMs-fgt_alpha-IR-omega!=0-b=0}
\eeqa
with solutions
\beqa
\tilde{f}_\alpha(r) &=& C_1 e^{-iq e_3  \tilde{r}} e^{-i \frac{\omega \ell_2^2}{r-r_\star}} \text{HeunD}\bigg[+4\tilde\omega,-4\big(\nu_\lambda^2+(1+2iq e_3 )\tilde\omega\big),0,4\big(\nu_\lambda^2-(1+2iq e_3 )\tilde\omega\big),-\frac{\tilde\omega-2iq e_3  \tilde{r}}{\tilde\omega+2iq e_3  \tilde{r}}\bigg] \nn\\
&+& C_2 e^{+iq e_3  \tilde{r}} e^{+i \frac{\omega \ell_2^2}{r-r_\star}} \text{HeunD}\bigg[-4\tilde\omega,-4\big(\nu_\lambda^2+(1+2iq e_3 )\tilde\omega\big),0,4\big(\nu_\lambda^2-(1+2iq e_3 )\tilde\omega\big),-\frac{\tilde\omega-2iq e_3  \tilde{r}}{\tilde\omega+2iq e_3  \tilde{r}}\bigg],\nn\\
\tilde{g}_\alpha(r) &=& C_3 e^{+iq e_3  \tilde{r}} e^{-i \frac{\omega \ell_2^2}{r-r_\star}} \text{HeunD}\bigg[+4\tilde\omega,-4\big(\nu_\lambda^2-(1-2iq e_3 )\tilde\omega\big),0,4\big(\nu_\lambda^2+(1-2iq e_3 )\tilde\omega\big),-\frac{\tilde\omega-2iq e_3  \tilde{r}}{\tilde\omega+2iq e_3  \tilde{r}}\bigg] \nn\\
&+& C_4 e^{-iq e_3  \tilde{r}} e^{+i \frac{\omega \ell_2^2}{r-r_\star}} \text{HeunD}\bigg[-4\tilde\omega,-4\big(\nu_\lambda^2-(1-2iq e_3 )\tilde\omega\big),0,4\big(\nu_\lambda^2+(1-2iq e_3 )\tilde\omega\big),-\frac{\tilde\omega-2iq e_3  \tilde{r}}{\tilde\omega+2iq e_3  \tilde{r}}\bigg], \nn
\eeqa
where $\tilde{r}\equiv {(r-r_\star)}/{r_\star}$, $\tilde\omega \equiv \sqrt{4 q e_3 \omega  /r_\star }\ell_2$ and HeunD is Heun doubleconfluent function $\text{HeunD}[a,b,c,d,z]$, which is solution to the ODE $y^{\prime\prime}(z)-[a^2 (z^2+1) - 2 z(z^2-1)]y^\prime(z)/(z^2-1)^2  + [b z^2 +(2a+c)z +d]y(z)/(z^2-1)^3=0$. For the charged case with both $(b_0,b_r)\ne 0$ and $q\ne 0$, it turns out that the general wave functions can also be expressed as linear combination of Heun doubleconflument function, but we do not intend to present them here.

\subsection{Fermi surface}

In this section, we try to extract the information on the Fermi surface for Dirac fermions by solving flow equation near the boundary.

\subsubsection{Flow equation and Fermi surface}

For the $b_r\ne 0$ case, in the infinite boundary, one has
\beqa
\partial_r \xi_{\alpha,\beta} =   - \beta b_r  (1+\xi_{\alpha,\beta}^2)  - 2 m_D  \frac{\ell}{r}\xi_{\alpha,\beta},
\eeqa
which has the solutions
\beqa
\xi_{\alpha,\beta}(r) = - \frac{J_{\frac{1}{2}+ m_D \ell}(\beta b_r r) C_1 + Y_{\frac{1}{2}+ m_D \ell}(\beta b_r r)}{J_{-\frac{1}{2}+ m_D \ell}(\beta b_r r) C_1 + Y_{-\frac{1}{2}+ m_D \ell}(\beta b_r r)}.
\eeqa
In fact, they are independent of subscript $\alpha=1,2$.

In the infinite boundary, according to Eq.(\ref{Eq:DiracEOM-flow-xi_alpha-UV-AdS(d+1)-omega!=0}), in the absence of $b_r$, the flow equation becomes
\beqa
\partial_r \xi_\alpha =    \frac{\ell^2}{r^2} ( \omega + q \mu )  (1+\xi_\alpha^2) +  \frac{\ell^2}{r^2} \bigg(  - \tilde{m}_\alpha \frac{r_\star}{\ell} (1-\xi_\alpha^2)  - 2m_D\frac{r}{\ell}\xi_\alpha \bigg),
\eeqa
where $\tilde{m}_\alpha$ is defined in Eq.(\ref{Eq:mt_alpha}) with $\alpha=1,2$. The analytical solution to the above flow equation turns out to be
\beqa
\xi_\alpha(r) \overset{r\to \infty}{=} \sqrt\frac{\tilde{m}_\alpha r_\star- \ell (\omega+ q\mu)}{\tilde{m}_\alpha r_\star +  \ell(\omega+ q\mu) }  N(r)
\equiv \sqrt\frac{\tilde{m}_\alpha^-}{\tilde{m}_\alpha^+} N(r), \nn
\eeqa
where $N(r)$ is a function
\beqa
N(r) = \frac{e^{i \ell m_D \pi} \Gamma[\frac{1}{2}- m_D \ell ] I_{+\frac{1}{2}- m_D \ell}\big( \ell \frac{r_\star}{r}\sqrt{\tilde{m}^+_\alpha\tilde{m}^-_\alpha}\big)C_1 -i \Gamma[\frac{3}{2}+ m_D \ell] I_{-\frac{1}{2}+ m_D \ell}\big( \ell \frac{r_\star}{r}\sqrt{\tilde{m}_\alpha^+\tilde{m}_\alpha^-} \big)}{e^{i  \ell m_D \pi} \Gamma[\frac{1}{2}- m_D \ell ] I_{-\frac{1}{2}- m_D \ell}\big( \ell \frac{r_\star}{r}\sqrt{\tilde{m}_\alpha^+\tilde{m}_\alpha^-}\big)C_1 -i \Gamma[\frac{3}{2}+ m_D \ell] I_{+\frac{1}{2}+ m_D \ell}\big(  \ell \frac{r_\star}{r} \sqrt{\tilde{m}^+_\alpha\tilde{m}^-_\alpha} \big)}, \nn
\eeqa
which in fact is independent of subscript $\alpha=1,2$, since the product $\tilde{m}^+_\alpha \tilde{m}^-_\alpha$ is independent of $\alpha$,
\beqa
  \tilde{m}^{\pm}_\alpha  &\equiv & \tilde{m}_\alpha \pm  \frac{\ell}{r_\star}(\omega+ q\mu) = [-(-)^\alpha \lambda \pm (\omega+ q\mu) ]\frac{\ell}{r_\star}.
\label{Eq:mt_pm}
\eeqa

For the vacuum case, i.e., $q=0$, $ \omega\ll \tilde{m}_\alpha $, the two mass spectra are degenerate.
In this case, $ \tilde{m}^+_\alpha \approx \tilde{m}^-_\alpha \approx \tilde{m}_\alpha  $,
\beqa
\ii
N(r) \! \overset{r\to \infty}{=} \!
\frac{1+2\nu_3}{C(\tilde{m}^\pm, \nu_3)}  \bigg( \frac{i}{2} \bigg)^{1+2\nu_3}  \bigg(  \ell \frac{ r_\star}{r}\sqrt{\tilde{m}^+\tilde{m}^-}\bigg)^{2\nu_3} \3i ,
\eeqa
where $C(\tilde{m}^\pm, \nu_3) $ is a function to be determined by near horizon boundary condition $\xi_\alpha = i  $ for the $\omega\ne 0  $ case as shown in Eq.(\ref{Eq:xi_alpha-horizon-omega!=0}),
or $\xi_\alpha = (m_D\ell_2-\nu_\lambda)/(q e_3  + \tilde{m}_\alpha \ell_2) $ for the $\omega = 0  $ case as shown in Eq.(\ref{Eq:xi_alpha-horizon-omega=0}),
\beqa
C(\tilde{m}^\pm, \nu_3)\left\{\begin{aligned}
& (-1)^{\frac{1}{2}+\nu_3}\frac{\Gamma[\frac{3}{2}+\nu_3]}{\Gamma[\frac{1}{2}-\nu_3]}  \frac{
 I_{\frac{1}{2}+\nu_3}\big(  \ell \sqrt{\tilde{m}^+ \tilde{m}^-} \big) +i \sqrt\frac{\tilde{m}_\alpha^-}{\tilde{m}_\alpha^+}  I_{-\frac{1}{2}+\nu_3}\big(  \ell \sqrt{\tilde{m}^+\tilde{m}^-} \big)}{   I_{-\frac{1}{2}-\nu_3}\big(  \ell  \sqrt{\tilde{m}^+\tilde{m}^-} \big) + i \sqrt\frac{\tilde{m}_\alpha^-}{\tilde{m}_\alpha^+} I_{\frac{1}{2}-\nu_3}\big(  \ell  \sqrt{\tilde{m}^+\tilde{m}^-} \big)}, \quad \omega\ne 0, \\
& (-1)^{\frac{1}{2}+\nu_3}\frac{\Gamma[\frac{3}{2}+\nu_3]}{\Gamma[\frac{1}{2}-\nu_3]}  \frac{
(m_D\ell_2 -\nu_\lambda) I_{\frac{1}{2}+\nu_3}\big(  \ell \sqrt{\tilde{m}^+ \tilde{m}^-} \big) - (q e_3  + \tilde{m}_\alpha \ell_2) \sqrt\frac{\tilde{m}_\alpha^-}{\tilde{m}_\alpha^+} I_{-\frac{1}{2}+\nu_3}\big(  \ell \sqrt{\tilde{m}^+\tilde{m}^-} \big)}{ (m_D\ell_2 -\nu_\lambda)  I_{-\frac{1}{2}-\nu_3}\big(  \ell \sqrt{\tilde{m}^+\tilde{m}^-} \big) - (q e_3  + \tilde{m}_\alpha \ell_2) \sqrt\frac{\tilde{m}_\alpha^-}{\tilde{m}_\alpha^+} I_{\frac{1}{2}-\nu_3}\big(  \ell  \sqrt{\tilde{m}^+\tilde{m}^-} \big)}, \quad \omega=0 .
\end{aligned}\right. \nn
\eeqa
where $I_{\nu}(z)$ is the modified/hyperbolic Bessel function of the first kind, defined by $I_\nu(z)=i^{-\nu}J_\nu(iz)$, which is exponentially growing function $I_\nu(z)\overset{z\to\infty}{\sim}e^z/\sqrt{2\pi z}$. It is one branch of the two linearly independent solutions to the modified Bessel's differential equation $z^2y^{\prime\prime}(z)+zy^\prime(z)-(z^2+\nu^2)y(z)=0$.
For the $\omega=0 $ case, one can solve the pole of $C(\tilde{m}_\alpha,\nu_3)=0$ near the Fermi surface, we expand $\tilde{m}_\alpha$ in the numerator of $C(\tilde{m}_\alpha,\nu_3)$ for the $\omega=0$ case, meanwhile remember that $\nu_\lambda=\sqrt{(m_D^2+\tilde{m}^2)\ell_2^2-q^2 e_3^2}$ also depends on $\tilde{m}$, then one has
\beqa
\tilde{m}_\alpha \overset{\lambda\to \lambda_F}{\approx}  \frac{\mu  \ell}{e_3  r_\star  - \mu  \ell \ell_2}\bigg( q e_3  + \big( \sqrt{m_D^2\ell_2^2-q^2 e_3^2}-m_D\ell_2 \big) \frac{J_{ m_D \ell + \frac{1}{2}}(q\mu  \ell^2/r_\star)}{J_{ m_D \ell - \frac{1}{2}}(q\mu  \ell^2/r_\star)}  \bigg),
\eeqa
where $J_{\nu}(z)$ is the first kind of Bessel functions, which is one of independent solutions of differential equation $z^2y^{\prime\prime}(z)+zy^\prime(z)+(z^2-\nu^2)y(z)=0$, and we have considered the result that $J_{\nu}(z)\overset{\nu\to \infty}{\sim} \sqrt\frac{2}{\pi z}\cos\big(-z + \frac{\pi}{4}(2\nu+1)\big) $, this implies that the Fermi momentum approaches to a maximum,
\beqa
\tilde{m}_\alpha \overset{m_D\to \infty}{=} \mu   \ell \frac{ q e_3 }{e_3  r_\star  - \mu  \ell   \ell_2} + {\mathcal O}(\frac{1}{m_D}).
\eeqa
Therefore the Fermi momentum can be approximately calculated through Eq.(\ref{Eq:lambda-k-b}),
\beqa
k_F \overset{m_D \ne 0}{\approx} \pm \sqrt{\lambda^2 + b^2} = \pm \sqrt{ \tilde{m}^2 \frac{r_\star^2}{\ell^2} +  b^2 }.
\eeqa
In the absence of chiral gauge field $b_i=0$, it turns out to be,
\beqa
k_F \overset{m_D \ne 0}{\approx} && \ii -(-1)^\alpha \frac{  \mu  r_\star}{e_3  r_\star  - \mu  \ell \ell_2  }\bigg(  q e_3  + (\sqrt{m_D^2\ell_2^2-q^2 e_3^2} -m_D\ell_2) \frac{J_{ m_D \ell + \frac{1}{2}}(q\mu  \ell^2/r_\star)}{J_{ m_D \ell - \frac{1}{2}}(q\mu  \ell^2/r_\star)}  \bigg).  \label{Eq:kF-massive}
\eeqa
In the limit of $ \mu \to 0 $, the Fermi momentum approaches to $ | q\mu| $.

\end{widetext}

\subsubsection{Retarded Green's function and fermi surface:massless fermion, vacuum and low frequency}

Thus one obtains the retarded Green's function for the standard quantization as
\beqa
G_{s\alpha}^R(\omega,\lambda) =  \frac{1+2\nu_3}{C(\tilde{m}^\pm_\alpha, \nu_3)} \bigg( \frac{i}{2} \bigg)^{1+2\nu_3}  \sqrt\frac{\tilde{m}^-_\alpha}{\tilde{m}^+_\alpha} \bigg(  r_\star \ell  \sqrt{\tilde{m}^+\tilde{m}^-} \bigg)^{2\nu_3} ,  \nn
\eeqa
note that $\tilde{m}_\alpha $ carries a subscript $\alpha$. Moreover, according to Eq.(\ref{Eq:GIt-property-1}),
\beqa
G_{a\alpha}(\omega,\lambda,m_D)  \equiv  - G_{s\alpha}^{-1}(\omega,\lambda,m_D) = G_{s\alpha}(\omega,-\lambda, -m_D), \nn
\eeqa
one obtains the retarded Green's function for the alternative quantization
\beqa
\3i && \3i G_{a\alpha}^R(\omega,\lambda,m_D) = G_{s\alpha}^R(\omega,-\lambda,-m_D) \nn\\
\3i && \3i = \frac{1-2\nu_3}{C(-\tilde{m}^\mp_\alpha, -\nu_3)} \sqrt\frac{\tilde{m}^+_\alpha}{\tilde{m}^-_\alpha} \bigg( \frac{i}{2} \bigg)^{1-2\nu_3} r_\star^{-2\nu_3}  \bigg(  \ell \sqrt{\tilde{m}^+\tilde{m}^-}\bigg)^{-2\nu_3}  . \nn
\eeqa

In the massless limit i.e., $\nu_3=0$ (or $m_D=0$), the retarded Green's functions become
\beqa
\ii && G_{s\alpha}^R(\omega,\lambda) = \frac{1}{C( \tilde{m}_\alpha^\pm,  \nu_3=0)} \frac{i}{2} \sqrt\frac{\tilde{m}^-_\alpha}{\tilde{m}^+_\alpha}, \nn\\
\ii && G_{a\alpha}^R(\omega,\lambda) = \frac{1}{C(-\tilde{m}_\alpha^\pm, -\nu_3=0)} \frac{i}{2} \sqrt\frac{\tilde{m}^+_\alpha}{\tilde{m}^-_\alpha},
\eeqa
where $ C(\tilde{m}^\pm_\alpha, \nu_3=0) $ turns out to be
\beqa
 \left\{\begin{aligned}
& \frac{i}{2}\frac{\tanh( \ell\sqrt{\tilde{m}^+\tilde{m}^-})+i\sqrt\frac{\tilde{m}_\alpha^-}{\tilde{m}_\alpha^+}}{1 + i\sqrt\frac{\tilde{m}_\alpha^-}{\tilde{m}_\alpha^+}\tanh( \ell\sqrt{\tilde{m}^+\tilde{m}^-})}, \quad \omega\ne 0\nn \\
& \frac{i}{2} \frac{\sqrt\frac{\tilde{m}_\alpha \ell_2 - q e_3 }{\tilde{m}_\alpha \ell_2 + q e_3} \tanh{( \ell\sqrt{\tilde{m}^+\tilde{m}^-})} + \sqrt\frac{\tilde{m}^-_\alpha}{\tilde{m}^+_\alpha} }{ \sqrt\frac{\tilde{m}_\alpha\ell_2-q e_3 }{\tilde{m}_\alpha\ell_2+q e_3 }+\sqrt\frac{\tilde{m}^-_\alpha}{\tilde{m}^+_\alpha} \tanh{( \ell\sqrt{\tilde{m}^+\tilde{m}^-})} },  \quad \omega=0 		 \end{aligned}\right.
\eeqa
where $\tilde{m}^+ \tilde{m}^- = (\lambda^2 -   (\omega+ q \mu)^2){\ell^2}/{r_\star^2} $ for the $\omega\ne 0$ case and $\tilde{m}^+ \tilde{m}^- = (\lambda^2 -   q^2 \mu^2){\ell^2}/{r_\star^2} $ for the $\omega= 0$ case, respectively. It is worth noticing that when $\lambda \to -\lambda $ and $ m_D \to -m_D $, then $ \tilde{m}^\pm \to -\tilde{m}^\mp $ and $ \nu_3 \to -\nu_3$.

\begin{figure}[H]
\begin{center}
  \includegraphics[scale=0.8]{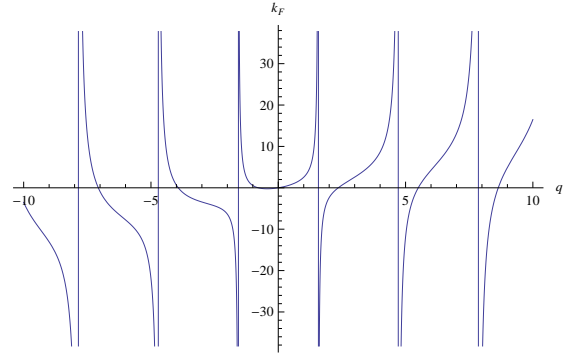}
\end{center}
  \caption{Fermi momentum $k_F$ vs. electric charge $q$ for massless Weyl fermion: the Fermi momentum is a periodic function of the charge with divergent poles.} \label{fig:kF_mD=0-q}
\end{figure}

For the $\omega=0$ case, one can solve the pole of $ C(\tilde{m}^\pm_\alpha, \nu_3=0) $ near the Fermi surface in the massless limit $\nu_3 = 0 $ (or $m_D = 0 $) but $q\ne 0$, which gives, respectively, $\tilde{m}_\alpha(\lambda\to \lambda_F)$ as
\beqa
\left\{\begin{aligned}
& \frac{  \ell (q\mu + \omega)}{r_\star} \bigg( 1 - i \tan\frac{  \ell^2 q\mu}{r_\star}\bigg), ~ \omega\ne 0 \nn \\
&      \frac{ \mu  \ell q e_3 }{ e_3  r_\star + i \mu  \ell  \ell_2 \tan \frac{  \ell^2 q\mu}{r_\star} }\bigg( 1 + i \tan\frac{  \ell^2 q\mu}{r_\star}\bigg),  ~ \omega = 0.
 		 \end{aligned}\right.
\eeqa
Moreover, in the vacuum, i.e., $\omega, \tilde{m}_\alpha \gg q $, or $q=0$ ($ \nu_\lambda = \tilde{m}_\alpha \ell_2 $), for the $\omega=0$ case,
\beqa
 C(\tilde{m}^\pm_\alpha, \nu_3=0)
 \overset{\omega=0}{=} \frac{i}{2}\frac{\tanh( \ell\sqrt{\tilde{m}^+\tilde{m}^-})+\sqrt\frac{\tilde{m}^-_\alpha}{\tilde{m}^+_\alpha}}{1+\sqrt\frac{\tilde{m}^-_\alpha}{\tilde{m}^+_\alpha}\tanh( \ell\sqrt{\tilde{m}^+\tilde{m}^-})}. \nn
\eeqa		
which gives
\beqa
\tilde{m}_\alpha \overset{\omega=0}{=} \frac{ \ell q \mu }{r_\star}\bigg( 1+  \tan \frac{  \ell^2 q \mu}{r_\star} \bigg).
\eeqa
Therefore the Fermi momentum in the massless limit near the Fermi surface,
\beqa
\lambda_F  \overset{m_D = 0}{\approx}   -(-1)^\alpha  q \mu \bigg( 1+  \tan \frac{  \ell^2 q \mu}{r_\star} \bigg),
\label{Eq:kF-massless}
\eeqa
which gives that $\lambda_F \! \overset{\mu \to 0}{\approx} \!  -(-1)^\alpha  q \mu$, where $\mu \equiv e_3  {r_\star}/{\ell_2^2}$.
One can check that for both massive case in Eq.(\ref{Eq:kF-massive}) and massless case in Eq.(\ref{Eq:kF-massless}), the Fermi momentum is a periodic function of charge $q$ with divergent poles, e.g., as shown in Fig.~\ref{fig:kF_mD=0-q} for massless Weyl fermion case.

In the vacuum limit, $q=0$, $ \omega \to 0$, one has $\tilde{m}^+_\alpha \approx \tilde{m}^-_\alpha \approx \tilde{m}_\alpha,  \sqrt{\tilde{m}^+\tilde{m}^-} \approx |\tilde{m}_\alpha|$, then
\beqa
 C(\tilde{m}^\pm_\alpha, \nu_3=0)  \left\{\begin{aligned}
& \frac{i}{2}\frac{\tanh( \ell|\tilde{m}|)+i}{1 + i\tanh( \ell|\tilde{m}|)}, \quad \omega\ne 0\nn \\
& \frac{i}{2},  \quad \omega=0. 		 \end{aligned}\right.
\eeqa
Thus the vacuum retarded Green's functions for the standard and alternative quantizations become, respectively,
\beqa
G_{s\alpha}^R(\omega,\lambda) =  \sqrt\frac{\tilde{m}^-_\alpha}{\tilde{m}^+_\alpha}, \quad \tilde{G}_{a\alpha}^R(\omega,\lambda) =  \sqrt\frac{\tilde{m}^+_\alpha}{\tilde{m}^-_\alpha}. \nn
\eeqa

According to Eq.(\ref{Eq:mt_pm}), for the standard and alternative quantizations, with a small frequency fluctuation, one obtains
\beqa
\3i && \3i G_{s1}^R(\omega,\lambda) =  \sqrt\frac{\tilde{m}^-_1}{\tilde{m}^+_1} = - \sqrt\frac{\lambda -  ( \omega+i\epsilon)}{\lambda+( \omega+i\epsilon)}, \nn\\
\3i && \3i G_{s2}^R(\omega,\lambda) =  \sqrt\frac{\tilde{m}^-_2}{\tilde{m}^+_2} = \sqrt\frac{\lambda + ( \omega+i\epsilon)}{\lambda-( \omega+i\epsilon)}; \\
\3i && \3i G_{a1}^R(\omega,\lambda) =  \sqrt\frac{\tilde{m}^+_1}{\tilde{m}^-_1} = - \sqrt\frac{\lambda + ( \omega+i\epsilon)}{\lambda-( \omega+i\epsilon)}, \nn\\
\3i && \3i G_{a2}^R(\omega,\lambda) =  \sqrt\frac{\tilde{m}^+_2}{\tilde{m}^-_2} = \sqrt\frac{\lambda- ( \omega+i\epsilon)}{\lambda+( \omega+i\epsilon)};
\eeqa
which are consistent with properties of retarded Green's function of massless spinor at UV fixed point, as shown in Eq.(\ref{Eq:GI-property-4-mD=0}) and Eq.(\ref{Eq:GIt-property-1-mD=0}). Consequently, this implies that both $\text{Im}G_{s1}^R(\omega,\lambda)$ and $\text{Im}G_{a1}^R(\omega,\lambda)$ have linear dispersion peak at $\omega \approx -  \lambda $, while both $\text{Im}G_{s2}^R(\omega,\lambda)$ and $\text{Im}G_{a1}^R(\omega,\lambda)$ have divergent peak at $\omega \approx +  \lambda $.

\subsubsection{Spectral function with finite frequency and charge}

Thus the general ARPES spectral function with finite frequency and charge is proportional to
\beqa
\ii && \3i \text{Im}G_{s\alpha}^R(\omega,\lambda) = -\frac{1}{2}\sqrt\frac{\tilde{m}^-_\alpha}{\tilde{m}^+_\alpha}\text{Re}\bigg( \frac{1}{C(\tilde{m}^\pm_\alpha, \nu_3)}\bigg) \nn\\
\ii &\overset{\omega\ne 0}{=}& \3i \frac{1}{2}\sqrt\frac{\tilde{m}^-_\alpha}{\tilde{m}^+_\alpha} \frac{4\sqrt{\tilde{m}^+_\alpha\tilde{m}^-_\alpha}}{(\tilde{m}^+_\alpha + \tilde{m}^-_\alpha)\cosh(2 \ell\sqrt{\tilde{m}^+\tilde{m}^-})-(\tilde{m}^+_\alpha-\tilde{m}^-_\alpha)}  \nn\\
\ii &=& \3i \frac{\tilde{m}_\alpha -  \frac{\ell}{r_\star}(\omega+ q\mu)}{ \tilde{m}_\alpha \cosh(2 \ell\sqrt{\tilde{m}^+\tilde{m}^-}) -  \frac{\ell}{r_\star}(\omega+ q\mu)}, \nn
\eeqa
where $\tilde{m}_\alpha^\pm$ is defined in Eq.(\ref{Eq:mt_pm}) and $\tilde{m}^+\tilde{m}^- \equiv \tilde{m}_\alpha^+\tilde{m}_\alpha^- = [\lambda^2 - (\omega+q\mu)^2] {\ell^2}/{r_\star^2}$.

\begin{figure}[H]
\begin{center}
   \includegraphics[scale=0.8]{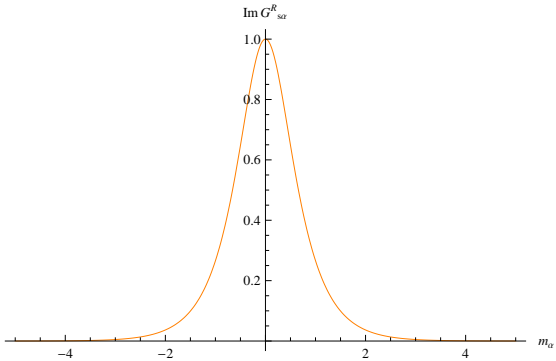}
\end{center}
  \caption{The shape of the spectral function Im$G^R_{s\alpha}(\omega)$: Imaginary part of retarded Green's function with nonzero frequency, i.e., $\omega = -10^{-3}$. We have chosen paprameters $\ell=r_\star=1$ and $q=0$.} \label{fig:ImGRs_alpha}
\end{figure}

Note that for the $ \omega=0 $ case, $\text{Re}C(\tilde{m},\nu_3=0)=0 $, as a result, $G^R(\omega=0,\lambda) $ has no imaginary part, $ \text{Im}G^R(\omega=0,\lambda)=0 $, this is consistent with Eq.(64) in ref.~\cite{Faulkner:2009wj}. Comparing with the vacuum limit case, the divergent peak of the ARPES at finite frequency and electric charge, is smoothed out into finite size peak, as shown in Fig.~\ref{fig:ImGRs_alpha}.

\subsubsection{Anomalous Hall effect of retarded Green's function for Weyl fermion: near Fermi surface}

\begin{figure}[H]
\begin{center}
  \includegraphics[scale=0.45]{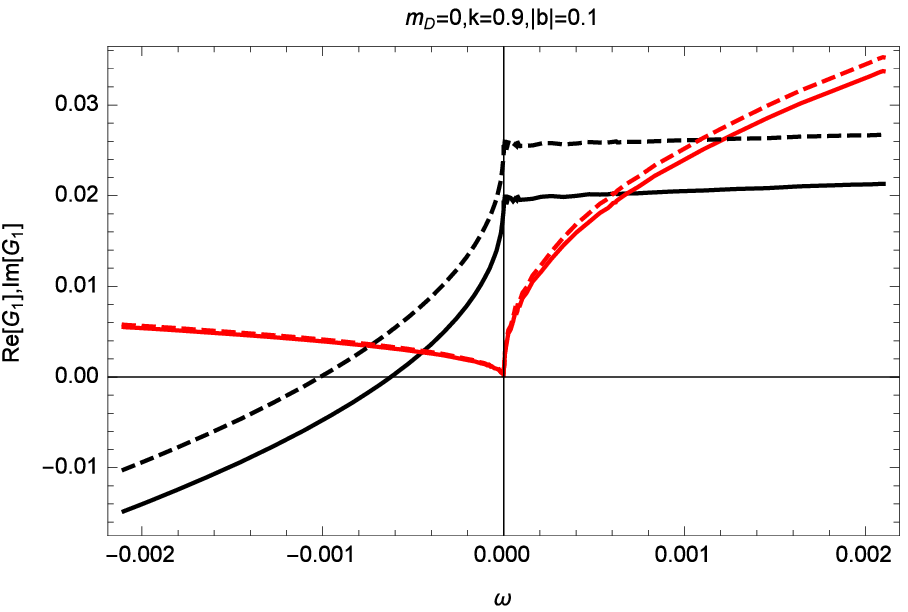} \includegraphics[scale=0.45]{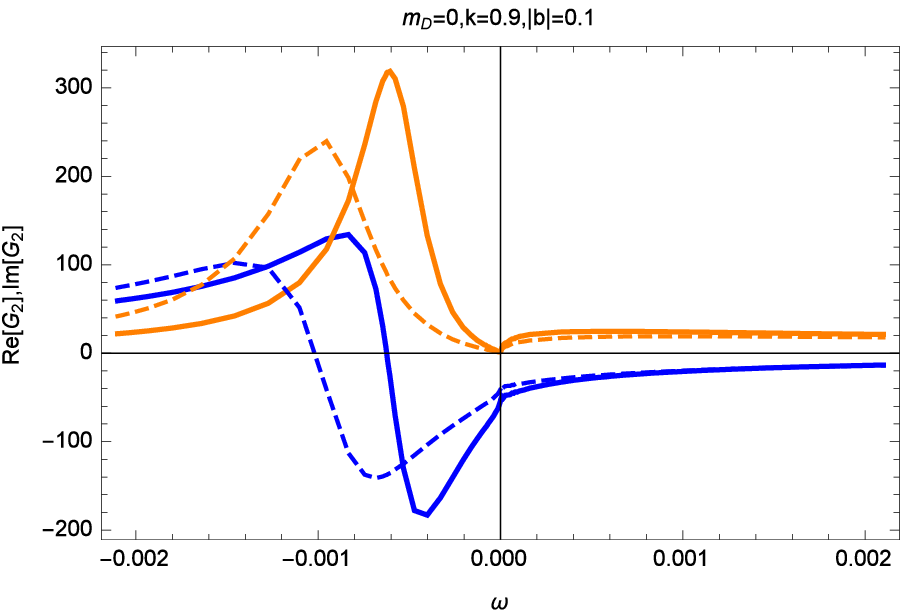}\\
  \includegraphics[scale=0.45]{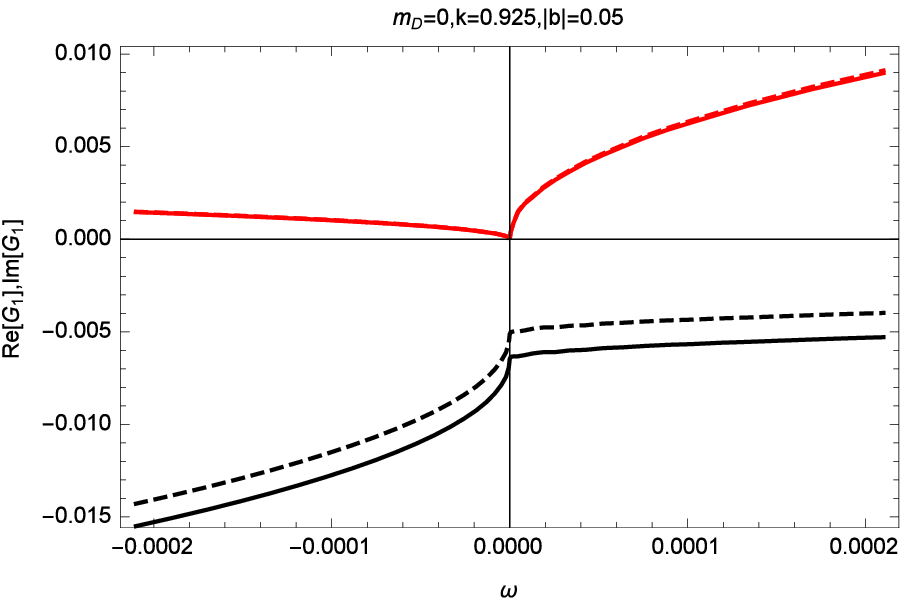} \includegraphics[scale=0.45]{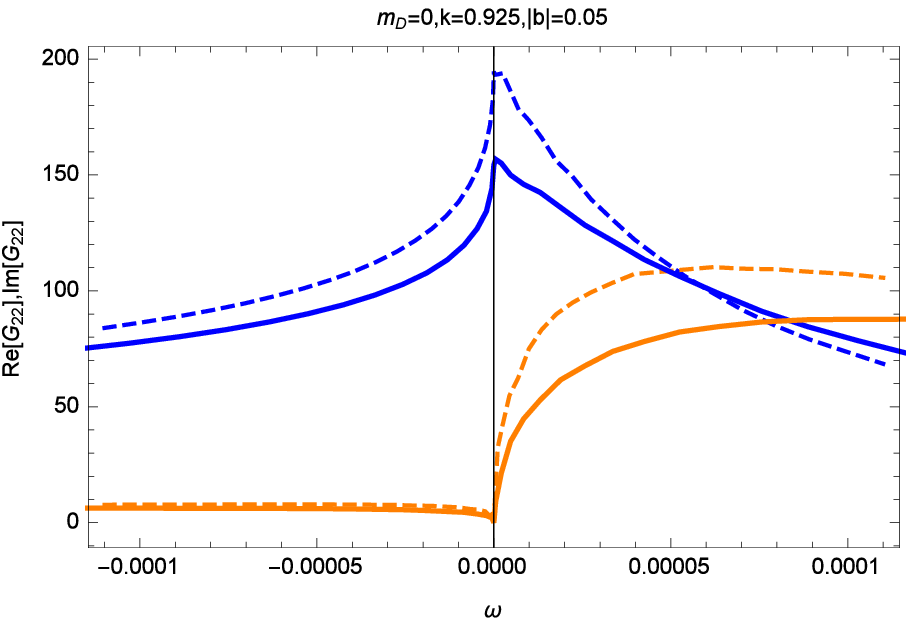}\\
\end{center}
  \caption{
  AHE of retarded Green's function for Weyl fermions ($m_D=0$) near Fermi surface from AdS$_4$ space-time: Re$G^R_{\alpha}(\omega)$ (black/blue curves), Im$G^R_{\alpha}(\omega)$ (red/orange curves) with $\alpha=1,2$, respectively, as functions of frequency $\omega$, given momentum at $k=0.9<k_F$ or $k=0.925>k_F$ and $|b|=0.1$. We have chosen parameters $r_\star=\ell=g_F=q=1$, the Fermi momentum turns out to be $k_F=0.9158$.
   } \label{fig:AHE_GR1-GR2_AdS}
\end{figure}

By observing Fig.~\ref{fig:AHE_GR1-GR2_AdS}, it is obvious that the spectral functions Im$G^R_{2}(\omega)$: (a) For the $k<k_F$ case, there is a quasi-particle like peak at $\omega<0$, while there is a tiny bump at $\omega>0$. As $k$ approaches $k_F$, the peak and the bump approach $\omega=0$ and their heights approach infinity. (b) For the $k>k_F$ case, there is a bump at $\omega>0$ and a smaller bump at $\omega<0$. As $k\to k_F$ from positive side, i.e., $(k-k_F)\to 0^+$, both bumps approach $\omega=0$ and their heights approach infinity. Without loss of generality, we have chosen a set of input parameters as $r_\star=\ell=g_F=q=1$.

For the special case without AHE, i.e., $|\vec{b}|=0$ for fermion in the AdS$_4$ space-time, the retarded Green's function Re$G^R_{2}(\omega)$ (solid blue curve) and Im$G^R_{2}(\omega)$ (solid orange curve), just recover the results in ref.~\cite{Liu:2009dm}, as shown in Fig.~3 and Fig.~5, respectively.

\subsubsection{Anomalous Hall effect of retarded Green's functions for Dirac Fermion: Far from Fermi Surface}

\begin{figure}[H]
\begin{center}
  \includegraphics[scale=0.45]{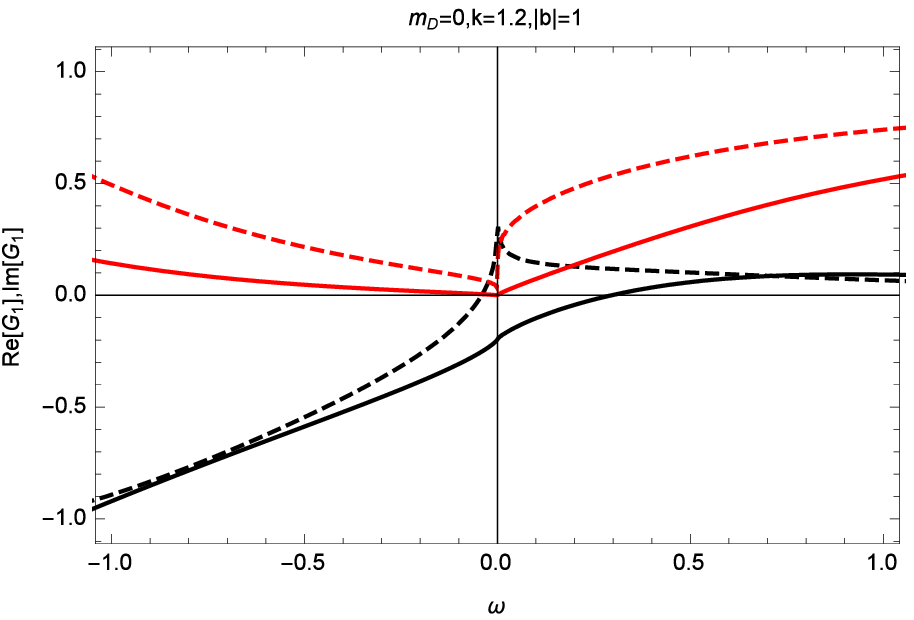} \includegraphics[scale=0.45]{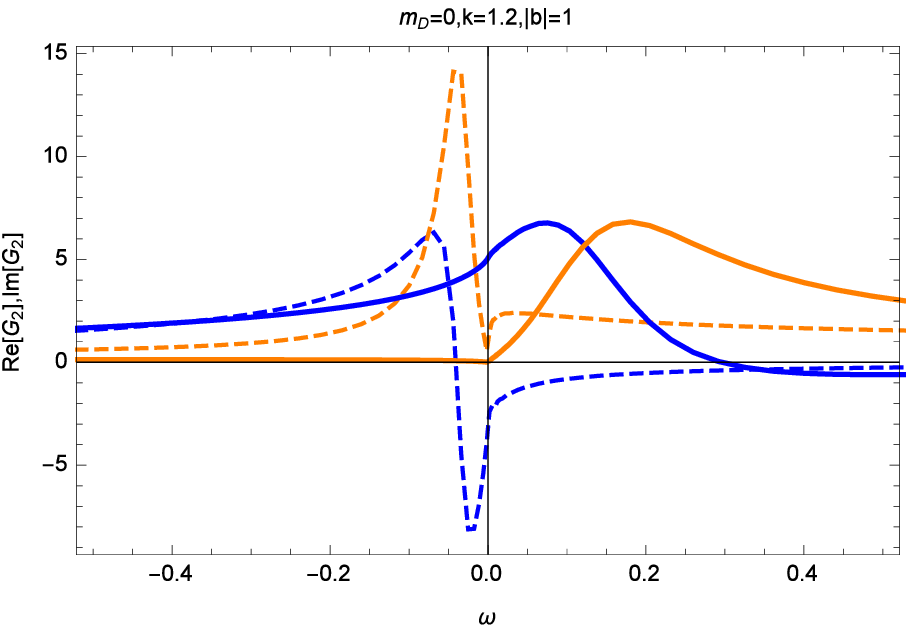}\\
  \includegraphics[scale=0.45]{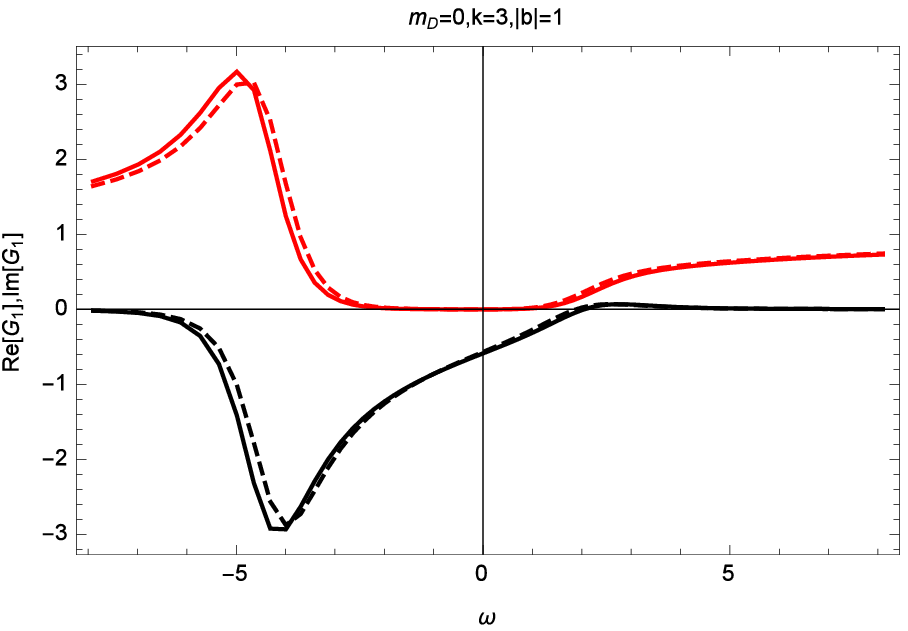} \includegraphics[scale=0.45]{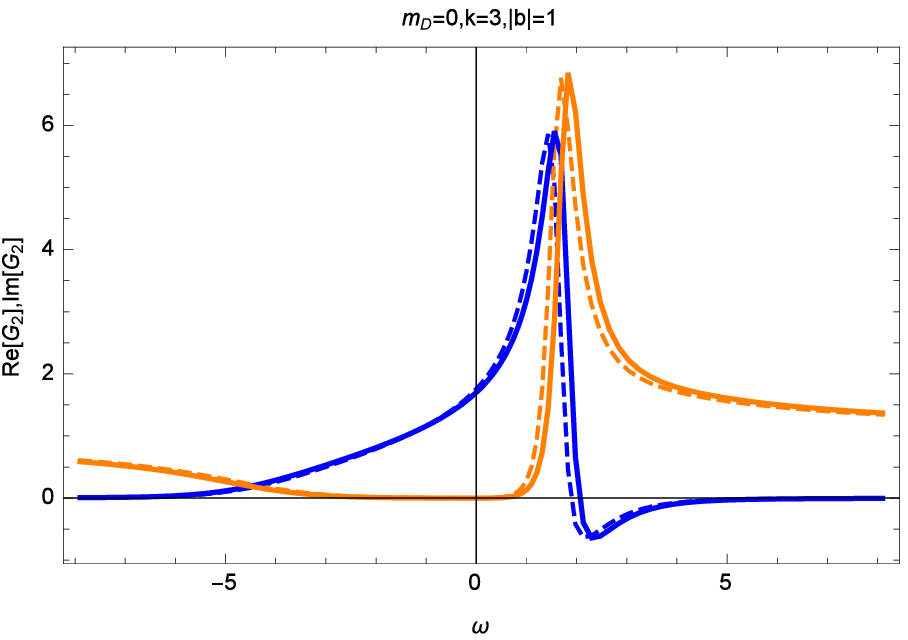}\\
\end{center}
  \caption{
  AHE of retarded Green's function for Weyl fermions ($m_D=0$) in AdS$_4$ space time: Re$G^R_\alpha(\omega)$ (blue curve) and Im$G^R_\alpha(\omega)$ (orange curve), given momentum $|\vec{k}|=1.2 > q \mu $ (small momentum) and $|\vec{k}|=3.0 > q \mu $ (large momentum). We have chosen input parameters, $r_\star=\ell=g_F=q=1$, $|\vec{b}| = 0$ (solid curve) and $|\vec{b}|=1$ (dashed curve).
   } \label{fig:massless_AHE_AdS}
\end{figure}

\begin{figure}[H]
\begin{center}
  \includegraphics[scale=0.45]{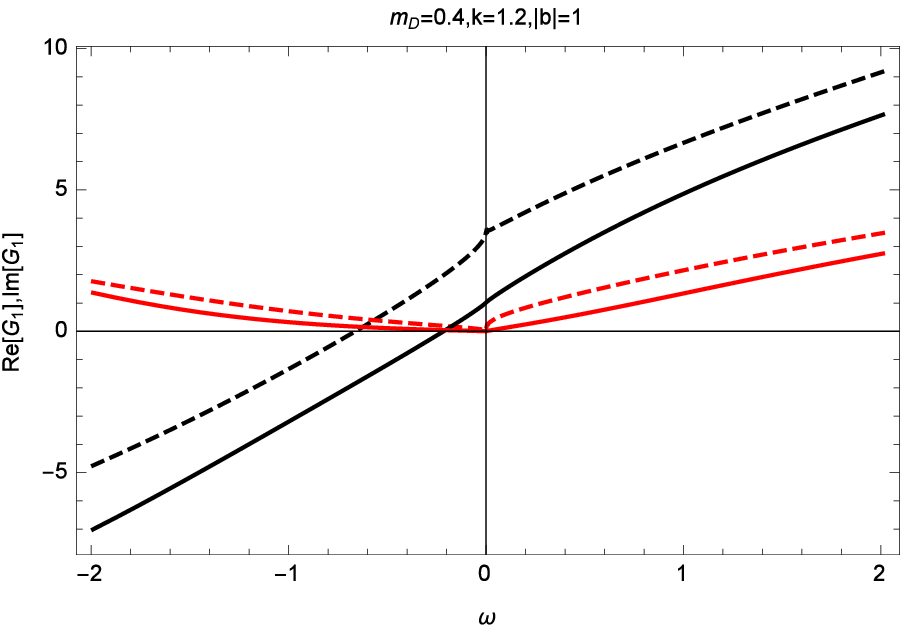} \includegraphics[scale=0.45]{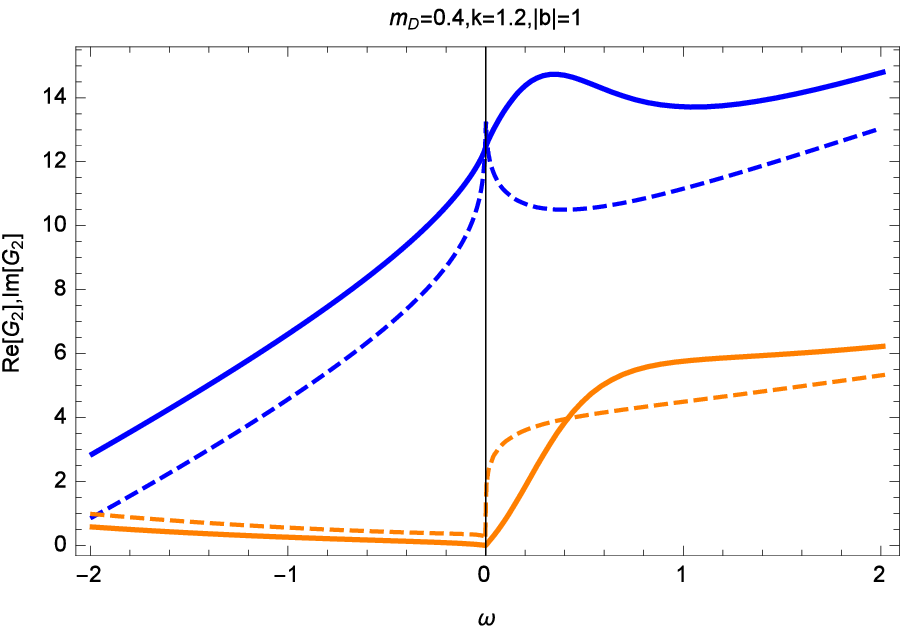}\\
  \includegraphics[scale=0.45]{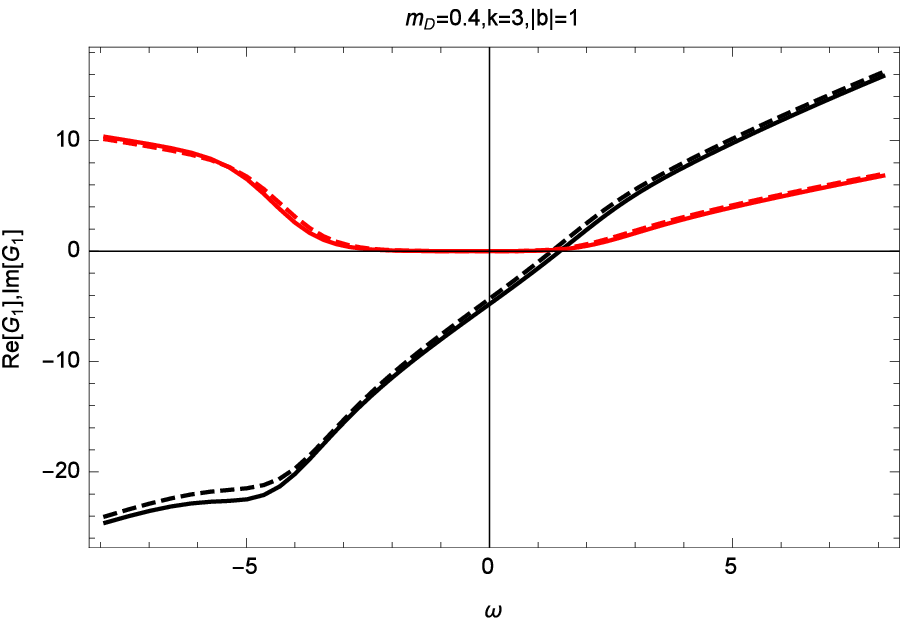} \includegraphics[scale=0.45]{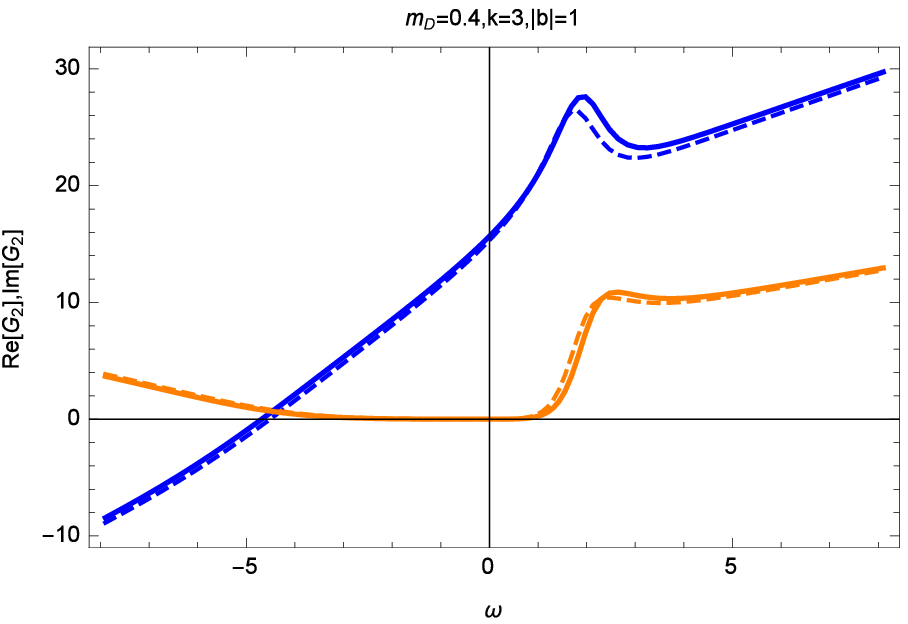}\\
\end{center}
  \caption{
  AHE of retarded Green's function for Dirac fermions ($m_D=0.4$) in AdS$_4$ space-time: Re$G^R_{\alpha}(\omega)$ (blue curve) and Im$G^R_{\alpha}(\omega)$ (orange curve), as functions of frequency $\omega$, given fixed momentum $|\vec{k}|=1.2 > q \mu $ (small momentum) and $|\vec{k}|=3.0 > q \mu $ (large momentum). We have chosen input parameters, $r_\star=\ell=g_F=q=1$, $|\vec{b}|=1$ (dashed curve), and $|\vec{b}| = 0$ (solid curve).
   } \label{fig:massive_AHE_AdS}
\end{figure}

In Fig.~\ref{fig:massless_AHE_AdS} and Fig.~\ref{fig:massive_AHE_AdS}, for spectral functions, i.e., Im$G_{i}$, they approach to unity $1$ in the $|\omega|\to\infty$ limit. Given a fixed large momentum, e.g., $k=3.0$, Im$G_{1}$ has a linearly-dispersing constant height peak at $\omega=-k-q\mu = -4.0$, while Im$G_{2}$ has a sharp peak due to divergence at $\omega = k-q\mu = 2.0$. Both Im$G_{i}$ are roughly zero between $\omega\in(-k-q\mu,k-q\mu)$ as in the vacuum case. Without loss of generality, we have chosen a set of parameters, as $r_\star=\ell=g_F=q=1$.

By observing the cases with larger momentum, e.g., $k=3.0$, it is obvious that the Green's function is affected by $|b|$ in the case that $|b|\sim |k|$ while it will not change too much in the case that $|b|\ll |k|$ when $|k|$ is far from the Fermi surface $k_F$, as expected.

For the special case without AHE, i.e., $|\vec{b}|=0$ for fermion in the AdS$_4$ space-time, the spectral function Im$G^R_{2}(\omega)$ (solid orange curve), just recovers the result in ref.~\cite{Liu:2009dm}, as shown in Fig.~1.

\subsubsection{Chiral magnetic effect for Dirac and Weyl Fermions}

\begin{figure}[H]
\begin{center}
  \includegraphics[scale=0.45]{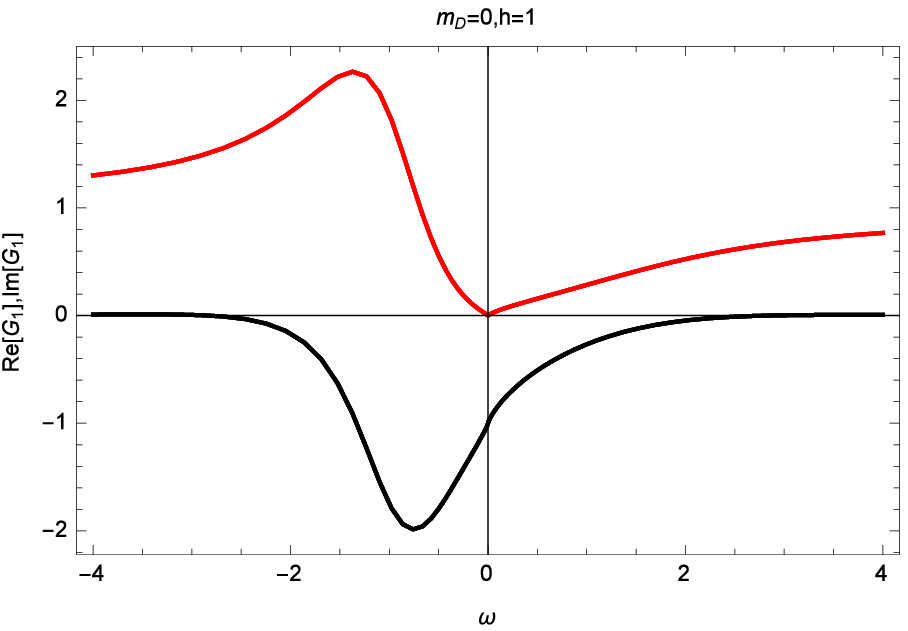} \includegraphics[scale=0.45]{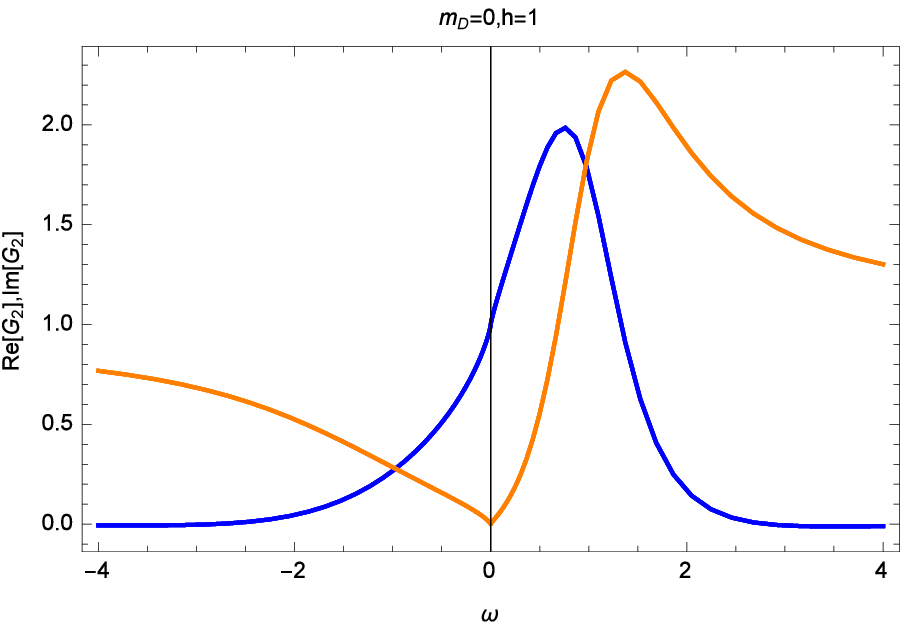}\\
  \includegraphics[scale=0.45]{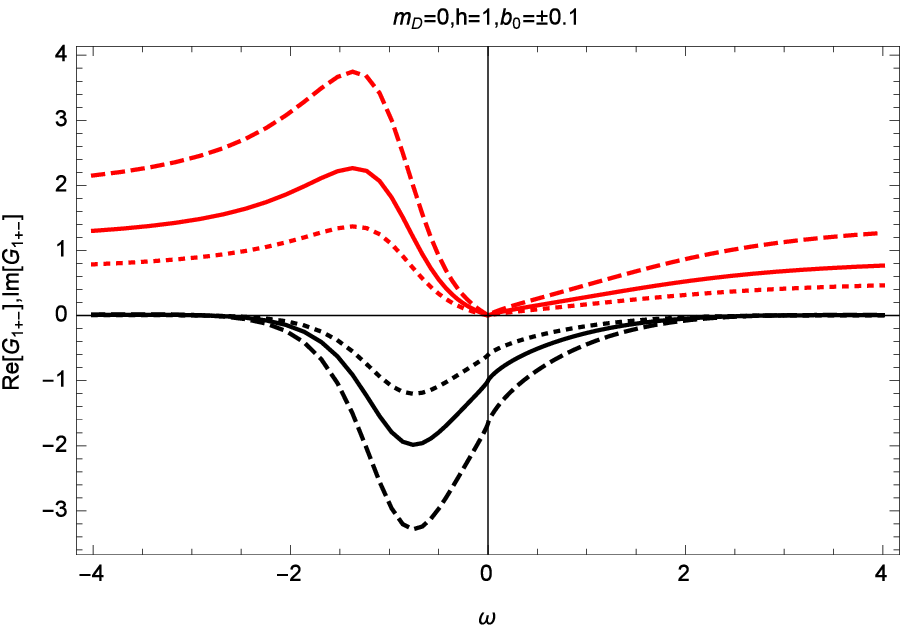} \includegraphics[scale=0.45]{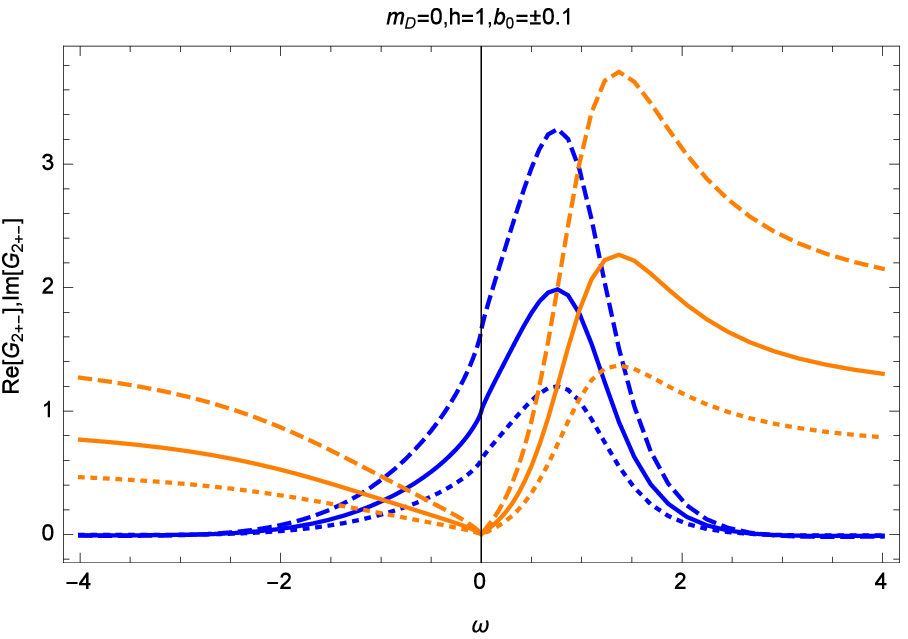}\\
\end{center}
  \caption{CME of retarded Green's function for Weyl fermions ($m_D=0.0$) in AdS$_4$ space-time: Re$G^R_{\alpha}(\omega)$ (black/blue curve), Im$G^R_{\alpha}(\omega)$ (red/orange curve) with $\alpha=1,2$, as functions of frequency $\omega$, given external magnetic field $|\vec{h}|=1$, and $b_{x,y}=0$. We have chosen parameters $b_0=0$ (solid thick curve),$b_0=0.1$ (dashed curve) and $b_0=-0.1$ (dotted curve), respectively.
   } \label{fig:CME_Weyl_AdS}
\end{figure}

\begin{figure}[H]
\begin{center}
  \includegraphics[scale=0.45]{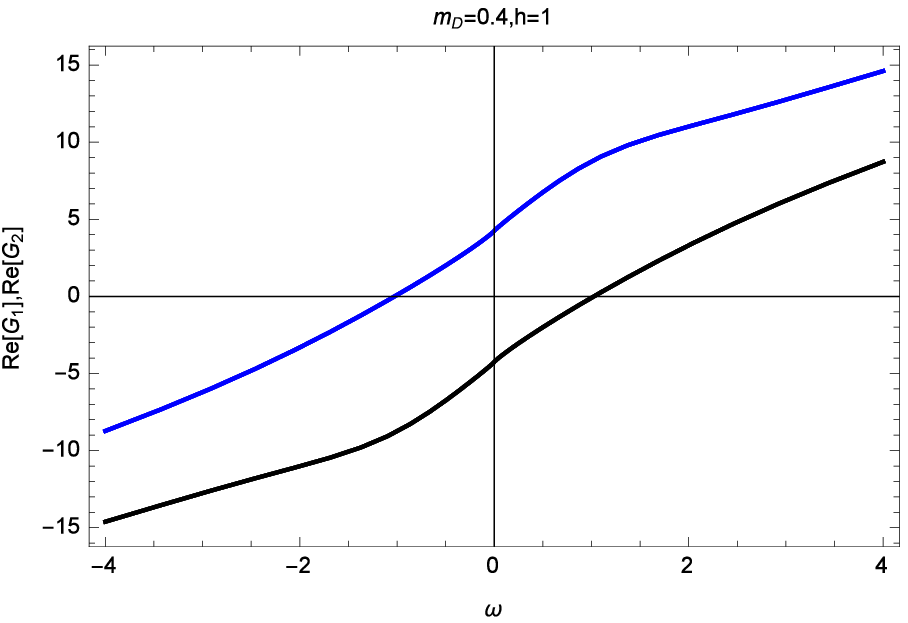} \includegraphics[scale=0.45]{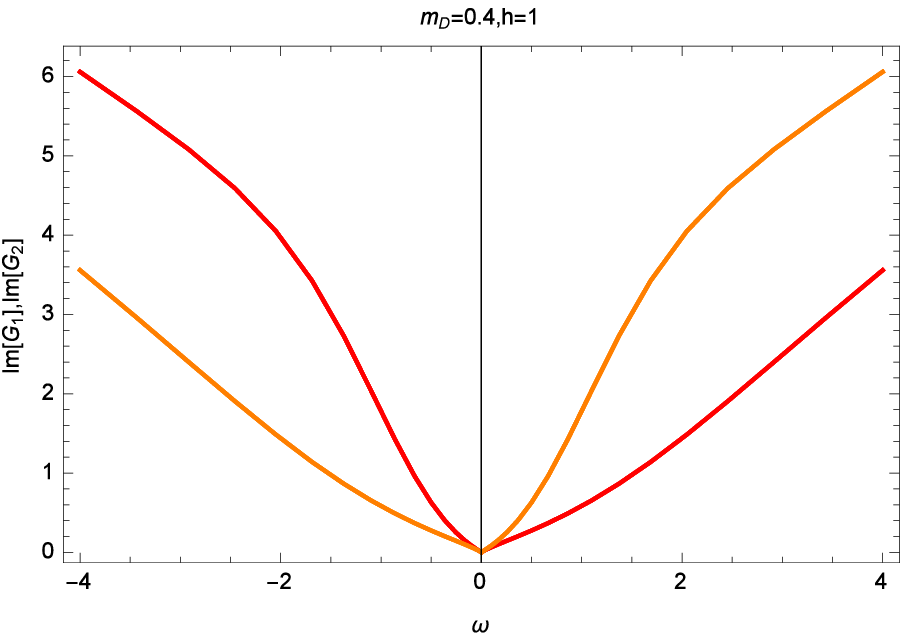}\\
  \includegraphics[scale=0.45]{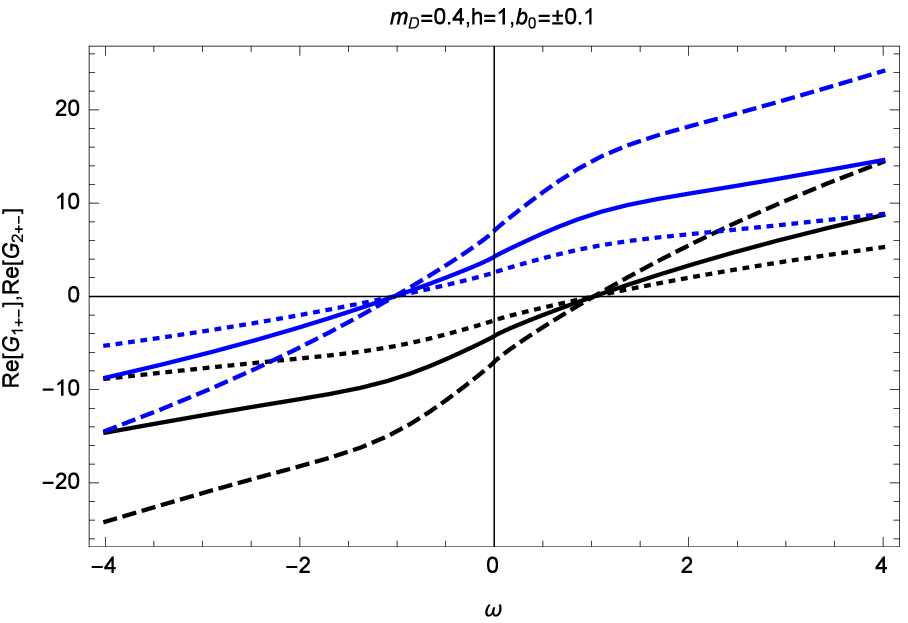} \includegraphics[scale=0.45]{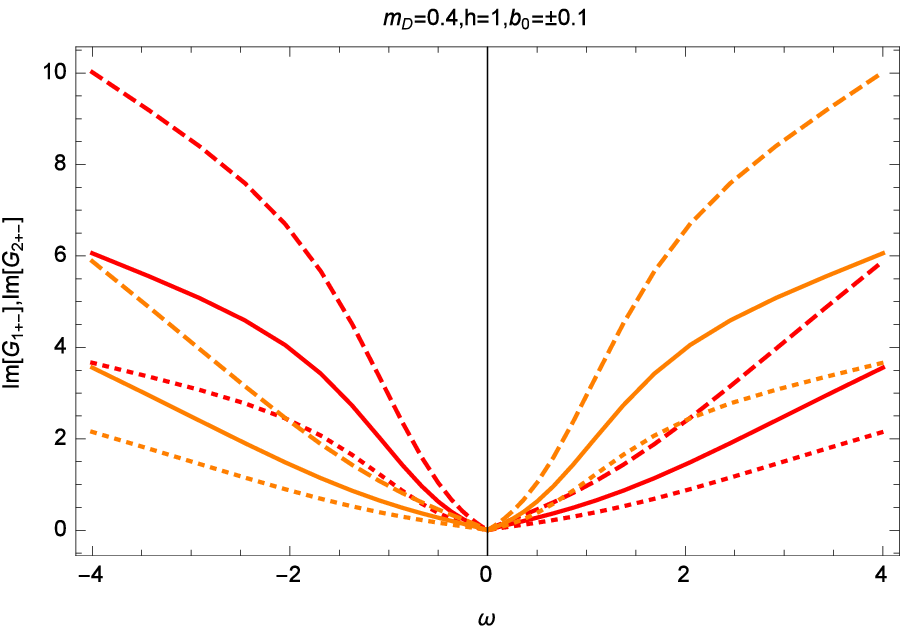}\\
\end{center}
  \caption{CME of retarded Green's function for Dirac fermions ($m_D=0.4$) in AdS$_4$ space-time: Re$G^R_{\alpha}(\omega)$ (black/blue curve), Im$G^R_{\alpha}(\omega)$ (red/orange curve) with $\alpha=1,2$, respectively, as functions of frequency $\omega$, given external magnetic field $|\vec{h}|=1$ and $b_{x,y}=0$. We have chosen parameter $b_0=0$ (solid thick curve), $b_0=0.1$ (dashed curve) and $b_0=-0.1$ (dotted curve), respectively.
   } \label{fig:CME_Dirac_AdS}
\end{figure}

In Fig.~\ref{fig:CME_Weyl_AdS} and Fig.~\ref{fig:CME_Dirac_AdS}, the retarded Green's function, i.e., $G_{\alpha,\beta\beta^\prime}$ are those corresponding to the energy eigenstate. Without loss of generality, we have chosen a set of parameters as, $r_\star=\ell=g_F$, $q_0=\mu r_\star = 0$ for Weyl and Dirac fermions with $m_D=0,0.4$, respectively. For CME case, we have turned off the external electromagnetic field, i.e., $\mu=0$. In the presence of external magnetic field, the momentum becomes that in Eq.(\ref{Eq:lambda_n-h-b}). For simplicity, we have set $b_{x,y}=0$, but $b_0 = \pm 0.1$ and considered only the ground state case, i.e., $n=0$. Therefore, $\lambda=\sqrt{qh}=\sqrt{h}$ with a magnetic field $\vec{h}=1$. The symmetry present in the above figures for $\alpha=1,2$, respectively, can be explained by observing the invariance in the flow equations as
\beqa
\omega \leftrightarrow -\omega, \quad \xi_\alpha \leftrightarrow - \xi_\alpha.
\eeqa

\section{IR Correlation Functions of Charged Fermion at Zero and Finite Temperatures}
\label{sec:IRgR-AdS(d+1)}

\subsection{In the Ricci flat case}
\label{sec:flat-k=0}

To extract the retarded Green's function for an operator ${\mathcal O}$, we need to solve the Dirac equations with the in-falling boundary conditions at the horizon. We identify $\psi_+$ as the source and its canonical momentum in terms of radial $r$ slicing as the vacuum expectation value.

At this step, let's consider a fermionic operator ${\mathcal O}_F$ for the boundary field theory,
by introducing a bulk fermion spinor field $\psi(t,x,u\sim 1/r)$ with mass $m_D$ and charge $q$.
The boundary field theory in the UV, e.g. CFT$_3$ in the UV is characterized completely and only by the dimension of the a CFT$_3$ operator, i.e., $\Delta_\pm$, which is given in terms of bulk quantities for neutral bulk Dirac field~\cite{Henningson:1998cd,Henneaux:1998ch,Son:2002sd,Iqbal:2009fd,Gubser:2012yb,Hinterbichler:2015pta},
\beqa
\ii
&& \psi_+  \overset{u\to 0}{\sim}  A u^{\Delta_-} + B u^{\Delta_+ + 1},  ~ \psi_- \overset{u\to 0}{\sim}  C u^{\Delta_- +1} + D u^{\Delta_+},   \nn\\
&& \Delta_- = \frac{3}{2} -   m_D \ell ,   ~  \Delta_+ = \frac{3}{2} +   m_D \ell. 
\eeqa

As investigated and summarized in appendix~\ref{app:TCBH_AdS}, for Einstein gravity in the AdS$_{4}$ space-time, the bulk gravity flows from the scale $u$ to AdS$_2\times {\mathbb R}^{2}$ in the near horizon region with scale $u_\star$ defined in Eq.(\ref{Eq:u_star}). The conformal field theory in the IR boundary, i.e., CFT$_1$, or $(0+1)$-dimensional conformal quantum mechanics, is characterized by the dimension of the operator $\Psi$. For $(3+1)$-dimensional bulk theory, it is asymptotic to AdS$_2\times {\mathbb R}^{2}$ at IR boundary. To be concrete, the background metric and gauge field are given by Eq.(\ref{Eq:ed-gF}) with $e_3=g_F\mu\ell_2/(\sqrt{2})$ with $\ell_2=\ell/\sqrt{6}$.


In terms of the near horizon coordinates $(t,x,\zeta)$ of AdS$_2$, without loss of generality, one can choose the above Gamma matrices in $D=(3+1)$-dimensions as shown in Eq.(\ref{Eq:Gamma-rep-1}),
\beqa
\ii\ii
\Gamma^{\underline{\zeta}} \! = \! \bigg(\begin{array}{cc}
\sigma^3 & \3i 0 \\
0 & \3i \sigma^3
\end{array}\bigg) , \Gamma^{\underline{t}} \! = \! \bigg( \begin{array}{cc}
i\sigma^1 & \3i 0 \\
0 & \3i i\sigma^1
\end{array}\bigg) , \Gamma^{\underline{1}} \! = \! \bigg(\begin{array}{cc}
-\sigma^2 & \3i 0 \\
0 & \3i \sigma^2
\end{array}\bigg) \! , \label{Eq:Gamma-rep-2}
\eeqa
which is compatible with the choice made in Eq.(\ref{Eq:Gamma-rep-1}), where the relative minus sign of the Gamma matrix for the radial coordinate is due to the change in the orientation between $r$ and $\zeta$. The bulk spinors can be decomposed as
\beqa
&& \Psi = (\Psi_+, \Psi_-)^T \equiv \Phi_+ + \Phi_- = (-gg^{\zeta\zeta})^{-\frac{1}{4}}\psi = \sqrt{\frac{\zeta}{r_\star}\frac{\ell}{\ell_2} } \psi , \nn\\
&& \Psi_\beta = P_\beta \Psi, \quad P_\beta \equiv  \frac{1}{2}(\textbf{1} - \beta \Gamma^{\underline{\zeta}}\Gamma^{\underline{t}}\Gamma^{\underline{1}}) ,
\eeqa
with $\beta=\pm$, and
\beqa
P_+ =  \text{diag}(\textbf{1}_{2},\textbf{0}_{2}), \quad P_- = \text{diag}(\textbf{0}_{2},\textbf{1}_{2}) , \quad P_+ + P_- = \textbf{1}_{2}, \nn
\eeqa
where the projection operator is consistent with that in Eq.(\ref{Eq:P_beta}).
The bulk Dirac equation is simplified to decoupled 1st order ODEs for its spinor components $\Psi_\alpha$,
\beqa
\ii \bigg[ \partial_\zeta \!+\! i\sigma^2 \bigg( \omega \!+\! q \frac{ e_3}{\zeta} \bigg) \!-\! \frac{\ell_2}{\zeta} ( m_D \sigma^3 \!+\! \tilde{m}_\alpha \sigma^1 ) \bigg] \Psi_\alpha = 0, \label{Eq:Dirac-Psi_alpha-AdS2}
\eeqa
where $\alpha=1,2$, are the index for energy eigenvalue, i.e., $\pm \lambda$, where $\lambda=\sqrt{k^2-b^2}$ can be induced by the momentum in ${\mathbb R}^{2}$ for Ricci flat cutoff hypersurface, as shown in Eq.(\ref{Eq:lambda-k-b}). By assuming that spatial directions are homogeneous, the parity-violating mass term $\tilde{m}_\alpha$ is defined in Eq.(\ref{Eq:mt_alpha}). It is worth noticing that for $\alpha=1,2$, the parity violating mass terms are, respectively,
\beqa
\tilde{m}_1 =  \sqrt{k^2-b^2} \frac{\ell}{r_\star},  \quad \tilde{m}_2 = - \sqrt{k^2-b^2} \frac{\ell}{r_\star},
\eeqa
which are related by a reversion of the momentum $\lambda=\sqrt{k^2-b^2}$, due to spatial inversion symmetry.
Therefore, the spatial inversion symmetry of the EOMs implies that, once one has obtained the solution of $\Psi_1$,
the solution of $\Psi_2$ can be obtained by flipping the sign of the momentum $k$, $\Psi_2(\lambda) = \Psi_1(-\lambda)$.
Near the boundary of AdS$_2$, in the infinite boundary limit with $\zeta \to 0$, the equations become 1st order ODEs as
\beqa
&& \partial_\zeta \Psi_\alpha - \frac{1}{\zeta}U(\zeta) \Psi_\alpha=0, \nn \\
&& U(\zeta) =  m_D \ell_2 \sigma^3 + \tilde{m}_\alpha \ell_2 \sigma^1 - i q e_3  \sigma^2, \label{Eq:U-zeta}
\eeqa
which can be rewritten to a 2nd order ODE,
\beqa
\bigg(\partial_\zeta^2  + \frac{1}{\zeta}\partial_\zeta  - \frac{(m_D^2 + \tilde{m}^2)\ell_2^2 - q^2 e_3^2}{\zeta^2} \bigg) \Psi_\alpha = 0. \label{Eq:EOM-AdS2-R(d-1)-fermions-asymptotic}
\eeqa
The solution turns out to be
\beqa
&& \Psi_1 = c_1 \zeta^{\nu_\lambda} + c_1^\star \zeta^{-\nu_\lambda}, \nn\\
&& \Psi_2  \sim \frac{c_1}{m_D\ell_2+ \nu_\lambda}\zeta^{\nu_\lambda} + \frac{c_1^\star}{m_D\ell_2 - \nu_\lambda}\zeta^{-\nu_\lambda}, \nn
\eeqa
where $c_1$ is an arbitrary complex constant and $c_1^\star$ is its conjugate, the scaling dimension and effective mass square are
\beqa
&& \nu_\lambda = \sqrt{(m_D^2+\tilde{m}^2)\ell_2^2-q^2 e_3^2} \equiv \sqrt{m_\lambda^2 \ell_2^2 - q^2 e_3^2 - i \epsilon },\nn\\
&& m_\lambda^2 \equiv m_D^2 + \tilde{m}^2 = m_D^2 + \lambda^2 \frac{\ell^2}{r_\star^2}, \label{Eq:nu_lambda-m_lambda}
\eeqa
where in the case without external magnetic field, $\lambda^2 = |\vec{k}|^2 - |\vec{b}^2|$ as shown in Eq.(\ref{Eq:lambda-k-b}).
The solution of $\Psi_\alpha$ near the boundary of AdS$_2$ region can be written as
\beqa
\Psi_\alpha = c_2 v_{\alpha,+} \zeta^{\nu_\lambda} + c_2^\star v_{\alpha,-} \zeta^{-\nu_\lambda}, \label{Eq:Psi-vpm}
\eeqa
where $c_2$ is another arbitrary complex constant and $c_2^\star$ is its conjugate. $v_\pm$ are real eigenvectors of $U(\zeta)$ with eigenvalues $\pm\nu_\lambda$, respectively, and the subscript $\pm$ stand for the signs of the corresponding eigenvectors,
\beqa
v_{\alpha,\pm} \sim \left(\begin{array}{c}
1  \\
\frac{m_D\ell_2 \mp \nu_\lambda}{q e_3 - \tilde{m}_\alpha\ell_2}
\end{array}
\right).
\eeqa
The relative normalization of $v_+$ and $v_-$ is a convention, which affects the normalization of the AdS$_2$ Green's function. By using the identity
\beqa
\frac{m_D^2\ell_2^2 - \nu_\lambda^2}{q e_3 - \tilde{m}_\alpha\ell_2} \equiv q e_3  + \tilde{m}_\alpha \ell_2,
\eeqa
one can make the choice of $v_{\alpha,\pm}$ as shown in Eq.(\ref{Eq:vpm})\footnote{To compare with the result in ref.~\cite{Faulkner:2009wj}, in the following we have chosen the same normalization conversion.}
\beqa
\ii
v_{\alpha,\pm} \equiv \left(\begin{array}{c}
m_D\ell_2 \pm \nu_\lambda  \\
\tilde{m}_\alpha\ell_2 + q e_3
\end{array}
\right). \label{Eq:vpm_2}
\eeqa
In summary, the asymptotic behavior of $\Psi$ near the boundary of AdS$_2$ region can be written as
\beqa
&& \Psi_\alpha = B v_{\alpha,+} \zeta^{\nu_\lambda}(1 + {\mathcal O}(\zeta)) + A v_{\alpha,-} \zeta^{-\nu_\lambda}(1+ {\mathcal O}(\zeta)) , \nn\\
&& B = \frac{c_2}{m_D\ell_2 + \nu_\lambda} , \quad A = \frac{c_2^\star}{m_D\ell_2 - \nu_\lambda}. \label{Eq:Psi-fermions-AdS2-asymptotic}
\eeqa
With the normalization defined above, the bottom components of $v_\pm$ are equal so that $A(k)$ and $B(k)$ can be extracted from the asymptotics of $u$ (or $\zeta$). While in the generic case, the frequency can not be neglected, the $\omega$ dependence can be scaled away by redefining $\zeta \to {q e_3 }/{\omega}$, in this case, the generic solutions to Eq.(\ref{Eq:Dirac-Psi_alpha-AdS2}) are
\beqa
\ii &&\ii \Psi_\alpha \!=\! B v_{\alpha,+} \bigg(\frac{\zeta}{\omega}\bigg)^{\nu_\lambda}(1 \!+\! {\mathcal O}(\zeta)) \!+\! A v_{\alpha,-} \bigg( \frac{\zeta}{\omega}\bigg)^{-\nu_\lambda}(1 \!+\! {\mathcal O}(\zeta)) , \nn \\
\ii&&\ii B = \frac{c_2 \omega^{\nu_\lambda}}{m_D\ell_2 + \nu_\lambda} , \quad A = \frac{c_2^\star\omega^{-\nu_\lambda}}{m_D\ell_2 - \nu_\lambda}. \label{Eq:Psi-I-omega!=0}
\eeqa
Thus we conclude that $A\sim \omega^{-\nu_\lambda}$, $B\sim \omega^{\nu_\lambda}$, then after imposing infalling boundary condition for $\Psi$ at the horizon, the retarded Green's function at IR for the boundary operator in the CFT$_1$ dual to $\Psi$ can then be estimated to be proportional to ${\mathcal G}^R(\omega) = {B}/{A} \sim \omega^{2\nu_\lambda}$. This implies a coordinate space correlation function by doing an inverse Fourier transformation, which gives ${\mathcal G}^R(t)\sim t^{-2\Delta^{\text{IR}}}$, with the scaling dimension $\Delta^{\text{IR}}$ of the IR CFT operator ${\mathcal O}_F$ given by $\Delta^{\text{IR}} = {1}/{2} + \nu_\lambda$.

However, there is an ambiguity in the above definition of the retarded Green's function, as the ratio of $B/A$ depends on the relative normalization of the eigenspinors $v_\pm$ in Eq.(\ref{Eq:vpm}), thus one should keep in mind that there is a relative normalization rescaling factor in front of regarded Green's function, i.e., $v_\pm \to c_\pm v_\pm, \Rightarrow {\mathcal G}^R \to {c_+}/{c_-}{\mathcal G}^R$, while the rescaling cancels when matching to the UV $G^R$ in the infinite boundary condition, thus will not affect the full Green's function $G^R$ computed. Physically this can be easily understood, since if one does not use low frequency expansion approach in ref.~\cite{Faulkner:2009wj}, the retarded Green function's $G^R$ is completely determined by the full bulk gravity in the infinite boundary limit, which will not depend on the IR normalization choice at all.

\subsubsection{IR correlation functions at zero temperature}

In the near horizon region, the geometric background is AdS$_2\times {\mathbb R}^{2}$. The metric and the gauge field are
\beqa
&& g_{tt}(\zeta)= \frac{\ell_2^2}{\zeta^2}, \quad g_{\zeta\zeta}(\zeta) = \frac{\ell_2^2}{\zeta^2}, \quad g_{xx}(\zeta) =\frac{r_\star^2}{\ell^2}, \nn\\
&& A_t(\zeta) = \frac{e_3}{\zeta},  \quad \text{and/or} \quad  A_x(y) = - h y,
\eeqa
where the spatial component of the $U(1)$ gauge field is present when an external magnetic field is applied, as shown in Eq.(\ref{Eq:A_M}).

Thus, one has
\beqa
 \lambda_1^\pm(\zeta) = \omega \pm \beta b_0 + \beta b_\zeta + q \frac{e_3}{\zeta}  , \quad \lambda_2(\zeta) = \frac{\ell_2}{\zeta} \tilde{m}_f, \nn
\eeqa
where $\tilde{m}_f$ is defined through $m_f(r)$ in Eq.(\ref{Eq:m_f(r)}), namely,
\beqa
\tilde{m}_f \equiv m_f(r_\star) = m_D + i \tilde{m}_\alpha, \label{Eq:mt_f}
\eeqa
with $\tilde{m}_\alpha$ being defined in Eq.(\ref{Eq:mt_alpha}) with $\alpha=1,2$.

When we make the redefinition that $b_r=-b_\zeta$, there is a relative minus sign between $b_\zeta$ and $b_r$, if one wants to compare with those results obtained in radial coordinate $r$.

The radial sector of the Dirac equation in the $\zeta$ coordinate is given in Eq.(\ref{Eq:fg_zeta_2nd}), with coefficient given in Eq.(\ref{Eq:gamma-1234-zeta}), which can be expressed more explicitly as
\begin{widetext}
\beqa
 \gamma_1(\zeta) & = &  2i \beta b_0 + \frac{1}{\zeta}, \quad \gamma_2(\zeta)  =   \bigg( (\omega \!+\!  q \frac{e_3 }{\zeta}) \!+\! \beta b_\zeta \bigg)^2 - b_0^2 - \frac{\ell_2^2}{\zeta^2}( m_D^2 + \tilde{m}_\alpha^2 )  - i \bigg( \frac{\omega - \beta b_0 + \beta b_\zeta}{\zeta} \bigg), \nn\\
 \gamma_3(\zeta) & = &  2i \beta b_0 + \frac{1}{\zeta}, \quad \gamma_4(\zeta)  =   \bigg( (\omega \!+\!  q \frac{e_3 }{\zeta}) \!+\! \beta b_\zeta \bigg)^2 - b_0^2 - \frac{\ell_2^2}{\zeta^2}( m_D^2 + \tilde{m}_\alpha^2 )  + i \bigg( \frac{\omega + \beta b_0 + \beta b_\zeta}{\zeta} \bigg),
\eeqa
where $\tilde{m}_\alpha \equiv -(-1)^\alpha \lambda {\ell}/{r_\star}$.

According to Eq.(\ref{Eq:gamma-lambda-zeta}) in the $\zeta$ coordinates, the two coupled 1st order ODEs can be transformed into two decoupled $2$-nd ODEs,
\beqa
&& \partial_\zeta^2 \tilde{f}_{\alpha,\beta} + \bigg( 2i \beta b_0 + \frac{1}{\zeta} \bigg) \partial_\zeta \tilde{f}_{\alpha,\beta} + \bigg[\bigg( \omega + \beta b_\zeta + \frac{q e_3 }{\zeta}\bigg)^2 - b_0^2 - |\tilde{m}_f|^2\frac{\ell_2^2}{\zeta^2} -  \frac{i(\omega - \beta b_0 + \beta b_\zeta)}{\zeta}\bigg]\tilde{f}_{\alpha,\beta} =0, \nn \\
&& \partial_\zeta^2 \tilde{g}_{\alpha,\beta} + \bigg( 2i \beta b_0 + \frac{1}{\zeta} \bigg) \partial_\zeta \tilde{g}_{\alpha,\beta} + \bigg[\bigg( \omega + \beta b_\zeta + \frac{q e_3 }{\zeta}\bigg)^2 - b_0^2 - |\tilde{m}_f|^2\frac{\ell_2^2}{\zeta^2} +  \frac{i(\omega + \beta b_0 + \beta b_\zeta)}{\zeta}\bigg]\tilde{g}_{\alpha,\beta} =0. \label{Eq:fgt_alpha-zeta-T=0}
\eeqa

In the infinite boundary with $\zeta\to 0$, the equations are dominated by
\beqa
 \partial_\zeta^2 \tilde{f}_\alpha +  \frac{1}{\zeta} \partial_\zeta \tilde{f}_\alpha - \frac{|\tilde{m}_f|^2 \ell_2^2 - q^2 e_3^2}{\zeta^2} \tilde{f}_\alpha =0, \quad \partial_\zeta^2 \tilde{g}_\alpha +  \frac{1}{\zeta} \partial_\zeta \tilde{g}_\alpha - \frac{|\tilde{m}_f|^2 \ell_2^2 - q^2 e_3^2}{\zeta^2} \tilde{g}_\alpha =0, \nn 
\eeqa
which have the solutions
\beqa
\tilde{f}_\alpha = \tilde{a}_{\alpha\uparrow} \zeta^{-\nu_\lambda} + \tilde{b}_{\alpha\uparrow} \zeta^{\nu_\lambda}, \quad \tilde{g}_\alpha = \tilde{a}_{\alpha\downarrow} \zeta^{-\nu_\lambda} + \tilde{b}_{\alpha\downarrow} \zeta^{\nu_\lambda}, \label{Eq:ft_alpha-gt_alpha-ba_alpha-zeta}
\eeqa
where the conformal weight $\nu_\lambda$ is defined in Eq.(\ref{Eq:nu_lambda-m_lambda}).

In the absence of external magnetic field, the eigenvalue of transverse momentum is defined in Eq.(\ref{Eq:lambda-k-b}), while in the presence of external magnetic field, the corresponding eigenvalues are quantized as defined in Eq.(\ref{Eq:lambda_n-h-b}).
\beqa
\tilde{m}_\alpha^2 = (k_x^2 - b_x^2)\frac{\ell^2}{r_\star^2} \equiv \tilde{m}_k^2 - \tilde{m}_{b_x}^2, \quad k_x \in {\mathbb R}, \quad \text{or} \quad  \tilde{m}_\alpha^2 = \bigg( 2qh\big(n+\frac{1}{2}\big)  - b_y^2 \bigg) \frac{\ell^2}{r_\star^2}\equiv \tilde{m}_n^2 - \tilde{m}_{b_y}^2, \quad n\in {\mathbb Z},
\eeqa
where we have defined $\tilde{m}_k \equiv k_x {\ell}/{r_\star}$, $\tilde{m}_{b_{x,y}}  \equiv b_{x,y} {\ell}/{r_\star}$, and $\tilde{m}_n \equiv \sqrt{2qh\big( n + \frac{1}{2} \big)}{\ell}/{r_\star}$ with $n\in {\mathcal Z}$.

By solving the general EOMs in Eq.(\ref{Eq:fgt_alpha-zeta-T=0}), one obtains
\beqa
\ii
\tilde{f}_{\alpha,\beta}(\zeta) \3i &=& \3i e^{ -i(\omega+\beta b_\zeta+\beta b_0)\zeta}\zeta^{\nu_\lambda}\big[C_1 U[1 + \nu_\lambda + iq e_3 , 1+ 2\nu_\lambda, 2i(\omega+\beta b_\zeta)\zeta] +  C_2 L_{-1-\nu_\lambda-i q e_3}^{2\nu_\lambda}[2i(\omega+\beta b_\zeta)\zeta] \big], \label{Eq:ft_alpha-zeta}\\
\ii
\tilde{g}_{\alpha,\beta}(\zeta) \3i &=& \3i e^{ -i(\omega+\beta b_\zeta+\beta b_0)\zeta}\zeta^{\nu_\lambda}\big[C_3 U[ \nu_\lambda + iq e_3 , 1+ 2\nu_\lambda, 2i(\omega+\beta b_\zeta)\zeta] +  C_4 L_{-\nu_\lambda-i q e_3}^{2\nu_\lambda}[2i(\omega+\beta b_\zeta)\zeta] \big], \label{Eq:gt_alpha-zeta}
\eeqa
where where $C_1$, $C_2$, $C_3$ and $C_4$ are all constants. $\nu_\lambda\equiv \sqrt{(m_D^2+\tilde{m}_\alpha^2)\ell_2^2 - q^2 e_3^2}$, $U(a,b,z)$ is the Tricomi confluent hypergeometric function, which is a degenerate form of a hypergeometric differential equation where two of the three regular singularities merge into an irregular singularity, and $L_n^a(z) $ is the generalized Laguerre polynomial. In the $\zeta\to \infty$ limit, they have the asymptotic behavior as

\beqa
 U[1+\nu_\lambda+iq e_3 , 1+ 2\nu_\lambda, 2i(\omega+\beta b_\zeta)\zeta] & \sim & \zeta^{-1 -\nu_\lambda - i q e_3}, \quad L_{-1-\nu_\lambda - i q e_3}^{2\nu_\lambda}[2i(\omega+\beta b_\zeta)\zeta] \sim e^{2i(\omega+\beta b_\zeta)\zeta}\zeta^{-\nu_\lambda + i q e_3 }. \nn \\
 U[\nu_\lambda + iq e_3 , 1+ 2\nu_\lambda, 2i(\omega+\beta b_\zeta)\zeta] & \sim & \zeta^{-\nu_\lambda - iq e_3}, \quad L_{-\nu_\lambda-i q e_3}^{2\nu_\lambda}[2i(\omega+\beta b_\zeta)\zeta] \sim e^{2i(\omega+\beta b_\zeta)\zeta}\zeta^{-1 -\nu_\lambda + i q e_3 }. \nn
\eeqa
It is obvious that the Laguerre will be unpredictable in the near horizon limit, $\zeta\to \infty$, since oscillation phase is infinity.
\beqa
\ii
\tilde{f}_{\alpha,\beta}(\zeta) \3i & \sim & \3i e^{ -i \beta b_0\zeta}\zeta^{\nu_\lambda}\big[C_1 e^{ -i(\omega+\beta b_\zeta)\zeta} \zeta^{-1-\nu_\lambda - iq e_3} +  C_2 e^{i(\omega+\beta b_\zeta)\zeta}\zeta^{-\nu_\lambda + i q e_3 } \big], \\
\ii
\tilde{g}_{\alpha,\beta}(\zeta) \3i & \sim & \3i e^{ -i \beta b_0\zeta}\zeta^{\nu_\lambda}\big[C_3 e^{ -i(\omega+\beta b_\zeta)\zeta} \zeta^{-\nu_\lambda - i q e_3} +  C_4  e^{i(\omega+\beta b_\zeta)\zeta}\zeta^{-1-\nu_\lambda + i q e_3 } \big].
\eeqa
Considering that the frequency sector as Fourier decomposition of wave function in Eq.(\ref{Eq:psi-Psi-Fourier}),
\beqa
&& e^{-i\omega t} e^{-i\beta b_0 \zeta} e^{-i(\omega+\beta b_\zeta)\zeta} \zeta^{-i q e_3 } = e^{-i\omega(t + \zeta + \beta \frac{b_0+b_\zeta}{\omega}\zeta + \frac{q e_3 }{\omega} \ln\zeta )} , \nn\\
&& e^{-i\omega t} e^{-i\beta b_0 \zeta} e^{i(\omega+\beta b_\zeta)\zeta} \zeta^{i q e_3 } = e^{-i\omega(t - \zeta + \beta \frac{b_0-b_\zeta}{\omega} \zeta - \frac{q e_3 }{\omega} \ln\zeta )}.
\eeqa
In the case with $b_0=b_r=0$ and $q=0$, the first wave solution corresponds to the in-falling wave one while the second solution corresponds to the out-going wave one in the near horizon limit, i.e., $\zeta\to \infty$. By imposing the in-falling wave condition to isolate the un-physical solutions, namely, one should drop the out-going wave by imposing $C_2=C_4=0$, but only keep the in-going solutions
\beqa
\ii
\tilde{f}_{\alpha,\beta}(\zeta) \3i &=& \3i C_1    [2i(\omega+\beta b_\zeta)]^{-\frac{1}{2}-\nu_\lambda} e^{ -i \beta b_0\zeta}\zeta^{-\frac{1}{2}} W_{-\frac{1}{2}-i q e_3 , \nu_\lambda}[2i(\omega+\beta b_\zeta)\zeta]  , \nn \\
\ii
\tilde{g}_{\alpha,\beta}(\zeta) \3i &=& \3i  C_3   [2i(\omega+\beta b_\zeta)]^{-\frac{1}{2}-\nu_\lambda} e^{ -i \beta b_0\zeta}\zeta^{-\frac{1}{2}} W_{+\frac{1}{2}-i q e_3 , \nu_\lambda}[2i(\omega+\beta b_\zeta)\zeta] , \label{Eq:tF_alpha-tG_alpha}
\eeqa
where $W_{k,m}[z]$ is Whittaker-$W$ function.

At this step, by using the definition of $\tilde{R}_\alpha$ in Eq.(\ref{Eq:fgt_alpha-transQ_FGt_alpha}), one can consider the asymptotic behavior of the in-falling wave solution at infinite boundary limit $\zeta\to 0$ as already explored in Eq.(\ref{Eq:ft_alpha-gt_alpha-ba_alpha-zeta}),
\beqa
\tilde\Psi_\alpha \sim \tilde{R}_\alpha \equiv (\tilde{F}_\alpha,\tilde{G}_\alpha) = Q^{-1}(\tilde{f}_\alpha,\tilde{g}_\alpha)^T \equiv \tilde{a}_\alpha \zeta^{-\nu_\lambda} + \tilde{b}_\alpha \zeta^{\nu_\lambda},  \label{Eq:Rt_alpha-b/a-zeta}
\eeqa
where $\tilde{a}_\alpha=(\tilde{a}_{\alpha\uparrow},\tilde{a}_{\alpha\downarrow})^T$ and $\tilde{b}_\alpha=(\tilde{b}_{\alpha\uparrow},\tilde{b}_{\alpha\downarrow})^T$ are $2\times 1$ column vectors with $\alpha=1,2$. By imposing the normalization conversion in Eq.(\ref{Eq:vpm}), we require that
\beqa
\tilde{b}_\alpha = \frac{\tilde{b}_{\alpha\uparrow}}{\tilde{b}_{\alpha\downarrow}} \sim \frac{v_{+\uparrow}}{v_{+\downarrow}} = \frac{m_D \ell_2 + \nu_\lambda}{\tilde{m}_\alpha\ell_2 + q e_3 }, \quad \tilde{a}_\alpha = \frac{\tilde{a}_{\alpha\uparrow}}{\tilde{a}_{\alpha\downarrow}} \sim \frac{v_{-\uparrow}}{v_{-\downarrow}} = \frac{m_D \ell_2 - \nu_\lambda}{\tilde{m}_\alpha\ell_2 + q e_3 }.
\eeqa
This can be obtained by considering asymptotic behavior of wave function in the infinite boundary condition, i.e., $\zeta\to 0$, Eq.(\ref{Eq:tF_alpha-tG_alpha}) becomes
\beqa
&& \tilde{f}_\alpha(\zeta) \sim C_1 \bigg( \zeta^{\nu_\lambda} \frac{\Gamma[-2\nu_\lambda]}{\Gamma[1-\nu_\lambda+i q e_3]} +    \zeta^{-\nu_\lambda} [2i(\omega+b_\zeta)]^{-2\nu_\lambda}\frac{\Gamma[2\nu_\lambda]}{\Gamma[1+\nu_\lambda+i q e_3 ]}  \bigg), \nn\\
&& \tilde{g}_\alpha(\zeta) \sim C_3 \bigg( \zeta^{\nu_\lambda} \frac{\Gamma[-2\nu_\lambda]}{\Gamma[-\nu_\lambda+i q e_3]} +    \zeta^{-\nu_\lambda} [2i(\omega+b_\zeta)]^{-2\nu_\lambda}\frac{\Gamma[2\nu_\lambda]}{\Gamma[\nu_\lambda+i q e_3 ]}  \bigg).
\eeqa
where we have dropped a phase factor $e^{-i\beta b_0 \zeta}$, which one should keep in mind, when dealing with the Green's function with different helicity $\beta^\prime \ne \beta$.

According to Eq.(\ref{Eq:fgt_alpha-transQ_FGt_alpha}), one has the asymptotic behavior
\beqa
\tilde{R}_{\alpha,\beta\beta^\prime} &=&
\left(
  \begin{array}{c}
    \tilde{F}_{\alpha,\beta} \\
    \tilde{G}_{\alpha,\beta^\prime} \\
  \end{array}
\right) = Q^{-1} \left(
  \begin{array}{c}
    \tilde{f}_{\alpha,\beta} \\
    \tilde{g}_{\alpha,\beta^\prime} \\
  \end{array}
\right) = \frac{1}{\sqrt{2}}\left(
            \begin{array}{cc}
              1 & 1 \\
             -i & i \\
            \end{array}
          \right) \left(
  \begin{array}{c}
    \tilde{f}_{\alpha,\beta} \\
    \tilde{g}_{\alpha,\beta^\prime} \\
  \end{array}
\right) = \frac{1}{\sqrt{2}} \left(
  \begin{array}{c}
    \tilde{f}_{\alpha,\beta} +  \tilde{g}_{\alpha,\beta^\prime} \\
    -i(\tilde{f}_{\alpha,\beta} -  \tilde{g}_{\alpha,\beta^\prime}) \\
  \end{array}
\right) \nn\\
&\overset{\zeta\to 0}{=}& \frac{\Gamma[-2\nu_\lambda]}{\Gamma[1-\nu_\lambda+i q e_3 ]} \frac{1}{\sqrt{2}}\left(
                             \begin{array}{c}
                                  C_1 -  C_3(\nu_\lambda-i q e_3 )  \\
                               -i\big( C_1 +  C_3(\nu_\lambda-i q e_3 ) \big) \\
                             \end{array}
                           \right) \zeta^{\nu_\lambda} \nn\\
&+& \frac{\Gamma[2\nu_\lambda]}{\Gamma[1+\nu_\lambda+i q e_3 ]} [2i(\omega+b_\zeta)]^{-2\nu_\lambda} \frac{1}{\sqrt{2}}\left(
                             \begin{array}{c}
                                  C_1 +  C_3(\nu_\lambda+i q e_3 ) \\
                               -i \big( C_1 -  C_3(\nu_\lambda+i q e_3 ) \big) \\
                             \end{array}
                           \right) \zeta^{-\nu_\lambda} ,
\eeqa
thus by choosing the conversion in Eq.(\ref{Eq:vpm}), and considering the definition of $\tilde{R}_\alpha\equiv \tilde{b}_\alpha \zeta^{\nu_\lambda} + \tilde{a}_\alpha \zeta^{-\nu_\lambda} $ in Eq.(\ref{Eq:Rt_alpha-b/a-zeta}), one obtains
\beqa
&& \tilde{b}_\alpha = \frac{\Gamma[-2\nu_\lambda]}{\Gamma[1-\nu_\lambda+i q e_3 ]} \frac{1}{\sqrt{2}} \left(
                             \begin{array}{c}
                                  C_1 -  C_3(\nu_\lambda-i q e_3 )  \\
                               -i\big( C_1 +  C_3(\nu_\lambda-i q e_3 ) \big) \\
                             \end{array}
                           \right) \sim v_+ \equiv \left(\begin{array}{c}
m_D\ell_2 + \nu_\lambda  \\
\tilde{m}_\alpha\ell_2 + q e_3
\end{array}
\right) , \nn\\
&& \tilde{a}_\alpha = \frac{\Gamma[2\nu_\lambda]}{\Gamma[1+\nu_\lambda+i q e_3 ]} [2i(\omega+b_\zeta)]^{-2\nu_\lambda} \frac{1}{\sqrt{2}} \left(
                             \begin{array}{c}
                                  C_1 +  C_3(\nu_\lambda+i q e_3 ) \\
                               -i\big( C_1 -  C_3(\nu_\lambda+i q e_3 ) \big) \\
                             \end{array}
                           \right) \sim v_- \equiv \left(\begin{array}{c}
m_D\ell_2 - \nu_\lambda  \\
\tilde{m}_\alpha\ell_2 + q e_3
\end{array}
\right),
\eeqa
which gives
\beqa
\frac{C_3}{C_1}  = -\frac{1}{\nu_\lambda- i q e_3 }\frac{\nu_\lambda - i q e_3 + (m_D-i\tilde{m}_\alpha)\ell_2 }{\nu_\lambda + i q e_3  + (m_D+i\tilde{m}_\alpha)\ell_2 } = \frac{1}{\nu_\lambda + i q e_3 }\frac{\nu_\lambda+i q e_3 - (m_D-i\tilde{m}_\alpha)\ell_2 }{\nu_\lambda-i q e_3  - (m_D+i\tilde{m}_\alpha)\ell_2 }.
\eeqa
By substituting back into $\tilde{R}_\alpha$, one obtains
\beqa
\tilde{R}_\alpha &=& \sqrt{2} C_1 \frac{1}{\nu_\lambda+ i q e_3 + (m_D+i\tilde{m}_\alpha)\ell_2 } \frac{\Gamma[-2\nu_\lambda]}{\Gamma[1-\nu_\lambda+i q e_3 ]}  v_+ \zeta^{\nu_\lambda} \nn\\
&& -\sqrt{2}C_1  [2i(\omega+b_\zeta)]^{-2\nu_\lambda} \frac{1}{\nu_\lambda-i q e_3  - (m_D+i\tilde{m}_\alpha)\ell_2} \frac{\Gamma[2\nu_\lambda]}{\Gamma[1+\nu_\lambda+i q e_3 ]} v_- \zeta^{-\nu_\lambda}. \nn
\eeqa
Thus, one finally obtains
\beqa
\ii\ii
\tilde \Psi_\alpha  \sim  \tilde{B}_\alpha v_+ \zeta^{\nu_\lambda}(1+ {\mathcal O}(\zeta)) + \tilde{A}_\alpha v_- \zeta^{-\nu_\lambda}(1+{\mathcal O}(\zeta)),
\eeqa
where
\beqa
\tilde{B}_\alpha &=& \frac{\sqrt{2}C_1}{\nu_\lambda + i q e_3  + (m_D+i\tilde{m}_\alpha)\ell_2} \frac{\Gamma[-2\nu_\lambda]}{\Gamma[1-\nu_\lambda+ i q e_3 ]}  \equiv \frac{\tilde{b}_{\alpha\uparrow}}{m_D \ell_2 + \nu_\lambda} = \frac{\tilde{b}_{\alpha\downarrow}}{\tilde{m}_\alpha\ell_2 + q e_3  } , \nn\\
\tilde{A}_\alpha &=& \frac{-\sqrt{2}C_1[2i(\omega+b_\zeta)]^{-2\nu_\lambda}}{\nu_\lambda - i q e_3  - (m_D + i \tilde{m}_\alpha) \ell_2 } \frac{\Gamma[2\nu_\lambda]}{\Gamma[1+\nu_\lambda + i q e_3 ]}  \equiv  \frac{\tilde{a}_{\alpha\uparrow}}{m_D \ell_2 - \nu_\lambda} = \frac{\tilde{a}_{\alpha\downarrow}}{\tilde{m}_\alpha\ell_2 + q e_3  }. \nn
\eeqa

In the presence of chiral gauge field, according to Eq.(\ref{Eq:tPsi_alpha-Psi_alpha}) and Eq.(\ref{Eq:transU1_tPsi-Psi}), one has $\Psi_\alpha = U_1^{-1} \tilde\Psi_\alpha$, namely
\beqa
\left(
  \begin{array}{c}
    \Psi_{\alpha,+} \\
    \Psi_{\alpha,-} \\
  \end{array}
\right) = U_1^{-1} \left(
  \begin{array}{c}
    \tilde\Psi_{\alpha,+} \\
    \tilde\Psi_{\alpha,-} \\
  \end{array}
\right)
&\sim & \frac{1}{\sqrt{2}}\left(
                                                                                           \begin{array}{cc}
                                                                                             1 & i \\
                                                                                             i & 1 \\
                                                                                           \end{array}
                                                                                         \right) \left(
  \begin{array}{c}
    \tilde{R}_{\alpha,+} \\
    \tilde{R}_{\alpha,-} \\
  \end{array}
\right) =  \frac{1}{\sqrt{2}} \left(
  \begin{array}{c}
    \tilde{R}_{\alpha,+} + i \tilde{R}_{\alpha,-} \\
    i( \tilde{R}_{\alpha,+} - i \tilde{R}_{\alpha,-}) \\
  \end{array}
\right).
\eeqa
Thus, the asymptotic behavior in the infinite boundary becomes
\beqa
\left(
  \begin{array}{c}
    \Psi_{\alpha,+} \\
    \Psi_{\alpha,-} \\
  \end{array}
\right) \sim \left(
  \begin{array}{c}
    (\tilde{B}_{\alpha,+}+i\tilde{B}_{\alpha,-})   \\
    i(\tilde{B}_{\alpha,+}-i\tilde{B}_{\alpha,-})   \\
  \end{array}
\right) v_+ \zeta^{\nu_\lambda} + \left(
  \begin{array}{c}
    (\tilde{A}_{\alpha,+}+i\tilde{A}_{\alpha,-})  \\
    i(\tilde{A}_{\alpha,+}-i\tilde{A}_{\alpha,-})  \\
  \end{array}
\right) v_- \zeta^{-\nu_\lambda},
\eeqa
from which, we obtain the correlation function for $\Psi_{\alpha}$ as\footnote{Note that $(-1)^{\nu_\lambda}=(-1+i\epsilon)^{\nu_\lambda}=e^{i\pi \nu_\lambda}$, to entail ${\mathcal G}_\alpha^A$ is analytical in the lower half complex $\omega$ plane.}, respectively,
\beqa
{\mathcal G}_{\alpha,\beta\beta^\prime}^A(\omega, \lambda)  \equiv  \frac{\tilde{B}_{\alpha,\beta} + i\tilde{B}_{\alpha,-\beta}}{ \tilde{A}_{\alpha,\beta^\prime} + i \tilde{A}_{\alpha,-\beta^\prime} } = \frac{\tilde{\mathcal G}^A_{\alpha,\beta\beta^\prime}+i \tilde{\mathcal G}^A_{\alpha,-\beta\beta^\prime}}{\tilde{\mathcal G}^A_{\alpha,-\beta-\beta^\prime}+i\tilde{\mathcal G}^A_{\alpha,-\beta\beta^\prime}} \tilde{\mathcal G}^A_{\alpha,-\beta-\beta^\prime}.
\eeqa
The results can be categorized into two cases:
\begin{itemize}
  \item If $\beta = \beta^\prime$, one has ${\mathcal G}_{\alpha}^A(\omega, \lambda) \equiv {\mathcal G}_{\alpha,\beta\beta}^A(\omega, \lambda)=\tilde{\mathcal G}_{\alpha,\beta\beta}^A(\omega, \lambda)\equiv \tilde{\mathcal G}_{\alpha}^A(\omega, \lambda)$.
  \item If $\beta = - \beta^\prime$, one has
  \beqa
  {\mathcal G}_{\alpha,\beta-\beta}^A(\omega, \lambda) = \frac{\tilde{\mathcal G}^A_{\alpha,\beta-\beta}+ i \tilde{\mathcal G}^A_{\alpha}}{\tilde{\mathcal G}^A_{\alpha}+i \tilde{\mathcal G}^A_{\alpha,\beta-\beta}} \tilde{\mathcal G}^A_{\alpha} = \frac{\cosh{(\beta \pi b_0 \zeta)} - i \sinh{(\beta \pi b_0 \zeta)}}{\cosh{(\beta \pi b_0 \zeta)} + i \sinh{(\beta \pi b_0 \zeta)}} \tilde{\mathcal G}^A_{\alpha},
  \eeqa
  where we have used that $\tilde{\mathcal G}^A_{\alpha,\beta-\beta}\tilde{\mathcal G}^A_{\alpha,-\beta\beta}=\tilde{\mathcal G}^A_{\alpha,\beta\beta}\tilde{\mathcal G}^A_{\alpha,-\beta-\beta}\equiv (\tilde{\mathcal G}^A_{\alpha})^2$.
\end{itemize}
Since we have chosen the conversion that $\partial_t\to  -i\omega$, the correlation functions obtained above should be identified as the advanced Green's function at the IR fixed point. Alternatively, by choosing the conversion that ($\partial_t\to i \omega$) and with the similar procedure, the retarded Green's function of the spinor operator ${\mathcal O}_F$ can be obtained as
\beqa
{\mathcal G}_{\alpha,\beta\beta^\prime}^R(\omega, \lambda) = \frac{\tilde{\mathcal G}^R_{\alpha,\beta\beta^\prime}+i \tilde{\mathcal G}^R_{\alpha,-\beta\beta^\prime}}{\tilde{\mathcal G}^R_{\alpha,-\beta-\beta^\prime}+i\tilde{\mathcal G}^R_{\alpha,-\beta\beta^\prime}} \tilde{\mathcal G}^R_{\alpha,-\beta-\beta^\prime}.
\eeqa
One would expect that
\beqa
{\mathcal G}_{\alpha,\beta-\beta}^R(\omega, \lambda) =   \frac{\cosh{(\pi \beta b_0 \zeta)} + i \sinh{(\pi \beta b_0 \zeta)}}{\cosh{(\pi \beta b_0 \zeta)} - i \sinh{(\pi \beta b_0 \zeta)}} \tilde{\mathcal G}^R_{\alpha}.
\eeqa
In the absence of chiral gauge fields $b_0$ and $b_r$, it is not necessary to do transformation between $\tilde\Psi_\alpha$ and $\Psi_\alpha$, i.e., the matrix $U_1$, as shown in Eq.(\ref{Eq:tPsi_alpha-Psi_alpha}). Therefore, $\Psi_\alpha \equiv \tilde\Psi_\alpha$, the wavefunctions reduce to be those without the tilde
\beqa
\ii\ii
\Psi_\alpha  = \tilde\Psi_{\alpha} \sim  \tilde{B}_\alpha v_+ \zeta^{\nu_\lambda}(1+ {\mathcal O}(\zeta)) + \tilde{A}_\alpha v_- \zeta^{-\nu_\lambda}(1+{\mathcal O}(\zeta)). \label{Eq:Psi_alpha-omega=0}
\eeqa
With this, we obtain the advanced correlation function for $\Psi_\alpha$,
\beqa
 \ii\ii {\mathcal G}_\alpha^A(\omega, \lambda) = \tilde{\mathcal G}_\alpha^A(\omega, \lambda) \equiv \tilde{B}_\alpha \tilde{A}_\alpha^{-1} = e^{i\pi\nu_\lambda}(2\omega)^{2\nu_\lambda}\frac{\Gamma[-2\nu_\lambda]}{\Gamma[2\nu_\lambda]}  \frac{\Gamma[1+\nu_\lambda + i q e_3 ]}{\Gamma[1-\nu_\lambda + i q e_3 ]}\frac{-\nu_\lambda + (m_D+i\tilde{m}_\alpha)\ell_2 + i q e_3 }{\nu_\lambda + (m_D+i\tilde{m}_\alpha)\ell_2 + i q e_3  }, \label{Eq:gA-I}
\eeqa
where $\tilde{m}_\alpha = -(-1)^\alpha \lambda \ell/r_\star$ and we have used that $(2i\omega)^{2\nu_\lambda} = (2\omega)^{2\nu_\lambda}e^{i\pi \nu_\lambda}$. Similarly as stated before, the retarded correlation function is given by
\beqa
 {\mathcal G}^R_\alpha(\omega,\lambda) = e^{-i\pi\nu_\lambda}(2\omega)^{2\nu_\lambda}\frac{\Gamma[-2\nu_\lambda]}{\Gamma[2\nu_\lambda]}\frac{\Gamma[1+\nu_\lambda - i q e_3 ]}{\Gamma[1-\nu_\lambda - i q e_3 ]} \frac{-\nu_\lambda + (m_D - i\tilde{m}_\alpha)\ell_2 - i q e_3 }{\nu_\lambda + (m_D - i\tilde{m}_\alpha)\ell_2 - i q e_3  }. \label{Eq:gR-I}
\eeqa
Note that $(-1)^{\nu_\lambda}=(-1-i\epsilon)^{\nu_\lambda}=e^{-i\pi \nu_\lambda}$, to entail ${\mathcal G}_\alpha^R(\omega,\lambda)$ is analytical in the upper half complex $\omega$ plane.


With the spatial inversion symmetry, by noticing that $\tilde{m}_2=-\tilde{m}_1$, one can expect
\beqa
{\mathcal G}_1(\omega,\lambda) = {\mathcal G}_2(\omega, -\vec{k}),
\eeqa
which implies that one can simply restrict to one of them, e.g., ${\mathcal G}_2(\omega,\lambda)$, without loss of generality. The Green's function at the IR fixed point can be diagonalized to take the form
\beqa
\ii
{\mathcal G}^R(\omega,\lambda) \2i = \2i \left(\begin{array}{cc}
\! {\mathcal G}_1^R(\omega,\lambda) &   \\
& \3i { \mathcal G}_2^R(\omega,\lambda)
\end{array}
\right),  \label{Eq:gR}
\eeqa
which is obtained from the expansion coefficients at the boundary. The subscript $\alpha=1,2$ indicates a multiplicity which arises in the boundary theory as a consequence of the Lorentz invariance.

The retarded Green's function can also be expressed as
\beqa
\ii {\mathcal G}_\alpha^R(\omega,\lambda)= g_f(\nu_\lambda)(-i\omega)^{2\nu_\lambda}, \quad g_f(\nu_\lambda) \equiv 2^{2\nu_\lambda}\frac{\Gamma[-2\nu_\lambda]}{\Gamma[2\nu_\lambda]} \frac{\Gamma[1+\nu_\lambda - i q e_3 ]}{\Gamma[1-\nu_\lambda - i q e_3 ]}\frac{-\nu_\lambda + (m_D - i\tilde{m}_\alpha)\ell_2 - i q e_3 }{\nu_\lambda + (m_D - i\tilde{m}_\alpha)\ell_2 - i q e_3  }. \label{Eq:Gk-IR-spinor}
\eeqa
It is worth noticing that there are two zeros for spinor's IR CFT green's function, namely,
\beqa
m_D \ell_2 = \nu_\lambda, \quad \text{and} \quad \tilde{m}_\alpha\ell_2 = - q e_3 ,
\eeqa
where the second identity is consistent with the first one by the definition of $\nu_\lambda$. When the second equality is true, then the matrix $U(\zeta)$ in Eq.(\ref{Eq:U-zeta}) becomes diagonal. If one restricts to $\alpha=1$, then for ${\mathcal G}_1(\omega,\lambda)$, the prefactor is
\beqa
 g_f(\nu_\lambda)  \equiv  2^{2\nu_\lambda}\frac{\Gamma[-2\nu_\lambda]}{\Gamma[2\nu_\lambda]}\frac{\Gamma[1+\nu_\lambda - i q e_3 ]}{\Gamma[1-\nu_\lambda - i q e_3 ]}  \frac{-\nu_\lambda + (m_D - i \lambda \frac{\ell}{r_\star})\ell_2 - i q e_3 }{\nu_\lambda + (m_D - i \lambda \frac{\ell}{r_\star})\ell_2 - i q e_3  },  \label{Eq:cnuk}
\eeqa
\end{widetext}
since, $\tilde{m}_\alpha = \lambda {\ell}/{r_\star} $ for $\alpha=1$, where in this case without external magnetic field, $\lambda=\sqrt{k^2-b^2}$ as shown in Eq.(\ref{Eq:lambda-k-b}).

The advanced Green's function at IR fixed point is given by ${\mathcal G}_\alpha^A(\omega,\lambda)  = g_f^\star(\nu_\lambda)(i\omega)^{2\nu_\lambda}$.

For a spinor, from the correlation functions above, one finds that
\beqa
\ii && \ii\ii \frac{{\mathcal G}^R(\omega,\lambda)}{{\mathcal G}^A(\omega,\lambda)} \! = \! e^{-2\pi \nu_\lambda i}\frac{g_f(\nu_\lambda)}{g_f^\star(\nu_\lambda)} \! = \! e^{-2\pi i \nu_\lambda} \frac{\sinh[\pi(q e_3  - i\nu_\lambda) ]}{\sinh[\pi(q e_3  + i\nu_\lambda )]} \quad \quad \nn\\
\3i &=& \3i -e^{-2\pi i \nu_\lambda} \frac{\sin[\pi(\nu_\lambda + i q e_3 ) ]}{\sin[\pi(\nu_\lambda - i q e_3 )]}  \! = \! \frac{e^{-2\pi \nu_\lambda i} - e^{-2\pi q e_3 }}{e^{2\pi \nu_\lambda i} - e^{-2\pi q e_3 }}.  \label{Eq:GRF-GAF}
\eeqa

For real $\nu_\lambda$, the equation in Eq.(\ref{Eq:GRF-GAF}) gives the phase of $g_f(\nu_\lambda)$, and for imaginary $\nu_\lambda$, the equation gives the modulus of $g_f(\nu_\lambda)$.

Alternative, by introducing
\beqa
{\mathcal G}^R(\omega,\lambda) \equiv c(\lambda)\omega^{2\nu_\lambda}, \nn
\eeqa
where $c(\lambda)$ denotes the {prefactor}
\beqa
 c(\lambda) = g_f(\nu_\lambda)(-i)^{2\nu_\lambda} \equiv |c(\lambda)|e^{i\gamma_\lambda} , \label{Eq:c(lambda)-g_f(nu_lambda)}
\eeqa
and $\gamma_\lambda$ is the phase
\beqa
\gamma_\lambda \equiv \text{arg}[c(\lambda)],
\eeqa
so that
\beqa
c^\star(\lambda) = g_f^\star(\nu_\lambda)i^{2\nu_\lambda} \equiv |c(\lambda)|e^{-i\gamma_\lambda}. \nn
\eeqa
Thus, one has
\beqa
\ii\3i
\frac{{\mathcal G}^R(\omega,\lambda)}{{\mathcal G}^A(\omega,\lambda)} \!  = \! \frac{c(\lambda)}{c^\star(\lambda)} \! = \!  e^{2i\gamma_\lambda} \! = \! \frac{(e^{-2\pi \nu_\lambda i} - e^{-2\pi q e_3 })^2}{|e^{2\pi \nu_\lambda i} - e^{-2\pi q e_3 }|^2}.\quad
\eeqa

In the following, we briefly discuss the physical consequence due to the presence of the scaling dimension $\nu_\lambda$ as a complex number in the complex plane:
\begin{enumerate}
\item For real $\nu_\lambda$: The ratio ${\mathcal G}^R(\omega,\lambda)/{\mathcal  G}^A(\omega,\lambda)$ becomes a pure phase and one has
\beqa
\gamma_\lambda = \text{arg}[ \Gamma[-2\nu_\lambda](e^{-2\pi \nu_\lambda i} - e^{-2\pi q e_3 })  ]. \label{Eq:gamma_lambda}
\eeqa
The factor $e^{i\gamma_\lambda}$ and thus $c(\lambda)$ always lie in the upper-half complex plane, and $e^{i\gamma_\lambda + 2\pi \nu_\lambda i}$ always lies in the upper-half complex plane. Therefore in both cases, the phase lies in the upper half complex plane, implying that there is no instability problem for spinors. Namely, for $\nu_\lambda\in(0,1/2)$,
\beqa
\gamma_\lambda + 2\pi \nu_\lambda < \pi, \quad \Rightarrow \quad \pi - \gamma_\lambda > 2\pi \nu_\lambda. \label{Eq:gamma_lambda-nu_lambda}
\eeqa
The causality of the boundary field theory requires that the retarded Green's function must be analytical in the upper-half complex $\omega$ plane. This is equivalent to that there should be no poles in the upper complex $\omega$ plane.

\item For pure imaginary $\nu_\lambda = - i\rho_\lambda (\rho_\lambda>0)$, the ratio becomes real and gives the modulus of $c(\lambda)$.
\beqa
 \frac{e^{-2\pi \rho_\lambda } - e^{-2\pi q e_3 }}{e^{2\pi \rho_\lambda} + e^{-2\pi q e_3 }} = e^{2i\gamma_\lambda} <  e^{-4\pi \rho_\lambda}  <1.
\eeqa
\item For generic $\nu_\lambda$, ${\mathcal G}_\alpha^R(\omega,\lambda)$ has a {logarithmic branch point} at $\omega=0$. One can choose the branch cut along the negative imaginary axis, i.e., the physical sheet to be $(-\pi/2,3\pi/2)$, which resolves into a line of poles along the branch cut when going to finite temperature.
\end{enumerate}

\subsubsection{IR correlation functions at finite temperature}

In the zero temperature case, the near horizon region of a charged black brane is AdS$_2\times {\mathbb R}^{2}$ space-time. The background metric and gauge field in the finite temperature case are given by
\beqa
&&   ds^2  = \frac{\ell_2^2}{\zeta^2}\bigg(  -h(\zeta)dt^2 + h(\zeta)^{-1} d\zeta^2  \bigg) + \frac{r_\star^2}{\ell^2} dx_{d-1}^2, \nn\\
&&   A_t(\zeta)  = \frac{e_3 }{\zeta}\bigg(1 - \frac{\zeta}{\zeta_0} \bigg),
\eeqa
from which, one has the metric components,
\beqa
g_{tt} = \frac{\ell_2^2}{\zeta^2}h(\zeta), \quad g_{\zeta\zeta} = \frac{\ell_2^2}{\zeta^2}h(\zeta)^{-1}, \quad g_{xx} = \frac{r_\star^2}{\ell^2}, \label{Eq:gt2-gzeta2-gx2-T!=0}
\eeqa
where $h(\zeta)$ is given in Eq.(\ref{Eq:ds2-T!=0-AdS2}), namely,
\beqa
h(\zeta)\equiv 1 - \frac{\zeta^2}{\zeta_0^2},
\eeqa
where $\zeta_0$ plays the role of finite temperature $ T = (2\pi\zeta_{0,k})^{-1} \ne 0$ as defined in Eq.(\ref{Eq:TH-zeta0-eta0-k!=0}).

In the representation of Eq.(\ref{Eq:Gamma-rep-2}), the Dirac equation in Eq.(\ref{Eq:Dirac-Psi_alpha-beta_zeta}), namely, Eq.(\ref{Eq:Dirac-EOMs-R1-R2}) in $\zeta$ coordinates becomes,
\begin{widetext}
\beqa
\bigg[    \textbf{1}_2 \bigg(  -\frac{\partial_\zeta}{\sqrt{g_{\zeta\zeta}}} \!-\! i \beta \frac{b_0}{\sqrt{g_{tt}}} \bigg) \!+\!  \sigma^3 m_D  \bigg] \tilde\Psi_{\alpha,\beta} = \bigg(  i\sigma^2 \omega_\beta(\zeta) \!+\! (-1)^\alpha \frac{\lambda}{\sqrt{g_{xx}}} \sigma^1 \bigg) \tilde\Psi_{\alpha,\beta} ,
\eeqa
where $\omega_\beta(\zeta)$ is defined in Eq.(\ref{Eq:omega(zeta)-m_alpha(zeta)}). The bulk spinors can be decomposed as
\beqa
\tilde\Psi_{\alpha,\beta} &=& P_{\beta} \Psi_{\alpha}, \quad \tilde\Psi
= (-gg^{\zeta\zeta})^{-\frac{1}{4}}\tilde\psi = \sqrt{\frac{\zeta}{\ell_2}\frac{\ell}{r_\star} } h(\zeta)^{-\frac{1}{4}}\tilde\psi, \nn
\eeqa
where $P_\beta$ with $\beta=\pm$ is the spin projection operator defined in Eq.(\ref{Eq:P_beta}). With $\tilde\Psi_{\alpha,\beta} \sim R_{\alpha,\beta}$, it turns out
\beqa
\textbf{1}_2  \frac{\partial_\zeta}{\sqrt{g_{\zeta\zeta}}} R_{\alpha,\beta} - \bigg(   m_D \sigma^3 \!-\! i \beta \frac{b_0}{\sqrt{g_{tt}}} \textbf{1}_2 -  \omega_\beta(\zeta) i\sigma^2 \!-\! (-1)^\alpha \frac{\lambda}{\sqrt{g_{xx}}} \sigma^1 \bigg) R_{\alpha,\beta} = 0.
\eeqa
One has
\beqa
\bigg[ \textbf{1}_2  \partial_\zeta -  \sqrt{g_{\zeta\zeta}} \bigg( m_D \sigma^3 \!-\! (-1)^\alpha \lambda \frac{1}{\sqrt{g_{xx}}}  \sigma^1 \bigg) \!+\! i \beta b_0 \sqrt\frac{g_{\zeta\zeta}}{g_{tt}} \textbf{1}_2 +  \bigg( \sqrt\frac{g_{\zeta\zeta}}{g_{tt}} (\omega \!+\!  q A_t) \!+\! \beta b_\zeta \bigg) i\sigma^2     \bigg] \tilde\Psi_{\alpha,\beta} = 0.
\eeqa
In the background metric and gauge field given in Eq.(\ref{Eq:gt2-gzeta2-gx2-T!=0}), the EOMs become
\beqa
\bigg[ \textbf{1}_2 \partial_\zeta -   \frac{\ell_2}{\zeta}\frac{1}{\sqrt{h(\zeta)}} \big( m_D \sigma^3 + m_\alpha(\zeta) \sigma^1 \big) \!+\! i \beta b_0 \frac{1}{h(\zeta)} \textbf{1}_2 +  \bigg( \frac{1}{h(\zeta)} \Big(\omega \!+\!  q e_3  \big( \frac{1}{\zeta} - \frac{1}{\zeta_0} \big)  \Big) \!+\! \beta b_\zeta \bigg) i\sigma^2     \bigg] \tilde\Psi_{\alpha,\beta} = 0,
\eeqa
where $m_\alpha(\zeta)$ is defined in Eq.(\ref{Eq:omega(zeta)-m_alpha(zeta)}).

The equations of motion become two coupled ODEs as shown in Eq.(\ref{Eq:fg_zeta_1st}) with the variables $\lambda_{1,2}(\zeta)$ given by Eq.(\ref{Eq:lambda_1^pm-lambda_2-zeta}), by choosing the conversion that $\partial_t\to  -i\omega$,
\beqa
&& \lambda_1^\pm(\zeta) \equiv \sqrt\frac{g_{\zeta\zeta} }{g_{tt}}[(\omega \pm \beta b_0) \!+\!  q A_t(\zeta)] \!+\! \beta b_\zeta = \frac{1}{h(\zeta)}\bigg(\omega \pm \beta b_0 + q e_3  \big( \frac{1}{\zeta} - \frac{1}{\zeta_0} \big) \bigg) + \beta b_\zeta , \nn\\
&& \lambda_2(\zeta)  \equiv \sqrt{g_{\zeta\zeta}}[m_D + i m_\alpha(\zeta)]  = \frac{\ell_2}{\zeta}\frac{1}{\sqrt{h(\zeta)}}(m_D + i \tilde{m}_\alpha),
\eeqa
where $m_\alpha(\zeta)$ is defined in Eq.(\ref{Eq:omega(zeta)-m_alpha(zeta)}) and $\tilde{m}_\alpha$ just recovers that defined in Eq.(\ref{Eq:mt_alpha}).

From the two coupled $1$-st order ODEs, one obtains two decoupled $2$-nd order ODEs as shown in Eq.(\ref{Eq:fg_zeta_2nd}), with the parameters given in Eq.(\ref{Eq:gamma-1234-zeta}),
\beqa
&& \partial_\zeta^2 \tilde{f}_{\alpha,\beta} + \frac{2\zeta^2-\zeta_0^2-2i\beta b_0\zeta\zeta_0^2}{\zeta(\zeta^2-\zeta_0^2)} \partial_\zeta \tilde{f}_{\alpha,\beta} + \bigg[ - \frac{\nu_\lambda^2+q^2 e_3^2}{\zeta^2(1-\frac{\zeta^2}{\zeta_0^2})} + \frac{ \Big( \omega + q e_3 \big(\frac{1}{\zeta} - \frac{1}{\zeta_0} \big) + \beta b_\zeta\big( 1 - \frac{\zeta^2}{\zeta_0^2} \big) \Big)^2 - b_0^2 }{\big( 1 - \frac{\zeta^2}{\zeta_0^2} \big)^2} \nn\\
 && -i\frac{\omega - \beta b_0 - \frac{q e_3}{\zeta_0}\big( 1 - \frac{\zeta}{\zeta_0} \big) + \beta b_\zeta \big( 1 - \frac{\zeta^2}{\zeta_0^2} \big) \big( 1 - 2 \frac{\zeta^2}{\zeta_0^2} \big)}{\zeta \big( 1 - \frac{\zeta^2}{\zeta_0^2} \big)^2} \bigg]\tilde{f}_{\alpha,\beta} = 0, \nn\\
 && \partial_\zeta^2 \tilde{g}_{\alpha,\beta} + \frac{2\zeta^2-\zeta_0^2-2i\beta b_0\zeta\zeta_0^2}{\zeta(\zeta^2-\zeta_0^2)} \partial_\zeta \tilde{g}_{\alpha,\beta} + \bigg[ - \frac{\nu_\lambda^2+q^2 e_3^2}{\zeta^2(1-\frac{\zeta^2}{\zeta_0^2})} + \frac{ \Big( \omega + q e_3 \big(\frac{1}{\zeta} - \frac{1}{\zeta_0} \big) + \beta b_\zeta\big( 1 - \frac{\zeta^2}{\zeta_0^2} \big) \Big)^2 - b_0^2 }{\big( 1 - \frac{\zeta^2}{\zeta_0^2} \big)^2} \nn\\
 && +i\frac{\omega + \beta b_0 - \frac{q e_3}{\zeta_0}\big( 1 - \frac{\zeta}{\zeta_0} \big) + \beta b_\zeta \big( 1 - \frac{\zeta^2}{\zeta_0^2} \big) \big( 1 - 2 \frac{\zeta^2}{\zeta_0^2} \big)}{\zeta \big( 1 - \frac{\zeta^2}{\zeta_0^2} \big)^2} \bigg] \tilde{g}_{\alpha,\beta} = 0, \label{Eq:fgt_alpha-zeta-T!=0}
\eeqa
where we have considered that $(m_D^2 + \tilde{m}_\alpha^2) \ell_2^2 = \nu_\lambda^2 + q^2 e_3^2$ due to Eq.(\ref{Eq:nu_lambda-m_lambda}).

Absorbing $b_\zeta$ into the redefinition of wave function, the equations can be simplified as
\beqa
&& \partial_\zeta^2 \tilde{f}_{\alpha,\beta} + \frac{2\zeta^2-\zeta_0^2-2i\beta b_0\zeta\zeta_0^2}{\zeta(\zeta^2-\zeta_0^2)} \partial_\zeta \tilde{f}_{\alpha,\beta} + \bigg[ - \frac{\nu_\lambda^2+q^2 e_3^2}{\zeta^2(1-\frac{\zeta^2}{\zeta_0^2})} + \frac{ \Big( \omega + q e_3 \big(\frac{1}{\zeta} - \frac{1}{\zeta_0} \big)  \Big)^2 - b_0^2 }{\big( 1 - \frac{\zeta^2}{\zeta_0^2} \big)^2}  -i\frac{\omega - \beta b_0 - \frac{q e_3}{\zeta_0}\big( 1 - \frac{\zeta}{\zeta_0} \big)  }{\zeta \big( 1 - \frac{\zeta^2}{\zeta_0^2} \big)^2} \bigg]f_{\alpha,\beta} = 0, \nn\\
&& \partial_\zeta^2 \tilde{g}_{\alpha,\beta} + \frac{2\zeta^2-\zeta_0^2-2i\beta b_0\zeta\zeta_0^2}{\zeta(\zeta^2-\zeta_0^2)} \partial_\zeta \tilde{g}_{\alpha,\beta} + \bigg[ - \frac{\nu_\lambda^2+q^2 e_3^2}{\zeta^2(1-\frac{\zeta^2}{\zeta_0^2})} + \frac{ \Big( \omega + q e_3 \big(\frac{1}{\zeta} - \frac{1}{\zeta_0} \big) \Big)^2 - b_0^2 }{\big( 1 - \frac{\zeta^2}{\zeta_0^2} \big)^2}  +i\frac{\omega + \beta b_0 - \frac{q e_3}{\zeta_0}\big( 1 - \frac{\zeta}{\zeta_0} \big)  }{\zeta \big( 1 - \frac{\zeta^2}{\zeta_0^2} \big)^2} \bigg] \tilde{g}_{\alpha,\beta} = 0. \nn
\eeqa
The above equations turn out to be analytically solvable.


By solving EOMs in Eq(\ref{Eq:fgt_alpha-zeta-T!=0}), one obtains
\beqa
\tilde{f}_{\alpha,\beta} &=&   \bigg(  \frac{\zeta - \zeta_0}{\zeta + \zeta_0 } \bigg)^{-\frac{1}{2}i(\omega-\beta b_0)\zeta_0 + i q e_3  } \bigg[ C_{1} \bigg( \frac{\zeta}{\zeta - \zeta_0} \bigg)^{\nu_\lambda}  {}_2 F_1[ \nu_\lambda - i q e_3,  \frac{1}{2} + \nu_\lambda - i q e_3 + i \omega \zeta_0, 1 + 2\nu_\lambda;  \frac{2\zeta}{\zeta-\zeta_0} ] \nn \\
&+& C_{2} (-2)^{-2\nu_\lambda}  \bigg(  \frac{\zeta}{\zeta - \zeta_0} \bigg)^{-\nu_\lambda}     {}_2 F_1[ -\nu_\lambda - i q e_3,  \frac{1}{2} - \nu_\lambda - i q e_3  + i \omega \zeta_0, 1 - 2\nu_\lambda;  \frac{2\zeta}{\zeta-\zeta_0} ] \bigg], \nn\\
\tilde{g}_{\alpha,\beta} &=&   \bigg(  \frac{\zeta - \zeta_0}{\zeta + \zeta_0 } \bigg)^{+\frac{1}{2}i(\omega+\beta b_0)\zeta_0 - i q e_3  } \bigg[ C_{3} \bigg( \frac{\zeta}{\zeta - \zeta_0} \bigg)^{\nu_\lambda}  {}_2 F_1[ \nu_\lambda + i q e_3,  \frac{1}{2} + \nu_\lambda + i q e_3  - i \omega \zeta_0, 1 + 2\nu_\lambda;  \frac{2\zeta}{\zeta-\zeta_0} ] \nn \\
&+& C_{4} (-2)^{-2\nu_\lambda}  \bigg(  \frac{\zeta}{\zeta - \zeta_0} \bigg)^{-\nu_\lambda}     {}_2 F_1[ -\nu_\lambda + i q e_3,  \frac{1}{2} - \nu_\lambda + i q e_3  - i \omega \zeta_0, 1 - 2\nu_\lambda;  \frac{2\zeta}{\zeta-\zeta_0} ] \bigg],
\eeqa
where ${}_2 F_1(a,b;c;z)$ is the Gauss's hypergeometric function, which is an independent solution of the hypergeometric differential equation $z(1-z)y^{\prime\prime}(z)+(c-(a+b+1)z)y^\prime(z)-ab y(z)=0$. In the near horizon region, the coefficients of the solution are chosen so that the solution corresponds to the physical in-falling wave one, which gives
\beqa
\ii
\frac{C_1}{C_2} \3i &=& \3i \frac{\Gamma[-2\nu_\lambda]}{\Gamma[2\nu_\lambda]}\frac{\Gamma[ \nu_\lambda - i q e_3  ]}{\Gamma[ -\nu_\lambda - i q e_3 ]}\frac{\Gamma[\frac{1}{2}+\nu_\lambda + i q e_3  - i \omega\zeta_0  ]}{\Gamma[\frac{1}{2}-\nu_\lambda + i q e_3  - i \omega\zeta_0]}, \nn\\
\frac{C_3}{C_4} \3i &=& \3i \frac{\Gamma[-2\nu_\lambda]}{\Gamma[2\nu_\lambda]}\frac{\Gamma[1 + \nu_\lambda - i q e_3  ]}{\Gamma[ 1 - \nu_\lambda - i q e_3 ]}\frac{\Gamma[\frac{1}{2} + \nu_\lambda + i q e_3  - i \omega\zeta_0  ]}{\Gamma[\frac{1}{2}-\nu_\lambda + i q e_3  - i \omega\zeta_0]},
\label{Eq:C1VsC2-C3VsC4}
\eeqa
the coefficients obey the identity,
\beqa
\frac{C_3}{C_4} = \frac{q e_3  + i \nu_\lambda}{q e_3  - i \nu_\lambda}\frac{C_1}{C_2}. \label{Eq:C1VsC2-C3VsC4-ratio}
\eeqa
In the infinite boundary, the wave function has the asymptotic behavior
\beqa
\tilde{f}_{\alpha,\beta} &=&   (-1)^{-\frac{1}{2}i(\omega-\beta b_0)\zeta_0 + i q e_3  +\nu_\lambda} \bigg[ C_{1}  \bigg( \frac{\zeta}{\zeta_0} \bigg)^{\nu_\lambda}   + C_{2} 2^{-2\nu_\lambda} \bigg(  \frac{\zeta}{\zeta_0} \bigg)^{-\nu_\lambda}   \bigg], \nn\\
\tilde{g}_{\alpha,\beta} &=&   (-1)^{+\frac{1}{2}i(\omega+\beta b_0)\zeta_0 - i q e_3  +\nu_\lambda}  \bigg[ C_{3} \bigg( \frac{\zeta}{ \zeta_0} \bigg)^{\nu_\lambda}   + C_{4} 2^{-2\nu_\lambda} \bigg(  \frac{\zeta}{\zeta_0} \bigg)^{-\nu_\lambda}   \bigg].
\eeqa
According to Eq.(\ref{Eq:fgt_alpha-transQ_FGt_alpha}), one has the asymptotic behavior
\beqa
\tilde{R}_{\alpha,\beta} &\equiv &
\left(
  \begin{array}{c}
    \tilde{F}_{\alpha,\beta} \\
    \tilde{G}_{\alpha,\beta} \\
  \end{array}
\right) = Q^{-1} \left(
  \begin{array}{c}
    \tilde{f}_{\alpha,\beta} \\
    \tilde{g}_{\alpha,\beta} \\
  \end{array}
\right) = \frac{1}{\sqrt{2}}\left(
            \begin{array}{cc}
              1 & 1 \\
             -i & i \\
            \end{array}
          \right) \left(
  \begin{array}{c}
    \tilde{f}_{\alpha,\beta} \\
    \tilde{g}_{\alpha,\beta} \\
  \end{array}
\right) = \frac{1}{\sqrt{2}} \left(
  \begin{array}{c}
    \tilde{f}_{\alpha,\beta} +  \tilde{g}_{\alpha,\beta} \\
    -i(\tilde{f}_{\alpha,\beta}-  \tilde{g}_{\alpha,\beta}) \\
  \end{array}
\right) \nn\\
&\overset{\zeta\to 0}{=}& \frac{1}{\sqrt{2}} \left(
                             \begin{array}{c}
                                C_1 + C_3    \\
                                -i(C_1 - C_3)\\
                             \end{array}
                           \right) \bigg(\frac{\zeta}{\zeta_0}\bigg)^{\nu_\lambda}    +           \frac{2^{-2\nu_\lambda}}{\sqrt{2}}   \left(
                             \begin{array}{c}
                               C_2 + C_4   \\
                               -i(C_2 - C_4) \\
                             \end{array}
                           \right) \bigg(\frac{\zeta}{\zeta_0}\bigg)^{-\nu_\lambda}.
\eeqa
On the other hand, as in Eq.(\ref{Eq:Rt_alpha-b/a-zeta}), one has
\beqa
\tilde\Psi_\alpha \sim \tilde{R}_\alpha \equiv \tilde{b}_\alpha\zeta^{\nu_\lambda} + \tilde{a}_\alpha \zeta^{-\nu_\lambda},
\eeqa
where $\tilde{B}_\alpha$ and $\tilde{A}_\alpha$ are two by two matrices.
By imposing the normalization conversion defined in Eq.(\ref{Eq:vpm}),
\beqa
\tilde{b}_\alpha = \frac{\tilde{b}_{\alpha\uparrow}}{\tilde{b}_{\alpha\downarrow}} \sim \frac{v_{\alpha,+\uparrow}}{v_{
\alpha,+\downarrow}} = \frac{m_D \ell_2 + \nu_\lambda}{\tilde{m}_\alpha \ell_2 + q e_3 }, \quad \tilde{a}_\alpha = \frac{\tilde{a}_{\alpha\uparrow}}{\tilde{a}_{\alpha\downarrow}} \sim \frac{v_{\alpha,-\uparrow}}{v_{\alpha,-\downarrow}} = \frac{m_D \ell_2 - \nu_\lambda}{\tilde{m}_\alpha\ell_2 + q e_3 },
\eeqa
one can obtain
\beqa
C_1 = \frac{\nu_\lambda+i q e_3  + (m_D + i \tilde{m}_\alpha)\ell_2}{\nu_\lambda - i q e_3  + (m_D - i \tilde{m}_\alpha)\ell_2}C_3, \quad C_2 = \frac{\nu_\lambda-i q e_3  - (m_D + i \tilde{m}_\alpha)\ell_2}{\nu_\lambda + i q e_3  - (m_D - i \tilde{m}_\alpha)\ell_2}C_4, \nn
\eeqa
From the equations above and by using the equality $(m_D^2+\tilde{m}^2)\ell_2^2 = \nu_\lambda^2 + (q e_3 )^2$ due to Eq.(\ref{Eq:nu_lambda-m_lambda}), one obtains
\beqa
\frac{C_1}{C_2} = \frac{q e_3  - i \nu_\lambda}{q e_3  + i \nu_\lambda}\frac{C_3}{C_4},
\eeqa
which is consistent with that in Eq.(\ref{Eq:C1VsC2-C3VsC4-ratio}). By substituting $C_1$ and $C_2$ back into the expansion of the solution, we finally obtain
\beqa
\ii\ii
\tilde\Psi_\alpha  \sim  \tilde{B}_\alpha v_{\alpha,+} \zeta^{\nu_\lambda} + \tilde{A}_\alpha v_{\alpha,-} \zeta^{-\nu_\lambda},
\eeqa
with
\beqa
\tilde{B}_\alpha &=& \frac{\zeta_0^{-\nu_\lambda}}{\nu_\lambda-i q e_3  +(m_D - i \tilde{m}_\alpha)\ell_2}C_3  \equiv  \frac{\tilde{b}_{\alpha\uparrow}}{m_D \ell_2 + \nu_\lambda} = \frac{\tilde{b}_{\alpha\downarrow}}{\tilde{m}_\alpha\ell_2 + q e_3  }, \nn\\
\tilde{A}_\alpha &=& 2^{-2\nu_\lambda}\frac{\zeta_0^{\nu_\lambda}}{-\nu_\lambda - i q e_3  + (m_D -i\tilde{m}_\alpha)\ell_2}C_4 \equiv  \frac{\tilde{a}_{\alpha\uparrow}}{m_D \ell_2 - \nu_\lambda} = \frac{\tilde{a}_{\alpha\downarrow}}{\tilde{m}_\alpha\ell_2 + q e_3  }. \nn
\eeqa
In the presence of chiral gauge field, according to Eq.(\ref{Eq:tPsi_alpha-Psi_alpha}) and Eq.(\ref{Eq:transU1_tPsi-Psi}), one has $\Psi_\alpha = U_1^{-1} \tilde\Psi_\alpha$, namely
\beqa
\left(
  \begin{array}{c}
    \Psi_{\alpha,+} \\
    \Psi_{\alpha,-} \\
  \end{array}
\right) = U_1^{-1} \left(
  \begin{array}{c}
    \tilde\Psi_{\alpha,+} \\
    \tilde\Psi_{\alpha,-} \\
  \end{array}
\right)
&\sim & \frac{1}{\sqrt{2}}\left(
                                                                                           \begin{array}{cc}
                                                                                             1 & i \\
                                                                                             i & 1 \\
                                                                                           \end{array}
                                                                                         \right) \left(
  \begin{array}{c}
    \tilde{R}_{\alpha,+} \\
    \tilde{R}_{\alpha,-} \\
  \end{array}
\right) =  \frac{1}{\sqrt{2}} \left(
  \begin{array}{c}
    \tilde{R}_{\alpha,+} + i \tilde{R}_{\alpha,-} \\
    i( \tilde{R}_{\alpha,+} - i \tilde{R}_{\alpha,-}) \\
  \end{array}
\right).
\eeqa
Thus the asymptotic behavior in the infinite boundary becomes
\beqa
\left(
  \begin{array}{c}
    \Psi_{\alpha,+} \\
    \Psi_{\alpha,-} \\
  \end{array}
\right) \sim \left(
  \begin{array}{c}
    (\tilde{B}_{\alpha,+}+i\tilde{B}_{\alpha,-})   \\
    i(\tilde{B}_{\alpha,+}-i\tilde{B}_{\alpha,-})   \\
  \end{array}
\right) v_+ \zeta^{\nu_\lambda} + \left(
  \begin{array}{c}
    (\tilde{A}_{\alpha,+}+i\tilde{A}_{\alpha,-})  \\
    i(\tilde{A}_{\alpha,+}-i\tilde{A}_{\alpha,-})  \\
  \end{array}
\right) v_- \zeta^{-\nu_\lambda},
\eeqa
from which, we obtain the correlation function for $\Psi_\alpha$,
\beqa
{\mathcal G}_{\alpha,\beta\beta^\prime}^T(\omega,\lambda) & \equiv & \frac{\tilde{B}_{\alpha,\beta} + i \tilde{B}_{\alpha,-\beta}}{\tilde{A}_{\alpha,\beta^\prime} + i \tilde{A}_{\alpha,-\beta^\prime}} = \frac{\tilde{\mathcal G}^T_{\alpha,\beta\beta^\prime}+i \tilde{\mathcal G}^T_{\alpha,-\beta\beta^\prime}}{\tilde{\mathcal G}^T_{\alpha,-\beta-\beta^\prime}+i\tilde{\mathcal G}^T_{\alpha,-\beta\beta^\prime}} \tilde{\mathcal G}^T_{\alpha,-\beta-\beta^\prime}.
\eeqa
The results can be categorized into two cases:
\begin{itemize}
  \item If $\beta = \beta^\prime$, one obtains ${\mathcal G}_{\alpha}^T(\omega, \lambda) \equiv {\mathcal G}_{\alpha,\beta\beta}^T(\omega, \lambda)=\tilde{\mathcal G}_{\alpha,\beta\beta}^T(\omega, \lambda)\equiv \tilde{\mathcal G}_{\alpha}^T(\omega, \lambda)$.
  \item If $\beta = - \beta^\prime$, one has
  \beqa
  {\mathcal G}_{\alpha,\beta-\beta}^T(\omega, \lambda) = \frac{\tilde{\mathcal G}^T_{\alpha,\beta-\beta}+ i \tilde{\mathcal G}^T_{\alpha}}{\tilde{\mathcal G}^T_{\alpha}+i \tilde{\mathcal G}^T_{\alpha,\beta-\beta}} \tilde{\mathcal G}^T_{\alpha} = \frac{\cosh{( \beta \pi b_0 \zeta_0/2)} + i\sinh{(\beta \pi b_0 \zeta_0/2)}}{\cosh{( \beta \pi b_0 \zeta_0/2)} - i\sinh{(\beta \pi b_0 \zeta_0/2)}}  \tilde{\mathcal G}^T_{\alpha},
  \eeqa
  where we have used that $\tilde{\mathcal G}^T_{\alpha,\beta-\beta}\tilde{\mathcal G}^T_{\alpha,-\beta\beta}=\tilde{\mathcal G}^T_{\alpha,\beta\beta}\tilde{\mathcal G}^T_{\alpha,-\beta-\beta}\equiv (\tilde{\mathcal G}^T_{\alpha})^2$.
\end{itemize}

In the absence of chiral gauge fields $b_0$ and $b_\zeta$, there is no rotation between $\Psi_\alpha$ and $\tilde\Psi_\alpha$, as shown in Eq.(\ref{Eq:Psi_alpha-omega=0}), the notation tilde associated with the wave functions can be neglected,
\beqa
{\mathcal G}_\alpha^T(\omega,\lambda) & = & \tilde{\mathcal G}_\alpha^T(\omega,\lambda) \equiv {\tilde{B}_\alpha}{\tilde{A}_\alpha^{-1}} = \bigg(\frac{\zeta_0}{2}\bigg)^{-2\nu_\lambda}\frac{-\nu_\lambda - i q e_3  + (m_D - i\tilde{m}_\alpha)\ell_2}{\nu_\lambda - i q e_3  +(m_D -i \tilde{m}_\alpha)\ell_2}\frac{C_3}{C_4} \nn\\
&=& \bigg(\frac{\zeta_0}{2}\bigg)^{-2\nu_\lambda}\frac{-\nu_\lambda - i q e_3  + (m_D - i\tilde{m}_\alpha)\ell_2}{\nu_\lambda - i q e_3  +(m_D -i \tilde{m}_\alpha)\ell_2} \frac{\Gamma[-2\nu_\lambda]}{\Gamma[2\nu_\lambda]}\frac{\Gamma[1 + \nu_\lambda - i q e_3  ]}{\Gamma[ 1 - \nu_\lambda - i q e_3 ]}\frac{\Gamma[\frac{1}{2} + \nu_\lambda + i q e_3  - i \omega\zeta_0  ]}{\Gamma[\frac{1}{2}-\nu_\lambda + i q e_3  - i \omega\zeta_0]}, \nn
\eeqa
where in the last equality, we have used Eq.(\ref{Eq:C1VsC2-C3VsC4}). By using the temperature definition $T = (2\pi \zeta_0)^{-1}$ as in Eq.(\ref{Eq:TH-zeta0-eta0-k!=0}), we have obtained the IR correlation functions at finite temperature from the AdS gravity in the near horizon region,
\beqa
{\mathcal G}_\alpha^T(\omega,\lambda) = ( 4\pi T)^{2\nu_\lambda}\frac{-\nu_\lambda - i q e_3  + (m_D -i\tilde{m}_\alpha)\ell_2}{\nu_\lambda - i q e_3  +(m_D -i \tilde{m}_\alpha)\ell_2}  \frac{\Gamma[-2\nu_\lambda]}{\Gamma[2\nu_\lambda]}\frac{\Gamma[1 + \nu_\lambda - i q e_3  ]}{\Gamma[ 1 - \nu_\lambda - i q e_3 ]}\frac{\Gamma[\frac{1}{2} + \nu_\lambda + i q e_3  - i \frac{\omega}{2\pi T}   ]}{\Gamma[\frac{1}{2}-\nu_\lambda + i q e_3  - i \frac{\omega}{2\pi T} ]}, \nn
\eeqa
where $\tilde{m}_\alpha$ is defined in Eq.(\ref{Eq:mt_alpha}), one can define
\beqa
{\mathcal G}^T(\omega,\lambda) \equiv T^{2\nu_\lambda}g_f\bigg(\frac{\omega}{2\pi T}, \nu_\lambda\bigg), \label{Eq:gR-T!=0}
\eeqa
where $g_f(x, \nu_\lambda)$ is a scaling function given by
\beqa
g_f(x,\nu_\lambda)\equiv (4\pi)^{2\nu_\lambda} \frac{-\nu_\lambda - i q e_3  + (m_D - i\tilde{m}_\alpha)\ell_2}{\nu_\lambda - i q e_3  +(m_D -i \tilde{m}_\alpha)\ell_2} \frac{\Gamma[-2\nu_\lambda]}{\Gamma[2\nu_\lambda]} \frac{\Gamma[1 + \nu_\lambda - i q e_3  ]}{\Gamma[ 1 - \nu_\lambda - i q e_3 ]} \frac{\Gamma[\frac{1}{2} + \nu_\lambda + i q e_3  - i x  ]}{\Gamma[\frac{1}{2}-\nu_\lambda + i q e_3  - i x]}. \label{Eq:gf}
\eeqa
For the whole solution of the Dirac equation, there is a useful relation, ${\mathcal G}_1^{T}(\omega,\lambda) = {\mathcal G}_2^{T}(\omega,-k)$, which is still kept as the zero temperature case. It implies that one can simply restrict to one of them without loss of generality, e.g., $\alpha=1$,
\beqa
 {\mathcal G}_1^T(\omega,\lambda) = ( 4\pi T )^{2\nu_\lambda}\frac{-\nu_\lambda - i q e_3  + (m_D - i \lambda \frac{\ell}{r_\star})\ell_2}{\nu_\lambda - i q e_3  + (m_D -i \lambda \frac{\ell}{r_\star})\ell_2}  \frac{\Gamma[-2\nu_\lambda]}{\Gamma[2\nu_\lambda]}\frac{\Gamma[1 + \nu_\lambda - i q e_3  ]}{\Gamma[ 1 - \nu_\lambda - i q e_3 ]}\frac{\Gamma[\frac{1}{2} + \nu_\lambda + i q e_3  - i \frac{\omega}{2\pi T}  ]}{\Gamma[\frac{1}{2}-\nu_\lambda + i q e_3  - i \frac{\omega}{2\pi T} ]}, \nn  
\eeqa
where $\lambda=\sqrt{|\vec{k}|^2 - |\vec{b}|^2}$ without external magnetic field. The finite temperature Green's function at the IR fixed point can be diagonalized to take the form
\beqa
{\mathcal G}^T(\omega,\lambda)
= \left( \begin{array}{cc}
{\mathcal G}_1^T(\omega,\lambda) & \\
    & {\mathcal G}_1^T(\omega,-\lambda)
\end{array}
\right).
\eeqa
The fermion self energy at finite temperature becomes
\beqa
{\mathcal G}^T(\omega,\lambda)  \sim (4\pi T)^{2\nu_\lambda}\frac{\Gamma[\frac{1}{2}+\nu_\lambda+ i q e_3  - i \frac{\omega}{2\pi T}]}{\Gamma[\frac{1}{2}-\nu_\lambda+ i q e_3  - i  \frac{\omega}{2\pi T}]}   \xrightarrow[]{T\to 0}   (4\pi T)^{2\nu_\lambda} \bigg(-i \frac{\omega}{2\pi T}\bigg)^{2\nu_\lambda} \sim c_\lambda \omega^{2\nu_\lambda}, \label{Eq:gT=0-gR}
\eeqa
where in the non-zero small temperature limit, the order $(\pi T)^{2\nu_\lambda}$ characterizes the thermal smearing effect of the Fermi surface. In the zero temperature limit, i.e., $T\to 0$, the line of discrete poles of the Gamma function at finite temperature merges as a branch cut for $\omega^{2\nu_\lambda}$ at $T=0$, which coincides with the near horizon geometry at zero temperature, which is AdS$_2$ along the normal vector of the cutoff hyper-surface. Therefore in the zero temperature limit, the finite temperature Green's function just reduces to the zero temperature retarded Green's function. Alternatively, it is worthy to notice that the original branch point at $\omega = 0$ of the zero temperature retarded Green's function in Eq.(\ref{Eq:Gk-IR-spinor}) disappears and the branch cut is replaced at finite temperature by a line of poles parallel to the next imaginary axis.

\subsection{Charged fermions in the bulk with $k\ne 0$}
\label{sec:topology-k!=0}

In the above sections, we have studied the bulk geometry with Ricci flat $k=0$ hypersurface, namely, AdS RN black brane with Ricci flat horizon. While it is also significant to understand the quantitative behavior of the charged fermion in AdS RN black brane with non-flat horizon. In the following, we will focus on topological charged black brane in AdS$_4$ gravity, whose spatial hypersurface is a spheries/hyperbolic surface with normalized constant Ricci scalar $k=1$ and $k=-1$ as their topological indexes, respectively. Assuming the generic metric of the topological charged black hole is
\beqa
 ds^2 = - f(r) dt^2 + \frac{dr^2}{f(r)} + r^2 d\Omega_{2,k}^2, \quad d\Omega_{2,1}^2 = d\theta^2 + \sin^2\theta d\phi^2, \quad d\Omega_{2,-1}^2 = d\theta^2 + \sinh^2\theta d\phi^2, \nn
\eeqa
with red-shift factor $f(r)$ given in Eq.(\ref{Eq:fr-gr}). The near horizon metric at zero and finite temperature limits are given in Eq.(\ref{Eq:ds2-T!=0-AdS2}) and Eq.(\ref{Eq:ds2-T=0-AdS2}), respectively, by defining $\zeta_k$ and $\zeta_{0,k}$ as those in Eq.(\ref{Eq:zeta-eta-k!=0}), where the coordinate $r_0 \le r<\infty$ has a one-to-one correspondence to the coordinates $\zeta_{0,k}\ge \zeta_k>0$.

\subsubsection{IR correlation functions at zero temperature}


As discussed above, due to topological quantization, the continuous momentum $|\vec{k}|^2$ in the Ricci flat case is replaced by discrete quantum number $\kappa^2=l(l+1)$ in the Ricci curved case with $k=\pm 1$.

Therefore, for a charged fermion at zero temperature, the near horizon geometry of a topological charged black brane is given by Eq.(\ref{Eq:ds2-T=0-AdS2}). The IR correlation function can be expressed more explicitly with the topological index $k=1$, namely
\beqa
{\mathcal G}_l^R(\omega)= c(\nu_l)(-i\omega)^{2\nu_l}, \quad c(\nu_l) \equiv 2^{2\nu_l}\frac{\Gamma[-2\nu_l]}{\Gamma[2\nu_l]}  \frac{\Gamma[1+\nu_l - i q e_{3,k}]}{\Gamma[1-\nu_l - i q e_{3,k}]}\frac{-\nu_l + (m_D- i\tilde{m}_{\alpha,l})\ell_{2,k} - i q e_{3,k}}{\nu_l + (m_D - i\tilde{m}_{\alpha,l})\ell_{2,k} - i q e_{3,k} }, \label{Eq:Gl-IR-spinor} 
\eeqa
where $\tilde{m}_{\alpha,l}$ is given below, and $ l \in {\mathbb Z}$. The scaling dimensions of the IR CFT for $s$,$p$,$d$, \ldots waves with $l=0,1,2, \ldots$ are
\beqa
 \Delta_{l\pm}^{\text{IR}}= \frac{1}{2}\pm \nu_l, \quad \nu_l = \sqrt{ (m_D^2 + \tilde{m}_{\alpha,l}^2)\ell_{2,k}^{2} - q^2 e_{3,k}^{2}}, \quad \tilde{m}_{\alpha,l} \equiv -(-1)^\alpha \kappa_l \frac{\ell}{r_\star}, \quad \kappa_l = l, -(l+1),  \label{Eq:nu_l-m_l}
\eeqa
with $\ell_{2,k}$ defined in Eq.(\ref{Eq:ell_2k}), $e_{3,k}$ defined in Eq.(\ref{Eq:edk-ed}).

\subsubsection{IR correlation functions of fermion at finite temperature}

For a charged fermion field at finite temperature, the near horizon geometry of a topological charged black brane in AdS$_4$ is given by Eq.(\ref{Eq:ds2-T!=0-AdS2}). The IR correlation function at finite temperature from the extreme horizon region becomes
\beqa
 {\mathcal G}_l^T(\omega) = (4\pi T)^{2\nu_l}\frac{-\nu_l - i q e_{3,k} + (m_D -i\tilde{m}_{\alpha,l})\ell_{2,k}}{\nu_l - i q e_{3,k} +(m_D -i \tilde{m}_{\alpha,l})\ell_{2,k}}  \frac{\Gamma[-2\nu_l]}{\Gamma[2\nu_l]}\frac{\Gamma[1 + \nu_l - i q e_{3,k} ]}{\Gamma[ 1 - \nu_l - i q e_{3,k}]}\frac{\Gamma[\frac{1}{2} + \nu_l + i q e_{3,k} - i \frac{\omega}{2\pi T}   ]}{\Gamma[\frac{1}{2}-\nu_l + i q e_{3,k} - i \frac{\omega}{2\pi T} ]}, \nn
\eeqa
\end{widetext}
with corresponding parameters given in Eq.(\ref{Eq:nu_l-m_l}) and finite temperature $T = (2\pi \zeta_{0,k})^{-1} \ne 0$ defined in Eq.(\ref{Eq:TH-zeta0-eta0-k!=0}), where $\ell_{2,k}$ defined in Eq.(\ref{Eq:ell_2k}) with $ l \in{\mathbb Z}$, and $e_{3,k}$ defined in Eq.(\ref{Eq:edk-ed}).

\subsubsection{Topological quantization of fermion liquid}

The difference between the correlation functions at IR fixed point for topological case with $k=\pm 1$ and those with $k=0$ is that, the corresponding effective length $\ell_2$ and the effective dimensionless couplings $e_3 $ are dressed by the topology of the black brane, and changed into $\ell_{2,k}$ and $e_{3,k}$ defined in Eq.(\ref{Eq:ell_2k}) and Eq.(\ref{Eq:edk-ed}), respectively. Moreover, the continuous momentum squares $|\vec{k}|^2$ for $k=0$ case, is replaced by the discrete momentum squares $\kappa^2$ for $k=\pm 1$ case, with $\kappa^2 = l(l+1)$. Take $k=1$ case as an example, one needs to change Dirac wave functions with plane wave expansion into those with spherical harmonics expansion, and $\kappa_l=l,-(l+1)$. Consequently, the continuous effective mass of fermions at IR CFT with $k=0$ case given in Eq.(\ref{Eq:nu_lambda-m_lambda}), becomes quantized at IR CFT$_3$ with $k=\pm 1$ case.

As physical consequence, the Fermi surface will be topological quantized, since the Fermi momentum at Fermi surface, namely $k_F$ has to be a quantum number $l_F\in {\mathbb Z}$. Consequently, it is expected that the Fermi momentum will have a discrete sparse distributed pattern.

\subsection{Non-Fermi liquid with anomalous Hall effect}

In this section, as a demo, we explore the topological non-Fermi liquid behavior of anomalous Hall effect in topological materials. The non-Fermi liquid phase in these topological metals, can be described by the spectral function, which is the physical observable through ARPES experiment.

\subsubsection{Oscillatory phase at zero frequency}

The oscillatory region is determined by the scaling dimension of the IR CFT in Eq.(\ref{Eq:nu_lambda-m_lambda}), i.e., $\nu_\lambda = \sqrt{(m_D^2+\tilde{m}^2)\ell_2^2-q^2 e_3^2}<0$, where $\tilde{m}_\alpha$, $\ell_2$ and $e_3$ are given in Eq.(\ref{Eq:mt_alpha}), Eq.(\ref{Eq:ell_2}) and Eq.(\ref{Eq:ed-gF}). Intuitively, the oscillate phase will be present given an arbitrary large charge $q$. The momentum in the oscillatory region is
\beqa
|\vec{k}| < \sqrt{ |\vec{b}|^2 + \bigg( q^2 \frac{g_F^2}{2} - m_D^2 \bigg) \frac{r_\star^2}{\ell^2} }.
\eeqa
It is obvious that the presence of the AHE, i.e., $|\vec{b}|\ne 0$, will increase the momentum in the oscillatory region, while the presence of the Dirac mass, i.e., $m_D \ne 0$, will decrease the momentum in the oscillatory region.

\begin{figure}[H]
\begin{center}
  \includegraphics[scale=0.45]{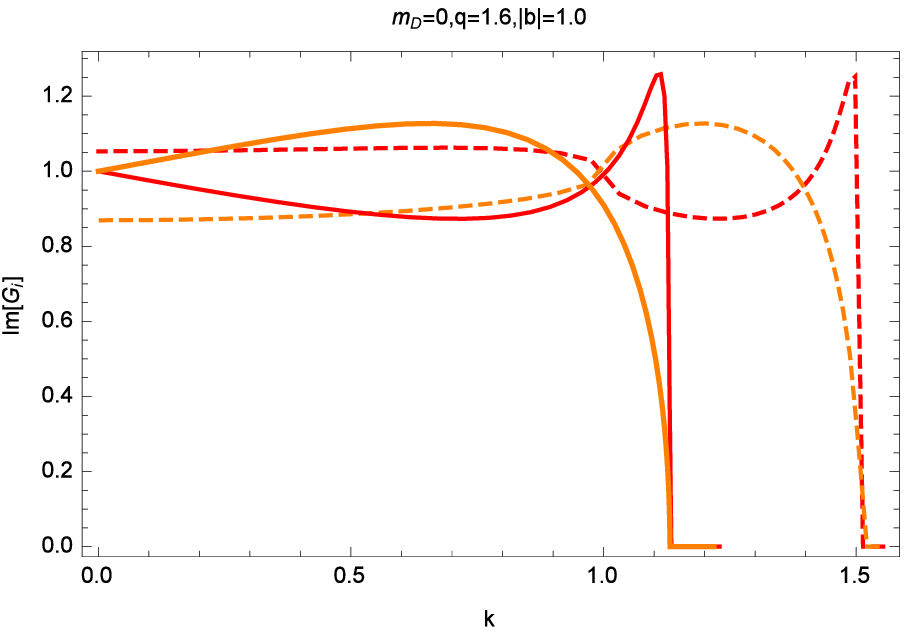} \includegraphics[scale=0.45]{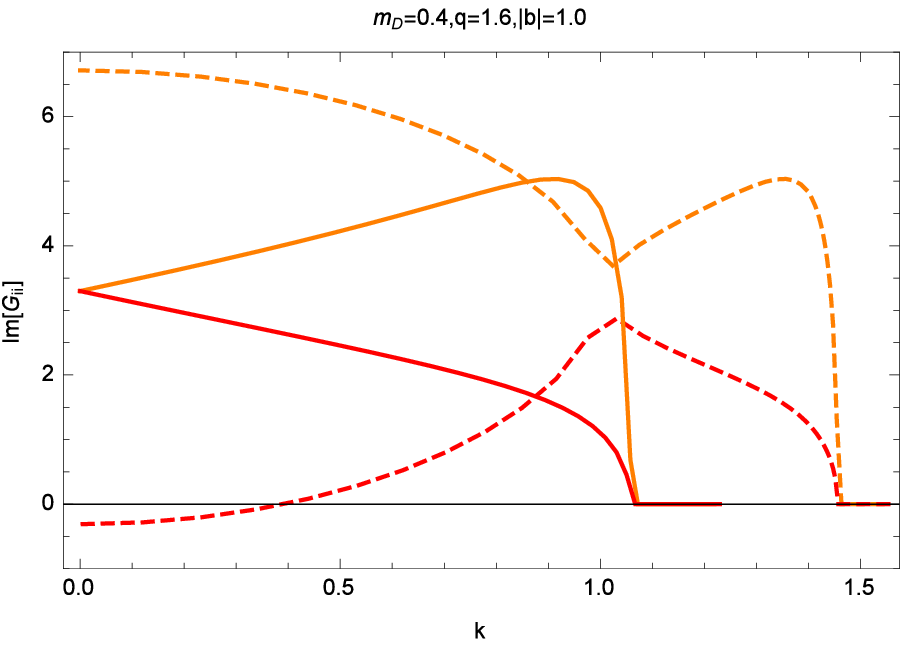} \\
\end{center}
  \caption{AHE of spectral functions with oscillatory phase at zero frequency ($\omega=0$) for Weyl ($m_D=0$) and Dirac ($m_D=0.4$) fermion in AdS$_4$ space-time: Im$G^R_{1}(k)$ (Red solid curve) and Im$G^R_{2}(k)$ (orange solid curve), as functions of momentum $|\vec{k}|$.
  For massless Weyl fermion, there is a bump near $k=qg_F/\sqrt{2}\approx 0.707$ in spectral function Im$G^R_2(k)$.
  For massive Dirac fermion, the Fermi momentum decreases with increasing of mass.
  We have chosen a slightly larger charge, i.e., $q=1.6>1$, and $|\vec{b}|=1$ (dashed curves).
  } \label{fig:massless_massive-charge_AdS}
\end{figure}

By observing Fig.~(\ref{fig:massless_massive-charge_AdS}), the Fermi surface crosses the boundary of the oscillatory region and the bump becomes a peak. With the increase of $|\vec{b}|$, the boundary of the oscillatory is shifted to larger Fermi momentum. Without loss of generality, we have chosen a set of parameters as, $r_\star=\ell=g_F=1$, a slightly larger charge $q=1.6$ and $|\vec{b}|=1$.

For the special case without AHE, i.e., $|\vec{b}|=0$ for fermion in the AdS$_4$ space-time, the spectral function Im$G^R_{1,2}(\omega)$ (solid orange curve) in Fig.~\ref{fig:massless_massive-charge_AdS} just reproduces the result in ref.~\cite{Liu:2009dm}, as shown in Fig.~10.

\subsubsection{Non-Fermi liquid phase at low frequency}

\begin{figure}[H]
\begin{center}
  \includegraphics[scale=0.45]{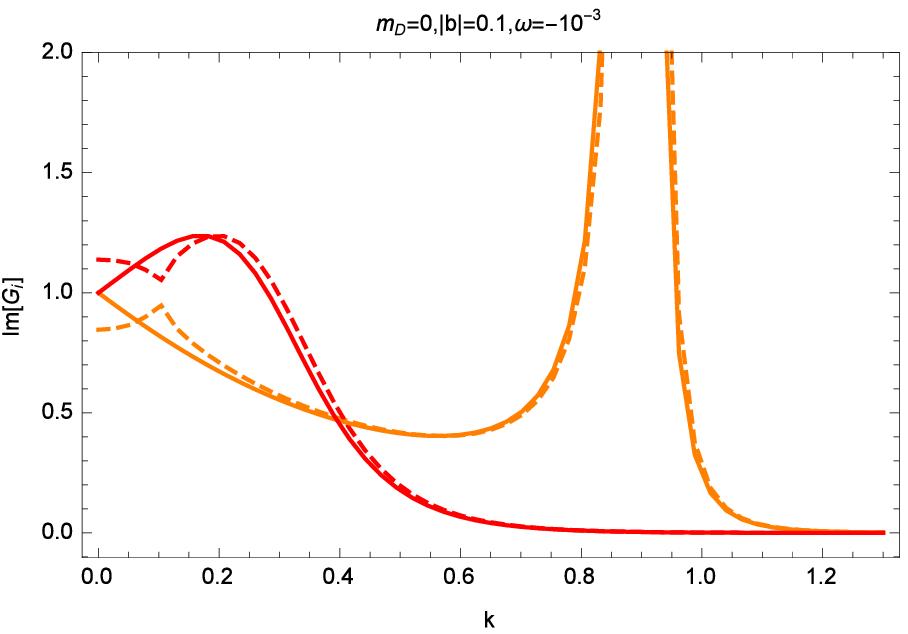} \includegraphics[scale=0.45]{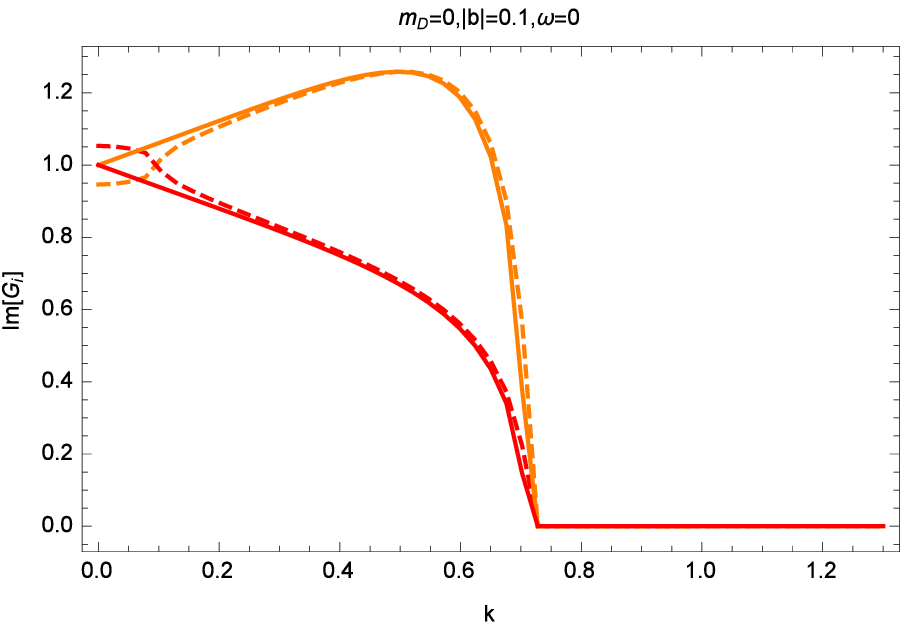}\\
\end{center}
  \caption{AHE of spectral function Im$G^R_{\alpha}(k)$ vs. momentum $k$ for Weyl fermion ($m_D=0$) in AdS$_4$ space-time: Im$G^R_{1}(k)$ (red curve) and Im$G^R_{2}(k)$ (orange curve), as functions of momentum $k$, at low frequency ($\omega=-10^{-3}$) or zero frequency ($\omega=0$). We have chosen the input parameter $r_\star=\ell=g_F=q=1$ and $|\vec{b}| = 0.1$ (dashed curve).
  There is a finite peak $\sim 42$ present in Im$G^R_{2}(k)$ near Fermi momentum $k_F\approx 0.9185$.
  (a)For $\omega=-10^{-3}$ case: By observing the AHE ($|\vec{b}|\ne 0$) of spectral function Im$G^R_{2}(k)$(orange curve), it is obvious that except for there is a relative small but new bump present at $k=|\vec{b}|=0.1$, the location of the original Fermi surface $k_F$ is also shifted to a new location at $\sqrt{k_F^2+|\vec{b}|^2}=\sqrt{0.9185^2+0.1^2} \approx 0.9239$.
  In the limit of zero frequency, i.e., $\omega \to 0^-$, namely from negative axis toward zero point, the height of the peak goes to infinity and the location of the peak approaches the exact value of Fermi momentum $k_F$.
  (b)For $\omega=0$ case: Both spectral function Im$G^R_{\alpha}(\omega=0)$ with $\alpha=1,2$, becomes identically zero in the region $k>q g_F /\sqrt{2}=1/\sqrt{2}\approx 0.7071$. In the presence of AHE, e.g., $|\vec{b}| = 0.1$. The spectral function Im$G^R_{\alpha}(k,\omega=0)$ becomes identically zero in the region $k=\sqrt{q^2g_F^2/2+|\vec{b}|^2}=\sqrt{1/2+0.1^2} \approx 0.7141$.
   } \label{fig:massless_Fermi_momentum_AdS}
\end{figure}

\begin{figure}[H]
\begin{center}
  \includegraphics[scale=0.45]{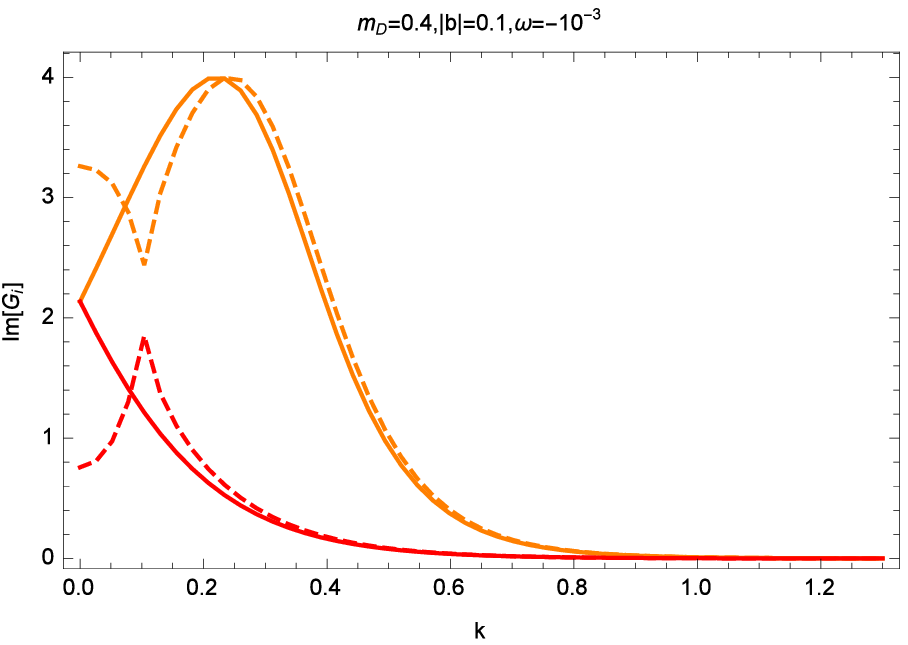} \includegraphics[scale=0.45]{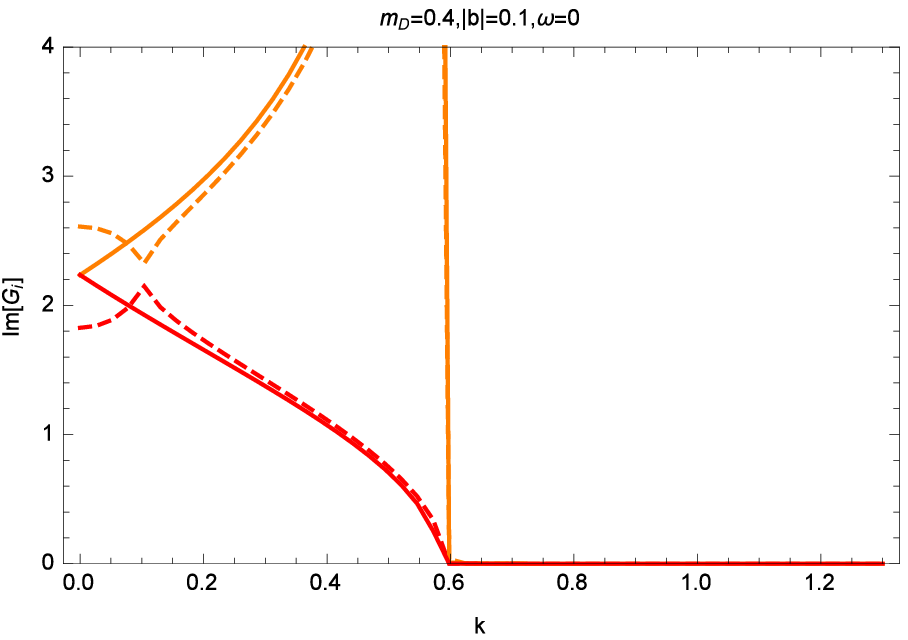}\\
\end{center}
  \caption{AHE of spectral function Im$G^R_{\alpha}(k)$ vs. momentum $k$ for Dirac fermion ($m_D=0.4$) in AdS$_4$ space-time: Im$G^R_{1}(k)$ (red curve) and Im$G^R_{2}(k)$ (orange curve), as functions of momentum $k$, at low frequency ($\omega=-10^{-3}$) or zero frequency ($\omega=0$).
  (a)For $\omega \ne 0$ case: There is no peak for the spectral function.
  (b)For $\omega=0$ case: There is a peak located at $k_F\approx 0.575$ for Im$G^R_{2}$. Meanwhile, both spectral function Im$G^R_{\alpha}(k)$ with $\alpha=1,2$, also becomes identically zero in the region $k=\sqrt{q^2g_F^2/2-m_D^2}=\sqrt{1/2-0.4^2}\approx 0.5831$. In the presence AHE, e.g., $|b|=0.1$(dashed orange curve), there will be a small bump for Im$G^R_{1}$(red curve), meanwhile, a small sinking for Im$G^R_{2}$(orange curve). Both spectral functions become identically zero in the $k=\sqrt{q^2g_F^2/2-m_D^2+|\vec{b}|^2}=\sqrt{1/2-0.4^2+0.1^2}\approx 0.5916$.
   } \label{fig:massive_Fermi_momentum_AdS}
\end{figure}

In Fig.~\ref{fig:massless_Fermi_momentum_AdS} and Fig.~\ref{fig:massive_Fermi_momentum_AdS}, we show the exact spectral function with respect to the momentum near zero frequency. At zero frequency, the Fermi momentum $k_F$ is revealed though the location with a peak for the spectral functions, which can be fitted according to Eq.(\ref{Eq:GR_FL}). Without loss of generality, we have chosen a set of parameters as, $r_\star=\ell=g_F=q=1$, and $|\vec{b}|=0.1$.

Thus, the parameters which character the non-Fermi liquid behavior of the retarded Green's function, namely the residue ${\mathcal Z}$, self-energy $\Sigma(\omega)$ and Fermi velocity $\gamma_\lambda$ can be determined.

For the special case without AHE, i.e., $|\vec{b}|=0$ for fermion in the AdS$_4$ space-time, the spectral function Im$G^R_{\alpha}(\omega)$ with $\alpha=1,2$ (solid red/orange curve) in Fig.~\ref{fig:massless_Fermi_momentum_AdS} just recovers the result in ref.~\cite{Liu:2009dm}, as shown in Fig.~4 there.

\section{Conclusion and Discussion}
\label{sec:concl-dis}

It is well known that the collective excitation of non-Fermi liquids, widely exists in heavy fermion systems with magnetism and exhibits a single-particle spectral function with sharp Fermi surface, whose form is distinct from those of the Landau's Fermi liquid theory~\cite{Landau}. Non-Fermi liquid behavior has widely been observed in many strongly correlated electronic systems such as heavy fermion compounds with large effective massive quasi-particle in the $3$d transition element series and the $4$f rare earth series in the periodic table~\cite{Varma:1989zz,Hewson,Stewart:2001zz}, or the strange metal phase of high critical temperature $T_c$ cuprates superconductors close to a quantum critical point~\cite{Hertz:1976zz,Si:2011hh}. The interactions between the underlying fundamental fermions are strong among many bodies but their collective excitation behavior can not be described by the ordinary quasi-particle picture in the framework of Fermi liquid theory~\cite{Anderson}, i.e, their low energy excitations near the Fermi surface do not behave as weakly interacting fermionic quasi-particles or holes. Phenomenologically, the typical non-Fermi liquid behavior includes but not limit to: with Fermi surface, no quasi-particle but a bump with broader peak, e.g., large decay width and short life time, much softer singularity of APRES spectrum near Fermi surface, logarithmic decayed residue etc.

Recently, a plenty of new experiments~\cite{Xu:2015,Xu:2015cga,Lu:2015zre,Lv:2015pya} in discovering novel topological materials such as topological metals~\cite{Haldane:2004zz,Burkov:2011} and topological insulators~\cite{Hasan:2010xy,Qi:2008ew}, have revealed the fact that chiral anomaly plays a universal and significant role in understanding the IR physics of strongly correlated electron system with topological phases.

In this paper, we aim at exploring the topological non-Fermi liquid with topological phases due to chiral anomaly, such as AHE and CME.
As a minimal model, we consider a Lagrangian with a new independent chiral gauge field $b_M$ induced by chiral anomaly in $(3+1)$-dimensional anti-de Sitter space-time dual to a $(2+1)$-dimensional conformal field theory. To be concrete, the scaling laws of the non-Fermi liquid at or near QCP, are described by a CFT$_3$ dual to a topological charged Reissner-Nordstr\"om  black hole in AdS$_4$ gravity. As a probe, a bulk Dirac fermion couples not only with the background $U(1)$ gauge field $A_M$, but also with the  chiral gauge field $b_M$, which serves as an essential ingredient to induce an intrinsic magnetic field and magnetization field in AHE, or an intrinsic electric polarization in CME. In this case, topological phases such as AHE and CME can be described in a unified framework of field theory. Within the framework, we calculate the boundary UV retarded Green's functions and IR correlation functions at zero and finite temperatures, respectively. The exact UV retarded Green's function can be obtained by solving flow equations, which shows us typical non-Fermi liquid behavior with AHE and CME at zero and low frequencies, respectively. In practice, the physical quantities to describe the non-Fermi liquid, such as Fermi surface, Fermi velocity, residues and self energy or decay width, can be determined by fitting with the spectral functions of the non-Fermi liquid at low frequency at zero or finite temperatures, respectively.

The presence of the chiral gauge field, has significantly changed the shape of the retarded Green's functions in the case without chiral anomaly, i.e., $G^R(\omega,\lambda,b_M=0)$, which we take them as the benchmark ones. For  simplicity, we have assumed that the chiral gauge fields are constants, which will modify the scaling dimension of IR correlation functions. For example, the discrepancy between the spectral functions of AHE with $|\vec{b}|=0.1$ and the benchmark ones at small frequency for Weyl and Dirac fermions cases, have been shown in Fig.~\ref{fig:massless_Fermi_momentum_AdS} and Fig.~\ref{fig:massive_Fermi_momentum_AdS}, respectively. In the non-Fermi liquid with AHE, the presence of the small non-vanishing $|\vec{b}|$ will induce a small bump in the spectral functions, compared to the benchmark ones. We have also validated and drawn figures on the non-Fermi liquid with CME at zero or small frequency for Dirac and Weyl fermion cases, respectively, while we do not intend to show them in this paper. The difference between the retarded Green's function with $b_0\ne 0$ and the benchmark one are that the former will be scaled with respect to the latter, which might be enlarged or shrunk, depending on the sign of the $b_0$. Moreover, the rescaling is larger at smaller momentum, but is smaller at larger momentum.
Here we have considered  a constant chiral gauge field. Certainly, it is interesting to consider an arbitrary chiral gauge field. In addition, we have also obtained the IR correlation functions in the case of Ricci non-flat horizon with $k\ne 0$. We have found that effective mass of bulk Dirac fermion at IR fixed point is topologically quantized, consequently the Fermi momentum presents a discrete sparse distributed pattern, in contrary to the case with the Ricci flat horizon.

In summary, we have investigated the topological non-Fermi liquid in $(2+1)$-dimensional space-time, by introducing not only an external gauge field, but also a new independent intrinsic chiral gauge field induced by chiral anomaly in $(3+1)$-dimensional space-time. As a probe, a bulk Dirac fermion couples to both the external gauge field and the induced chiral gauge field. We have studied the topological non-Fermi liquid with topological phases, such as AHE and CME. At or near the QCP, the behavior of the $(2+1)$-dimensional topological non-Fermi liquid, is described by a strongly coupled CFT$_3$ dual to a topological charged black brane in AdS$_{4}$ bulk gravity. We obtained the conformal scaling dimension of the spinor operators in both UV CFT$_3$ and IR CFT$_1$. We obtained the exact UV retarded Green's functions by solving the flow equations numerically, from which the Fermi momentum, residue and self energy or the decay width of the spectral functions can be determined given input parameters. We also obtained the analytical IR correlation functions at zero temperature and finite temperatures around the QCP, from which the non-analytic low frequency behavior of the topological non-Fermi liquid is exhibited. In the absence of chiral anomaly, our results reproduce and exactly reduce to those obtained in the RN AdS background~\cite{Liu:2009dm} without the chiral gauge field.

Here we would like to make some comments on the theory based upon the gauge/gravity duality. The AdS/CFT correspondence~\cite{Maldacena:1997re,Gubser:1998bc,Witten:1998qj} relates  the infinite boundary of the bulk to high energy at short distance in the dual field theory  and near horizon boundary in the bulk to low energy at large distance. In this case, the asymptotic behavior and corresponding UV physical effect in infinite boundary for $G^R$ from AdS$_4$ are totally reflected in $\Delta_\pm =3/2 \pm \nu_3$. By using AdS/CFT correspondence, one needs $G^R$ at UV fixed point, e.g., it can be obtained directly from asymptotic behavior of bulk gravity in infinite boundary. The reason that one needs correlation function ${\mathcal G}^R$ at IR fixed point is that: the UV $G^R$ obtained does not have good behavior in the small frequency limit at IR fixed point, namely can not make series expansion in the IR limit. Thus from the beginning, one solves ${\mathcal G}^R$ directly in the near horizon region, which has good behavior in the small frequency limit~\cite{Faulkner:2009wj}. From the IR side, we can match ${\mathcal G}^R$ in the IR to that in the UV, and give the so called master equation for $G^R$ in the UV, this is in formalism different from but equivalent to that obtained directly from UV side. The corresponding IR physical effect in the near horizon boundary for ${\mathcal G}^R$ from AdS$_2$ is totally reflected in $\Delta_\pm^{\text{IR}} = 1/2 \pm \nu_\lambda$. In a word, the low frequency behavior of the non-Fermi liquid is sensitive to the near horizon limit of the bulk. It has been found that through the ADM decomposition, the Brown-York tensor of a bulk gravity theory can induce a boundary field theory with local energy and momentum, e.g., when in the near horizon region, the boundary field theory is nothing but hydrodynamics on the cutoff boundary, with horizon radius as a cutoff scale, as an effective classical field theory of a gauge theory of bulk gravity~\cite{Son:2007vk,Iqbal:2008by,Bredberg:2011jq,Cai:2012mg}. While in high curvature and high energy situation, an ultraviolet completed quantum gravity theory is still required, which is hopefully perturbative approachable in the ultra-violet high energy. Therefore, it is expected that the near horizon region behavior of a quantum gravity theory with well understanding behavior at shot distance, would naturally provide a more insightful understanding on the IR CFT close to the QCP for non-Fermi liquid.

In the main context of the paper, we have started from an action of bulk Dirac field as shown in Eq.(\ref{Eq:S_fermion}), the bulk Dirac fermion couples with the $U(1)$ gauge field through a minimal couplings, i.e., $q \bar\psi \Gamma^M A_M \psi $, as well as the chiral gauge field term $q \bar\psi \Gamma^M b_M \Gamma^{5} \psi $. The chiral gauge field $b_M$ is induced by the chiral anomaly in $(3+1)$-dimensional space-time, which origins from triangle loops of chiral fermions in the bulk as discussed in more detail in appendix~\ref{app:ChiralAnomaly_ChiralGaugeField}. Therefore, the physical consequence of $b_M$ accounts for one loop level effect from perturbative aspect of quantum field theory. Meanwhile the other important non-minimal but physical couplings such as anomalous magnetic moment couplings~\cite{Edalati:2010ww,Edalati:2010ge}, will also be present at one loop level. The Pauli dipole interaction term in the bulk, will modify the fermion retarded Green's function on the boundaries, and derives the dynamical formation of a Mott insulator at strong coupling. We have also noticed that a holographic model for undoped Weyl semimetal phase through a relativistic neutral CFT for a four-dimensional chiral fermions in Lifshitz gravity with critical exponent $z$~\cite{Kachru:2008yh,Griffin:2012qx}, especially $z = 2$ case~\cite{Xu:2010eg}, has been studied in Ref.~\cite{Gursoy:2012ie}.

\vspace{0.2in}  \centerline{\bf{Acknowledgements}} \vspace{0.2in}

Y.-H.~Qi thanks valuable discussions with Wen-Hui Duan, Eun-Ah Kim, Wan Li, Jun-Wei Liu, Jun Ni, Xiao-Liang Qi, Yang Qi, Su-Fei Shi, Tong-Hao Tu, S.-S.-H. Henry Tye, Jing Wang, Ya-Yu Wang, Xiao-Gang Wen, Yong Xu, Anthony Zee, Long Zhang, and Yang Zhang on different condensed matter physics on different occasions. He would like to thank Bang-Fen Zhu and Yong-Gang Zhao for teaching him solid state physics, Guang-Ming Zhang for instructing him quantum statistics, and Yong-Shi Wu for instructing him QFT in condensed matter physics. This work is supported in part by the National Natural Science Foundation of China (No.11121064, No.11375247, No.11435006, and No.11475237).



\appendix

\section{Chiral Anomaly and Chiral Gauge Field}
\label{app:ChiralAnomaly_ChiralGaugeField}

\subsection{Chiral anomaly in high dimensional space-time}

In an arbitrary $D=(d+1)$-even dimensions, consider that the massless Dirac field coupled to the gauge field is invariant under the global, e.g., $U(1)$ chiral rotation,

\beqa
\psi^\prime= e^{i\theta\gamma^{D+1}}\psi , \quad \bar\psi^\prime = \psi^\dag e^{-i\theta\gamma^{D+1}} \gamma^0 = \bar\psi e^{i\theta\gamma^{D+1}},  \label{Eq:Chiral_Rotation}
\eeqa
where $e^{i\theta\gamma^{D+1}}=\textbf{1}\cos\theta+i\gamma^{D+1}\sin\theta$ and $\gamma^{D+1}$ is the chiral operator defined as
\beqa
\gamma^{D+1} \equiv \eta \prod_{n=1}^{D}\gamma^n = (\gamma^{D+1})^\dag, \quad (\gamma^{D+1})^2=1,
\eeqa
where the phase $\eta=i^p$ in Euclidean space and $i^{p-1}$ in Minkowski space-time (Alternatively, one can use $\eta=(-i)^{p}$ in Euclidean space and $i^{p+1}$ in Minkowski spacetime).

\begin{widetext}
In $D$-even Euclidean space,
\beqa
\gamma^{D+1} & \equiv & i^p \gamma^1 \gamma^2 \ldots \gamma^{D} = i^{p} (-1)^{(D-1)+(D-2)+ \ldots + 2 + 1} \gamma^{D}\gamma^{D-1}\ldots \gamma^2 \gamma^1, \quad p =[\frac{D}{2}]. \nn\\
&=& i^p (-1)^{p(2p-1)}\gamma^{D}\gamma^{D-1}\ldots \gamma^2 \gamma^1 = (-i)^p \gamma^{D}\gamma^{D-1}\ldots \gamma^2 \gamma^1,
\eeqa
while in Minkowski space-time,
\beqa
\gamma^{D+1} & \equiv & i^{p-1} \gamma^0 \gamma^1 \ldots \gamma^{D-1} = i^{p-1} (-1)^{(D-1)+(D-2)+ \ldots + 2 + 1} \gamma^{D-1}\gamma^{D-2}\ldots \gamma^1 \gamma^0, \quad p =[\frac{D}{2}]. \nn\\
&=& i^{p-1} (-1)^{p(2p-1)}\gamma^{D-1}\gamma^{D-2}\ldots \gamma^1 \gamma^0 = (-i)^{p+1} \gamma^{D-1}\gamma^{D-2}\ldots \gamma^1 \gamma^0.
\eeqa

Then the gauge current interaction term becomes
\beqa
\bar\psi i\cancel{D} \psi \to \bar\psi^\prime i\cancel{D} \psi^\prime = \psi^\dag e^{-i\theta\gamma^{D+1}} \gamma^0 i\cancel{D} e^{i\theta\gamma^{D+1}}\psi= \psi^\dag \gamma^0 i \cancel{D}\psi = \bar\psi i\cancel{D}\psi.
\eeqa
Considering that $\bar\psi i\cancel{D}\psi = \bar\psi_L i\cancel{D}\psi_R  + \bar\psi_R i\cancel{D}\psi_L $, the chiral rotation transformation is equivalent to
\beqa
\bar\psi^\prime_L = \bar \psi e^{-i\theta}, \quad \psi_R^\prime = \psi_R e^{i\theta}, \quad \bar\psi^\prime_R = \bar\psi e^{i\theta}, \quad \psi_L^\prime = \psi_L e^{-i\theta}.
\eeqa

The axial current at tree level (classical level) is conserved,
\beqa
\partial_\mu j^\mu_A = \partial_\mu (\bar\psi \gamma^\mu \gamma^{D+1} \psi) =0.
\eeqa
While the massless Dirac field is coupled to the gauge field under the local, e.g., $U(1)$ chiral rotation $\psi^\prime= e^{i\theta(x)\gamma^{D+1}}\psi$, or gauge invariant Lie algebra generator chiral rotation $\psi^\prime= e^{i\theta^a(x)t^a\gamma^{D+1}}\psi$, then the Lagrangian density of gauge current interaction term is no longer invariant,
\beqa
\bar\psi i\cancel{D} \psi & \to & \psi^\dag e^{-i\theta(x)\gamma^{D+1}} \gamma^0 i\cancel{D} e^{i\theta(x)\gamma^{D+1}}\psi = \psi^\dag \gamma^0 i\cancel{\partial} (i\theta(x)\gamma^{D+1}) \psi \nn\\
&=& - \bar\psi  \gamma^\mu  (\partial_\mu\theta(x)) \gamma^{D+1} \psi = -(\partial_\mu\theta(x)) j_A^\mu.
\eeqa
Under the gauge invariant chiral rotation, the generating function in the path integral, transforms as
\beqa
Z= \int {\mathcal D}\psi {\mathcal D} \bar\psi e^{i\int d^Dx\bar\psi i\cancel{D}\psi } & \overset{}{\to} & \int {\mathcal D}\psi^\prime {\mathcal D} \bar\psi^\prime e^{i\int d^Dx\bar\psi^\prime i\cancel{D}\psi^\prime } = \int {\mathcal D}\psi^\prime {\mathcal D} \bar\psi^\prime e^{i \int d^Dx [\bar\psi i\cancel{D}\psi -(\partial_\mu\theta(x))j_A^\mu ] } \nn\\
&=& \int {\mathcal D}\psi^\prime {\mathcal D} \bar\psi^\prime e^{i \int d^Dx \bar\psi i\cancel{D}\psi   }\bigg(  1 - i \int d^Dx  (\partial_\mu\theta(x))j_A^\mu +  {\mathcal O}(\theta^2) \bigg).  \label{Eq:Local_Chiral_Rotation}
\eeqa
If the fermion measure doesn't change under the chiral rotation, namely, ${\mathcal D}\psi^\prime {\mathcal D} \bar\psi^\prime={\mathcal D}\psi {\mathcal D} \bar\psi$, we have
\beqa
\int {\mathcal D}\psi {\mathcal D} \bar\psi e^{i\int d^Dx\bar\psi i\cancel{D}\psi } \bigg(  1 + i \int d^Dx  \theta(x)  \partial_\mu j_A^\mu +  {\mathcal O}(\theta^2) \bigg), \quad \Rightarrow \quad \partial_\mu j_A^\mu =0.
\eeqa
However, the fermion measure is not invariant under the chiral rotation, due to the Jacobian for the Grassmann integral
\beqa
&& {\mathcal D}\psi^\prime  = {\mathcal D} e^{i\theta(x)\gamma^{D+1}} \psi = {\mathcal D}\psi (\det e^{i\theta(X) \gamma^{D+1}} )^{-1} = {\mathcal D}\psi e^{-i \text{Tr} [ \theta(x)\gamma^{D+1} ] }, \nn\\
&& {\mathcal D}\bar\psi^\prime  = {\mathcal D}\bar\psi  e^{i\theta(x)\gamma^{D+1}} = {\mathcal D}\bar\psi (\det e^{i\theta(X) \gamma^{D+1}} )^{-1} = {\mathcal D}\bar\psi e^{-i \text{Tr} [ \theta(x)\gamma^{D+1} ] },
\eeqa
where we have used that $\text{Tr}A = \ln[ \det \exp{(A)}]$ or $\ln \det M = \text{Tr} \ln M $. Thus, one has
\beqa
{\mathcal D}\psi^\prime {\mathcal D}\bar\psi^\prime = {\mathcal D}\psi {\mathcal D}\bar\psi e^{- 2i \text{Tr} [ \theta(x)\gamma^{D+1} ] },
\eeqa
where the trace $\text{Tr}$ indicates that it is applied in both the spinor flavor space, i.e., $\text{tr}_f$ and the gauge group space, i.e., $\text{tr}_a$. The trace can be calculated and it turns out
\beqa
\text{Tr} [ \theta(x)\gamma^{D+1} ]
&\overset{p=[\frac{D}{2}]}{=}& \frac{1}{p!}\frac{(-1)^p}{(4\pi)^p}\int d^Dx  (2i)^p \text{tr}_a[ i\theta(x) F \wedge F\wedge \ldots \wedge F], \nn\\
&=& \frac{1}{p!}\frac{(-i)^{p}}{(2\pi)^p}\int d^Dx \text{tr}_a [i\theta(x) F^p] \overset{\text{U(1)}}{=} \frac{1}{p!}\frac{(-i)^{p}}{(2\pi)^p}\int d^Dx i\theta(x) \text{tr}_a [ F^p], \label{Eq:Tr-gamma(D+1)-Fp}
\eeqa
where $p \equiv [{D}/{2}]$, and we have used
\beqa
(i\cancel{D})^2 &=& -\cancel{D}\cancel{D} = - D_\mu D_\nu \gamma^\mu \gamma^\nu = - D_\mu D_\nu \frac{1}{2}\bigg( \{ \gamma^\mu , \gamma^\nu  \} + [ \gamma^\mu , \gamma^\nu ] \bigg) = - D_\mu D_\nu \frac{1}{2}\bigg( 2g^{\mu\nu} -2i \sigma^{\mu\nu} \bigg), \nn\\
&=& -D^2 + i \sigma^{\mu\nu}D_\mu D_\nu = -D^2 + i \sigma^{\mu\nu}(-\frac{i}{2}F_{\mu\nu}) =  -D^2 + \frac{1}{2}\sigma^{\mu\nu}F_{\mu\nu},
\eeqa
\end{widetext}
where $\sigma^{\mu\nu}\equiv {i}[ \gamma^\mu , \gamma^\nu ]/{2}$. One can take the differential forms of gauge field and its curvature tensor,
\beqa
\ii && \ii A\equiv -i A_\mu dx^\mu = - i A_\mu^a t^a dx^\mu, \nn\\
\ii && \ii F\equiv - i \frac{1}{2}F_{\mu\nu}dx^\mu \wedge dx^\nu = - i \frac{1}{2}F_{\mu\nu}^a t^a dx^\mu \wedge dx^\nu, \qquad
\eeqa
and $D\equiv d + A$, $F\equiv dA + A^2$, $DF=dF+[A,F]=(dA A - A dA)+(AdA-dA A)=0$. It is worth noticing that
\beqa
\ii\ii \text{tr}_f[\gamma^{D+1}\sigma^{\mu_1\mu_2}\ldots \sigma^{\mu_{D-1}\mu_{D}}] = i^{p} \epsilon^{\mu_1 \mu_2 \ldots \mu_{D-1}\mu_{D}}.
\eeqa
While above calculations are done in the Euclidean space, for the case in the Minkowski space-time, one must note that there are two missing $i^2$ due to that two $\gamma^0$s are present, thus
\beqa
\ii\ii
\text{tr}_f[\gamma^{D+1}\sigma^{\mu_1\mu_2}\ldots \sigma^{\mu_{D-1}\mu_{D}}]
= - i^{p} \epsilon^{\mu_1 \mu_2 \ldots \mu_{D-1}\mu_{D}}.
\eeqa
Consequently, for the case in the Minkowski space-time, there is minus sign in front of those corresponding expressions in the Euclidean space. This can be understood more intuitively as
\beqa
\text{tr}_f (\gamma^5 \gamma^\mu\gamma^\nu \gamma^\rho\gamma^\sigma) = \left\{\begin{aligned}
& 4\epsilon^{\mu\nu\rho\sigma} , \quad \text{Euclidean} \\
& \frac{4}{i}\epsilon^{\mu\nu\rho\sigma} = -4i \epsilon^{\mu\nu\rho\sigma} . \quad \text{Minkowski}
\end{aligned}\right.
\eeqa
Considering Eq.(\ref{Eq:Local_Chiral_Rotation}), one obtains
\begin{widetext}
\beqa
Z & \overset{}{\to} &
\int {\mathcal D}\psi^\prime {\mathcal D} \bar\psi^\prime e^{i \int d^Dx \bar\psi i\cancel{D}\psi   }\bigg(  1 - i \int d^Dx  (\partial_\mu\theta(x))j_A^\mu +  {\mathcal O}(\theta^2) \bigg) \nn\\
&=& \int {\mathcal D}\psi {\mathcal D} \bar\psi e^{- 2i \text{Tr} [ \theta(x)\gamma^{D+1} ]} e^{i \int d^Dx \bar\psi i\cancel{D}\psi   }\bigg(  1 - i \int d^Dx  (\partial_\mu\theta(x))j_A^\mu +  {\mathcal O}(\theta^2) \bigg) \nn\\
&=& \int {\mathcal D}\psi {\mathcal D} \bar\psi \bigg( 1 - 2i \text{Tr} [ \theta(x)\gamma^{D+1} ] + {\mathcal O}(\theta(x)^2) \bigg) e^{i \int d^Dx \bar\psi i\cancel{D}\psi   }\bigg(  1 - i \int d^Dx  (\partial_\mu\theta(x))j_A^\mu +  {\mathcal O}(\theta(x)^2) \bigg).
\eeqa
Thus
\beqa
\int {\mathcal D}\psi {\mathcal D} \bar\psi  e^{i \int d^Dx \bar\psi i\cancel{D}\psi   } \bigg(  - 2i \text{Tr} [ \theta(x)\gamma^{D+1} ] - i \int d^Dx  (\partial_\mu\theta(x))j_A^\mu +  {\mathcal O}(\theta(x)^2) \bigg)=0.
\eeqa
\end{widetext}
The axial current becomes non-conserved in the local chiral rotations through Eq.(\ref{Eq:Chiral_Rotation}) and Eq.(\ref{Eq:Tr-gamma(D+1)-Fp})
\beqa
\int d^D x  i \theta(x) \partial_\mu j_A^\mu &=& 2i \text{Tr}[\theta(x)\gamma^{D+1}] \nn\\
&=& \frac{2}{p!}\frac{(-i)^{p}}{(2\pi)^p}\int d^Dx i \text{tr}_a [\theta(x) F^p]. \nn
\eeqa
Assuming that we are considering a $U(1)$ chiral rotation, i.e., $\theta(x)$, but not a covariant chiral rotation $\theta(x)\equiv \theta^a(x)t^a$, then we obtain the non-conserved $U(1)$ axial current as~\cite{Zumino:1983rz,Zumino:1983ew,Wu:1983kz},
\beqa
\partial_\mu j_A^\mu = \frac{2}{p!}\frac{(-i)^{p}}{(2\pi)^p} \text{tr}_a F^p,
\eeqa
and $\text{tr}_a F^p$ is just the $p$-th Chern character $\Omega_{2p}$,
\beqa
\Omega_{2p} \equiv \text{tr}_a F^p = d {\mathcal L}_{2p-1}^{\text{CS}}(A).  \label{Eq:L-Chern-Simons}
\eeqa
where ${\mathcal L}_{2p-1}^{\text{CS}}(A)$ are the corresponding Chern-Simons terms in $(2p-1)=(D-1)$-odd-dimensions~\cite{Zumino:1983ew},
\beqa
{\mathcal L}_{2p-1}^{\text{CS}}(A) = p \int_0^1 dt  \,\text{tr}[A(tdA+t^2 A^2)^{p-1} ].
\eeqa

\subsubsection{Chiral anomaly in Euclidean space}

According to the general result above, the axial current violation in the Euclidean space is given by
\beqa
\ii\ii\ii \partial_\mu j^{\mu}_A  = \frac{2}{p!}\frac{(-1)^p}{(4\pi)^p}q^p\epsilon^{\mu_1\ldots \mu_{D}} \text{tr}_a[F_{\mu_1\mu_2}\ldots F_{\mu_{D-1}\mu_{D}}], \label{Eq:pjV(d+1)_Euclidean}
\eeqa
where $p\equiv [{D}/{2}]$ and $q$ is the charge coupling of the gauge field.

For example, in $D=(1+1)$-dimensions, $p=1$, Eq.(\ref{Eq:pjV(d+1)_Euclidean}) gives the $1$-st Chern character
\beqa
\ii\ii \partial_\mu j^{\mu}_A  = - 2\frac{1}{2\pi}q \epsilon^{\mu_1\mu_2 } \text{tr}_a[F_{\mu_1\mu_2}] = - \frac{q}{2\pi} \epsilon^{\mu \nu } F_{\mu \nu}.
\eeqa
According to Eq.(\ref{Eq:L-Chern-Simons}), the CS$_3$ term is
\beqa
\text{tr}_a F = d {\mathcal L}_{1}^{\text{CS}}(A) = d\text{tr} A .
\eeqa
In $D=(3+1)$-dimensions, $p=2$, Eq.(\ref{Eq:pjV(d+1)_Euclidean}) gives the $2$-nd Chern character
\beqa
\partial_\mu j^{\mu}_A  
=\frac{ q^2}{16\pi^2}\epsilon^{\mu\nu \rho \sigma} F_{\mu\nu} F_{\rho\sigma}, \nn
\eeqa
and the CS$_3$ term is
\beqa
\text{tr}_a F^2 = d {\mathcal L}_{3}^{\text{CS}}(A) = d\text{tr}\bigg(AdA + \frac{2}{3}A^3 \bigg). \nn
\eeqa
In $D=(5+1)$-dimensions, $p=3$, the $3$-rd Chern character is
\beqa
\partial_\mu j^{\mu}_A  
&=& - \frac{ q^3}{ 192 \pi^3}\epsilon^{\mu \nu \rho \sigma \beta \gamma } F_{\mu \nu} F_{ \rho \sigma}F_{ \beta \gamma},
\eeqa
while the CS$_5$ term is
\beqa
\text{tr}_a F^3 = d {\mathcal L}_{5}^{\text{CS}}(A) = d\text{tr}\bigg( A(dA)^2 + \frac{3}{2}A^3 dA + \frac{3}{5}A^5 \bigg). \nn
\eeqa

\subsubsection{Chiral anomaly in Minkowski space-time}

As discussed above, there is a relative minus sign for the chiral anomaly in the Minkowski space-time, compared to that in the Euclidean space.
The axial current violation in Minkowski space-time is
\beqa
\ii && \ii \partial_\mu j^{\mu}_A  = \frac{2}{p!}\frac{(-1)^{p+1}}{(4\pi)^p}q^p\epsilon^{\mu_1\mu_2\ldots \mu_{D-1}\mu_{D}} \text{tr}_a[F_{\mu_1\mu_2}\ldots F_{\mu_{D-1}\mu_{D}}], \nn\\
&&\label{Eq:pjV(d+1)_Minkowski}
\eeqa
where $p\equiv [{D}/{2}]$ and $q$ is the charge coupling of the gauge field. For example, in $D=(1+1)$-dimensions, $p=1$, Eq.(\ref{Eq:pjV(d+1)_Minkowski}) gives
\beqa
\partial_\mu j^{\mu}_A  = 2\frac{1}{2\pi}q \epsilon^{\mu_1\mu_2 } \text{tr}_a[F_{\mu_1\mu_2}] = \frac{q}{2\pi} \epsilon^{\mu \nu } F_{\mu \nu}.
\eeqa
In $D=(3+1)$-dimensions, $p=2$, Eq.(\ref{Eq:pjV(d+1)_Minkowski}) gives
\beqa
\partial_\mu j^{\mu}_A  &=& - 2\frac{1}{2!}\frac{ q^2}{(4\pi)^2}\epsilon^{\mu_1 \mu_2 \mu_3 \mu_4} F_{\mu_1\mu_2} F_{\mu_3\mu_4} \nn\\
&=& -\frac{ q^2}{16\pi^2}\epsilon^{\mu\nu \rho \sigma} F_{\mu\nu} F_{\rho\sigma}.
\eeqa
which agrees with the Adler-Bell-Jackiw anomaly~\cite{Adler:1969gk,Bell:1969ts}.
In $D=(5+1)$-dimensions, $p=3$, it becomes
\beqa
\partial_\mu j^{\mu}_A  
&=& \frac{ q^3}{ 192 \pi^3}\epsilon^{\mu \nu \rho \sigma \beta \gamma } F_{\mu \nu} F_{ \rho \sigma}F_{ \beta \gamma}.
\eeqa

\subsection{Chiral anomaly in $(3+1)$-dimensional space-time}

Chiral anomaly means that the original chiral symmetry is no longer respected at quantum loop level due to triangle diagrams of chiral fermions. To impose the requirement that the current desnity
\beqa
j^\mu \equiv \bar\psi  \gamma^\mu  \psi = j^{\mu R} + j^{\mu L}, \label{Eq:jV}
\eeqa
is still conserved at quantum loop level, i.e., $\partial_\mu j^\mu =0$, the chiral current density
\beqa
j^{\mu 5} \equiv \bar\psi  \gamma^\mu \gamma^5 \psi = j^{\mu R} - j^{\mu L}, \label{Eq:jV5}
\eeqa
is not conserved, where $\gamma^5\equiv i \gamma^0\gamma^1\gamma^2\gamma^3$ is the Dirac chiral matrix that anti-commutes with all other ones, i.e., $\{\gamma^5, \gamma^\mu \}=0$. Namely,
\beqa
\partial_\mu j^{\mu 5}  = - \frac{q^2}{8\pi^2}F_{\mu\nu}\tilde{F}^{\mu\nu}. \label{Eq:pjV5}
\eeqa
where $\tilde{F}^{\mu\nu}$ is a dual tensor of the electromagnetic field strength tensor,
\beqa
\tilde{F}^{\mu\mu} \equiv \frac{1}{2} \epsilon^{\mu\nu\rho\sigma} F_{\rho\sigma}.
\eeqa
The physical meaning of the chiral anomaly term $F\tilde{F}$ implies that if electric fields are applied in paralleled with magnetic field or vice verse, the chiral field is not conserved, since $\tilde{F}F \sim \vec{E}\cdot\vec{B}$.

\subsection{Chiral gauge field in $(3+1)$-dimensional space-time}

Consider the quantum field theory for a Dirac fermion, e.g., quantum electrodynamics (QED) in $(3+1)$-dimensional space-time with topological term $F\tilde{F}\sim\vec{E}\cdot\vec{B}$.
The generating function is
\beqa
Z = \int {\mathcal D}\psi {\mathcal D}\bar\psi \exp{ \bigg( i\int dt \int dx^3 {\mathcal L} \bigg) },
\eeqa
with the Lagrangian
\beqa
\ii
&& {\mathcal L} = \bar\psi ( i\gamma^\mu D_\mu - m_D ) \psi  - \frac{1}{4}F_{\mu\nu}F^{\mu\nu} + \frac{q^2}{16\pi^2}\theta(x) F_{\mu\nu} \tilde{F}^{\mu\nu}. \nn\\
&& \label{Eq:L-FFt}
\eeqa
where $\psi$ is 4-component Dirac spinor, $F_{\mu\nu}=\partial_\mu A_\nu - \partial_\nu A_\mu$ is the curvature tensor of $U(1)$ gauge field $A_\mu$, and the covariant derivative is
\beqa
D_\mu \equiv  \partial_\mu + i q A_\mu  .
\eeqa

By using the chiral current relation in Eq.(\ref{Eq:jV5}) and Eq.(\ref{Eq:pjV5}), the Lagrangian in Eq.(\ref{Eq:L-FFt}) can be described as an effective Lagrangian\footnote{Note the conversion of the sign: if Eq.(\ref{Eq:pjV5}) becomes $\partial_\mu j^{\mu 5} = q^2/8\pi^2 F\tilde{F}$, then Eq.(\ref{Eq:L-b-theta}) should be ${\mathcal L} \sim \bar\psi[ i\gamma^\mu( \partial_\mu  + i q A_\mu + i b_\mu \gamma_5) ] \psi$. $\mu =1,2,3,4$, $b_4=-i b_0$,$\gamma^4=-i \gamma^0$. All $\gamma^\mu$ are antihermitian while $\gamma^5=-i \gamma^0 \gamma^1 \gamma^2 \gamma^3$ is Hermitian, and $\{\gamma^5, \gamma^\mu\}=0$},
\beqa
\ii\ii\ii
{\mathcal L} = \bar\psi [ i\gamma^\mu( \partial_\mu  + i q A_\mu - i b_\mu \gamma_5) - m_D ] \psi \!-\! \frac{1}{4}F_{\mu\nu}F^{\mu\nu} . \label{Eq:L-b-theta}
\eeqa
where $b_\mu$ is a new dynamical constant vector field defined through
\beqa
b_\mu \equiv \frac{1}{2} \partial_\mu \theta(x), \, \mu=0,1,2,3,  \label{Eq:bu-theta}
\eeqa
which can be viewed as a chiral gauge field. Assuming that $b_\mu\ne 0$, it means that $\theta(x)$ is not a constant term, thus can not be dropped as boundary term in $(4+1)$-dimension. While since $F\tilde{F}$ is a CP violating term, thus so does $\theta$. Consequently, $b_\mu(x)$ can be viewed as a source term breaking time inversion symmetry of the bulk.

In QFT, $\theta$ vacuum term is a boundary term present at $(4+1)$-dimensions, which can not be renormalized by interactions. While it can be promoted to be a dynamic field in $(3+1)$-dimensions, i.e., $\theta(x^\mu)$ depends on the space-time, which can be realized in condensed matter systems with topological properties, the dynamics of which can be described by topological quantum field theory (TQFT). For example, the $\theta$ term can be realized as the source of the longitudinal magnetoelectric effect or equivalently integer quantum Hall effect (IQHE) conductivity on its surface in topological insulators (TIs)~\cite{Hasan:2010xy,Qi:2008ew}, by imposing inhomogeneous magnetic fields to the bulk of TIs state, so that $\theta(x^\mu)$ depends on the position.

It is worthy to notice that the Lagrangian in Eq.(\ref{Eq:L-b-theta}) is invariant under a chiral gauge transformation,
\beqa
\psi \to e^{i \theta(x)\gamma^5/2}\psi, \quad \bar\psi \to \bar\psi e^{i\theta(x)\gamma^5/2},
\eeqa
which implies that both the vector current and the chiral current are conserved under the chiral gauge transformation,
\beqa
j^\mu &=&  \bar\psi \gamma^\mu  \psi \to \bar\psi e^{i\theta(x)\gamma^5/2} \gamma^\mu  e^{i\theta(x)\gamma^5/2} \psi = \bar\psi \gamma^\mu \psi, \nn\\
j^{\mu 5} &=& \bar\psi \gamma^\mu \gamma^5 \psi  \to \bar\psi e^{i\theta(x)\gamma^5/2}   \gamma^\mu \gamma^5  e^{i\theta(x)\gamma^5/2} \psi =  \bar\psi \gamma^\mu \gamma^5 \psi, \nn
\eeqa
while the Lagrangian in Eq.(\ref{Eq:L-b-theta}) transforms into
\beqa
{\mathcal L} &=& \bar\psi [ i\gamma^\mu( \partial_\mu  + i q A_\mu - i b_\mu \gamma_5) - m_D ] \psi \nn\\
& \to & \bar\psi e^{i\theta(x)\gamma^5/2}   [ i\gamma^\mu( \partial_\mu  + i q A_\mu - i b_\mu \gamma_5) - m_D ] e^{i\theta(x)\gamma^5/2}   \psi \nn \\
&=& \bar\psi   [ i\gamma^\mu( \partial_\mu  + i q A_\mu ) - m_D e^{i\theta(x)\gamma^5} ]   \psi. \label{Eq:L-chiraltrans}
\eeqa
where $b_\mu$ is eliminated from the action by using Eq.(\ref{Eq:bu-theta}), and it is obvious that the Dirac mass term violates the chiral gauge symmetry at tree level. In addition, the measure of the Fermion's path integral gives a non-trivial term due to Jacobian of the chiral transformation, which can be interpreted as an additional term in the Dirac action,
\beqa
{\mathcal D}\psi {\mathcal D}\bar\psi \to {\mathcal D}\psi {\mathcal D}\bar\psi \det{[e^{i\theta(x)\gamma^5}]} \equiv  {\mathcal D}\psi {\mathcal D}\bar\psi e^{-S_\theta/\hbar}, \nn
\eeqa
where the equivalent action term is
\beqa
S_\theta = -i \hbar \text{Tr}{[\theta(x) \gamma^5]}, \label{Eq:S-theta-Tr-gamma5}
\eeqa
which has been calculated in Eq.(\ref{Eq:Tr-gamma(D+1)-Fp}).
Combing with the transformed Lagrangian in Eq.(\ref{Eq:L-chiraltrans}), as well as the pure gauge field term, one obtains
\beqa
\ii\ii\ii S = \int d^4x  \bar\psi   [ i\gamma^\mu( \partial_\mu  + i q A_\mu ) - m_D e^{i\theta(x)\gamma^5} ]   \psi + S_\theta,  \label{Eq:S-theta}
\eeqa
where $e^{i\theta(x)\gamma^5}=\textbf{1}\cos\theta(x)+i\gamma^5\sin\theta(x)$.
In conclusion, the dynamics of action with $\theta$ term as shown in Eq.(\ref{Eq:S-theta}), can be described by an effective Lagrangian in Eq.(\ref{Eq:L-b-theta}) with $b_\mu$ introduced, and a phase associated with the Dirac mass $m_D \to m_D e^{-i\theta(x)\gamma^5}$.

The trace term in the action in Eq.(\ref{Eq:S-theta-Tr-gamma5}) can be computed by summing over a complete basis of fermion state, e.g., eigen states of the Dirac operator $\cancel{D} = \cancel{p} + i q \gamma^\mu A_\mu $ with appropriate regularization. As a special case shown in Eq.(\ref{Eq:Tr-gamma(D+1)-Fp}), the final result in $(3+1)$-dimensions turns out to be one~\cite{Fujikawa:1979ay,Fujikawa:1980eg} in terms of the electromagnetic fields, an action in the Euclidean space with imaginary time, i.e.,
\beqa
S_\theta = \frac{i q^2}{32\pi^2} \int d^4x \theta(x)\epsilon^{\mu\nu\rho\lambda}F_{\mu\nu}F_{\rho\lambda} = \frac{i q^2}{8\pi^2} \int d^4x \theta(x) \vec{E}\cdot \vec{B}. \nn
\eeqa
The action in real time/Minkowski space-time after Wick rotation $\tau \to i t$ becomes\footnote{Where we have taken the natural units $\hbar=c=1$ and henceforth. e.g., Eq.(\ref{Eq:S-CS}), }
\beqa
S_\theta &=& \frac{q^2}{32\pi^2}\int dt d\vec{r} \theta(\vec{r},t) \epsilon^{\mu\nu\rho\sigma} F_{\mu\nu}F_{\rho\sigma} \nn\\
&=& \frac{q^2}{8\pi^2}\int dt d\vec{r} \theta(\vec{r},t) \vec{E}\cdot\vec{B}, \label{Eq:S-theta}
\eeqa
where $\theta(x)$ can be viewed as an ``axion'' field, which has the form
\beqa
\ii
\theta(\vec{r},t) = 2 (\vec{b}\cdot \vec{r}  - b_0 t), \quad b_\mu = (-b_0, \vec{b}),   \label{Eq:theta-brt}
\eeqa
where $\vec{b} = (b_x, b_y, b_z)$ and $2\vec{b}$ is the separation between the Weyl nodes in momentum space, while $2b_0$ is the separation between the nodes in energy. In momentum space, $b_0$ and $\vec{b}$ denote the shift in energy and momentum, respectively.
The broken time reversal (TR) symmetry (${\mathcal T}$)  allows for a non-trivial dependence of $\theta$ on spatial coordinates, thus without loss of generality, one can set $\vec{b}=(0,0,b)$; while broken inversion symmetry (${\mathcal P}$) allows for a non-trivial time dependence coordinates, i.e., $b_0=b$.

\subsection{Topological effects due to chiral anomaly}

The $\theta$ term in the action in Eq.(\ref{Eq:S-theta}) can be re-expressed in the Chern-Simons form,
\beqa
S_\theta &=& - \frac{q^2}{8\pi^2}\int dt d\vec{r} \partial_\mu \theta \epsilon^{\mu\nu\rho\lambda}A_\nu \partial_\rho A_\lambda.  \label{Eq:S-CS}
\eeqa
The currents are given through variation of the action with respect to the vector potential,
\beqa
j_\nu = \frac{q^2}{4\pi^2}\partial_\mu \theta \epsilon^{\mu\nu\rho\lambda} \partial_\rho A_\lambda = \frac{q^2}{2\pi^2} b_\mu \epsilon^{\mu\nu\rho\lambda} \partial_\rho A_\lambda.
\eeqa

In terms of components of $b_\mu$, one has
\beqa
j_\nu = \frac{q^2}{2\pi^2} b_k \epsilon^{k\nu\rho\lambda} \partial_\rho A_\lambda; \quad j_\nu = -\frac{q^2}{2\pi^2} b_0 \epsilon^{0\nu\rho\lambda} \partial_\rho A_\lambda, \nn
\eeqa
where $k =1,2,3$, the first term describes the anomalous Hall effect (AHE)~\cite{Haldane:1988,Nagaosa:2003,Nagaosa:2010ahe,Chang:2013}, while the second term describes the chiral magnetic effect (CME)~\cite{Fukushima:2008xe}. Equivalently, above current can be separated into time and spatial currents,
\beqa
j_0 &=& \frac{q^2}{2\pi^2} b_k \epsilon^{k0ij} \partial_i A_j, \nn\\
j_i &=& \frac{q^2}{2\pi^2} b_k \epsilon^{kij0} \partial_j A_0 - \frac{q^2}{2\pi^2} b_0 \epsilon^{0ijk} \partial_j A_k.
\eeqa
In terms of electromagnetic field, $(E^k,h^k) \equiv ( \partial^k A_0, \epsilon^{kij}\partial_i A_j)$ and $(\rho,\vec{j})\equiv(j_0,j_i)$, one has
\beqa
\rho &=& \frac{q^2}{2\pi^2} \vec{b} \cdot \vec{h}, \label{Eq:rho}\\
\vec{j} &=& \frac{q^2}{2\pi^2}\vec{b} \times \vec{E} + \frac{q^2}{2\pi^2} b_0 \vec{h}.\label{Eq:j}
\eeqa
where Eq.(\ref{Eq:rho}) and the first term in Eq.(\ref{Eq:j}) encode the AHE in topological materials with broken time reversal symmetry ${\mathcal T}$, while the second term in Eq.(\ref{Eq:j}) describes the CME in topological materials with broken parity/inversion symmetry ${\mathcal P}$.

In $(3+1)$-dimensions, by imposing the electric field $A_M\equiv(A_r,A_\mu)=(0,A_t(r),-hy,0)$ or $A_M=(0,A_t(r),0,hx)$, thus the external electric field is $E_M \equiv (E_r,E_\mu) = (\partial_r A_t(r),0,0,0)$ and external magnetic field is $h_M=(h_r,h_\mu)=(h,0,0,0)$, one has
\beqa
&&\rho = \frac{q^2}{2\pi^2} b_r h , \quad j^r = \frac{q^2}{2\pi^2} b_0 h, \nn\\
&& j^x = \frac{q^2}{2\pi^2} b_y E_r , \quad j^y =-\frac{q^2}{2\pi^2} b_x E_r ,
\eeqa
where nonvanishing $j^{r}$ leads to chiral magnetic effect (CME) with $b_0\ne 0$, and nonvanishing $j^{x,y}$ leads to anomalous Hall effect (AHE) with $b_y\ne 0$ or $b_x\ne 0$, respectively, as will be clear in the following.

\subsubsection{Anomalous Hall effect}
\label{sec:AHE}

For anomalous Hall effect, one has
\beqa
&& \rho = \frac{q^2}{2\pi^2} \vec{b} \cdot \vec{h} = \frac{q^2}{4\pi^2} \vec{h} \cdot \nabla \theta, \nn\\
&& \vec{j} = \frac{q^2}{2\pi^2}\vec{b} \times \vec{E} = \frac{q^2}{4\pi^2}  \nabla \theta \times \vec{E},
\eeqa
by identifying the charge density $\rho$ with the divergence of a magnetic field $\vec{H}$ and the current $\vec{j}$ with the curl of a magnetization field $\vec{M}$ in the bulk of an insulator,
\beqa
\rho = \nabla \cdot \vec{H}, \quad \vec{j} = \nabla \times \vec{M},
\eeqa
one obtains
\beqa
\nabla \cdot \vec{H} = \frac{q^2}{4\pi^2} \vec{h} \cdot \nabla \theta, \quad \nabla \times \vec{M} = \frac{q^2}{4\pi^2}  \nabla \theta \times \vec{E}.
\eeqa
For time reversal invariant case, e.g., the quantized topological magnetoelectric effect in TIs with $\theta=\pi$, one obtains the polarization and magnetization vector in the bulk,
\beqa
\vec{H} = \frac{q^2}{4\pi} \vec{h}, \quad \vec{M} = \frac{q^2}{4\pi} \vec{E}, \quad \theta =\pi,
\eeqa
which means that the magnetic field $\vec{H}$ is induced as a response to an applied external magnetic field $\vec{h}$, and the magnetization field $\vec{M}$ is induced as a response to an applied external electric field $\vec{E}$, respectively.

\subsubsection{Chiral magnetic effect}
\label{sec:CME}

For chiral magnetic effect, one has
\beqa
\rho = 0, \quad \vec{j} = \frac{q^2}{2\pi^2} b_0 \vec{h} = - \frac{q^2}{4\pi^2} \partial_t \theta \vec{h},
\eeqa
where $\vec{j}$ is the electromagnetic current, and $b_0$ denotes the change of the chirality, i.e., the difference between chiral fermion zero mode ($E=0$) with right and left handness. It can be parameterized as a chemical potential for an asymmetry between the number of right and left chiral fermions, e.g. $b_0 \equiv \mu_5 N_c$, with $N_c$ the color degree of freedom for the same $f$-th generation flavor fermion. To be concrete, in the chiral limit $m_D\to 0$, $\mu_5 = \int d^4x \partial_\mu j^{\mu}_5 = \int d^3x j^0_5 = 2N_f (n_R - n_L)$, where $n_R$ includes particle and antiparticles with right handed helicity while $n_L$ includes those with left-handed helicity.
The observation of the non-vanishing of $b_0$, implies both ${\mathcal P}$ and ${\mathcal CP}$ violations from chiral anomalies.

By identifying the current $\vec{j}$ with the time gradient of a polarization field in the bulk of an insulator,
\beqa
\vec{j} = \partial_t \vec{P},
\eeqa
one obtains
\beqa
\partial_t \vec{P} = - \frac{q^2}{4\pi^2} \partial_t \theta \vec{h}.
\eeqa
For the time reversal invariant case, e.g, the quantized topological magnetoelectric effect in TI with $\theta =\pi$, one obtains the polarization vector in the bulk,
\beqa
\vec{P} = - \frac{q^2}{4\pi} \vec{h}, \quad \theta =\pi,
\eeqa
which means that the electric polarization field $\vec{P}$ is induced as a response to an applied external magnetic field.

\subsubsection{Topological magnetoelectric effect}
\label{sec:TMEE}

In the absence of external magnetic field $\vec{B}$, AHE is observable, but CME will be absent,
\beqa
\text{AHE}:&& ~\rho = 0, ~ \vec{j} = \frac{q^2}{2\pi^2}\vec{b} \times \vec{E}, \Rightarrow \, \vec{M} = \frac{q^2}{4\pi} \vec{E}, \nn\\
\text{CME}:&& ~\rho = 0, ~ \vec{j} = 0.
\eeqa
While in the absence of external electric field $\vec{E}$, both AHE and the CME are observable, since one has
\beqa
\text{AHE}:&& ~\rho = \frac{q^2}{2\pi^2} \vec{b} \cdot \vec{h}, ~ \vec{j} = 0, \Rightarrow \, \vec{H} = \frac{q^2}{4\pi} \vec{h}, \nn\\
\text{CME}:&& ~\rho = 0, ~ \vec{j} =  \frac{q^2}{2\pi^2} b_0 \vec{h}, \Rightarrow \, \vec{P} = - \frac{q^2}{4\pi} \vec{h},
\eeqa
where we have chosen a realization of topological magnetoelectric effect with $\theta=\pi$.

\section{Dirac representation in $4$-dimensional space-time}
\label{app:DiracRep_RotationalInvariance}

\subsection{Dirac representation in $(3+1)$-dimensional space-time with spatial rotational invariance}

In $(3+1)$-dimensions, by assuming the rotational invariance, it is convenient to choose the basis of bulk Gamma matrices in $(3+1)$-dimensions~\cite{Wilczek:1981iz,Polchinski:1998rr}, so that the Dirac equation is real for real $\omega$ and $k$:\footnote{The basis is chosen so that the Dirac equation is real for real $\omega$ and $k$, thus one may call the real basis for Dirac equation in $D=(3+1)$-dimensions.}

In this case\footnote{In the following, we have used the direct multiplication rule from left to right, which is a little different from conversion of the direct product defined in Matrix theory in some literature.}, $\mu= (t,1,2)$, $\Gamma^{3} = \Gamma^r$,
\beqa
&& \Gamma^{t} =  i\sigma^1 \otimes \textbf{1}_{2}
=  \left(\begin{array}{cc}
  i\sigma^1   & 0\\
0 &  i\sigma^1
\end{array}
\right),  \nn\\
&& \Gamma^r = (-\sigma^3) \otimes  \textbf{1}_{2}
= \left(\begin{array}{cc}
-\sigma^3  & 0\\
0 & -\sigma^3
\end{array}
\right) , \nn\\
&& \Gamma^1 = \sigma^2 \otimes (-\sigma^3)
= \left(\begin{array}{cc}
-\sigma^2  & 0\\
0 & \sigma^2
\end{array}
\right) ,  \nn\\
&& \Gamma^{2} = \sigma^2 \otimes (-i\gamma^t) = \sigma^2 \otimes \sigma^1
= \left(\begin{array}{cc}
0 & \sigma^2 \\
\sigma^2  & 0
\end{array}
\right) ,
\label{Eq:Gamma-rep-1}
\eeqa
where $\Gamma^{m} =  \sigma^2 \otimes \gamma^{m-1}$, $\gamma^t\equiv i\sigma^1$, $\gamma^1 \equiv  -\sigma^3$, and
$\gamma^{t,1}$ are $2\times 2$ Pauli matrices in $2$ dimensions, and $\textbf{1}_{2}$ is a $2 \times 2$ identity matrix, with $p=[D/2]=2$.

\begin{itemize}
  \item[a.]
In $D=(3+1)$-even dimensions, the fermions carry chirality, and the chirality operator is defined as
\beqa
\ii
\Gamma^{2p+1} \2i &=& \2i  \Gamma^{D+1} = \Gamma^{5} \equiv i^{p-1} \Gamma^{\underline{t}} \Gamma^{\underline{1}} \Gamma^{\underline{2}} \Gamma^{\underline{r}}  \qquad \nn\\
\2i &=& \2i  - \sigma^2 \otimes \sigma^2 = \left(\begin{array}{cc}
0 & i\sigma^2   \\
- i\sigma^2& 0
\end{array}  \right) .   \label{Eq:Gamma-rep-1_Chirality}
\eeqa

Consequently, the chirality projection operator is defined through
\beqa
\ii
\Gamma^\pm \equiv \frac{1}{2}(\textbf{1}_{2^p} \pm \Gamma^{2p+1}) \!=\!  \left(\begin{array}{cc}
\textbf{1}_{2}  &  \pm i\sigma^2  \\
\mp i\sigma^2   &  \textbf{1}_{2}
\end{array} \right) .  \label{Eq:chirality-projection}
\eeqa
One has,
\beqa
&& \Gamma^\pm \Gamma^{\underline{r}} = \Gamma^{\underline{r}} \Gamma^\mp, \quad \Gamma^\pm \Gamma^{\underline{\mu}} =\Gamma^{\underline{\mu}}\Gamma^\mp, \quad \Gamma^\pm \Gamma^{5} = \pm \Gamma^\pm, \nn\\ &&  \quad \Gamma^{\underline{\mu}}\Gamma^{\underline{r}}= - \Gamma^{\underline{r}}\Gamma^{\underline{\mu}}, \quad \Gamma^{\underline{\mu}} \Gamma^{5}= - \Gamma^{5} \Gamma^{\underline{\mu}}.
\eeqa
In the representation, the spin projection operator along the spatial direction $k^i$ becomes
\beqa
P_\pm^{\hat{k}^i} \equiv \frac{1}{2}(\textbf{1}_{4} \pm \Sigma^i \hat{k}^i), ~ \Sigma^{i} \equiv \Gamma^{\underline{5}} \Gamma^{\underline{t}} \Gamma^{\underline{i}}, ~ \hat{k}\equiv \frac{\vec{k}}{|\vec{k}|}, \label{Eq:spin-projection_ki}
\eeqa
where $\hat{k}^i$ is the i-th component of the unit vector $\hat{k}$.


\item[b.] In $D=(2+1)$-odd dimensions, the chirality operator is defined as projector $\tilde\Gamma^d=\tilde\Gamma^r$,
\beqa
\Gamma^t=i\sigma^1, \quad \Gamma^r= -\sigma^3, \quad \Gamma^1 = \sigma^2.
\eeqa
In this case, the projection operator is defined through
\beqa
\Gamma_\pm \equiv \frac{1}{2}(\textbf{1}_{2} \pm \Gamma^{\underline{r}}) \!=\!  \textbf{1}_{2} \mp \sigma^3 .  \label{Eq:helicity-projection}
\eeqa
In the homogeneous spatial space, one has $k=(k^1,0)$, the Gamma matrices in $D=(3+1)$-dimensions ($p=[D/2]=2$), reduce to those in $D=(2+1)$-dimensions ($p=[3/2]=1$). In this case, the representations of bulk Dirac fermion in an arbitrary $D$-dimensional space-time, can always be simplified by choosing the representation in $D$-odd dimensional space-time, under the assumption of homogeneous spatial invariance.

\end{itemize}

Note that the assumption of rotational invariance is essential for the choice of Dirac fermion representation, since it guarantees that the spacial dimension is homogeneous, so that in momentum space, the problems in $2$-dimensional spatial dimension can be reduced to the problem in $1$-dimensional spatial dimension.
Then the equation in Eq.(\ref{Eq:Dirac-Psi-k-2}) depends on three Gamma matrices $\Gamma^{\underline{t}}$, $\Gamma^{\underline{r}}$ and $\Gamma^{\underline{1}}$. The projection operator can be defined as
\beqa
\Sigma^{\underline{1}} \equiv \Gamma^{\underline{t}}\Gamma^{\underline{1}}\Gamma^{\underline{r}} = \Gamma^{\underline{r}}\Gamma^{\underline{t}}\Gamma^{\underline{1}}.
\eeqa
In the ultra-relativistic limit, i.e., $m_D\ll |\vec{k}|$, the spin projection operator becomes ordinary chirality operator for massive particles~\cite{Itzykson:1980rh}, i.e.,  $P_\pm^{\hat{k}^i} \overset{}{=} (\textbf{1}_{4}\pm \tilde\Gamma^{\underline{5}} )/2$.

Assuming the rotational invariance is kept in the $(d-1)$-dimensional spatial space as in Eq.(\ref{Assump:rotational-invariance}). In this case, the Dirac equation in the $D=(d+1)$-dimensional bulk, can be reduced in analogy to that in $D=(2+1)$-dimensional bulk, with only three coordinates $(t,r,x)$. Therefore, we just need to solve the Driac equation in $D=(2+1)$-dimensions, i.e., $d=2$. Vice verse, one can generalize the representation of Dirac fermion chosen for $(2+1)$-dimensional case and extend to the $(d+1)$-dimensional case. In the chosen representation in Eq.(\ref{Eq:Gamma-rep-1}), the spin projection operator can be expressed more explicitly,
\beqa
\Sigma^{\underline{1}}  \equiv  \Gamma^{\underline{r}}\Gamma^{\underline{t}}\Gamma^{\underline{1}} =  \textbf{1}_2  \otimes (-\sigma^3)=  \left(\begin{array}{cc}
 -\textbf{1}_{2} & 0\\
0 & \textbf{1}_{2}
\end{array}
\right). \nn
\eeqa


Thus, the spin projection operator $P^{\hat{k}^i}_\pm$ becomes
\beqa
&& P_\pm^{\hat{k}^1} = \frac{1}{2}(\textbf{1}_{4} \mp \Sigma^{\underline{1}}), \quad P_+^{\hat{k}^1} + P_-^{\hat{k}^1} = \textbf{1}_{4}, \quad \hat{k}\equiv \frac{\vec{k}}{|\vec{k}|}, \nn\\
&& P_+^{\hat{k}^1} = \left(\begin{array}{cc}
\textbf{1}_{2} & 0\\
0 & \textbf{0}_{2}
\end{array}
\right), \quad P_-^{\hat{k}^1} =  \left(\begin{array}{cc}
\textbf{0}_{2} & 0\\
0 & \textbf{1}_{2}
\end{array}
\right),  \label{Eq:P_pm}
\eeqa
where we have normalized $\hat{k}^1=1$.

In the homogeneous spatial space, the projectors in Eq.(\ref{Eq:spin-projection_ki}) can be simplified to be Eq.(\ref{Eq:P_pm}).  By choosing the representation in Eq.(\ref{Eq:Gamma-rep-1}), one can reexpressed them as
\beqa
P_{\beta} \equiv \frac{1}{2}(\textbf{1}_{4} - \beta \Sigma^{\underline{1}}), \quad P_+ + P_- = \textbf{1}_{4},   \label{Eq:P_beta}
\eeqa
where we have dropped the superscript $\hat{k}^1$, i.e., $P_\pm \equiv P_\pm^{\hat{k}^1}$, namely
\beqa
 P_+ = \textbf{1}_{2}\otimes\left(\begin{array}{cc}
1 & 0\\
0 & 0
\end{array}
\right), \quad  P_- = \textbf{1}_{2} \otimes \left(\begin{array}{cc}
0 & 0\\
0 & 1
\end{array}
\right),
\eeqa
which are all commute with the Dirac operator in Eq.(\ref{Eq:Dirac-Psi-k}).

If one denotes the bulk Dirac fermion in terms of upper half and lower half component spinors $\Psi_\beta$, $\beta=\pm $, i.e., $\Psi_{+}$ and $\Psi_{-}$, which are just eigen-states of the projection operator for bulk fermion, as defined below for $p=2$ case (or in $D=4,5$-dimensions), $\Psi$ is a Dirac spinor with four components, which can be decomposed into two Weyl fermions with two components,
\beqa
\Psi \equiv \left(
              \begin{array}{c}
                \Psi_{+} \\
                \Psi_{-} \\
              \end{array}
            \right) = P_+ \Psi + P_- \Psi \equiv \Phi_+ + \Phi_-,  \label{Eq:Psi}
\eeqa
where we have introduced the notation for the eigenstate of projection operator~\footnote{
Generally speaking, they are different from the eigenstate of the bulk chirality operators introduced in Eq.(\ref{Eq:chirality-projection}), i.e., $\Psi^\pm \equiv \Gamma^\pm \Psi = (\textbf{1}_4 \pm \Gamma^{5})/2 \Psi$, unless in the ultra-relativistic limit. }
\beqa
\Phi_\beta \equiv P_\beta \Psi,  \label{Eq:Phipm}
\eeqa
with $\beta=\pm $. Therefore, in the $D=(3+1)$ dimensions, a $4$-components bulk Dirac spinor $\Psi$ has two projected $(3+1)$-dimensional Weyl spinor $P_\pm\Psi\equiv \Phi_\pm$ as its components, each transforms as a $2$-components Weyl spinors with opposite helicity.

According to Eq.(\ref{Eq:Phipm}), $\Phi_\pm$ can be expressed more explicitly,
\beqa
\ii\ii \Phi_+ \equiv P_+ \Psi = \left(
                               \begin{array}{c}
                                 \Psi_+ \\
                                 0 \\
                               \end{array}
                             \right), ~ \Phi_- \equiv P_- \Psi = \left(
                               \begin{array}{c}
                                 0 \\
                                 \Psi_- \\
                               \end{array}
                             \right).
\eeqa
It is worthy to notice that in this representation, the upper and lower half component $\Phi$ just decouple from each other. While one should keep in mind that, in an arbitrary chosen representation, it is possible that the projection operator could mix the upper half and lower half components of the bulk Dirac spinors.

\subsection{Dirac equation in generic spatial homogeneous space-time}
\label{sec:BulkDiracEOM-Homogeneous}

Consider a generic diagonal background metric with rotationally invariance along $x^i$ directions,
\beqa
\dr s^2 & = & - g_{tt}(r)\dr t^2 + g_{rr}(r)\dr r^2 + g_{xx}(r)\dr x^i \dr x_i \nn\\
& \equiv & \eta_{ab} {e^a}_M {e^b}_N \dr x^M \dr x^N, \label{Eq:ds2-generic}
\eeqa
where $i = 1, 2, \ldots d-1$. The boundary is taken at $r = \infty$, various components of the metric have the asymptotic behavior of AdS$_{d+1}$ with unit radius, namely $\dr s^2 \overset{r\to \infty}{\to} - {r^2}/{\ell^2}\dr t^2 + {\ell^2}/{r^2}\dr r^2 + {r^2}/{\ell^2} \dr x_{d-1}^2$.
From which, we have ${e^a}_M = \text{diag}(g_{tt}^{1/2},g_{rr}^{1/2},g_{xx}^{1/2}, \ldots)$, where $\ldots$ denote the remaining $(d-2)$-dimensional spatial components.


Then the non-vanishing components of spin connection are given by\footnote{In the following, for the briefness, we will neglect the variables of the metric function, unless to be specified later on.}
\beqa
 \omega_{\underline{t}\,\underline{r}\,t} &=& -\omega_{\underline{r}\,\underline{t}\,t} = - \frac{g_{tt}^\prime}{2\sqrt{g_{rr}}\sqrt{g_{tt}}}, \nn \\
 \omega_{\underline{r}\,\underline{x}\,x} &=& -\omega_{\underline{x}\,\underline{r}\,x} = - \frac{g_{xx}^\prime}{2\sqrt{g_{rr}}\sqrt{g_{xx}}}.
\eeqa
Thus, the connection matrix $\Omega_M\equiv\omega_{abM}\Gamma^{ab}/4$, which is the product of spin connection and the spinor representation of the homogeneous Lorentz group in the geometric background in Eq.(\ref{Eq:ds2-generic}), becomes,
\beqa
&& \Omega_t = -\frac{g_{tt}^\prime}{4\sqrt{g_{rr}}\sqrt{g_{tt}}}\Gamma^{\underline{t}}\Gamma^{\underline{r}}, \quad \Omega_r = 0, \nn\\
&& \Omega_i =  \frac{g_{xx}^\prime}{4\sqrt{g_{rr}}\sqrt{g_{xx}}}\Gamma^{\underline{i}}\Gamma^{\underline{r}},
\eeqa
and the Gamma matrix in the curved space-time becomes
\beqa
&& \Gamma^t = \ea^t \Gamma^a = \frac{1}{\sqrt{g_{tt}}}\Gamma^{\underline{t}}, \quad \Gamma^r = \ea^r \Gamma^a = \frac{1}{\sqrt{g_{rr}}}\Gamma^{\underline{r}}, \nn\\
&& \Gamma^i =\ea^i \Gamma^a = \frac{1}{\sqrt{g_{xx}}}\Gamma^{\underline{i}}, \label{Eq:Gamma-Gamma-tan}
\eeqa
and the chirality operator is
\beqa
\Gamma^{2p+1} &\equiv & \Gamma^{D+1} = \frac{1}{(2p)!} \sqrt{-g}\varepsilon_{\mu_1\mu_2\ldots \mu_{D}} \Gamma^{\mu_1}\Gamma^{\mu_2}\ldots \Gamma^{D}\nn\\
&=& i^{\frac{D}{2}-1} \sqrt{-g} \Gamma^{t} \Gamma^1 \ldots \Gamma^{d-2}\Gamma^{d-1}\Gamma^r, \nn\\
&=& i^{\frac{D}{2}-1} \sqrt{-g}\frac{1}{\sqrt{g_{tt}g_{xx}^{d-1}g_{rr}}}  \Gamma^{\underline{t}} \Gamma^{\underline{1}} \ldots \Gamma^{\underline{d-2}}\Gamma^{\underline{d-1}}\Gamma^{\underline{r}}, \nn\\
&=& \Gamma^{\underline{D+1}} = \Gamma^{\underline{2p+1}},
\eeqa
where $\epsilon_{\mu_1\ldots \mu_{D}}$ and $\epsilon^{\mu_1\ldots \mu_{D}}$ are the co-variant and contra-variant Levi-Civita tensors, respectively. They are both defined through the Levi-Civita tensor density $\varepsilon$ with the conversion $\varepsilon^{12\ldots D}=1$,
\beqa
\epsilon_{\mu_1\ldots \mu_{D}}=\sqrt{-g}\varepsilon_{\mu_1\ldots \mu_{D}}, \quad \epsilon^{\mu_1\ldots \mu_{D}}=\frac{1}{\sqrt{-g}}\varepsilon^{\mu_1\ldots \mu_{D}}. \nn
\eeqa
And $\Gamma^{\underline{D+1}}$ is the chirality gamma matrix in flat Euclidean space,
\beqa
\Gamma^{\underline{2p+1}} = \Gamma^{\underline{D+1}} \equiv i^{\frac{D}{2}-1} \Gamma^{\underline{t}}\Gamma^{\underline{1}} \ldots \Gamma^{\underline{d-2}}\Gamma^{\underline{d-1}}\Gamma^{\underline{r}}.
\eeqa
Therefore, the chirality operator in curved space-time, i.e., $\Gamma^{2p+1}$ will not be affected by the curvature of the space-time, and is the same as that in flat space-time. In other words, the degenerate of the chirality state of Dirac fermion will not be split by any gravitational field, whatever strong.

Then the spin connection term due to spinor field in curved space-time in the Dirac equation turns out,
\beqa
\Gamma^M \Omega_M &=& (\ec^M \Gamma^c)(\frac{1}{4}\omega_{abM}\Gamma^{ab}) = \frac{\Gamma^{\underline{r}}}{\sqrt{g_{rr}}}[\log( g_{tt}g_{xx})^{1/4}]^\prime \nn\\
&=& \frac{\Gamma^{\underline{r}}}{\sqrt{g_{rr}}}[\log(-gg^{rr})^{1/4}]^\prime,
\eeqa
where we have used the identities in Eq.(\ref{Eq:Grassmanian-Algebra}). It is worthy to notice that all of the spin connection terms are present with the radial coordinate direction $r$. In the generic background in Eq.(\ref{Eq:ds2-generic}), the Dirac equation in bulk space-time $(\Gamma^M D_M - m_D) \psi =0 $, combining with the spin connection term becomes
\beqa
0 &=& \Gamma^M (\partial_M + \Omega_M - i q A_M - i b_M \Gamma^{2p+1})\psi - m_D  \psi \nn\\
&=&  \bigg[ \frac{\Gamma^{\underline{t}}}{\sqrt{g_{tt}}} \bigg( \partial_t - i q A_t + i b_0 \Gamma^{\underline{2p+1}} \bigg) \nn\\
&& +  \frac{\Gamma^{\underline{r}}}{\sqrt{g_{rr}}} \bigg(\partial_r +[\log(-gg^{rr})^{\frac{1}{4}}]^\prime - i b_r \Gamma^{\underline{2p+1}}  \bigg) \nn\\
&& +  \frac{\Gamma^{\underline{j}}}{\sqrt{g_{xx}}} \bigg(\partial_j \!-\! i q A_j - i b_j \Gamma^{\underline{2p+1}} \bigg) - m_D \bigg]\psi.
\eeqa
where $m_D$ is the mass of the Dirac fermion and the chiral gauge field $b_M=(b_\mu,b_r)$ where $b_\mu=(-b_0,b_j)$. $\Gamma^{2p+1}$ is the Chirality operator defined in $2p$-even dimensions, where $p\equiv [{D}/{2}]$.
By observing the equation above, one can define a new fermion field by using a gauge transformation $\psi = e^{\alpha(r)}\Psi$, where the phase angle
$\alpha(r) \equiv \log(-gg^{rr}(r))^{-\frac{1}{4}}$, so that the spinor connection term will be completely absorbed as an radial function and separated from the original fermion field $\psi$,
\beqa
\psi &\equiv& \exp[-\log(-gg^{rr})^{\frac{1}{4}}+i\Gamma^{\underline{2p+1}}\int_{r_h}^r b_{r}(s)ds ] \Psi\nn\\
&=& (-gg^{rr})^{-\frac{1}{4}}e^{i\Gamma^{ b(r) \underline{2p+1}}} \Psi,  \label{Eq:psi-Psi-x} \label{Eq:b(r)}
\eeqa
where the phase factor $b(r) \equiv \int_{r_h}^r b_r(s) ds$, which is a function with radial coordinate $r$ as variable. By assuming that $b_r$ is a constant, e.g., the vacuum expectation value of a field along radial direction, one can obtain $b(r)=b_r(r-r_h)$. In the absence of the chiral gauge field, e.g., $b_r=0$, the phase factor reduces to the identity matrix $\textbf{1}_4$.
In this case, the Dirac equation for the new defined fermion field $\Psi$ will behave like that there is no spinor connection's contribution at all, as the case in a flat space-time,
\beqa
&& \bigg(\frac{\Gamma^{\underline{t}}}{\sqrt{g_{tt}}} ( \partial_t \!-\! i q A_t \!+\! i b_0 \Gamma^{\underline{2p+1}} )
\!+\!  \frac{\Gamma^{\underline{r}}}{\sqrt{g_{rr}}} (\partial_r - i b_r \Gamma^{\underline{2p+1}})  \nn\\
&\!+\! &  \frac{\Gamma^{\underline{j}}}{\sqrt{g_{xx}}} (\partial_j \!-\! i q A_j \!-\! i b_j \Gamma^{\underline{2p+1}} )  \!-\! m_D \bigg)\Psi \!=\! 0,
\eeqa
where $\Psi \equiv (-gg^{rr})^{{1}/{4}} \exp{\big(i\Gamma^{\underline{2p+1}}\int_{r_h}^r b_{r}(s)ds\big)} \psi$.
The EOM can be re-expressed as
\beqa
\ii
&& i\bigg(\frac{\sqrt{g_{xx}}}{\sqrt{g_{tt}}}\Gamma^{\underline{t}}( -i\partial_t \!-\!  q A_t \!+\! b_0 \Gamma^{\underline{2p+1}}) \nn\\
&& \!+\!   \Gamma^{\underline{j}} (-i\partial_j \!-\!  q A_j \!-\!  b_j \Gamma^{\underline{2p+1}} ) \bigg) \Psi \nn\\
\ii && +  \frac{\sqrt{g_{xx}}}{\sqrt{g_{rr}}} [\Gamma^{\underline{r}}(\partial_r \!-\!  ib_r \Gamma^{\underline{2p+1}}) \!-\! \sqrt{g_{rr}} m_D]  \Psi = 0.
\eeqa
Assuming that we are considering a system with broken time reversal symmetry (or charge parity violation, i.e., $\cancel{{\mathcal CP}}$, assuming ${\mathcal CPT}$ is conserved), but still with unbroken inversion symmetry, i.e., P is unbroken. In this case, the problem can be simplified by setting $\vec{b}=(b_1,b_2,\ldots b_{d-1},b_r) = (b_1,0,\ldots,0) \ne \vec{0}$, meanwhile $b_0=0$. The physical consequences of chiral gauge field $b_M$ are discussed in more detail in appendix~\ref{app:ChiralAnomaly_ChiralGaugeField}.

\subsection{Dirac equation in bulk with Ricci flat surface}
\label{sec:BulkDiracEOM-k=0}

For simplicity, let's consider firstly the bulk gravity with Ricci flat brane, e.g., Eq.(\ref{Eq:ds2-fr}) with $k=0$ case. Assuming that the theory is translationally invariant along the ordinary space-time $x^\mu$ direction on the brane, i.e., all the metric components depend on radial coordinate $r$ only, then the field $\Psi(r,x^\mu)$ can be expressed in momentum space along the $x^\mu$ direction, namely
\beqa
 \Psi(r,x^\mu) &=& \Psi(r,k^\mu)e^{-i\omega t + i\vec{k}\cdot \vec{x}},  \label{Eq:Psi-x2k}\\
 \partial_\mu &=& (\partial_t,\partial_j)\to (-i\omega, ik_j), \quad k_\mu  = (-\omega,\lambda). \nn
\eeqa
Combining Eq.(\ref{Eq:psi-Psi-x}) and Eq.(\ref{Eq:Psi-x2k}), we have
\beqa
\psi(r,x^\mu) &=& (-gg^{rr})^{-\frac{1}{4}}e^{-i\omega t + i k_j x^j}\Psi(r,k^\mu). 
\label{Eq:psi-Psi-x-2}
\eeqa
In the end, we obtain the EOM of Dirac fermion for $\Psi(r,k_\mu)$ in the momentum space, by assuming that $b_r=0$,
\beqa
&&  i \Gamma^{\underline{\mu}} ( K_\mu - B_\mu \Gamma^{\underline{2p+1}} )\Psi + \frac{\sqrt{g_{xx}}}{\sqrt{g_{rr}}} (\Gamma^{\underline{r}}\partial_r - \sqrt{g_{rr}} m_D )  \Psi = 0 ,  \nn\\
&& K_\mu  = (-K_0, K_j) \equiv \bigg(-\frac{\sqrt{g_{xx}}}{\sqrt{g_{tt}}}( \omega +  q A_t ), (k_j -q A_j)  \bigg) , \nn \\
&& B_\mu = (-B_0, B_j) \equiv \bigg(-\frac{\sqrt{g_{xx}}}{\sqrt{g_{tt}}} b_0 , b_j \bigg), \label{Eq:Dirac-Psi-k}
\eeqa
where $p = [D/2]$. In summary, the translation invariance in $(\vec{x},t)$ turns the Drac equation into an ordinary differential equation of $r$, and rotation invariance allows one to set $k_i=\delta_i^1 k_1$ without loss of generality. For bulk gravity with $U(1)$ gauge field as $A_M=(A_r,A_\mu)=(0,A_t(r),A_i(x_j))$, the $K_{0,i}$ and $B_{0,i}$ are given by
\beqa
K_0(r) \2i &=& \2i \frac{\sqrt{g_{xx}(r)}}{\sqrt{g_{tt}(r)}} [ \omega \!+\!  q A_t(r) ], ~ K_i = k_i - q A_i(x_j), \qquad\quad \label{Eq:K0} \\
B_0(r) \2i &=& \2i \frac{\sqrt{g_{xx}(r)}}{\sqrt{g_{tt}(r)}} b_0, ~ B_j = b_j. \label{Eq:B0}
\eeqa
It is convenient to decompose $\Psi$ in terms of eigenvalue of spinor representation along radial direction $\Gamma^{r}$,
\beqa
\ii\ii
\Psi = \Psi_+ + \Psi_-, ~ \Psi_\pm = \Gamma_\pm \Psi, ~ \Gamma_\pm \equiv \frac{1}{2}(\textbf{1}_{2^p}\pm \Gamma^{\underline{r}}), \label{Eq:Psipm}
\eeqa
where $p=[D/2]$, $\Psi_+$ or $\Psi_-$ carries only half of the components of $\Psi$. In this case, according to Eq.(\ref{Eq:psi-Psi-x-2}), one has
\beqa
\ii
\psi_\pm(r,x^\mu) & \equiv & \Gamma_\pm \psi(r,x^\mu) \nn\\
&=& (-gg^{rr})^{-\frac{1}{4}}e^{-i\omega t + i k_i x^i}\Psi_\pm(r,k^\mu). \label{Eq:psi-pm-Psi-pm}
\eeqa

\section{Generic Solution to Radial Sector of Dirac Equation}
\label{app:radial_Dirac_EOMs}

\begin{widetext}

\subsection{Master equation for the radial sector}
\label{sec:Master_EOMs-r}

The radial sector of the EOMs of Dirac equation in Eq.(\ref{Eq:Dirac-EOMs-R1-R2}) can be reexpressed as
\beqa
 \bigg[ \textbf{1}_2  \frac{\partial_r}{\sqrt{g_{rr}}} +    m_D \sigma^3 - i \beta \frac{b_0}{\sqrt{g_{tt}}} \textbf{1}_2 -  \bigg( \frac{(\omega +  q A_t)}{\sqrt{g_{tt}}} - \beta\frac{b_r}{\sqrt{g_{rr}}} \bigg) i\sigma^2  - (-1)^\alpha \frac{\lambda}{\sqrt{g_{xx}}} \sigma^1 \bigg] \tilde{R}_{\alpha,\beta} = 0.
\eeqa
By introducing
\beqa
\omega_\beta(r) \equiv \frac{(\omega \!+\!  q A_t)}{\sqrt{g_{tt}}} \!-\! \beta\frac{b_r}{\sqrt{g_{rr}}}, \quad m_\alpha(r) \equiv \!-\! (-1)^\alpha \frac{\lambda}{\sqrt{g_{xx}}}, \label{Eq:omega(r)-m_alpha(r)}
\eeqa
one obtains
\beqa
\textbf{1}_2  \frac{\partial_r}{\sqrt{g_{rr}}} \tilde{R}_{\alpha,\beta} + \bigg(   m_D \sigma^3 \!-\! i \beta\frac{b_0}{\sqrt{g_{tt}}} \textbf{1}_2 -  \omega_\beta(r) i\sigma^2 + m_\alpha(r) \sigma^1 \bigg) \tilde{R}_{\alpha,\beta} = 0. \nn
\eeqa
Assuming
\beqa
\tilde{R}_{\alpha,\beta} = \left(
             \begin{array}{c}
               \tilde{F}_{\alpha,\beta}(r) \\
               \tilde{G}_{\alpha,\beta}(r) \\
             \end{array}
           \right), \label{Eq:Rt_alpha}
\eeqa
one obtains
\beqa
  \textbf{1}_2  \frac{\partial_r}{\sqrt{g_{rr}}} \left(
             \begin{array}{c}
               \tilde{F}_{\alpha,\beta}(r) \\
               \tilde{G}_{\alpha,\beta}(r) \\
             \end{array}
           \right) + \left(
                       \begin{array}{cc}
                         m_D\!-\! i \beta\frac{b_0}{\sqrt{g_{tt}}} & -\omega_\beta(r) + m_\alpha(r) \\
                         \omega_\beta(r)+m_\alpha(r) & -m_D\!-\! i \beta\frac{b_0}{\sqrt{g_{tt}}} \\
                       \end{array}
                     \right) \left(
             \begin{array}{c}
               \tilde{F}_{\alpha,\beta}(r) \\
               \tilde{G}_{\alpha,\beta}(r) \\
             \end{array}
           \right) = 0 . \nn
\eeqa
We find that it is useful to impose the following transformation for the radial sector of the wave function, $Q$,
\beqa
\left(
  \begin{array}{c}
    \tilde{f}_{\alpha,\beta} \\
    \tilde{g}_{\alpha,\beta} \\
  \end{array}
\right)  \equiv Q \left(
  \begin{array}{c}
    \tilde{F}_{\alpha,\beta} \\
    \tilde{G}_{\alpha,\beta} \\
  \end{array}
\right) = \frac{1}{\sqrt{2}}\left(
            \begin{array}{cc}
              1 & i \\
              1 & -i \\
            \end{array}
          \right) \left(
  \begin{array}{c}
    \tilde{F}_{\alpha,\beta} \\
    \tilde{G}_{\alpha,\beta} \\
  \end{array}
\right),  \label{Eq:fgt_alpha-transQ_FGt_alpha}
\eeqa
which leads to
\beqa
 \frac{\partial_r}{\sqrt{g_{rr}}} \left(
             \begin{array}{c}
               \tilde{f}_{\alpha,\beta}(r) \\
               \tilde{g}_{\alpha,\beta}(r) \\
             \end{array}
           \right) + \left(
                       \begin{array}{cc}
                         i\big(\omega_\beta(r) - \beta\frac{b_0}{\sqrt{g_{tt}}} \big) & m_D + i m_\alpha(r) \\
                         m_D - i m_\alpha(r) & -i\big(\omega_\beta(r) + \beta\frac{b_0}{\sqrt{g_{tt}}} \big) \\
                       \end{array}
                     \right) \left(
             \begin{array}{c}
               \tilde{f}_{\alpha,\beta}(r) \\
               \tilde{g}_{\alpha,\beta}(r) \\
             \end{array}
           \right) = 0.
\eeqa
We define
\beqa
&& \lambda_{1}^\pm(r) \equiv \sqrt{g_{rr}}\big(\omega_\beta(r) \pm \beta \frac{b_0}{\sqrt{g_{tt}}} \big) = \sqrt{g_{rr}}\bigg( \frac{(\omega \!+\!  q A_t) \pm \beta b_0 }{\sqrt{g_{tt}}} \!-\! \beta\frac{b_r}{\sqrt{g_{rr}}} \bigg) =  \sqrt\frac{g_{rr} }{g_{tt}}[(\omega \pm \beta b_0) \!+\!  q A_t] \!-\! \beta b_r , \quad \beta = \pm\nn\\
&& \lambda_{2}(r) \equiv \sqrt{g_{rr}}[m_D + i m_\alpha(r)] = \sqrt{g_{rr}}\bigg( m_D \!-\! i(-1)^\alpha \frac{\lambda}{\sqrt{g_{xx}}} \bigg), \quad \alpha=1,2. \label{Eq:lambda_1^pm-lambda_2-r}
\eeqa
where for the briefness, we have dropped the subscript $\alpha$ dependence upon $\lambda_{1}^\pm(r)$ and $\lambda_2(r)$, i.e., $\lambda_{1,\beta}^{\pm}(r)$ and $\lambda_{2,\alpha}(r)$.

It is worthy to notice that there is an inversion symmetry between $\lambda_1^\pm$, namely, $b_0\to -b_0$, and there is also an inversion symmetry between $\lambda_{2}^\alpha$ too, i.e., $\lambda \to -\lambda$. The field equations become two coupled $1$-st order ODEs,
\beqa
 -\partial_r \tilde{f}_{\alpha,\beta} = +i\lambda_1^-(r) \tilde{f}_{\alpha,\beta} + \lambda_2(r)\tilde{g}_{\alpha,\beta}, \quad -\partial_r \tilde{g}_{\alpha,\beta} = -i\lambda_1^+(r)    \tilde{g}_{\alpha,\beta} + \lambda_2^{\star}(r)\tilde{f}_{\alpha,\beta}, \label{Eq:fg_r_1st}
\eeqa
where $\lambda_2^{\star}(r)=\sqrt{g_{rr}}(r)[m_D-im_\alpha(r)]$ is the conjugate of $\lambda_2(r)$. From above equations, one can obtain two decoupled $2$-nd order ODEs as
\beqa
 \tilde{f}_{\alpha,\beta}^{\prime\prime}(r) + \gamma_1(r) \tilde{f}_{\alpha,\beta}^{\prime}(r) + \gamma_2(r)\tilde{f}_{\alpha,\beta}(r) =0, \quad \tilde{g}_{\alpha,\beta}^{\prime\prime}(r) + \gamma_3(r) \tilde{g}_{\alpha,\beta}^{\prime}(r) + \gamma_4(r)\tilde{g}_{\alpha,\beta}(r) =0,  \label{Eq:fg_r_2nd}
\eeqa
with
\beqa
 \gamma_1(r) & \equiv &  -i[\lambda_1^+(r)-\lambda_1^-(r)] - [\ln{\lambda_2(r)}]^\prime, \quad \gamma_2(r)  \equiv  \lambda_1^+(r)\lambda_1^-(r)- |\lambda_2(r)|^2 + i\lambda_1^-(r) \bigg(\ln\frac{\lambda_1^-(r)}{\lambda_2(r)}\bigg)^\prime, \nn\\
 \gamma_3(r) & \equiv & -i[\lambda_1^+(r)-\lambda_1^-(r)] - [\ln{\lambda_2^{\star}(r)}]^\prime, \quad \gamma_4(r)  \equiv  \lambda_1^+(r)\lambda_1^-(r)- |\lambda_2(r)|^2 - i\lambda_1^+(r)\bigg(\ln\frac{\lambda_1^+(r)}{\lambda_2^{\star}(r)}\bigg)^\prime. \label{Eq:gamma-lambda-r}
\eeqa

\subsubsection{$b_r$ is absent, $b_0$ is present}

The $b_r$ component can be absorbed into the radial sector of the wave functions, in this case, one has
\beqa
 \gamma_1(r) & = &  -2i \beta \sqrt\frac{g_{rr}}{g_{tt}}b_0 -  \ln^\prime\bigg[ \sqrt{g_{rr}}\bigg(m_D - i (-1)^\alpha \frac{\lambda}{\sqrt{g_{xx}}} \bigg) \bigg] , \nn \\
 \gamma_2(r) & = & \frac{g_{rr}}{g_{tt}}[(\omega \!+\!  q A_t)^2-b_0^2] - g_{rr}\bigg( m_D^2 + \frac{\lambda^2}{g_{xx}} \bigg) +  i  \sqrt\frac{g_{rr} }{g_{tt}}[(\omega - \beta b_0) \!+\!  q A_t]  \bigg[\ln\frac{\frac{(\omega - \beta b_0) \!+\!  q A_t}{\sqrt{g_{tt}}} }{ m_D \!-\! i(-1)^\alpha \frac{\lambda}{\sqrt{g_{xx}}}}\bigg]^\prime, \nn\\
 \gamma_3(r) & = &  -2i \beta \sqrt\frac{g_{rr}}{g_{tt}}b_0 -  \ln^\prime\bigg[ \sqrt{g_{rr}}\bigg(m_D + i (-1)^\alpha \frac{\lambda}{\sqrt{g_{xx}}} \bigg) \bigg] , \nn \\
 \gamma_4(r) & = &  \frac{g_{rr}}{g_{tt}}[(\omega \!+\!  q A_t)^2 - b_0^2] - g_{rr}\bigg( m_D^2 + \frac{\lambda^2}{g_{xx}} \bigg) -  i  \sqrt\frac{g_{rr} }{g_{tt}}[(\omega + \beta b_0) \!+\!  q A_t]   \bigg[\ln\frac{\frac{(\omega + \beta b_0) \!+\!  q A_t}{\sqrt{g_{tt}}} }{ m_D \!+\! i(-1)^\alpha \frac{\lambda}{\sqrt{g_{xx}}}}\bigg]^\prime.
\eeqa

\subsubsection{Both $b_0$ and $b_r$ are absent}

In the case that chiral gauge fields are all absent, i.e., $b_0=b_r=0$, then
\beqa
 \gamma_1(r) & = &  -  \ln^\prime\bigg[ \sqrt{g_{rr}}\bigg(m_D - i (-1)^\alpha \frac{\lambda}{\sqrt{g_{xx}}} \bigg) \bigg] , \nn \\
 \gamma_2(r) & = & g_{rr}\bigg[  \frac{(\omega \!+\! q A_t)^2}{g_{tt}} - \bigg( m_D^2 + \frac{\lambda^2}{g_{xx}} \bigg) \bigg] + i  \sqrt\frac{g_{rr} }{g_{tt}}(\omega \!+\!  q A_t)  \bigg[\ln\frac{\frac{\omega \!+\!  q A_t}{\sqrt{g_{tt}}} }{ m_D \!-\! i(-1)^\alpha \frac{\lambda}{\sqrt{g_{xx}}}}\bigg]^\prime, \nn\\
 \gamma_3(r) & = &  -  \ln^\prime\bigg[ \sqrt{g_{rr}}\bigg(m_D + i (-1)^\alpha \frac{\lambda}{\sqrt{g_{xx}}} \bigg) \bigg] , \nn \\
 \gamma_4(r) & = & g_{rr}\bigg[  \frac{(\omega \!+\! q A_t)^2}{g_{tt}} - \bigg( m_D^2 + \frac{\lambda^2}{g_{xx}} \bigg) \bigg] - i  \sqrt\frac{g_{rr} }{g_{tt}}(\omega  \!+\! q A_t)   \bigg[\ln\frac{\frac{\omega \!+\!  q A_t}{\sqrt{g_{tt}}} }{ m_D \!+\! i(-1)^\alpha \frac{\lambda}{\sqrt{g_{xx}}}}\bigg]^\prime.
\eeqa
It is worthy to notice that in this case, there is a symmetry between the coefficient
\beqa
 m_2(r) = -m_1(r),
\eeqa
where $m_\alpha(r)$ is defined in Eq.(\ref{Eq:m_alpha(r)-mt_alpha}) with $\alpha=1,2$. Thus if one exchanges the subscript $1$ and $2$, one has
\beqa
1 \leftrightarrow 2 \quad \Rightarrow \quad (\gamma_3(r),\gamma_4(r)) \leftrightarrow (\gamma_1(r), \gamma_2(r)).
\eeqa
This implies the identical relations between the wave functions since they satisfy the same EOMs in Eq.(\ref{Eq:fg_r_2nd}),
\beqa
\left(
  \begin{array}{c}
    \tilde{f}_1 \\
    \tilde{g}_1 \\
  \end{array}
\right) = \left(
                          \begin{array}{c}
                            \tilde{g}_2 \\
                            \tilde{f}_2 \\
                          \end{array}
                        \right), \quad \left(
  \begin{array}{c}
    \tilde{f}_2 \\
    \tilde{g}_2 \\
  \end{array}
\right) = \left(
                          \begin{array}{c}
                            \tilde{g}_1 \\
                            \tilde{f}_1 \\
                          \end{array}
                        \right).  \label{Eq:ft-gt-m_alpha-symmetry}
\eeqa
In this case, according Eq.(\ref{Eq:fgt_alpha-transQ_FGt_alpha}), one has
\beqa
\left(
  \begin{array}{c}
    \tilde{F}_{\alpha} \\
    \tilde{G}_{\alpha} \\
  \end{array}
\right) = Q^{-1} \left(
  \begin{array}{c}
    \tilde{f}_\alpha \\
    \tilde{g}_\alpha \\
  \end{array}
\right) = \frac{1}{\sqrt{2}}\left(
            \begin{array}{cc}
              1 & 1 \\
             -i & i \\
            \end{array}
          \right) \left(
  \begin{array}{c}
    \tilde{f}_\alpha \\
    \tilde{g}_\alpha \\
  \end{array}
\right),
\eeqa
where we have dropped the subscript $\beta=\pm$, since in the absence of $b_0$ and $b_r$, the two helicity eigen wave functions are degenerate, i.e., the EOMs are of the same form for both $\beta=\pm$ cases. To be more explicit, one obtains
\beqa
 \left(
  \begin{array}{c}
    \tilde{F}_{1} \\
    \tilde{G}_{1} \\
  \end{array}
\right)  = \frac{1}{\sqrt{2}} \left(
  \begin{array}{c}
    \tilde{f}_1 + \tilde{g}_1 \\
    i(\tilde{g}_1 - \tilde{f}_1) \\
  \end{array}
\right), \quad \left(
  \begin{array}{c}
    \tilde{F}_{2} \\
    \tilde{G}_{2} \\
  \end{array}
\right)  = \frac{1}{\sqrt{2}} \left(
  \begin{array}{c}
    \tilde{f}_2 + \tilde{g}_2 \\
    i(\tilde{g}_2 - \tilde{f}_2) \\
  \end{array}
\right) = \frac{1}{\sqrt{2}} \left(
  \begin{array}{c}
    \tilde{g}_1 + \tilde{f}_1 \\
    i(\tilde{f}_1 - \tilde{g}_1) \\
  \end{array}
\right) ,
\eeqa
which implies that
\beqa
\tilde{F}_2 = \tilde{F}_1, \quad \tilde{G}_2 = - \tilde{G}_1. \label{Eq:Ft_12-Gt_12}
\eeqa

\subsection{Master equation in the $\zeta$ coordinate}
\label{sec:Master_EOMs-zeta}

For field equations in $\zeta$ coordinate, they are similar to those in Eq.(\ref{Eq:fg_r_2nd}),
\beqa
 \tilde{f}_{\alpha,\beta}^{\prime\prime}(\zeta) + \gamma_1(\zeta) \tilde{f}_{\alpha,\beta}^{\prime}(\zeta) + \gamma_2(\zeta)\tilde{f}_{\alpha,\beta}(\zeta) =0, \quad \tilde{g}_{\alpha,\beta}^{\prime\prime}(\zeta) + \gamma_3(\zeta) \tilde{g}_{\alpha,\beta}^{\prime}(\zeta) + \gamma_4(\zeta)\tilde{g}_{\alpha,\beta}(\zeta) =0, \label{Eq:fg_zeta_2nd}
\eeqa
but with different parameters
\beqa
 \gamma_1(\zeta) & = &  2i \beta \sqrt\frac{g_{\zeta\zeta}}{g_{tt}}b_0 -  \ln^\prime\bigg[ \sqrt{g_{\zeta\zeta}}\bigg(m_D - i (-1)^\alpha \frac{\lambda}{\sqrt{g_{xx}}} \bigg) \bigg] , \nn \\
 \gamma_2(\zeta) & = & \bigg[  \bigg( \sqrt\frac{g_{\zeta\zeta}}{g_{tt}}[\omega \!+\!  q A_t(\zeta)] \!+\! \beta b_\zeta \bigg)^2 - \frac{g_{\zeta\zeta}}{g_{tt}}b_0^2 \bigg]- g_{\zeta\zeta}\bigg( m_D^2 + \frac{\lambda^2}{g_{xx}} \bigg)\\
             & - &  i \bigg( \sqrt\frac{g_{\zeta\zeta} }{g_{tt}}[(\omega - \beta b_0) \!+\!  q A_t(\zeta)] \!+\! \beta b_\zeta \bigg) \bigg[\ln  \frac{\sqrt\frac{g_{\zeta\zeta}}{g_{tt}}[(\omega - \beta b_0) \!+\!  q A_t(\zeta)] \!+\! \beta b_\zeta  }{ \sqrt{g_{\zeta\zeta}}\big( m_D \!-\! i(-1)^\alpha \frac{\lambda}{\sqrt{g_{xx}}} \big)    }  \bigg]^\prime, \nn\\
 \gamma_3(\zeta) & = &  2i \beta \sqrt\frac{g_{\zeta\zeta}}{g_{tt}}b_0 -  \ln^\prime\bigg[ \sqrt{g_{\zeta\zeta}}\bigg(m_D + i (-1)^\alpha \frac{\lambda}{\sqrt{g_{xx}}} \bigg) \bigg] , \nn \\
 \gamma_4(\zeta) & = & \bigg[  \bigg( \sqrt\frac{g_{\zeta\zeta}}{g_{tt}}[\omega \!+\!  q A_t(\zeta)] \!+\! \beta b_\zeta \bigg)^2 - \frac{g_{\zeta\zeta}}{g_{tt}}b_0^2 \bigg]- g_{\zeta\zeta}\bigg( m_D^2 + \frac{\lambda^2}{g_{xx}} \bigg) \nn\\
             & + &  i \bigg( \sqrt\frac{g_{\zeta\zeta} }{g_{tt}}[(\omega + \beta b_0) \!+\!  q A_t(\zeta)] \!+\! \beta b_\zeta \bigg) \bigg[\ln\frac{ \sqrt\frac{g_{\zeta\zeta} }{g_{tt}}[(\omega + \beta b_0) \!+\!  q A_t(\zeta)] \!+\! \beta b_\zeta}{ \sqrt{g_{\zeta\zeta}}\big( m_D \!+\! i(-1)^\alpha \frac{\lambda}{\sqrt{g_{xx}}}  \big) }\bigg]^\prime, \label{Eq:gamma-1234-zeta}
\eeqa
where we have replaced $r$ with $-\zeta$ and $b_r$ with $-b_\zeta$, to compare with the results in the $\zeta$ coordinate.

One can obtain the EOMs of Dirac equation in the $\zeta$ coordinates straightly, by making the following replacement,
\beqa
\lambda_1(r) \to -\lambda_1(\zeta), \quad \lambda_2(r) \to - \lambda_2(\zeta).
\eeqa
Thus
\beqa
\gamma_\alpha(\zeta) = \gamma_\alpha(r) (\lambda_\alpha(r) \to -\lambda_\alpha(\zeta) ), \quad  \alpha=1,2.
\eeqa

This can be checked by observing that the field equation of two coupled $1$-st order ODEs,
\beqa
 \partial_\zeta \tilde{f}_{\alpha,\beta} = +i\lambda_1^-(\zeta) \tilde{f}_{\alpha,\beta} + \lambda_2(\zeta)\tilde{g}_{\alpha,\beta}, \quad \partial_\zeta \tilde{g}_{\alpha,\beta} = -i\lambda_1^+(\zeta) \tilde{g}_{\alpha,\beta} + \lambda_2^\star(\zeta)\tilde{f}_{\alpha,\beta},  \label{Eq:fg_zeta_1st}
\eeqa
which are similar to Eq.(\ref{Eq:fg_r_2nd}) except that $r\to -\zeta$. This is equivalent to that
\beqa
\lambda_\alpha^\pm(r) \to -\lambda_\alpha^\pm(\zeta), \quad \alpha=1,2.
\eeqa
Thus
\beqa
\gamma_\alpha(r) \to \gamma_\alpha(\zeta)[ \, \lambda_\alpha(r) \to - \lambda_\alpha(\zeta) ], \quad \alpha=1,2.
\eeqa
Namely,
\beqa
 \gamma_1(\zeta) & \equiv &  i[\lambda_1^+(r)-\lambda_1^-(r)] - [\ln{\lambda_2(r)}]^\prime, \quad \gamma_2(\zeta)  \equiv  \lambda_1^+(r)\lambda_1^-(r)- |\lambda_2(r)|^2 - i\lambda_1^-(r) \bigg[\ln\frac{\lambda_1^-(r)}{\lambda_2(r)}\bigg]^\prime, \nn\\
 \gamma_3(\zeta) & \equiv & i[\lambda_1^+(r)-\lambda_1^-(r)] - [\ln{\lambda_2^\star(r)}]^\prime, \quad \gamma_4(\zeta)  \equiv  \lambda_1^+(r)\lambda_1^-(r)- |\lambda_2(r)|^2 + i\lambda_1^+(r)\bigg[\ln\frac{\lambda_1^+(r)}{\lambda_2^\star(r)}\bigg]^\prime,  \label{Eq:gamma-lambda-zeta}
\eeqa
with $r\to \zeta$ and
\beqa
&& \lambda_{1}^\pm(\zeta) \equiv \sqrt{g_{\zeta\zeta}}\bigg(\omega_\beta(\zeta) \pm \beta \frac{b_0}{\sqrt{g_{tt}}} \bigg)  =  \sqrt\frac{g_{\zeta\zeta} }{g_{tt}}[(\omega \pm \beta b_0) \!+\!  q A_t] \! + \! \beta b_\zeta , \nn\\
&& \lambda_2(\zeta) \equiv \sqrt{g_{\zeta\zeta}}[m_D + i m_\alpha(\zeta)] = \sqrt{g_{\zeta\zeta}}\bigg( m_D \!-\! i(-1)^\alpha \frac{\lambda}{\sqrt{g_{xx}}} \bigg), \label{Eq:lambda_1^pm-lambda_2-zeta}
\eeqa
where
\beqa
\omega_\beta(\zeta) \equiv \frac{(\omega \!+\!  q A_t)}{\sqrt{g_{tt}}} \!+\! \beta \frac{b_\zeta}{\sqrt{g_{\zeta\zeta}}}, \quad
m_\alpha(\zeta) \equiv \!-\! (-1)^\alpha \frac{\lambda}{\sqrt{g_{xx}}},  \label{Eq:omega(zeta)-m_alpha(zeta)}
\eeqa
and we have used the same definition as shown in Eq.(\ref{Eq:omega(r)-m_alpha(r)}) in the $r$ coordinate. From the above definitions, we obtain Eq.(\ref{Eq:gamma-1234-zeta}).

\end{widetext}

\section{Topological Charged Black Holes}
\label{app:TCBH_AdS}

In this section, we briefly review and list the main results on the topological charged black holes in the AdS$_{d+1}$ gravity as well as their asymptotic behavior in both the infinite boundary and near horizon boundary. For more details, refer to refs.~\cite{Li:2014fsa}

\subsection{Abelian gauge field in the $(3+1)$-dimensional bulk and effective IR gauge coupling}

The full action of bulk fields with Einstein gravity and Maxwell fields can be expressed as
\beqa
S  =  \frac{1}{2\kappa_G} \! \int \! d^{4}x \sqrt{-g} \bigg( R \!-\! 2\Lambda  \!-\! \frac{\ell^2}{g_F^2}F_{MN}F^{MN} \bigg),  \nn
\eeqa
where the first two terms comes from Einstein gravity with $\kappa_G=8\pi G_N$ and $G_N$ being the Newton's gravitational constant. $[R]=[\Lambda]=[L^{-2}]=2$ and $[F^2]=4$. The cosmological constant can be defined as $\Lambda \equiv - {3}/{\ell^2}$ in AdS$_4$ spacetime, where $\ell$ is the curvature radius of AdS$_4$ bulk.

In addition, we have introduced an effective dimensionless gauge coupling $g_F$ as a measure of the relative strength of the electromagnetic and gravitational interactions,
\beqa
 \frac{1}{4g_{em}^2} &=& \frac{1}{2\kappa_G} \frac{\ell^2}{g_F^2}, \quad \Rightarrow  \quad   g_{em} = \frac{\sqrt{\kappa_G} g_F}{\sqrt{2}\ell}, \quad \text{or} \quad\nn\\
   g_F &=& \frac{\sqrt{2}\ell}{\sqrt{\kappa_G}}g_{em}, \quad [g_F] = 0,\label{Eq:gF-effective}
\eeqa
where $g_{em}^2$ is the electromagnetic charge coupling of the $U(1)$ gauge field, which is controlled by the electromagnetic field action,
\beqa
 S^{EM} \3i &=& \3i \int d^{4}x  \sqrt{-g} {\mathcal L}_{em}, \nn\\
{\mathcal L}_{em} \3i &=& \3i - \frac{1}{4g_{em}^2}  g^{MN}g^{PQ}F_{MP}F_{NQ}.
\eeqa
By observing Eq.(\ref{Eq:gF-effective}), it is obvious that $g_F$ is becoming stronger as the gravitational interactions is becoming weaker, and it becomes divergent in the $\kappa_G \to 0$ limit. Thus the effective gauge coupling $g_F$ characterizes the relative strength of the $U(1)$ gauge interaction and gravitational interaction.

\subsection{Topological charged black holes in AdS$_{4}$}
\label{sec:AdS(d+1)}

The metric of topological charged black brane in Einstein gravity in AdS$_{4}$ space-time, turns out to be~\cite{Li:2014fsa}
\beqa
ds^2 = -  f(r) dt^2 + \frac{dr^2}{f(r)} + r^2 d\Omega_{2,k}^2,  \label{Eq:ds2-fr}
\eeqa
where $d\Omega_{2,k}^2$ denotes the line element of a $2$-dimensional manifold with constant scalar curvature $2k$. And $k=0,\pm 1$ indicate different topology of the $2$-dimensional manifold,
\beqa
d\Omega_{2,k}^2 = \gamma_{mn} dx^m dx^n,
\eeqa
where $\gamma_{mn}$ is the metric of the $2$-dimensional manifold. To be concrete, one has
\beqa
d\Omega_{2,k=1}^2 = d\theta^2 + \sin^2\theta d\phi^2, \quad d\Omega_{2,-1}^2 = d\theta^2 + \sinh^2\theta d\phi^2, \nn
\eeqa
where $k=\pm 1$ are topology indexes of black brane, for a sphere and hyperbolic surface, respectively.

For the topological charged black brane in AdS$_{4}$ space, namely, the AdS-Reissner-Nordstr\"om (RN) metric, the metric factor in Eq.(\ref{Eq:ds2-fr}) is given by
\beqa
\3i
 f(r) \! & \equiv & \! k +  \frac{r^2}{\ell^2} g(r)\overset{r\to \infty}{=} k + \frac{r^2}{\ell^2} , \nn\\
\3i
 g(r) \! &\equiv & \! 1 - \frac{c_1 \ell^2}{r^3} + \frac{\kappa_G}{g_{em}^2}\frac{1}{2}\frac{q_0^2+h_0^2}{r^{4}}, \label{Eq:fr-gr}
\eeqa
where the parameters $c_1\equiv 2M \ell^{-2}$ and $q_0$ can be parameterized by the mass $M$ and the charge $Q$ of RN black hole~\cite{Li:2014fsa},
\beqa
Q^2 = \frac{\kappa_G}{g_{em}^2}\frac{q_0^2+h_0^2}{2} = \frac{\ell^2}{g_F^2}(q_0^2+h_0^2) \equiv q^2 + h^2. \label{Eq:M-ell-Q-q0}
\eeqa

One of the non-vanishing elements of the associated gauge field is given by
\beqa
A_t(r) \overset{}{=} \mu  - \frac{q_0}{r}, \quad A_x(y) = - h y. 
\label{Eq:At-AdS(3+1)}
\eeqa
One has
\beqa
q_0 \2i & \equiv & \2i   \mu r_0,
\label{Eq:q0-mu}
\eeqa
so that $A_t$ vanishes at the horizon, where $r_0$ is the horizon radius determined by the largest positive root
of the redshift factor $f(r_0)=0$.

The charge $Q$ can be expressed with charactering length $\ell_F$,
\beqa
Q =  \ell_F \sqrt{q_0^2+h_0^2} , \quad \ell_F \equiv \frac{\ell}{g_F},
\label{Eq:Q-mu-AdS(d+1)}
\eeqa
with $g_F$ being dimensionless, i.e., $[g_F]=0$, thus $[Q] = [L^{2}]$ and $[q_0]=1$. The charge is defined by character length $\ell_F$ via Eq.(\ref{Eq:Q-mu-AdS(d+1)}), which essentially reflects the competition through relative strength of gravity and electromagnetic forces.

The mass parameter of the charged black hole can be expressed as
\beqa
2M  =
r_0^3 \bigg( 1 \!+\! \frac{Q^2}{r_0^{4}} \bigg) \!+\! k \ell^2 r_0 \3i . \quad \label{Eq:2M-AdS(d+1)}
\eeqa

$g(r)$ defined in Eq.(\ref{Eq:fr-gr}) can be re-expressed as
\beqa
g(r) &=& 1 - \frac{r_0^3}{r^3} j(r),  \label{Eq:gr-Q-epsilon} \\
j(r) &=& 1 + k \frac{\ell^2}{r_0^2} + \frac{Q^2}{r_0^{4}}\bigg( 1 - \frac{r_0}{r} \bigg), \nn
\eeqa
from which it is straightforward to check that $r_0$ is the horizon radius determined by $f(r_0)=k+ r_0^2g(r_0)/\ell^2 =0$.

The Hawking temperature of a topological black hole turns out to be
\beqa
T = \frac{3r_0}{4\pi \ell^2}\bigg( 1 + \frac{1}{3} k \frac{\ell^2}{r_0^2} -  \frac{1}{3}\frac{Q^2}{r_0^{4}}  \bigg).   \label{Eq:TH-AdS(d+1)-k!=0}
\eeqa
According to Eq.(\ref{Eq:TH-AdS(d+1)-k!=0}), the extremal radius of the charged black holes
is achieved at a length scale where the temperature is vanishing. Thus, one can express the specific value of $Q=Q(r_\star)$ in terms of the extremal horizon radius $r_\star$,
\beqa
Q^2 = 3 r_\star^{4} \bigg( 1 + \frac{1}{3} k\frac{\ell^2}{r_\star^2} \bigg). \label{Eq:Qstar-AdS(d+1)}
\eeqa
The near extremal topological charged black brane can be expressed with horizon radius $r_0$ and extremal black horizon radius $r_\star$, assuming $r_0 \gtrsim r_\star$, $g(r)$ can be re-expressed as
\beqa
&& g(r) =  1 - \frac{r_0^3}{r^3} \bigg[ 1 + 3\frac{r_\star^{4}}{r_0^{4}} j(r) + k \frac{\ell^2}{r_0^2}\bigg( 1 + \frac{r_\star^{2}}{r_0^{2}}j(r) \bigg) \bigg], \nn\\
&& j(r) \equiv  1 - \frac{r_0}{r} .  \label{Eq:gr-At-star-AdS(d+1)-T!=0-k!=0}
\eeqa
And the temperature of the topological charged black brane can be expressed as
\beqa
\ii
T \3i = \frac{3r_0}{4\pi \ell^2}\bigg[ 1 + k \frac{1}{3}\frac{\ell^2}{r_0^2} -  \frac{r_\star^{4}}{r_0^{4}} \bigg( 1 + k \frac{1}{3}\frac{\ell^2}{r_\star^2} \bigg)  \bigg]. \label{Eq:TH-AdS-k!=0}
\eeqa

\subsection{Ricci non-flat case with $k\ne 0$}

\subsubsection{Finite temperature}

By expanding the temperature defined in Eq.(\ref{Eq:TH-AdS-k!=0}) around extremal horizon region $r\approx r_\star$, one can expresse the finite temperature as
\beqa
T \overset{}{=} \frac{1}{2\pi\zeta_{0,k}} = \frac{1}{2\pi \zeta_0} \frac{\ell_2^2}{\ell_{2,k}^2} . \label{Eq:TH-zeta0-eta0-k!=0}
\eeqa
where
\beqa
 \ell_{2,k} \overset{k=0}{=} \frac{\ell}{\sqrt{6}} \equiv \ell_2,  \label{Eq:ell_2}
\eeqa
and $\ell_{2,k}$ is the re-scaling length depending on the topology of the $2$-dimensional manifold with radial direction as its normal vector,
\beqa
\ii\ii
\ell_{2,k} \equiv  \frac{\ell}{\sqrt{ 6 +  k \frac{\ell^2}{r_\star^2}}} \overset{k=0}{\equiv} \ell_2. \label{Eq:ell_2k}
\eeqa
We have expressed the $r$ (or $u\sim 1/r$) coordinate in terms of new parameter $\zeta_k$ as
\beqa
 \zeta_k & \equiv & \frac{\ell_{2,k}^{2}}{r-r_\star},  \quad \zeta_k \equiv  \frac{\ell_{2,k}^2}{\ell^2}\frac{u_\star^2}{u_\star-u};
\label{Eq:zeta-eta-k!=0} \\
 \zeta_{0,k} & \equiv & \frac{\ell_{2,k}^{2}}{r_0-r_\star},  \quad \zeta_{0,k} \equiv  \frac{\ell_{2,k}^2}{\ell^2}\frac{u_\star^2}{u_\star-u_0}.
\label{Eq:zeta0-eta0-k!=0}
\eeqa
where $u_\star$ is defined through the extreme horizon $r_\star$,
\beqa
u_\star \equiv \frac{\ell^2}{r_\star}. \label{Eq:u_star}
\eeqa
$u_0\equiv \ell^2/r_0$ and $u\equiv \ell^2/r$.

By using Eq.(\ref{Eq:Qstar-AdS(d+1)}), the charactering length $\ell_F$ in Eq.(\ref{Eq:Q-mu-AdS(d+1)}) can be re-expressed as,
\beqa
\ii\ii \ell_F  \equiv \frac{\ell_k}{g_F}  = \frac{Q}{\sqrt{q_0^2+h_0^2}} = \frac{1}{\sqrt{q_0^2+h_0^2}} r_\star^2 \sqrt{3 + k\frac{\ell^2}{r_\star^2}}.   \label{Eq:ell_F-ell_k}
\eeqa
The above equation leads to that in the bulk gravity with $k=0$ case,
\beqa
\ell_F  = \sqrt\frac{3}{q_0^2+h_0^2} r_\star^2 . \label{Eq:gF-r_0}
\eeqa

The near extremal horizon metric of topological charged black holes in AdS$_4$ space and the corresponding gauge field turn out to be,
\beqa
\ii && \ii ds^2 \!=\! \frac{\ell_{2,k}^{2}}{\zeta_k^2} \bigg[ \!-\! \bigg( 1 \!-\! \frac{\zeta_k^2}{\zeta_{0,k}^2} \bigg)  dt^2 \!+\! \bigg( 1 \!-\! \frac{\zeta_k^2}{\zeta_{0,k}^2} \bigg)^{-1} \ii  d\zeta_k^2 \bigg] \!+\!  r_\star^2  d\Omega_{2,k}^2, \label{Eq:ds2-T!=0-AdS2} \qquad  \\
\ii && \ii A_t(\zeta_k) = \frac{e_{3,k}}{\zeta_k} \bigg( 1 - \frac{\zeta_{k}}{\zeta_{0,k}} \bigg), \quad A_x(y) = -hy, \label{Eq:At-T!=0-AdS2}
\eeqa
where we have introduced a new dimensionless effective IR gauge coupling $e_{3,k}$ defined as
\beqa
e_{3,k} \equiv q_0 \frac{\ell_{2,k}^2}{r_0^2} = \mu \frac{\ell_{2,k}^2}{r_0}. \label{Eq:edk}
\eeqa
Further, one obtains
\beqa
\ii\ii \frac{e_{3,k}}{\ell_{2,k}^2} = \frac{q_0}{r_0^2} = \frac{\mu}{r_0} = \frac{e_{3,0}}{\ell_2^2}, \quad \Rightarrow \quad
e_{3,k} =  e_{3,0} \frac{\ell_{2,k}^2}{\ell_2^2},  \label{Eq:edk-ed0}
\eeqa
where $\ell_{2,k}$ is given in Eq.(\ref{Eq:ell_2k}).

\subsubsection{Zero temperature}

In the near extremal horizon limit $r_0 \to r_\star$ (or $u_0 \to u_\star$), Eq.(\ref{Eq:zeta0-eta0-k!=0}) becomes
\beqa
 \zeta_{0,k} \to \infty, \quad \text{or} \quad  \zeta_{0,k} \to \infty
\eeqa
which stands for the zero temperature limit with
\beqa
T  \overset{\zeta_{0,k}\to \infty}{=}0.
\eeqa

The charactering length $\ell_F$ is given by Eq.(\ref{Eq:ell_F-ell_k}), which leads to
\beqa
g_F   = \frac{\ell}{r_\star^2} \sqrt\frac{q_0^2+h_0^2}{3 + k\frac{\ell^2}{r_\star^2} } = \sqrt{2}\ell \frac{g_{em}}{\sqrt{\kappa_G}}.  \label{Eq:gF-r_star}
\eeqa
Therefore, the effective couplings of gauge field $g_F$, can be re-expressed as a function of radius of extremal black brane
\beqa
\frac{g_{em}}{\sqrt{\kappa_G}} =  \frac{1}{\sqrt{2} r_\star^2 } \sqrt\frac{q_0^2+h_0^2}{3 + k\frac{\ell^2}{r_\star^2} }.
\eeqa

The factor $g(r)$ in Eqs.(\ref{Eq:gr-At-star-AdS(d+1)-T!=0-k!=0}) and (\ref{Eq:gr-At-star-AdS(d+1)-T!=0-k!=0}) becomes
\beqa
g(r) \!\overset{r=r_\star}{=}\! 1 \!-\! \frac{r_\star^d}{r^d} \bigg( 1 \!+\! 3  j(r) \!+\! k \frac{\ell^2}{r_\star^2}[ 1 \!+\! j(r) ] \bigg), \label{Eq:gr-At-star-AdS(d+1)-T=0-k!=0}
\eeqa
with $ j(r)\equiv 1 - {r_\star}/{r}$. The gauge field in Eq.(\ref{Eq:At-AdS(3+1)}) becomes
\beqa
A_t(r) = \mu j(r), \quad A_x(y) = -hy,  \label{Eq:At-fr-AdS(d+1)-T=0}
\eeqa
with $q_0 \equiv \mu r_\star$.

Thus, the near extremal horizon metric of topological charged black holes in AdS$_{4}$ gravity and corresponding gauge field at zero temperature limit are
\beqa
 &&  ds^2   =  \frac{\ell_{2,k}^{2}}{\zeta_k^2} (  -    dt^2  +    d\zeta_k^2 )  +   r_\star^2  d\Omega_{2,k}^2,  \label{Eq:ds2-T=0-AdS2} \\
 &&  A_t(\zeta_k)  =  \frac{e_{3,k}}{\zeta_k} , \quad A_x(y) = - hy, \label{Eq:At-T=0-AdS2}
\eeqa
where $e_{3,k}$ is defined as in Eq.(\ref{Eq:edk}) except that $r_0=r_\star$,
\beqa
e_{3,k} \equiv  q_0 \frac{\ell_{2,k}^2}{r_\star^2} = \mu \frac{\ell_{2,k}^2}{r_\star}. \label{Eq:edk-r_star}
\eeqa
From Eq.(\ref{Eq:edk-r_star}), one obtains
\beqa
\frac{e_{3,k}}{\ell_{2,k}^2} = \frac{q_0}{r_\star^2} = \frac{e_{3}}{\ell_2^2}, \quad \Rightarrow \quad
e_{3,k} = e_3 \frac{\ell_{2,k}^2}{\ell_2^2},  \label{Eq:edk-ed}
\eeqa
In the Ricci flat case and absence of magnetic field, Eq.(\ref{Eq:gF-r_star}) gives
\beqa
g_F =  \frac{\ell}{r_\star^2} \frac{q_0}{\sqrt{3}}.
\eeqa
By observing Eqs.(\ref{Eq:edk-r_star}), one obtains
\beqa
e_3 = q_0 \frac{\ell_{2}^2}{r_\star^2} = \sqrt{3 } g_F \frac{\ell_{2}^2}{\ell}  . \label{Eq:ed-gF-mu}
\eeqa
Thus, the dimensionless effective IR gauge couplings $e_3$ can be defined with bulk gauge coupling $g_F$ as
\beqa
e_3 \equiv \frac{g_F\ell_2}{\sqrt{2}} = \frac{g_F}{2\sqrt{3}}\ell. \label{Eq:ed-gF}
\eeqa
which can be viewed as an effective charge in the near horizon region by observing the near horizon behavior of gauge field in Eq.(\ref{Eq:ds2-T!=0-AdS2}) and Eq.(\ref{Eq:ds2-T=0-AdS2}).

\subsection{Ricci flat case with $k=0$}

\subsubsection{Finite temperature}

For the Ricci flat case with $k=0$, the mass parameter in Eq.(\ref{Eq:2M-AdS(d+1)}) becomes
\beqa
2M  \overset{k=0}{=}  r_0^3 + \frac{Q^2}{r_0},
\label{Eq:2M-AdS(d+1)}
\eeqa
and Eq.(\ref{Eq:gr-Q-epsilon}) becomes
\beqa
g(r) = 1 - \frac{r_0^3}{r^3} - \frac{Q^2}{r^3}\bigg( \frac{1}{r_0} - \frac{1}{r}\bigg).   \label{Eq:gr-Q}
\eeqa

In this case, the temperature in Eq.(\ref{Eq:TH-AdS-k!=0}) becomes
\beqa
T \overset{k=0}{=}  \frac{3 r_0}{4\pi \ell^2}\bigg( 1 - \frac{1}{3}\frac{Q^2}{r_0^{4}} \bigg) = \frac{3 r_0}{4\pi \ell^2}\bigg( 1 -\frac{r_\star^4}{r_0^{4}} \bigg), \label{Eq:TH-AdS(d+1)}
\eeqa
where the charge of the black hole, according to Eq.(\ref{Eq:TH-AdS(d+1)}), can be measured by its extremal horizon radius $r_\star$ as
\beqa
\ii\ii
Q  =  \sqrt{3} r_\star^{2}. \label{Eq:Qstar-AdS(d+1)}
\eeqa
Alternatively, this is equivalent to that the extremal radius of horizon is defined by the charge of black brane,
\beqa
\ii\ii
r_\star^{4} \equiv \frac{1}{3}Q^2, \, \Rightarrow \, u_\star^{4} \overset{k=0}{\equiv} 3 \ell^{8} Q^{-2},
\label{Eq:r_star-Q:u_star-Q}
\eeqa
where we have introduced the conformal coordinate $u_\star\equiv {\ell^2}/{r_\star}$. From Eq.(\ref{Eq:r_star-Q:u_star-Q}), we have
\beqa
r_\star = r_0^{\text{min}} \le r_0, \quad \Leftrightarrow \quad
u_\star = u_0^{\text{max}} \ge u_0,
\eeqa
where $u_0$ is the horizon radius and $u_\star$ is the horizon radius for an extremal black brane. Thus the singularity at $r_\star$ is covered by the horizon radius $r_0$. In this case, the temperature in Eq.(\ref{Eq:TH-AdS-k!=0}) becomes\footnote{For briefness, we neglected the expression for functions of $f(u)$,$A_t(u)$, etc, in the conformal coordinate $u$, which can be obtained by making the changes such as $r_{0}/r \to u/u_{0}$ and $r_{\star}/r \to u/u_{\star}$.}
\beqa
T
&=& \left\{ \begin{aligned}
& \frac{3 r_0}{4\pi \ell^2}\bigg( 1 - \frac{r_\star^{4}}{r_0^{4}} \bigg), \quad r_0 \ge r_\star; \\
& \frac{3}{4\pi u_0}\bigg(1 -   \frac{u_0^{4}}{u_\star^{4}} \bigg), \quad u_0 \le u_\star.  \label{Eq:TH-AdS-k=0}
           \end{aligned} \right.
\eeqa

By using Eqs.(\ref{Eq:2M-AdS(d+1)}) and (\ref{Eq:Qstar-AdS(d+1)}) the mass of the black hole in Eq.(\ref{Eq:2M-AdS(d+1)}) becomes
\beqa
 4 r_\star^3 \le 2M = r_0^3 \bigg( 1 + 3  \frac{r_\star^{4}}{r_0^{4}} \bigg)\le 4 r_0^3. \nn
\eeqa
The lower bound of the mass just corresponds to the zero temperature case when $r_0=r_\star$.

For the Ricci flat case with $k=0$, by using Eq.(\ref{Eq:Qstar-AdS(d+1)}), the charged black brane in Eq.(\ref{Eq:gr-Q}) can be re-expressed as
\beqa
g(r)  =  1 - \frac{r_0^3}{r^3} - 3\frac{r_\star^{3}}{r^3}\bigg( \frac{r_\star}{r_0} - \frac{r_\star}{r}\bigg).  \label{Eq:gr-At-star-AdS(d+1)-T!=0}
\eeqa
At the finite temperature case, the new coordinates defined in Eq.(\ref{Eq:zeta0-eta0-k!=0}) become
\beqa
 \zeta & \equiv & \frac{\ell_{2}^{2}}{r-r_\star},  \quad \zeta \equiv  \frac{\ell_{2}^2}{\ell^2}\frac{u_\star^2}{u_\star-u};
\label{Eq:zeta-eta-k=0} \\
 \zeta_{0} & \equiv & \frac{\ell_{2}^{2}}{r_0-r_\star},  \quad \zeta_{0} \equiv  \frac{\ell_{2}^2}{\ell^2}\frac{u_\star^2}{u_\star-u_0},
\label{Eq:zeta0-eta0-k=0}
\eeqa
where $\ell_2$ is defined in Eq.(\ref{Eq:ell_2}). The metric near the horizon in Eq.(\ref{Eq:ds2-T!=0-AdS2}) becomes a black brane in AdS$_2\times{\mathbb R}^{2}$,
\beqa
\ii
ds^2 \! = \! \frac{\ell_2^2}{\zeta^2} \!\bigg[ \! - \! \bigg(1 \!-\! \frac{\zeta^2}{\zeta_0^2} \bigg)dt^2 \! + \! \bigg(1 \!-\! \frac{\zeta^2}{\zeta_0^2} \bigg)^{-1} \ii d\zeta^2 \! \bigg] \! + \! \frac{r_\star^2}{\ell^2} dx_{2}^2,    ~~
\label{Eq:ds2-T!=0-AdS2-k=0}
\eeqa
where $d\Omega_{2,0}^2 \equiv dx_{2}^2/\ell^2  $ is used. The gauge field near the horizon in Eq.(\ref{Eq:At-T!=0-AdS2}) is expressed as
\beqa
A_t(\zeta) = \frac{e_{3,0}}{\zeta} \bigg( 1 - \frac{\zeta}{\zeta_0} \bigg), \quad A_x(y) = - hy, \label{Eq:At-T!=0-AdS2-k=0}
\eeqa
where $e_{d,0}$ is given by Eq.(\ref{Eq:edk}),
\beqa
e_{d,0} = q_0 \frac{\ell_{2}^2}{r_0^2} = \mu \frac{\ell_{2}^2}{r_0}.  \label{Eq:ed0}
\eeqa
The temperature (with respect to $t$) in Eq.(\ref{Eq:TH-zeta0-eta0-k!=0}) becomes
\beqa
T = \frac{1}{2\pi \zeta_0}.  \label{Eq:TH-zeta0-eta0-k=0}
\eeqa
It implies that at finite charge density, the bulk geometry near horizon becomes AdS$_2\times {\mathbb R}^{2}$. The scale invariance of the AdS$_2$ implies that at low energies the corresponding dynamics on the boundary will be controlled by CFT$_1$, namely conformal quantum mechanics.

\subsubsection{Zero temperature}

Let us consider the zero temperature case with $T=0$, where $r_0 = r_\star$ (or in conformal coordinates $u_0 = u_\star$). In this case, $g(r)$ in Eq.(\ref{Eq:gr-At-star-AdS(d+1)-T!=0}) can be re-expressed as
\beqa
g(r) = 1 - 4 \frac{r_\star^3}{r^3} + 3 \frac{r_\star^{4}}{r^{4}}.  \label{Eq:gr-At-star-AdS(d+1)-T=0}
\eeqa
In the near extremal horizon limit $r_0 \to r_\star$ (or $u_0 \to u_\star$), Eq.(\ref{Eq:zeta0-eta0-k=0}) becomes
\beqa
 \zeta_{0} \to \infty,
\eeqa
which stands for the zero temperature limit, since by definition the temperature in Eq.(\ref{Eq:TH-zeta0-eta0-k=0}) becomes
\beqa
T  \overset{\zeta_0\to \infty}{=}0.
\eeqa

If $h_0=0$ and $q_0=\mu r_0 = \mu r_\star$, Eq.(\ref{Eq:gF-r_star}) and Eq.(\ref{Eq:u_star}) become
\beqa
\ii\ii
r_\star  = \frac{1}{\sqrt{3}} \frac{\ell\mu}{g_F}, \quad
u_\star \equiv \frac{\ell^2}{r_\star} =  \sqrt{3} g_F \frac{\ell}{\mu}, \label{Eq:u_star-gF}
\eeqa
where $\mu$ is the chemical potential, $u_\star$ corresponds to a single scale, which is proportional to the effective dimensionless gauge coupling $g_F$.

In the zero temperature limit, the metric in Eq.(\ref{Eq:ds2-T!=0-AdS2-k=0}) reduces to
\beqa
ds^2 = \frac{\ell_2^2}{\zeta^2}\big( - dt^2 + d\zeta^2 \big) + \frac{r_\star^2}{\ell^2}dx_{2}^2, \label{Eq:ds2-T=0-AdS2-k=0}
\eeqa
and the corresponding gauge field in Eq.(\ref{Eq:At-T!=0-AdS2-k=0}) becomes
\beqa
A_t(\zeta) = \frac{e_3}{\zeta}, \quad A_x(y) = - h y, \label{Eq:At-T=0-AdS2-k=0}
\eeqa
where $e_{3}$ is given by Eq.(\ref{Eq:edk}), or the extremal limit case of Eq.(\ref{Eq:ed0})
\beqa
e_3 = \lim_{r_0 \to r_\star} e_{3,0} = \mu \frac{\ell_{2}^2}{r_\star}. \label{Eq:ed-r_star}
\eeqa
By using Eq.(\ref{Eq:u_star-gF}), $e_d$ can be expressed with the gauge couplings $g_F$, and the effective IR curvature radius $\ell_2$, namely
\beqa
e_3 = \frac{g_F\ell_2}{\sqrt{2}} = \frac{g_F}{2\sqrt{3}}\ell .  \label{Eq:ed-gF}
\eeqa
Therefore, $e_3$ can be viewed as the effective dimensionless IR gauge coupling in the near horizon region.

\end{CJK}

\newpage
\bibliography{000}

\end{document}